\documentclass[12pt]{article}
\pdfoutput=1
\usepackage{draft,tikz-cd}

\newcommand{\cd}{\cdots}
\usepackage{float}
\usepackage{booktabs} 
\usepackage{array}    
\usepackage{caption} 
\usepackage{tikz,tikz-3dplot}
\usepackage{pgfplots}
\pgfplotsset{compat=1.17}
\usetikzlibrary{shapes,arrows,cd,chains,decorations.markings,decorations.pathmorphing,calc}
\tikzset{
	->-/.style args={#1rotate#2}{decoration={markings, mark=at position #1 with {\arrow[scale=1.5,rotate = #2 ]{stealth}}}, postaction={decorate}}
}
\tikzstyle{GraphNode}=[circle, draw=black, fill=black, inner sep=2pt, minimum size=5pt]
\tikzstyle{GraphEdge}=[black]
\pgfmathsetmacro{\gS}{1}
\definecolor{bmnblue}{RGB}{29,78,137}
\definecolor{bmnteal}{RGB}{18,126,131}
\definecolor{bmngold}{RGB}{184,134,11}
\definecolor{bmngray}{RGB}{70,70,70}
\definecolor{softbg}{RGB}{244,247,250}
\definecolor{softgreen}{RGB}{233,245,238}
\definecolor{softorange}{RGB}{251,243,230}
\definecolor{softblue}{RGB}{235,243,252}

\begin{document}

\begin{titlepage}

\begin{center}

\title{Finite-$N$ BMN index across all vacuum sectors}

\author{Chi-Ming Chang$^{a,b,c}$, Sarthak Duary$^{a}$, Kangning Liu$^{a,d}$}

\address{${}^a$Yau Mathematical Sciences Center (YMSC), Tsinghua University, Beijing, China}

\address{${}^b$Beijing Institute of Mathematical Sciences and Applications (BIMSA) Beijing, China}

\address{${}^c$Peng Huanwu Center for Fundamental Theory, Hefei, Anhui, China}

\address{${}^d$Department of Mathematical Sciences, Tsinghua University, Beijing, China}

\email{cmchang@tsinghua.edu.cn, sarthakduary@tsinghua.edu.cn, lkn22@mails.tsinghua.edu.cn}

\end{center}


\begin{abstract}
We compute the finite-$N$ Witten index of BMN matrix quantum mechanics after
summing over all partition-labeled supersymmetric vacuum sectors.  Starting from
the unitary-matrix integral for each sector, we adapt two complementary
evaluation methods to the BMN index: a symmetric-group character expansion,
which reduces each
fixed fugacity order to a finite combinatorial sum, and a residue expansion in
which the contributing poles are organized by rooted trees, with a colored-tree
generalization for multi-partition sectors.  Where practical, direct integration
and extraction of the constant term in the expanded integrand give independent
coefficient-by-coefficient checks.  We
evaluate every vacuum sector for $N\leq 9$.  In the equal-fugacity expansion,
the coefficients near charges $j\sim N^2$ exhibit substantial cancellations
between vacuum sectors, which tend to reduce the effective $N^2$ coefficient
of the total index entropy.  The finite-$N$ data also reveal
a nontrivial sectoral organization: near $j=N^2$, the sector giving the largest
contribution changes with $N$, from single-partition sectors at small rank to
double-partition sectors starting at $N=6$.  We call this phenomenon dominance
switching.  These results provide quantitative finite-$N$ input for using the
BMN index as a diagnostic of protected plane-wave black-hole sectors and suggest
in the controlled type-IIA regime a possible D2 dressed black-hole
interpretation in which D0 black-hole sectors are accompanied by macroscopic
spherical D2-brane degrees of freedom, analogous to dual dressed black holes in
$AdS_5\times S^5$.
\end{abstract}

\vfill

\end{titlepage}

\tableofcontents
\newpage

\section{Introduction}

The BMN matrix quantum mechanics \cite{Berenstein:2002jq} is a massive deformation of the BFSS matrix quantum mechanics \cite{Banks:1996vh,Taylor:2001vb} describing the DLCQ of M-theory on the maximally supersymmetric plane-wave background. The deformation lifts the flat directions, yielding a weak-coupling regime, a discrete spectrum, and a family of supersymmetric vacua described by fuzzy-sphere configurations labeled by partitions of the rank $N$ \cite{Berenstein:2002jq,Maldacena:2002rb}. Early perturbative analyses clarified the deformed symmetry structure and identified protected short multiplets in the discrete spectrum \cite{Dasgupta:2002hx,Kim:2002if,Dasgupta:2002ru}. 

The BMN model also admits rich brane and holographic interpretations. In one large-$N$ scaling, where the multiplicities are held fixed and the sizes of the irreducible blocks are taken large, its partition-labeled fuzzy-sphere vacua describe spherical M2-branes \cite{Berenstein:2002jq}. In the complementary large-multiplicity scaling, the same partition data admit transverse M5-brane and Lin-Maldacena geometric interpretations \cite{Maldacena:2002rb,Lin:2005nh,Asano:2017nxw}. The BMN model also sits inside a broader web of theories with $SU(2|4)$ symmetry, including consistent truncations related to $\mathcal{N}=4$ SYM on ${\mathbb R}\times S^3$ \cite{Kim:2003rza,Lin:2005nh,Ishiki:2006rt,Ishiki:2006yr}. Exact partition-function and localization methods for this family of theories were further developed in \cite{Asano:2012zt,Asano:2014vba}. Related BMN-like matrix models for more general pp-wave-like backgrounds and their supersymmetric black-object sectors were recently discussed in \cite{Lee:2026xoo}. At strong coupling, the BMN model itself exhibits a confined phase connected to the Lin-Maldacena description and a deconfined phase described by a black-hole geometry \cite{Costa:2014wya}. This makes it natural to ask whether finite-$N$ protected sectors in the BMN model already retain signatures of black-hole microstates.

Recent work on finite-$N$ protected sectors in related holographic settings has emphasized that black-hole microstates are not visible in naive large-$N$ graviton counting. Explicit constructions of non-graviton operators, finite-$N$ black-hole cohomologies, and the monotone/fortuitous distinction suggest that such states can instead appear as intrinsically finite-$N$ protected sectors \cite{Chang:2022mjp,Choi:2022caq,Choi:2023znd,Choi:2023vdm,Chang:2024zqi,Chang:2025rqy,deMelloKoch:2025ngs,Gaikwad:2025doublegauge}. The BMN model therefore provides a controlled and especially rich setting in which to study how the large-$N$ behavior of protected sectors can differ substantially across its many supersymmetric sectors. A recent Hamiltonian analysis derived the BMN Witten index in a weak-coupling expansion around all fuzzy-sphere vacua and found a black-hole-like growth of entropy scaling as $\mathcal{O}(N^2)$ at charges of order $N^2$ already in the trivial vacuum sector \cite{Chang:2024lkw}. In this paper we ask how that sector-by-sector picture changes after the sum over sectors: do different vacua reinforce this growth, or can cancellations suppress the total index at charges of order $N^2$?

More precisely, each partition
\begin{equation}
N=\sum_{k=1}^{K} n_k N_k
\end{equation}
with distinct positive integers $N_k$ defines a distinct weak-coupling sector with unbroken gauge group $\prod_k U(n_k)$, and the full index is obtained only after summing over all such sectors. Determining how this sector sum reshapes the total index is a necessary step toward using the BMN index as a diagnostic of plane-wave black-hole entropy, and it is also a prerequisite for isolating genuinely finite-$N$ states of the fortuitous type discussed in \cite{Chang:2022mjp,Choi:2023znd,Chang:2024zqi,Choi:2023vdm,Chang:2025rqy}. Throughout this paper we work at finite mass deformation, where the weak-coupling expansion around each vacuum is under control; we do not attempt to resolve the singular BFSS limit.

Our main technical goal is to make the finite-$N$ BMN index tractable across vacuum sectors. Starting from the matrix-integral formula of \cite{Chang:2024lkw}, we develop two complementary reorganizations. First, we rewrite the gauge projection in terms of symmetric-group characters, following \cite{Dolan:2007rq,Murthy:2020scj,Gaiotto:2021xce}, so that at each fixed fugacity order the unitary-matrix integral becomes a finite combinatorial sum. Second, we derive a residue representation of the same index and recast the contributing residues as sums over trees. These two descriptions provide independent computational handles on the same protected quantity. We also use two additional methods in computationally practical sectors: direct integration and extraction of the constant term in the expanded integrand. In the common domains of applicability, all available methods agree coefficient by coefficient.

Using these methods, we compute the index in every vacuum sector for $N\leq 9$; in this range, single-, double-, and triple-partition sectors already cover all partitions of $N$. Specializing to equal fugacities and writing
\begin{equation}
\mathcal{I}^{\mathrm{BMN}}(t)=\sum_j d_j t^j,
\end{equation}
we find that, at the largest ranks studied, cancellations between vacuum sectors suppress the total index relative to the larger sector contributions at charges $j\sim N^2$, with a tendency to reduce its effective $N^2$ coefficient as $N$ increases. The absence of $\mathcal{O}(N^2)$ entropy growth in the total index at fixed $j/N^2$ is emphasized in \cite{Komatsu:2026holography}, while our available ranks do not by themselves establish the asymptotic scaling. We also uncover a more refined finite-$N$ phenomenon: the sector giving the largest contribution to $|d_j|$ is not fixed once and for all, but changes with $N$. In particular, near $j=N^2$ the dominant contribution comes from single-partition sectors for small $N$, while double-partition sectors take over starting at $N=6$. We refer to this pattern as \emph{dominance switching}.

This sectoral structure also suggests a possible bulk interpretation.  If the
dominant sector at large $N$ contains an irreducible block whose size grows
without bound, then the corresponding fuzzy sphere is a macroscopic plane-wave
giant graviton.  In the controlled type-IIA region, the same object is a
spherical D2-brane carrying D0-brane charge.  As argued in
section~\ref{sec:dressed-bh}, comparing its radial scale
$U_{\rm D2}\sim\mu N_{\max}$ with the D0 horizon scale
$U_h\sim\mu [g_{\rm eff}^2(\mu)]^{1/5}$ gives the outside-horizon condition
$g_{\rm eff}^2(\mu)\lesssim N_{\max}^5$.  When this condition is met inside the
type-IIA window, it motivates interpreting the protected states counted by such
a dominant BMN index sector not as a pure D0 black hole, but as a D0 black-hole
sector dressed by large spherical D2-brane degrees of freedom.  This picture is
closely analogous in spirit to dual dressed black holes in
$AdS_5\times S^5$, where a black-hole core is dressed by large dual giant
gravitons, namely D3-branes wrapping an $S^3$ in $AdS_5$
\cite{Choi:2024ddbh,Choi:2025sgg}.  We also recall the related brane-fused
black-hole cohomologies of \cite{deMelloKoch:2024pcs}.  The analogy is
heuristic, but it gives a concrete way to interpret the sector dependence of
the BMN index and may help explain why, to our knowledge, no pure BPS black
object in eleven-dimensional supergravity asymptotic to the plane wave is
currently known, whether localized as a black hole or extended along the
M-theory circle as a black string.

The rest of the paper is organized as follows. In section~\ref{prelim} we review the BMN Hamiltonian and the weak-coupling expression for the index. Section~\ref{charexp} develops the symmetric-group character expansion.  The single-partition case is treated in section~\ref{singlepart}, while the general multi-partition index is discussed in section~\ref{multipart}. Section~\ref{resexp} presents the residue expansion of the BMN index. In particular, section~\ref{treesres} explains how the relevant residues can be organized as a sum over trees.  Section~\ref{meaningtrees} discusses the meaning of the contribution of the trees. Section~\ref{genpartsec} extends the construction to a general partition sector. Section~\ref{concheck} compares the resulting formulas in explicit multi-partition examples using direct integration and constant-term extraction. Section~\ref{datan} uses the computed series to analyze the growth of $\log |d_j|$ and the onset of dominance switching. Section~\ref{sec:dressed-bh} develops the type-IIA D2 dressed black-hole interpretation suggested by the sector dependence, and section~\ref{sec:conclusion} summarizes the conclusions and further directions. Appendix~\ref{partnot} sets up the partition notation and records a basic exponential identity used in section~\ref{charexp}. Appendix~\ref{sunbmn} collects the corresponding $SU(N)$ BMN indices.  The single-partition, double-partition, triple-partition, and total-index cases are presented separately in Appendices~\ref{spbmn}, \ref{dpbmn}, \ref{tpbmn}, and \ref{totalbmnin}, respectively. Appendix~\ref{closedform} derives closed-form expressions for the single-partition $SU(N)$ BMN index.  More specifically, Appendix~\ref{cf} focuses on the trivial vacuum and presents the closed-form $SU(N)$ BMN indices for $N=2,\ldots,7$, obtained by evaluating the residues as a sum over trees as described in section~\ref{treesres}.  Appendix~\ref{single-partitionclosedform} then treats the remaining single-partition sectors.  Appendix~\ref{itreseval} develops an iterative residue-evaluation procedure for the single-partition BMN index.  Finally, Appendix~\ref{exbmn} gives a detailed illustration of the character expansion method by working out the explicit sector $N_1=1,\ N_2=2,\ n_1=1,\ n_2=1$.


\section{Preliminaries of BMN matrix quantum mechanics}
\label{prelim}

In section \ref{prelim}, we briefly review the Hamiltonian of BMN matrix quantum mechanics, and summarize the Witten index computation using Hamiltonian formulation from \cite{Chang:2024lkw}. 
\subsection{The BMN Hamiltonian}
The total Hamiltonian for the BMN matrix quantum mechanics can be decomposed into two components, $H_0$ and $H_{\text{int}}$, expressed as \cite{Maldacena:2002rb}
\begin{equation}
\label{H_decomp}
H = H_0 + H_{\text{int}}.
\end{equation}
Here, $H_0$ corresponds to the BFSS matrix quantum mechanics limit, while $H_{\text{int}}$ contains the mass and deformation terms dependent on the parameter $\mu$. Explicitly, these terms are given by
\begin{equation}
\label{H_parts}
\begin{split}
H_0 &= R\,\mathrm{Tr}\left[\frac{1}{2}\sum_{I=1}^{9}(P^I)^2 - \frac{1}{4\ell_P^6}\sum_{I,J=1}^{9}[X^I,X^J]^2 - \frac{1}{2\ell_P^3}\Psi^T\gamma^I[X^I,\Psi]\right], \\
H_{\text{int}} &= \frac{R}{2}\,\mathrm{Tr}\left[\left(\frac{\mu}{3R}\right)^2\sum_{i=1}^{3}(X^i)^2 + \left(\frac{\mu}{6R}\right)^2\sum_{m=4}^{9}(X^m)^2 \right. \\
&\quad \left. +\, i\frac{\mu}{4R}\Psi^T\gamma^{123}\Psi + i\frac{2\mu}{3R\ell_P^3}\epsilon_{ijk}X^iX^jX^k\right].
\end{split}
\end{equation}
Here, the system involves nine bosonic $N \times N$ matrices denoted by $X^I$ (with $I = 1, \dots, 9$) and sixteen fermionic $N \times N$ matrices represented by $\Psi$ (with spinor indices omitted). The conjugate momenta associated with $X^I$ are $P^I$. The index $I$ is split into $i = 1, 2, 3$ and $m = 4, \dots, 9$. The model is defined with a $U(N)$ gauge symmetry.

The parameters $R$, $\ell_P$, and $1/\mu$ carry dimensions of length. Through the rescalings: $X^I = \ell_P \widetilde{X}^I$, $P^I = \ell_P^{-1} \widetilde{P}^I$, and $H_{0/{\rm int}}=R\ell_P^{-2}\widetilde{H}_{0/{\rm int}}$, such that $\widetilde{X}^I$, $\widetilde{P}^I$ and $\widetilde{H}_{0/{\rm int}}$ are dimensionless, 
one isolates a single dimensionless coupling constant
\begin{equation}
\label{coupling}
g^2 = \frac{R^3}{\mu^3 \ell_P^6}.
\end{equation}
In the strong coupling limit $g \to \infty$ (corresponding to $\mu \to 0$), the interaction term $H_{\text{int}}$ vanishes, and the theory becomes to the BFSS matrix quantum mechanics described by $H_0$. In this regime, the system exhibits $SO(9)$ symmetry, with $X^I$ and $P^I$ belonging to the vector representation and $\Psi$ to the spinor representation. The terms in $H_{\text{int}}$ break this symmetry, reducing it to $SO(3)$ $\times$ $SO(6)$.

\subsection{Witten index review}
\label{wittenindexrev}
The BMN matrix quantum mechanics is characterized by $SU(2|4)$ supersymmetry. The maximal bosonic subgroup of this symmetry is $SO(3) \times SO(6)$. The supersymmetry includes sixteen supercharges, denoted as $Q^m_\alpha$ and their Hermitian conjugates $(Q^m_\alpha)^\dagger$, where $\alpha = \pm$ represents the spinor index associated with $SU(2) \cong SO(3)$. These supercharges obey the following anti-commutation relations \cite{Dasgupta:2002hx}
\begin{equation}
\begin{split}
\{ Q^m_\alpha, (Q^n_\beta)^\dagger \} &= 2 \delta_n^m \delta_\beta^\alpha H + \frac{\mu}{3} \epsilon_{ijk} (\sigma^k)_\alpha^{{\beta}} \delta_n^m M^{ij} - \frac{2\mu}{3} \delta_\beta^{{\alpha}} R_n^m \,,
\end{split}
\end{equation}
where $M^{ij}$ and $R_n^m$ represent the generators for rotations in $SO(3)$ and $SU(4)$, respectively.
We select a specific supercharge, $Q := Q^4_-$. The anti-commutator of this charge with its Hermitian conjugate defines the operator $2\Delta$
\begin{equation}
\begin{split}
2\Delta := \{Q, Q^\dagger\} &= 2H - \frac{2\mu}{3} M^{12} - \frac{2\mu}{3} R_4^4 \\
&= 2H - \frac{2\mu}{3} M^{12} - \frac{\mu}{3} (M^{45} + M^{67} + M^{89}) \,.
\end{split}
\end{equation}
In the second line, we have decomposed the $SU(4)$ generator $R_4^4$ into the Cartan generators of $\mathrm{SO}(6)$ specifically $M^{45}$, $M^{67}$, and $M^{89}$ which correspond to rotations in three mutually orthogonal planes within $\mathbb{R}^6$. We consider the thermal partition function defined as
\begin{equation}
\begin{split}
Z = \text{Tr}\, \Omega \,, \quad \text{with} \quad \Omega := e^{-\beta \Delta - 2\omega M^{12} - \sum_{i=1}^3 \Delta_i M^{2i+2, 2i+3}} \,.
\end{split}
\end{equation}
For the Boltzmann factor $\Omega$ to anti-commute with the supercharge $Q$ (i.e., $\{Q, \Omega\}=0$), the chemical potentials must satisfy a specific linear constraint
\begin{equation}
\label{cons}
\begin{split}
\sum_{i=1}^3 \Delta_i - 2\omega = 2\pi i \pmod{4\pi i} \,.
\end{split}
\end{equation}
Imposing this condition yields the Witten index for the BMN matrix quantum mechanics \cite{Chang:2024lkw}
\begin{equation}
\mathcal{I}^{\mathrm{BMN}} = \text{Tr} \left[ (-1)^F \exp\left( -\beta \Delta - \sum_{i=1}^3 \Delta_i \left( M^{12} + M^{2i+2, 2i+3} \right) \right) \right].
\end{equation}
Here, we identify the fermion number $F$ with $2M^{12}$, given that the fermionic fields $\Psi$ possess half-integer eigenvalues under the rotation generator $M^{12}$. Consequently, the fermion parity operator is expressed as $(-1)^F = e^{2\pi i M^{12}}$. This structure ensures that bosonic and fermionic states contribute with opposite signs to the trace. States with $\Delta > 0$ form supersymmetric doublets ($|\Psi\rangle$ and $Q|\Psi\rangle$) whose contributions cancel exactly. Therefore, the Witten index receives contributions solely from BPS states (where $\Delta = 0$). As a result, the Witten index is independent of the inverse temperature $\beta$. The Witten index remains invariant under variations of the coupling constant $g$, except potentially at the singular limit of infinite coupling ($g \to \infty$).

\subsection{Witten index in the weak coupling limit}
\label{susyindex}
Since the index is coupling-independent, we evaluate it in the weak coupling limit (large $\mu$ limit, where the steep bosonic potential allows for an expansion around classical supersymmetric vacua. The potential $V$ is expressed as
\begin{equation}
V = \frac{R}{2} \text{Tr} \left[ \left( \frac{\mu}{3R} X^i + \ell_P^{-3} i \epsilon^{ijk} X^j X^k \right)^2 + \frac{1}{2 \ell_P^2} (i [X^m, X^n])^2 + \frac{1}{\ell_P^2} (i [X^m, X^i])^2 + \left( \frac{\mu}{6R} \right)^2 (X^m)^2 \right].
\end{equation}
\paragraph{Vacua in BMN matrix quantum mechanics.}
For classical supersymmetric vacua, each of these four positive-definite terms must vanish independently. The BMN matrix quantum mechanics admits a class of fuzzy sphere vacua \cite{Berenstein:2002jq}: 
\begin{equation}
\begin{split}
X^m = 0,\quad
X^i = \frac{\mu \ell_P^3}{3R} J^i ,
\end{split}
\end{equation}
where $J^i$ are $N\times N$ matrices forming a representation of $\mathrm{SU}(2)$ satisfying $[J^i,J^j] = i\epsilon^{ijk}J^k$. Any $N$-dimensional representation of $\mathrm{SU}(2)$ can be decomposed into irreducible components classified by a partition of the integer $N$ 
\begin{equation}
\begin{split}
N = \sum_{k=1}^{K} n_k N_k,
\end{split}
\end{equation}
where $n_k,\,N_k\in{\mathbb Z}_{>0}$ and $N_k$ are all distinct.
The trivial vacuum ($K=1,\, n_1=N,\, N_1=1$) preserves the full $U(N)$ gauge symmetry, while generic fuzzy sphere vacua break the symmetry to $\prod_k U(n_k)$. Expanding the theory around a given vacuum and taking the weak-coupling limit $g\ll1$, the Hamiltonian reduces to a set of harmonic oscillators organized into superselection sectors labeled by partitions of $N$. The BPS states are constructed from BPS letters that satisfy the condition $\Delta = 0$. Here, $\iota^{(m)}_{kl}(\Delta_i)$ is the $m$-th single-letter index derived from the BPS oscillator spectrum \cite{Chang:2024lkw} 
\begin{equation}
\label{eq:single_letter_index}
\iota^{(m)}_{kl}(\Delta_i)= \sum_{j=\frac{1}{2}|N_k - N_l|}^{\frac{1}{2}(N_k + N_l) - 1} (-1)^{2jm+1} e^{-jm(\sum_{i=1}^3 \Delta_i)} \prod_{a=1}^3(1 - e^{-m\Delta_a}) + \delta_{N_k, N_l}.
\end{equation}
The Witten index for a specific superselection sector takes the form
\begin{equation}
\label{dsum}
\mathcal{I}_{n_i; N_i}^{\mathrm{BMN}} = \int  \prod_{k=1}^K \left[d U_k \right] \exp \left[ \sum_{m=1}^\infty  \sum_{k,l=1}^K \frac{1}{m} \iota^{(m)}_{kl}(\Delta_i) \text{Tr} U_k^{\dagger m} \text{Tr} U_l^{m} \right].
\end{equation}
The total Witten index is given by summing the indices over all possible superselection sectors, which are determined by the values of $n_i$ and $N_i$. The integration over the $ n_k \times n_k $ unitary matrix $ U_k $ enforces gauge invariance under the unbroken subgroup $\prod_k U(n_k)$ i.e., remaining gauge symmetry associated with the $ \{n_i, N_i\} $ sector. The total Witten index is given by
\begin{equation}
\mathcal{I}^{\mathrm{BMN}} = \sum_{\substack{n_i, N_i \in \mathbb{Z}_{>0},\,N_i\neq N_j\,(i\neq j)
\\
N=\sum_{k=1}^{K}n_kN_k}} \mathcal{I}_{n_i; N_i}^{\mathrm{BMN}}.
\end{equation}

\section{Symmetric-group character expansion of the BMN index}
\label{charexp}

In this section, we recast the BMN index in terms of symmetric-group characters, following \cite{Dolan:2007rq, Murthy:2020scj, Gaiotto:2021xce}. The central ingredient is the Frobenius formula, which expresses multitraces of unitary matrices in terms of products of unitary-group characters and symmetric-group characters. The Haar integral in \eqref{dsum} is then evaluated using character orthogonality. We shall refer to this reorganization as the \emph{symmetric-group character expansion}. We stress that this does not alter the index; it is merely a different, more algebraic way of presenting the same quantity. Its main advantage is computational: the original gauge projection, written as a unitary-matrix integral in \eqref{dsum}, is transformed into a finite combinatorial problem involving integer partitions and symmetric-group characters. 

To keep the notation light, we specialize to equal fugacities
\begin{equation}
\Delta_1=\Delta_2=\Delta_3=\Delta,
\qquad
t^2=e^{-\Delta}.
\end{equation}
The single-letter indices \eqref{eq:single_letter_index} simplify to
\begin{equation}
\iota_{aa}(t^m)=1-(1-t^{2m})^3\sum_{j=0}^{N_a-1} t^{6jm},
\end{equation}
for diagonal blocks, and
\begin{equation}
\iota_{ab}(t^m)=\iota_{ba}(t^m)
=(1-t^{2m})^3
\sum_{j=\frac12|N_a-N_b|}^{\frac12(N_a+N_b)-1}(-1)^{2jm+1} t^{6jm},
\qquad a\neq b.
\end{equation}

\subsection{Single partition index}
\label{singlepart}
We begin with the simplest sector $N=n_1N_1$, in which the unbroken gauge group is $U(n_1)$. The index is
\begin{equation}
\mathcal{I}_{n_1;N_1}^{\mathrm{BMN}}(t)
=\int [dU]\, \exp\!\left[\sum_{m=1}^\infty \frac{1}{m}\,\iota_{11}(t^m)\,\Tr(U^m)\Tr(U^{\dagger m})\right].
\end{equation}
Expanding the plethystic exponential mode by mode gives
\begin{equation}
\begin{split}
\mathcal{I}_{n_1; N_1}^{\text{BMN}}(t)
&=\int [dU] \prod_{j=1}^\infty \sum_{k_j=0}^\infty \frac{1}{k_j!} \left( \frac{1}{j} \iota_{11}(t^j) \, \text{Tr}(U^j) \text{Tr}(U^{\dagger j}) \right)^{k_j}.
\end{split}
\end{equation}
It is convenient to package the nonnegative integers $\{k_m\}_{m\ge 1}$ into an integer partition
\begin{equation}\label{eqn:bi-intpar-multitrace}
P\equiv (1^{k_1}2^{k_2}3^{k_3}\cdots),
\end{equation}
with weight and length
\begin{equation}
|P|:=\sum_{m\ge 1} m k_m,
\qquad
\ell(P):=\sum_{m\ge 1} k_m.
\end{equation}
Such a partition \(P\) determines a multitrace operator built from \(|P|\) matrices:
\ie
\mathcal{O}_P(U):=\prod_{m\ge 1}(\Tr U^m)^{k_m}\,.
\fe
In terms of this notation, the expansion becomes
\begin{equation}
\mathcal{I}_{n_1;N_1}^{\mathrm{BMN}}(t)
=\int [dU]\,\sum_P \frac{1}{z_P}\,\iota_{11}(t)_P\,\mathcal{O}_P(U)\mathcal{O}_P(U^{\dagger}),
\end{equation}
where
\begin{equation}
z_P:=\prod_{m\ge 1} k_m!\,m^{k_m},
\qquad
\iota_{kl}(t)_P:=\prod_{m\ge 1} \iota_{kl}(t^m)^{k_m}.
\end{equation}

The next step is to expand the multitrace operator ${\cal O}_P(U)$ in irreducible characters. For each partition $\lambda$ of integer $|P|$ (denoted as $\lambda\vdash |P|$) with $\ell(\lambda)\le n_1$, let $\widetilde\chi_{\lambda}(U)$ denote the $U(n_1)$ character and let $\chi^{\lambda}(P)$ denote the character of the symmetric group $S_{|P|}$ on the conjugacy class $P$. The Frobenius formula gives
\begin{equation}\label{eqn:Frobenius_formula}
\mathcal{O}_P(U)
=\sum_{\substack{\lambda\vdash |P|\\ \ell(\lambda)\le n_1}}
\widetilde\chi_{\lambda}(U)\,\chi^{\lambda}(P).
\end{equation}
Combining this with the orthogonality of unitary-group characters,
\begin{equation}\label{eqn:ortho_chi}
\int [dU]\,\widetilde\chi_{\lambda}(U)\widetilde\chi_{\lambda'}(U^{\dagger})
=\delta_{\lambda\lambda'},
\end{equation}
we obtain
\begin{equation}
\int [dU]\,\mathcal{O}_P(U)\mathcal{O}_P(U^{\dagger})
=\sum_{\substack{\lambda\vdash |P|\\ \ell(\lambda)\le n_1}}
\chi^{\lambda}(P)^2.
\end{equation}
Therefore the single-partition index admits the character expansion
\begin{equation}
\label{spartionchar}
\boxed{
\mathcal{I}_{n_1;N_1}^{\mathrm{BMN}}(t)
=\sum_P \frac{\iota_{11}(t)_P}{z_P}
\sum_{\substack{\lambda\vdash |P|\\ \ell(\lambda)\le n_1}}
\chi^{\lambda}(P)^2
 .}
\end{equation}
At any fixed order in $t$, only finitely many partitions $P$ contribute, so evaluating the index reduces to a finite sum over symmetric-group characters.

\subsection{Multi-partition index}
\label{multipart}
We now generalize the same method to a vacuum labelled by the partition
\begin{equation}
N=\sum_{r=1}^{K} n_r N_r.
\end{equation}
The corresponding index is
\begin{equation}
\mathcal{I}_{n_i;N_i}^{\mathrm{BMN}}(t)
=\int \prod_{r=1}^{K}[dU_r]\,
\exp\!\left[
\sum_{m=1}^{\infty}\sum_{k,l=1}^{K}
\frac{1}{m}\,\iota_{kl}(t^m)\,\Tr(U_k^{\dagger m})\Tr(U_l^m)
\right].
\end{equation}
Expanding each factor as before gives
\begin{equation}
\mathcal{I}_{n_i;N_i}^{\mathrm{BMN}}(t)
=\sum_{\{P_{kl}\}} C_{\{P_{kl}\}}
\int \prod_{r=1}^{K}[dU_r]\,
\prod_{k,l=1}^{K}
\mathcal{O}_{P_{kl}}(U_l)\mathcal{O}_{P_{kl}}(U_k^{\dagger}),
\end{equation}
with
\begin{equation}
C_{\{P_{kl}\}}
:=\prod_{k,l=1}^{K}\frac{\iota_{kl}(t)_{P_{kl}}}{z_{P_{kl}}}.
\end{equation}

For later convenience, we define the sum of two partitions componentwise:
\begin{equation}
(1^{k_1}2^{k_2}\cdots)+(1^{k_1'}2^{k_2'}\cdots)
:=(1^{k_1+k_1'}2^{k_2+k_2'}\cdots).
\end{equation}
Then for each node $r$ we introduce the incoming and outgoing partitions
\begin{equation}
P_r^{\mathrm{L}}:=\sum_{k=1}^{K} P_{kr},
\qquad
P_r^{\mathrm{R}}:=\sum_{l=1}^{K} P_{rl}.
\end{equation}
Using $\mathcal{O}_P(U)\mathcal{O}_{P'}(U)=\mathcal{O}_{P+P'}(U)$, the integrand factorizes as
\begin{equation}
\prod_{k,l=1}^{K}
\mathcal{O}_{P_{kl}}(U_l)\mathcal{O}_{P_{kl}}(U_k^{\dagger})
=
\prod_{r=1}^{K}
\mathcal{O}_{P_r^{\mathrm{L}}}(U_r)
\mathcal{O}_{P_r^{\mathrm{R}}}(U_r^{\dagger}).
\end{equation}
For each $r$, applying the Frobenius formula and the orthogonality relation yields
\begin{equation}
\int [dU_r]\,\mathcal{O}_{P_r^{\mathrm{L}}}(U_r)\mathcal{O}_{P_r^{\mathrm{R}}}(U_r^{\dagger})
=
\delta_{|P_r^{\mathrm{L}}|,|P_r^{\mathrm{R}}|}
\sum_{\substack{\lambda_r\vdash |P_r^{\mathrm{L}}|\\ \ell(\lambda_r)\le n_r}}
\chi^{\lambda_r}(P_r^{\mathrm{L}})\chi^{\lambda_r}(P_r^{\mathrm{R}}).
\end{equation}
The Kronecker delta imposes conservation of total cycle weight at each node. When $n_r\ge |P_r^{\mathrm{L}}|$, the restriction $\ell(\lambda_r)\le n_r$ is inactive and this expression reduces to the usual symmetric-group orthogonality relation, which is nonzero only when \(|P_r^{\mathrm{L}}|=|P_r^{\mathrm{R}}|\).

Putting everything together, we obtain the general character expansion
\begin{equation}
\boxed{
\mathcal{I}_{n_i;N_i}^{\mathrm{BMN}}(t)
=
\sum_{\substack{\{P_{kl}\}\\ |P_r^{\mathrm{L}}|=|P_r^{\mathrm{R}}|\ \forall r}}
\left(\prod_{k,l=1}^{K}\frac{\iota_{kl}(t)_{P_{kl}}}{z_{P_{kl}}}\right)
\prod_{r=1}^{K}
\sum_{\substack{\lambda_r\vdash |P_r^{\mathrm{L}}|\\ \ell(\lambda_r)\le n_r}}
\chi^{\lambda_r}(P_r^{\mathrm{L}})\chi^{\lambda_r}(P_r^{\mathrm{R}})
.}
\end{equation}
This expresses the BMN index entirely in terms of partitions and character tables of symmetric groups.

Let us record the first two nontrivial special cases explicitly.

\subsubsection*{Double partition $N=n_1N_1+n_2N_2$, with $N_1\neq N_2$}

In this case the only nontrivial conservation law is $|P_{12}|=|P_{21}|$, and the general formula becomes
\begin{equation}
\label{dpbmnformula}
\begin{aligned}
\mathcal{I}_{n_1,N_1;n_2,N_2}^{\mathrm{BMN}}(t)
=
&\sum_{\substack{P_{11},P_{12},P_{21},P_{22}\\ |P_{12}|=|P_{21}|}}
\frac{(\iota_{11})_{P_{11}}(\iota_{12})_{P_{12}}(\iota_{21})_{P_{21}}(\iota_{22})_{P_{22}}}
{z_{P_{11}}z_{P_{12}}z_{P_{21}}z_{P_{22}}}
\\[2mm]
&\times
\sum_{\substack{\lambda_1\vdash |P_{11}+P_{21}|\\ \ell(\lambda_1)\le n_1}}
\chi^{\lambda_1}(P_{11}+P_{21})\chi^{\lambda_1}(P_{11}+P_{12})
\\[1mm]
&\times
\sum_{\substack{\lambda_2\vdash |P_{12}+P_{22}|\\ \ell(\lambda_2)\le n_2}}
\chi^{\lambda_2}(P_{12}+P_{22})\chi^{\lambda_2}(P_{21}+P_{22}).
\end{aligned}
\end{equation}

\subsubsection*{Triple partition $N=n_1N_1+n_2N_2+n_3N_3$, with $N_k\neq N_l$ for $k\neq l$}

The character expansion becomes
\begin{equation}
\label{tpbmnformula}
\begin{aligned}
\mathcal{I}_{n_1,N_1;n_2,N_2;n_3,N_3}^{\mathrm{BMN}}(t)
=
&\sum_{\substack{P_{11},P_{12},\ldots,P_{33}\\
|P_{12}+P_{13}|=|P_{21}+P_{31}|\\
|P_{21}+P_{23}|=|P_{12}+P_{32}|\\
|P_{31}+P_{32}|=|P_{13}+P_{23}|}}
\frac{\prod_{k,l=1}^{3}(\iota_{kl})_{P_{kl}}}{\prod_{k,l=1}^{3} z_{P_{kl}}}
\\[2mm]
&\times
\sum_{\substack{\lambda_1\vdash |P_{11}+P_{21}+P_{31}|\\ \ell(\lambda_1)\le n_1}}
\chi^{\lambda_1}(P_{11}+P_{21}+P_{31})\chi^{\lambda_1}(P_{11}+P_{12}+P_{13})
\\[1mm]
&\times
\sum_{\substack{\lambda_2\vdash |P_{12}+P_{22}+P_{32}|\\ \ell(\lambda_2)\le n_2}}
\chi^{\lambda_2}(P_{12}+P_{22}+P_{32})\chi^{\lambda_2}(P_{21}+P_{22}+P_{23})
\\[1mm]
&\times
\sum_{\substack{\lambda_3\vdash |P_{13}+P_{23}+P_{33}|\\ \ell(\lambda_3)\le n_3}}
\chi^{\lambda_3}(P_{13}+P_{23}+P_{33})\chi^{\lambda_3}(P_{31}+P_{32}+P_{33}).
\end{aligned}
\end{equation}

\section{Residue expansion of the BMN index}
\label{resexp}
In this section, we develop an alternative approach to computing the index based on the residue theorem, building on residue-based treatments of finite-$N$ index integrals in \cite{Imamura:2021ytr,Gadde:2025yoa,vanLeuven:2025gwr}. The unitary-matrix integral \eqref{dsum} over the matrices \(U_k\) can first be reduced to an integral over their eigenvalues \(y_{k,a}\), \(a=1,\cdots,n_k\), as
\begin{equation}
\begin{split}\label{eqn:BMN_indx_eigenvalues}
\mathcal{I}_{n_k; N_k}^{\mathrm{BMN}} 
&=\prod_{r=1}^K \frac1{n_r!}\prod_{a=1}^{n_r}
\oint \frac{dy_{r,a}}{2\pi i y_{r,a}}\,
\prod_{1\le a<b\le n_r}
(y_{r,a}-y_{r,b})(y^{-1}_{r,a}-y^{-1}_{r,b})
\\
&\quad \times
\exp \left[
\sum_{m=1}^\infty
\sum_{k,l=1}^K
\frac{1}{m}\,
\iota^{(m)}_{kl}(\Delta_i)
\sum_{a=1}^{n_k}\sum_{b=1}^{n_l}
y_{l,b}^m y_{k,a}^{-m}
\right]
\\
&=\prod_{r=1}^K \frac1{n_r!}\prod_{a=1}^{n_r}
\oint \frac{dy_{r,a}}{2\pi i y_{r,a}}\,
\prod_{k,l=1}^K \prod_{a=1}^{n_k}\prod_{b=1}^{n_l}
\sideset{}{'}\prod^{\frac{1}{2}(N_k + N_l) - 1}_{j=\frac{1}{2}|N_k - N_l|}
\\
&\qquad \times
\frac{
\left(1-\epsilon_j\,y_{l,b}y_{k,a}^{-1}e^{-j{\bf \Delta}}\right)
\prod_{i=1}^3
\left(1-\epsilon_j\,y_{l,b}y_{k,a}^{-1}
e^{-(j+1){\bf \Delta}+\Delta_i}\right)
}{
\left(1-\epsilon_j\,y_{l,b}y_{k,a}^{-1}e^{-(j+1){\bf \Delta}}\right)
\prod_{i=1}^3
\left(1-\epsilon_j\,y_{l,b}y_{k,a}^{-1}
e^{-j{\bf \Delta}-\Delta_i}\right)
}\,,
\end{split}
\end{equation}
where
\begin{equation}
\iota^{(m)}_{kl}(\Delta_i)
=
\sum_{j=\frac{1}{2}|N_k-N_l|}^{\frac{1}{2}(N_k+N_l)-1}
(-1)^{2jm+1}
e^{-jm(\Delta_1+\Delta_2+\Delta_3)}
\prod_{i=1}^3(1-e^{-m\Delta_i})
+\delta_{N_k,N_l},
\qquad
\epsilon_j:=(-1)^{2j},
\end{equation}
where \({\bf \Delta}=\Delta_1+\Delta_2+\Delta_3\), and the prime indicates that the factor \((1-\epsilon_j y_{l,b}y_{k,a}^{-1}e^{-j{\bf \Delta}})\) is omitted when \(j=0\) and \((k,a)=(l,b)\). The integration contour for each \(y_{k,a}\) is the unit circle \(|y_{k,a}|=1\) in the complex \(y_{k,a}\)-plane.
For the residue expansion, it is better to keep the pairwise pole-zero data
explicit from the outset. Introduce a multi-index
\[
I=(k,a),\qquad 1\le k\le K,\qquad 1\le a\le n_k,
\]
and write
\[
c(I)=k,\qquad y_I\equiv y_{k,a},\qquad V=\{(k,a)\}.
\]
After extracting the \(I=J\) factors in \eqref{eqn:BMN_indx_eigenvalues} into
an overall constant \(C_{\vec n,\vec N}\), and shifting the auxiliary pole from
\(0\) to a point \(u\) inside the unit circle, the index can be written as
\begin{equation}\label{eq:general_sector_integrand}
\mathcal I^{\mathrm{BMN}}_{\vec n;\vec N}
=
\frac{C_{\vec n,\vec N}}{\prod_{k=1}^K n_k!}
\oint \prod_{I\in V}\frac{dy_I}{2\pi i (y_I-u)}
\prod_{I\neq J}
\frac{\displaystyle\prod_{g\in \widehat Z_{IJ}}
\left(1-g\frac{y_I}{y_J}\right)}
{\displaystyle\prod_{f\in P_{IJ}}
\left(1-f\frac{y_I}{y_J}\right)} .
\end{equation}
The finite multisets \(P_{IJ}\) and \(Z_{IJ}\) are defined by
\begin{equation}
\frac{\prod_{g\in Z_{IJ}}(1-gx)}{\prod_{f\in P_{IJ}}(1-fx)}
=
\sideset{}{'}\prod_{j=\frac12|N_{c(I)}-N_{c(J)}|}^{\frac12(N_{c(I)}+N_{c(J)})-1}
\left[
\frac{(1-xe^{-j{\bf \Delta}})
\prod_{i=1}^3(1-xe^{-(j+1){\bf \Delta}+\Delta_i})}
{(1-xe^{-(j+1){\bf \Delta}})
\prod_{i=1}^3(1-xe^{-j{\bf \Delta}-\Delta_i})}
\right]^{(-1)^{2j}} .
\end{equation}
For equal colors we also include the Vandermonde zero,
\[
\widehat Z_{IJ}=
\begin{cases}
Z_{IJ}\sqcup\{1\}, & c(I)=c(J),\\
Z_{IJ}, & c(I)\neq c(J).
\end{cases}
\]
All products over multisets are understood with multiplicity.

The contour integral can then be evaluated by the residue theorem, which converts the integral into a sum over residues. One subtlety of this method is the organization of the poles. Although the final result is independent of the order of integration, the set of poles that appears at intermediate steps does depend on that order.

Let us illustrate this point with the factor
\begin{equation}
(1-f\tfrac{z_I}{z_J})^{-1}(1-f'\tfrac{z_K}{z_J})^{-1}
\end{equation}
in the denominator of the integrand. Suppose we perform the contour integrals in the order \(z_I\), \(z_J\), \(z_K\), and assume \(|f'|<|f|<1\). Then this factor does not produce any pole in the \(z_I\)-integral. By contrast, if we integrate in the order \(z_J\), \(z_I\), \(z_K\), then the \(z_J\)-integral may pick up the pole at
\begin{equation}
z_J^* = f' z_K,
\end{equation}
which in turn induces a pole in the \(z_I\)-integral at
\begin{equation}
    z_I^* = \frac{f'}{f} z_K\,.
\end{equation}
This is what is called a \emph{mixed pole} in the terminology of \cite{Imamura:2021ytr}. A useful criterion proposed in that paper is that the contributions from mixed poles cancel among themselves. Thus, the factor \((1-f \frac{z_I}{z_J})^{-1}\) effectively contributes only a pole in the \(z_J\)-integral.

Let us illustrate this with the integrand
\begin{equation}
    \frac{1}{(1-f_{12}\frac{z_1}{z_2}) (1-f_{21}\frac{z_2}{z_1})(1-f_{13}\frac{z_1}{z_3})(1-f_{31}\frac{z_3}{z_1})(1-f_{23}\frac{z_2}{z_3})(1-f_{32}\frac{z_3}{z_2})}\,.
\end{equation}
Suppose we integrate over \(z_1\). There are then two contributions,
\begin{equation}
\begin{aligned}
    (z_1^* = f_{21} z_2 )~ & : ~ \frac{f_{21}z_2}{(1-f_{12}f_{21})(1-f_{13} \frac{f_{21}z_2}{z_3} )(1-f_{31}\frac{z_3}{f_{21}z_2})(1-f_{23}\frac{z_2}{z_3})(1-f_{32}\frac{z_3}{z_2})}  \\
    (z_1^* = f_{31} z_3 )~ & : ~  \frac{f_{31} z_3}{(1-f_{12}\frac{f_{31} z_3}{z_2}) (1-f_{21}\frac{z_2}{f_{31} z_3})(1-f_{13}f_{31})(1-f_{23}\frac{z_2}{z_3})(1-f_{32}\frac{z_3}{z_2})}\,.
\end{aligned}
\end{equation}
We now examine the mixed-pole contributions. Suppose \(|f_{31}|<|f_{21}|\). Then there is a mixed pole in the \(z_2\)-integral. In each of the two channels above, there is exactly one such mixed-pole contribution, located at
\begin{equation}
    z_2^* = \frac{f_{31}}{f_{21}} z_3\,,
\end{equation}
and the two contributions cancel each other.

In \cite{Gadde:2025yoa}, an arrow notation was introduced to encode this residue structure. If the residue prescription picks the factor \((1-f\frac{z_I}{z_J})^{-1}\) in the \(z_J\)-integral, we draw an arrow
\begin{equation}
    I \stackrel{f}{\longrightarrow} J\,.
\end{equation}
We will make extensive use of this notation in the rest of this section, and we will review it in more detail in the next subsection.

\subsection{Residues as a sum over trees}
\label{treesres}
We first explain the combinatorics in the single-partition sector. In this case
the pairwise data are independent of \(I,J\), and the index takes the form
\begin{equation}
\mathcal I_N
=
\frac{1}{N!}
\left(\frac{\prod_{g\in Z}(1-g)}{\prod_{f\in P}(1-f)}\right)^N
\oint \prod_{i=1}^N \frac{dz_i}{2\pi i (z_i-u)}
\prod_{i\neq j}
\left(1-\frac{z_i}{z_j}\right)
\frac{\prod_{g\in Z}\left(1-g\frac{z_i}{z_j}\right)}
{\prod_{f\in P}\left(1-f\frac{z_i}{z_j}\right)} ,
\end{equation}
where \(P\) and \(Z\) are finite sets satisfying
\(|P|-|Z|=1\)\footnote{The full \(\mathcal N=4\) SYM index does not satisfy
this condition because of the derivative letter, whereas the BMN sectors
considered here do.} and \(P\cap Z=\emptyset\). The extra Vandermonde zero
precisely compensates the difference \(|P|-|Z|=1\), so \(z_i=0\) is not a pole
of the shifted integrand.

We organize the residue theorem as a sum of simple-pole data. In particular, we
do not introduce separate subscripted variables such as \(f_{ij}\). Instead, we
sum directly over rooted trees and separate colliding poles by a small
\(\epsilon\)-deformation. The admissible trees are characterized by the
following rules.
\begin{enumerate}
    \item A root edge
    \[
    u\to i
    \]
    means that the \(z_i\)-contour is evaluated at the pole \(z_i=u\) coming
    from \((z_i-u)^{-1}\).

    \item A non-root edge
    \[
    i\stackrel{f}{\longrightarrow}j,\qquad f\in P,
    \]
    means that the residue of the \(z_j\)-integral is taken at the pole
    \(z_j=fz_i\).

    \item Each physical vertex has one and only one incoming edge. This is the
    simple-pole prescription.

    \item The root has exactly one outgoing edge. Indeed, if both \(u\to i\)
    and \(u\to j\) with \(i\neq j\) were present, then \(z_i=z_j=u\), and the
    Vandermonde factor
    \[
    \left(1-\frac{z_i}{z_j}\right)\left(1-\frac{z_j}{z_i}\right)
    \]
    would kill the residue.

    \item More generally, for a fixed parent \(i\) and a fixed label \(f\in P\),
    there is at most one outgoing \(f\)-edge from \(i\). If
    \[
    i\stackrel{f}{\longrightarrow}j,\qquad
    i\stackrel{f}{\longrightarrow}k,\qquad
    j\neq k,
    \]
    then \(z_j=z_k=fz_i\), and the same Vandermonde zero again forces the
    residue to vanish.
\end{enumerate}
The root edge encodes the pole $z_i=u$, while each labeled non-root edge
$i \stackrel{f}{\longrightarrow} j$ encodes the pole $z_j=f z_i$. 

Thus every non-vanishing contribution is represented by a rooted tree. Once a
canonical traversal order is fixed, each unlabeled tree represents \(N!\)
equivalent labelings of the eigenvalues, and this cancels the prefactor \(1/N!\)
in the integral. In the formulas below, \(\mathfrak T_N\) denotes the set of
canonically labeled admissible trees, and we always choose the unique child of
the root to be \(z_1\). The canonical labeling can be defined according to the lexicographical order of the nodes of a tree. Fix an ordering of the multiset
\[
P=\{f_1,\ldots,f_{n_P}\}.
\]
Each node can be uniquely connected to the root, for example
\begin{equation}
    u \stackrel{f_{a_1}}{\longrightarrow} \cdot \stackrel{f_{a_2}}{\longrightarrow} \cdots \stackrel{f_{a_m}}{\longrightarrow} \cdot
\end{equation}
Then, one can use an ordered set $(a_1,\cd,a_m)$ to label this node. To compare the order of two nodes, one just compares the labels alphabetically. For instance, if the labels of two nodes are $(a_1,\cd,a_m)$ and $(b_1,\cd , b_n)$. Then $(a_1,\cd,a_m)<(b_1,\cd , b_n)$ if there exists a $k$ such that $a_1=b_1,\cd ,a_{k-1}=b_{k-1}$, but $a_{k}< b_k$. It is clear that this is a total order of the nodes. Using this canonical ordering, one can generate the trees recursively by adding a largest node to a tree. From now on, we will always use this canonical lexicographical order. 

For \(\Gamma\in \mathfrak T_N\), let \(\mathrm{Path}_\Gamma(u,j)\) be the unique
path from the root to the vertex \(j\). The undeformed pole location is
\[
z_j^*
=
u\prod_{e\in \mathrm{Path}_\Gamma(u,j)} f^e .
\]

To separate colliding trees, we use a prime-number regulator attached to the
undeformed evaluation \(z_j^*/u\) itself. Define a multiplicative map \(\mathrm{Num}\) by
\begin{equation}
\begin{aligned}
\mathrm{Num}(f_i)&=p_i,\\
\mathrm{Num}(f_{a_1}\cdots f_{a_m})&=\mathrm{Num}(f_{a_1})\cdots \mathrm{Num}(f_{a_m}),
\end{aligned}
\end{equation}
where \(p_i\) is the \(i\)-th prime number. For a fixed tree \(\Gamma\), every
denominator whose target is \(j\) is deformed by the same factor
\[
(f)^{(\epsilon)}_{j,\Gamma}
:=
f\left(1+\epsilon\,\mathrm{Num}\!\left(\frac{z_j^*}{u}\right)\right),
\qquad f\in P .
\]
The deformed pole locations are then defined recursively by
\begin{equation}\label{neweval}
z_j^\star=
\begin{cases}
u, & u\to j,\\[2mm]
(f)^{(\epsilon)}_{j,\Gamma}\,z_i^\star,
& i\stackrel{f}{\to}j\in \Gamma .
\end{cases}
\end{equation}
By construction, \(z_j^\star\to z_j^*\) as \(\epsilon\to0\). The important
point is that the deformation depends only on the undeformed evaluations of the
vertices, not on an additional code for the full path. Nevertheless, the
recursive definition still distinguishes different branches when their
intermediate undeformed evaluations differ. For example, the two paths
\[
u\to \cdot \stackrel{a}{\to}\cdot \stackrel{b}{\to}j,
\qquad
u\to \cdot \stackrel{b}{\to}\cdot \stackrel{a}{\to}j
\]
have the same final monomial \(ab\), but they give
\[
z_j^\star = ab\bigl(1+\epsilon\,\mathrm{Num}(a)\bigr)
\bigl(1+\epsilon\,\mathrm{Num}(ab)\bigr)u
\]
and
\[
z_j^\star = ab\bigl(1+\epsilon\,\mathrm{Num}(b)\bigr)
\bigl(1+\epsilon\,\mathrm{Num}(ab)\bigr)u,
\]
so they remain distinct at finite \(\epsilon\).

It is convenient to package the Jacobian factors from the selected non-root
poles into
\[
F_\Gamma^{(\epsilon)}
=
\prod_{\substack{i\stackrel{f}{\to}j\in \Gamma\\ i\neq u}}
\left(
1-(f)^{(\epsilon)}_{j,\Gamma}\frac{z_i^\star}{z_j^\star}
\right)
(f)^{(\epsilon)}_{j,\Gamma}z_i^\star .
\]
The tree expansion of the index is then
\begin{equation}\label{newIndFormula}
\mathcal I_N
=
\lim_{\epsilon\to 0}
\sum_{\Gamma\in \mathfrak T_N}
\left(\frac{\prod_{g\in Z}(1-g)}{\prod_{f\in P}(1-f)}\right)^N
\prod_{j=2}^N \frac{1}{z_j^\star-u}
\prod_{i\neq j}
\left(1-\frac{z_i^\star}{z_j^\star}\right)
\frac{\prod_{g\in Z}\left(1-g\frac{z_i^\star}{z_j^\star}\right)}
{\prod_{f\in P}\left(1-(f)^{(\epsilon)}_{j,\Gamma}\frac{z_i^\star}{z_j^\star}\right)}
\,F_\Gamma^{(\epsilon)} .
\end{equation}

The existence and order-independence of the limit are the standard facts about
divided differences. If \(h\) is holomorphic near \(z=z^*\), then
\[
\operatorname{Res}_{z=z^*}\frac{h(z)}{(z-z^*)^n}
=
\lim_{z_1,\ldots,z_n\to z^*}
\sum_{a=1}^n
\frac{h(z_a)}{\prod_{b\neq a}(z_a-z_b)} .
\]
Thus a higher-order residue is recovered as the coincident-point limit of
separated simple poles.

\subsubsection*{Equivalent classes}

For practical evaluation it is useful to group trees before taking the final
fugacity limits. In the trivial vacuum sector, two trees are called
\(t\)-equivalent if their undeformed evaluations \(\{z_i^*\}\) become equal, up
to permutation, after the specialization
\[
g_i\to t^4,\qquad f_{1,2,3}\to t^2,\qquad f_4\to t^6 .
\]
They are called BMN-equivalent if the same happens, again up to permutation,
after imposing the BMN relations
\[
g_{1}\to  f_2 f_3,\qquad
g_{2}\to  f_3 f_1,\qquad
g_{3}\to  f_1 f_2,\qquad
f_4 \to f_1 f_2 f_3 .
\]
Each \(t\)-equivalence class is a union of BMN-equivalence classes. In
practice, the \(\epsilon\)-limit inside a BMN-equivalence class is usually mild;
the more expensive step is the final \(t\)-expansion, where it is advantageous
to combine terms with the same denominator structure before expanding.

\subsection{Meaning of the trees}
\label{meaningtrees}

We now explain the geometric meaning of the residue trees.  The main point is
that a tree is not a new physical object by itself.  Rather, it is a
prescription for a residue channel: it tells us which denominator factors are
used to localize the eigenvalue variables.

For simplicity, first consider the single-partition case.  A tree has one root
edge
\[
u\to r,
\]
which selects the pole
\[
z_r=u,
\]
and every non-root edge
\[
i\stackrel{f}{\longrightarrow}j
\]
selects the pole
\[
z_j=fz_i .
\]
Therefore the tree fixes all eigenvalues recursively:
\[
z_j^*
=
u\prod_{e\in \mathrm{Path}_{\Gamma}(u,j)} f^e .
\]
Equivalently, the tree selects \(n\) denominator hypersurfaces whose
intersection is the pole \(p_\Gamma=(z_1^*,\ldots,z_n^*)\).

When this pole is isolated in the full divisor arrangement, the picture is
simple.  Around each selected denominator one draws a small circle in the
normal direction.  The product of these circles is a small \(n\)-torus,
\[
T_\Gamma:
\qquad
|h_1|=\delta_1,\quad \ldots,\quad |h_n|=\delta_n,
\]
and the tree contribution is the corresponding period of the meromorphic
\(n\)-form,
\[
\operatorname{Cont}(\Gamma)
=
\frac{1}{(2\pi i)^n}\int_{T_\Gamma}\omega .
\]
In this clean situation, the tree is literally represented by a small residue
torus.

The subtlety is that the BMN integrand contains many denominator factors, not
only the \(n\) factors selected by a given tree.  Thus several trees may land on
the same undeformed pole.  At such a collision point, the individual small
tori attached to the different trees are no longer separated inside the
complement of the full divisor.  They should not be regarded as independent
contours of the original undeformed integral.

It is useful to phrase this in terms of a local moduli space of hyperplane
arrangements.  Let
\[
D_\lambda=\bigcup_A H_A(\lambda)
\]
be the union of all denominator hypersurfaces, where \(\lambda\) denotes the
parameters controlling their relative positions.  In our application,
\(\lambda\) is represented by the small \(\epsilon\)-deformation.  We write
\[
M_\lambda=\mathbb C^n\setminus D_\lambda .
\]
For generic \(\lambda\), the relevant intersections are separated, and each
tree gives a genuine cycle
\[
[T_\Gamma(\lambda)]\in H_n(M_\lambda).
\]
At special values of \(\lambda\), several intersections collide.  Then some of
these \(n\)-cycles can merge or disappear.  In this sense, the top homology
seen by the residue tori is a function on the arrangement moduli space:
\[
\lambda \longmapsto H_n(M_\lambda),
\]
or more concretely
\[
\lambda \longmapsto b_n(M_\lambda):=\dim H_n(M_\lambda).
\]
This function is locally constant on the smooth strata of the moduli space, but
it can jump on the discriminant locus where hyperplanes collide.

The lower-dimensional linking data are more robust.  The one-cycle linking a
single divisor component is still well-defined locally.  What fails is the
ability to combine \(n\) such one-cycles into a canonical product torus
associated with one individual tree.  Thus the collision is not a problem with
the existence of elementary divisor links; it is a problem with separating the
higher-dimensional residue cycles.

A two-dimensional local model makes the picture transparent.  Consider three
lines through the same point,
\[
H_1=\{x=0\},\qquad
H_2=\{y=0\},\qquad
H=\{x-y=0\}.
\]
If we try to assign a small torus to the pair \((H_1,H_2)\), the natural
candidate is
\[
T_{12}(a,b)=\{|x|=a,\ |y|=b\}.
\]
This contour is allowed only when it does not hit the third line \(H\).  Using
\[
s=y,\qquad t=\frac{x}{y},
\]
we get
\[
T_{12}(a,b):
\qquad
|s|=b,\qquad |t|=\frac{a}{b}.
\]
The third line is simply
\[
t=1.
\]
Thus the ratio \(a/b\) matters.  If \(a<b\), the circle \(|t|=a/b\) lies inside
\(t=1\).  If \(a>b\), it lies outside \(t=1\).  To move from one choice to the
other, the contour must cross the wall
\[
a=b,
\]
where it hits the extra divisor \(H\).  Therefore, in the undeformed central
arrangement, there is no canonical small torus attached only to the pair
\((H_1,H_2)\).  The extra line passing through the same point obstructs the
separation of residue channels. This can also be seen directly by computing the homology group:
\begin{equation}
    H_2(\mathbb{C}^2-H_1-H_2-H)\cong \mathbb Z^2
\end{equation}

Now deform the third line slightly:
\[
\widetilde H=\{x-y=\epsilon\},
\qquad
\epsilon\neq 0.
\]
The three pairwise intersections are separated:
\[
p_{12}=(0,0),\qquad
p_{1\widetilde H}=(0,-\epsilon),\qquad
p_{2\widetilde H}=(\epsilon,0).
\]
Near each point, there is now an honest small residue torus linking only the
two relevant lines. Indeed, we have
\begin{equation}
    H_2(\mathbb{C}^2-H_1-H_2-\widetilde H)\cong \mathbb Z^3
\end{equation}
Thus, the deformation resolves one degenerate pole into
several separated simple residue channels.

This is exactly the role of the \(\epsilon\)-deformation in the tree formula.
At finite \(\epsilon\), the colliding trees move to nearby but distinct points
in the arrangement moduli space.  Each tree then has a well-defined local
residue contour.  In the limit \(\epsilon\to0\), these contours degenerate and
the individual tree contributions may become singular.  However, the periods
are meromorphic functions of the deformation parameter, and the original
BMN contour integral is nonsingular.  Therefore the singularities cancel only
after summing over all trees in the same collision class:
\[
\lim_{\epsilon\to0}
\sum_{\Gamma\ \mathrm{in\ a\ collision\ class}}
\operatorname{Cont}_{\epsilon}(\Gamma).
\]
The individual terms are regulator-dependent residue channels; the sum is the
intrinsic contribution of the undeformed pole.

This gives the physical interpretation of the trees.  A non-colliding tree is a
genuine small residue torus.  A colliding tree is one branch of a resolved
degenerate residue.  The \(\epsilon\)-deformation should therefore be viewed as
a contour regulator: it moves the hyperplane arrangement away from the
discriminant locus, computes the separated periods, and then takes the
coincident limit after summing over the colliding branches.

\subsection{Generalization to a general partition sector}
\label{genpartsec}
We now return to the full BMN sector \(\{n_k;N_k\}_{k=1}^K\). Since the
pairwise notation has already been introduced in
\eqref{eq:general_sector_integrand}, only the tree rules need to be
generalized.

A physical vertex is \(I=(k,a)\) with color \(c(I)=k\). A non-root edge
\[
I\stackrel{f}{\longrightarrow}J
\]
is allowed only if \(f\in P_{IJ}\), in which case it imposes
\[
y_J=fy_I .
\]
A root edge \(u\to J\) corresponds to the pole \(y_J=u\). The admissible
colored trees satisfy the following rules.
\begin{enumerate}
    \item Each physical vertex has one and only one incoming edge.

    \item The root may have several outgoing edges, but at most one to each
    color. If \(u\to J\) and \(u\to J'\) with \(c(J)=c(J')\) and \(J\neq J'\),
    then \(y_J=y_{J'}=u\), and the same-color Vandermonde zero kills the
    residue.

    \item More generally, fix a parent vertex \(I\), a target color \(r\), and
    a label \(f\). Then there is at most one outgoing edge
    \[
    I\stackrel{f}{\longrightarrow}J
    \qquad\text{with}\qquad c(J)=r .
    \]
    If two such edges existed, the two target vertices would coincide, and the
    same-color Vandermonde factor would again force the residue to vanish.

    \item Before adjoining the auxiliary root \(u\), the diagram is therefore a
    colored forest; after adjoining \(u\), it becomes a rooted colored tree.

    \item After fixing a canonical order within each color class, each
    unlabeled colored tree represents \(\prod_k n_k!\) equivalent labelings,
    which cancel the prefactor in \eqref{eq:general_sector_integrand}. We
    denote by \(\mathfrak T_{\vec n,\vec N}\) the set of canonically labeled
    admissible rooted colored trees.
\end{enumerate}

For \(\Gamma\in \mathfrak T_{\vec n,\vec N}\), let \(\mathrm{Path}_\Gamma(u,J)\)
be the unique path from the root to \(J\). The undeformed evaluation is
\[
y_J^*
=
u\prod_{e\in \mathrm{Path}_\Gamma(u,J)} f^e .
\]

To define the \(\epsilon\)-deformation, fix once and for all an ordering of the
distinct basic pole factors that appear in the various \(P_{IJ}\),
\[
\mathcal P_{\mathrm{all}}=\{\varphi_1,\ldots,\varphi_M\},
\]
and define \(\mathrm{Num}(\varphi_\alpha)=p_\alpha\),
extended multiplicatively to arbitrary monomials. If the same fugacity monomial
appears in different \(P_{IJ}\), it is treated as the same generator in
\(\mathcal P_{\mathrm{all}}\), because the deformation is meant to depend only
on the monomial \(y_J^*/u\).

For a fixed tree \(\Gamma\), every denominator whose target is \(J\) is
deformed by
\[
(f)^{(\epsilon)}_{J,\Gamma}
:=
f\left(1+\epsilon\,\mathrm{Num}\!\left(\frac{y_J^*}{u}\right)\right),
\qquad f\in P_{IJ}.
\]
The deformed pole locations are defined recursively by
\[
y_J^\star=
\begin{cases}
u, & u\to J \in \Gamma,\\[2mm]
(f)^{(\epsilon)}_{J,\Gamma}\,y_I^\star,
& I\stackrel{f}{\to}J\in \Gamma .
\end{cases}
\]
Again, the deformation depends only on the undeformed evaluations of the
vertices, not on an additional code for the full colored path. Distinct trees
are separated whenever their intermediate undeformed evaluations differ.

Let \(R(\Gamma)\) be the set of vertices directly attached to the root. The
Jacobian factors from the selected non-root poles are packaged into
\[
F_\Gamma^{(\epsilon)}
=
\prod_{\substack{I\stackrel{f}{\to}J\in \Gamma\\ I\neq u}}
\left(
1-(f)^{(\epsilon)}_{J,\Gamma}\frac{y_I^\star}{y_J^\star}
\right)
(f)^{(\epsilon)}_{J,\Gamma}y_I^\star .
\]
The contribution of \(\Gamma\) is
\[
\operatorname{Cont}_\epsilon(\Gamma)
=
C_{\vec n,\vec N}
\left[
\prod_{J\notin R(\Gamma)}\frac{1}{y_J^\star-u}
\prod_{I\neq J}
\frac{\displaystyle\prod_{g\in \widehat Z_{IJ}}
\left(1-g\frac{y_I^\star}{y_J^\star}\right)}
{\displaystyle\prod_{f\in P_{IJ}}
\left(1-(f)^{(\epsilon)}_{J,\Gamma}\frac{y_I^\star}{y_J^\star}\right)}
\right]
F_\Gamma^{(\epsilon)} .
\]
Finally,
\begin{equation}\label{eq:general_sector_tree_sum}
\mathcal I^{\mathrm{BMN}}_{\vec n;\vec N}
=
\lim_{\epsilon\to0}
\sum_{\Gamma\in \mathfrak T_{\vec n,\vec N}}
\operatorname{Cont}_\epsilon(\Gamma).
\end{equation}
For \(K=1\), the root has a single child and
\eqref{eq:general_sector_tree_sum} reduces to \eqref{newIndFormula}.
A colored rooted tree for the \textit{triple partition sector} namely $(n_1,n_2,n_3)=(3,2,2)$ is shown in Fig.~\ref{coloredex}.
\begin{figure}[H]
\centering
\begin{tikzpicture}[
    >=Stealth,
    line cap=round,
    line join=round,
    every node/.style={inner sep=0pt},
    edge/.style={->, semithick, draw=black!70},
    bigedge/.style={->, very thick, draw=black!70},
    root/.style={
        circle,
        draw=black!55,
        fill=gray!20,
        minimum size=1.45cm,
        line width=0.9pt
    },
    k1/.style={
        circle,
        draw=green!50!black,
        fill=green!20,
        minimum size=1.45cm,
        line width=1pt
    },
    k2/.style={
        circle,
        draw=yellow!50!black!80,
        fill=yellow!25,
        minimum size=1.45cm,
        line width=1pt
    },
    k3/.style={
        circle,
        draw=blue!65!black,
        fill=blue!20,
        minimum size=1.45cm,
        line width=1pt
    },
    edgelabel/.style={font=\scriptsize, fill=white, inner sep=1pt},
    leg/.style={font=\normalsize}
]

\node[root] (u)  at ( 0.0,  2.8) {$u$};

\node[k1]   (x1) at (-4.2,  1.2) {$x_1$};
\node[k2]   (y1) at ( 0.0,  0.8) {$y_1$};
\node[k3]   (z1) at ( 4.2,  1.2) {$z_1$};

\node[k2]   (y2) at (-6.1, -1.0) {$y_2$};
\node[k1]   (x2) at (-2.8, -1.0) {$x_2$};
\node[k3]   (z2) at ( 1.2, -1.0) {$z_2$};
\node[k1]   (x1b) at ( 4.2, -1.0) {$x_1$};

\draw[edge]    (u) -- (x1);
\draw[bigedge] (u) -- (y1);
\draw[edge]    (u) -- (z1);

\draw[edge] (x1) -- node[pos=0.48, above, sloped, edgelabel, xshift=1pt]
    {$\left(f\right)^{(\epsilon)}_{(2,2),\Gamma}$} (y2);
    
\draw[edge] (x1) -- node[pos=0.56, right, edgelabel, xshift=8pt, yshift=1pt]
    {$\left(f\right)^{(\epsilon)}_{(1,2),\Gamma}$} (x2);

\draw[edge] (y1) -- node[pos=0.52, above right, edgelabel, xshift=6pt, yshift=2pt]
    {$\left(f\right)^{(\epsilon)}_{(3,2),\Gamma}$} (z2);

\draw[edge] (z1) -- node[pos=0.52, right, edgelabel, xshift=8pt]
    {$\left(f\right)^{(\epsilon)}_{(1,3),\Gamma}$} (x1b);

\node[font=\large] at (0,-2.6)
    {Example $(n_1,n_2,n_3)=(3,2,2)$; colors encode $k$, node label is $a$.};
\node[k1, minimum size=0.75cm] (L1) at (-5.6,-4.0) {};
\node[leg, anchor=west] at (-4.9,-4.0) {$k=1:\ x_a$};

\node[k2, minimum size=0.75cm] (L2) at (-1.2,-4.0) {};
\node[leg, anchor=west] at (-0.5,-4.0) {$k=2:\ y_a$};

\node[k3, minimum size=0.75cm] (L3) at ( 3.4,-4.0) {};
\node[leg, anchor=west] at ( 4.1,-4.0) {$k=3:\ z_a$};

\end{tikzpicture}
\caption{A colored rooted tree for the example $(n_1,n_2,n_3)=(3,2,2)$.}
\label{coloredex}
\end{figure}

For general partition sectors, we do not push to higher $N$ in numerical computation. In order to test the validity of the formula \eqref{eq:general_sector_tree_sum}, we compute the following two cases using the tree method in double partition sectors: $(N_1,N_2,n_1,n_2)=(1,2,1,1)$ and $(N_1,N_2,n_1,n_2)=(1,2,2,1)$. The case $(N_1,N_2,n_1,n_2)=(1,2,1,1)$ is also directly computed in \eqref{N1211}, which agrees with the tree method. For, the case $(N_1,N_2,n_1,n_2)=(1,2,2,1)$, the tree method agrees with the character table expansion in \eqref{2112}.

\subsection{Computation of the BMN index for multi-partition sectors: consistency check}
\label{concheck}
In this section, we will compute the BMN index for multi-partition sectors using direct integral methods. These results match the symmetric-group character expansion results of the BMN indices for the same sectors, providing a non-trivial consistency check of our results. We express the double and triple partition sector indices in Appendices \ref{dpbmn} and \ref{tpbmn} using equations \eqref{dpbmnformula} and \eqref{tpbmnformula}. To illustrate the character expansion approach, Appendix \ref{exbmn} provides a detailed walkthrough of the computation for a particular example $N_1 = 1, N_2 = 2, n_1 = 1, n_2 = 1$.
\subsection*{$N=N_1+N_2$}

The BMN index is
\begin{equation}
\mathcal{I}_{1,1; N_1,N_2}^{\mathrm{BMN}}(t)
= \int_0^{2\pi} \frac{d\phi_1}{2\pi} \int_0^{2\pi} \frac{d\phi_2}{2\pi} \,
\exp\!\left[ \sum_{m=1}^\infty \frac{1}{m} \Big( \iota_{11}(t^m) + \iota_{22}(t^m) 
+ 2\,\iota_{12}(t^m) \cos\!\big(m(\phi_2 - \phi_1)\big) \Big) \right],
\end{equation}
where 
\[
\begin{aligned}
\iota_{aa}(t^m) &= 1 - (1 - t^{2m})^3 \sum_{j=0}^{N_a - 1} t^{6jm}, \qquad a = 1,2, \\
\iota_{12}(t^m) &=  (1 - t^{2m})^3 
\sum_{j=\frac{1}{2}|N_1 - N_2|}^{\frac{1}{2}(N_1 + N_2) - 1} 
(-1)^{2jm+1}t^{6jm}.
\end{aligned}
\]


The BMN index $\mathcal{I}_{1,1; N_1=1,N_2=1}^{\mathrm{BMN}}(t)$ is
\begin{equation}
\mathcal{I}_{1,1; N_1=1,N_2=2}^{\mathrm{BMN}}(t)
= \int_0^{2\pi} \frac{d\phi_1}{2\pi} \int_0^{2\pi} \frac{d\phi_2}{2\pi} \, \mathcal{Z}_{N_1=1,N_2=2}(t, \phi_1, \phi_2),
\end{equation}
where the reduced integrand $\mathcal{Z}_{N_1=1,N_2=2}(t, \phi_1, \phi_2)$, after performing the $m$-sum, is given by
\begin{equation}
\begin{split}
\mathcal{Z}_{N_1=1,N_2=2}(t,\phi_1,\phi_2)
&= \Big[
(1+t^2)^6
\left(1+e^{-i(\phi_1-\phi_2)}t^3\right)
\left(1+e^{i(\phi_1-\phi_2)}t^3\right)
\left(-1+t^{10}\right)^3 \\
&\quad \times
\left(1+t^{14}+2t^7\cos(\phi_1-\phi_2)\right)^3
\Big]\bigg/ \Big[
\left(-1+t^6\right)^2
\left(1+t^6\right)
\left(-1+t^8\right)^3 \\
&\quad \times
\left(1+t^{10}+2t^5\cos(\phi_1-\phi_2)\right)^3
\left(1+t^{18}+2t^9\cos(\phi_1-\phi_2)\right)
\Big].
\end{split}
\end{equation}

For $SU(N)$ index we divide the $U(N)$ index by the $U(1)$ part coming from $\mathcal{I}^{\mathrm{BMN}}_{n_1=1;N_1=1}(t)$
\begin{equation}
\mathcal{I}^{\mathrm{BMN}~U(1)}(t)
=\frac{(1+t^{2})^{3}}{1-t^{6}},
\nonumber
\end{equation}
and get
\begin{equation}\label{N1211}
\begin{split}
\mathcal{I}_{N_1=1; N_2=2}^{\mathrm{BMN}~SU(N)}(t)
= & \big( (1+t^2)^6(-1+t^{10})^3 \\
&\quad \times
\big(1+t^2(1+t^2)\big(8+28t^2+94t^4+246t^6+580t^8+1222t^{10}+2372t^{12} \\
&\quad +4275t^{14}+7223t^{16}+11524t^{18}+17452t^{20}+25192t^{22}+34790t^{24} \\
&\quad +46071t^{26}+58667t^{28}+71944t^{30}+85102t^{32}+97209t^{34}+107329t^{36} \\
&\quad +114601t^{38}+118415t^{40}+118392t^{42}+114574t^{44}+107266t^{46} \\
&\quad +97134t^{48}+85011t^{50}+71845t^{52}+58564t^{54}+45982t^{56}+34701t^{58} \\
&\quad +25125t^{60}+17401t^{62}+11487t^{64}+7200t^{66}+4258t^{68}+2363t^{70} \\
&\quad +1217t^{72}+575t^{74}+245t^{76}+91t^{78}+29t^{80}+7t^{82}+t^{84}\big)\big)\big) \Big/ \\
&\big( (-1+t^6)^2(1+t^6)(-1+t^8)^3
(1+t^2+t^4+t^6+t^8)^5(1+t^6+t^{12}) \\
&\quad \times
(1+t^2+t^4+t^6+t^8+t^{10}+t^{12})^3 \big).
\end{split}
\end{equation}
The reduced integrand $\mathcal{Z}_{N_1=1,N_2=3}(t, \phi_1, \phi_2)$, after performing the $m$-sum, is given by
\begin{equation}
\begin{split}
\mathcal{Z}_{N_1=1,N_2=3}(t,\phi_1,\phi_2)
&=
\Big[
(1+t^2)^6
\left(1-e^{-i(\phi_1-\phi_2)}t^6\right)
\left(1-e^{i(\phi_1-\phi_2)}t^6\right) \\
&\quad \times
\left(1-t^{10}\right)^3
\left(1-e^{-i(\phi_1-\phi_2)}t^{10}\right)^3
\left(1-e^{i(\phi_1-\phi_2)}t^{10}\right)^3
\left(1-t^{16}\right)^3
\Big] \Big/ \\
&\quad
\Big[
\left(-1+t^6\right)^2
\left(1-t^8\right)^3
\left(1-e^{-i(\phi_1-\phi_2)}t^8\right)^3
\left(1-e^{i(\phi_1-\phi_2)}t^8\right)^3 \\
&\quad \times
\left(1+t^6+t^{12}\right)
\left(1-e^{-i(\phi_1-\phi_2)}t^{12}\right)
\left(1-e^{i(\phi_1-\phi_2)}t^{12}\right)
\left(1-t^{14}\right)^3
\Big].
\end{split}
\end{equation}
All results obtained for $N = N_1 + N_2$ are consistent with the character expansion method noted in \ref{dpbmn}.

To optimize the double-partition integral, we introduce the relative fugacity projection, exploiting the fact that the integrand depends strictly on the phase difference. Because the integrand of the BMN index depends exclusively on the relative phase difference. We exploit this by introducing a relative fugacity, $x = e^{i(\phi_1 - \phi_2)}$. This allows us to map the continuous Haar measure integrals to a purely algebraic constant term extraction acting on the Taylor-expanded plethystic exponential.

\subsection*{Relative fugacity projection approach $N=N_1+N_2$}
The exact integration of the double-partition BMN index can be vastly optimized by transitioning to a fugacity basis and employing constant-term extraction. Because the integrand depends exclusively on the phase difference $\phi_2 - \phi_1$, the overall $U(1)$ gauge phase decouples, allowing us to reduce the two-dimensional integral to a single projection. We define a relative fugacity $x = e^{i(\phi_1 - \phi_2)}$. The cosine term naturally reorganizes into pairs of reciprocal variables
\begin{equation}
    2 \cos\big(m(\phi_2 - \phi_1)\big) = x^m + x^{-m}.
\end{equation}
This allows us to define the single-letter index $f(t, x)$ as a finite polynomial
\begin{equation}
    f(t, x) = \iota_{11}(t) + \iota_{22}(t) + \iota_{12}(t)\left(x + \frac{1}{x}\right).
\end{equation}
The full multi-particle index prior to integration is the plethystic exponential of this single-letter index. By Taylor expanding with respect to $t$ to a finite order $\mathcal{O}(t^K)$ \textit{before} integrating, we circumvent the generation of transcendental functions entirely. Finally, the remaining integration over the Haar measure projects the index onto the gauge-invariant zero-mode sector. By residue theorem, this is exactly equivalent to extracting the constant term (the coefficient of $x^0$) from the expanded polynomial
\begin{equation}
\mathcal{I}_{1,1; N_1,N_2}^{\mathrm{BMN}}(t) 
= \oint \frac{dx}{2\pi i x} \exp\left[ \sum_{m=1}^\infty \frac{1}{m} f(t^m, x^m) \right] 
= \Bigg[ \exp\left[ \sum_{m=1}^\infty \frac{1}{m} f(t^m, x^m) \right] \Bigg]_{x^0}.
\end{equation}
This purely algebraic operation is exact and avoids the severe computational overhead associated with evaluating symbolic multivariable integrals.

\subsection*{$N=N_1+N_2+N_3$}
We consider
\begin{equation}
n_1 = n_2 = n_3 = 1, \qquad U_a = e^{i\phi_a} \in U(1), \quad a = 1,2,3.
\end{equation}
The BMN index is
\begin{equation}
\mathcal{I}^{\mathrm{BMN}}_{1,1,1;N_1,N_2,N_3}(t)
=
\int \prod_{a=1}^3 \frac{d\phi_a}{2\pi} \,
\exp\left[
\sum_{m=1}^\infty \frac{1}{m}
\sum_{k,l=1}^3
\iota_{kl}(t^m) \,
e^{im(\phi_l - \phi_k)}
\right].
\end{equation}
We have
\begin{equation}
\iota_{aa}(t^m)
=
1 - (1 - t^{2m})^3 \sum_{j=0}^{N_a - 1} t^{6jm}.
\end{equation}

For $a < b$,
\begin{equation}
\iota_{ab}(t^m)
=
 (1 - t^{2m})^3
\sum_{j = \frac{|N_a - N_b|}{2}}^{\frac{N_a + N_b}{2} - 1} (-1)^{2jm+1}t^{6jm}.
\end{equation}
With
\begin{equation}
\theta_{12} = \phi_2 - \phi_1, \quad
\theta_{13} = \phi_3 - \phi_1, \quad
\theta_{23} = \phi_3 - \phi_2.
\end{equation}
We get
\begin{equation}
\begin{aligned}
\mathcal{I}^{\mathrm{BMN}}_{1,1,1;N_1,N_2,N_3}(t)
&=
\int_0^{2\pi} \frac{d\phi_1}{2\pi}
\int_0^{2\pi} \frac{d\phi_2}{2\pi}
\int_0^{2\pi} \frac{d\phi_3}{2\pi} \\
&\quad \times
\exp\Bigg[
\sum_{m=1}^\infty \frac{1}{m}
\Big(
\iota_{11}(t^m) + \iota_{22}(t^m) + \iota_{33}(t^m) \\
&\qquad\qquad
+ 2\iota_{12}(t^m)\cos(m\theta_{12})
+ 2\iota_{13}(t^m)\cos(m\theta_{13})
+ 2\iota_{23}(t^m)\cos(m\theta_{23})
\Big)
\Bigg].
\end{aligned}
\end{equation}
The reduced integrand $\mathcal{Z}_{N_1=1,N_2=2,N_3=3}(t, \phi_1, \phi_2, \phi_3)$, after performing the $m$-sum, is given by
\begin{equation}
\small
\begin{split}
\mathcal{Z}_{N_1=1,N_2=2,N_3=3}(t,\phi_1,\phi_2,\phi_3)
&=\Big[
\left(e^{i\phi_2}+e^{i\phi_1}t^3\right)
\left(e^{i\phi_1}+e^{i\phi_2}t^3\right)
\left(e^{i\phi_3}+e^{i\phi_2}t^3\right) \\
&\quad \times
\left(e^{i\phi_2}+e^{i\phi_3}t^3\right)
\left(1-t^4\right)^9
\left(e^{i\phi_3}-e^{i\phi_1}t^6\right)
\left(e^{i\phi_1}-e^{i\phi_3}t^6\right) \\
&\quad \times
\left(e^{i\phi_2}+e^{i\phi_1}t^7\right)^3
\left(e^{i\phi_1}+e^{i\phi_2}t^7\right)^3
\left(e^{i\phi_3}+e^{i\phi_2}t^7\right)^3 \\
&\quad \times
\left(e^{i\phi_2}+e^{i\phi_3}t^7\right)^3
\left(1-t^{10}\right)^6
\left(e^{i\phi_3}-e^{i\phi_1}t^{10}\right)^3 \\
&\quad \times
\left(e^{i\phi_1}-e^{i\phi_3}t^{10}\right)^3
\left(e^{i\phi_3}+e^{i\phi_2}t^{13}\right)^3
\left(e^{i\phi_2}+e^{i\phi_3}t^{13}\right)^3
\left(1-t^{16}\right)^3
\Big] \Big/ \\
&\quad
\Big[
\left(1-t^2\right)^9
\left(e^{i\phi_2}+e^{i\phi_1}t^5\right)^3
\left(e^{i\phi_1}+e^{i\phi_2}t^5\right)^3
\left(e^{i\phi_3}+e^{i\phi_2}t^5\right)^3 \\
&\quad \times
\left(e^{i\phi_2}+e^{i\phi_3}t^5\right)^3
\left(1-t^6\right)
\left(1-t^8\right)^6
\left(e^{i\phi_3}-e^{i\phi_1}t^8\right)^3 \\
&\quad \times
\left(e^{i\phi_1}-e^{i\phi_3}t^8\right)^3
\left(e^{i\phi_2}+e^{i\phi_1}t^9\right)
\left(e^{i\phi_1}+e^{i\phi_2}t^9\right) \\
&\quad \times
\left(e^{i\phi_3}+e^{i\phi_2}t^{11}\right)^3
\left(e^{i\phi_2}+e^{i\phi_3}t^{11}\right)^3
\left(1-t^{12}\right) \\
&\quad \times
\left(e^{i\phi_3}-e^{i\phi_1}t^{12}\right)
\left(e^{i\phi_1}-e^{i\phi_3}t^{12}\right)
\left(1-t^{14}\right)^3 \\
&\quad \times
\left(e^{i\phi_3}+e^{i\phi_2}t^{15}\right)
\left(e^{i\phi_2}+e^{i\phi_3}t^{15}\right)
\left(1-t^{18}\right)
\Big].
\end{split}
\end{equation}

\subsection*{Relative fugacity projection approach $N=N_1+N_2+N_3$}
We can vastly improve efficiency by mapping the problem to finite polynomial algebra. We introduce complex fugacities $z_a = e^{i\phi_a}$. The integration measure transforms as
\begin{equation}
    \int_0^{2\pi} \frac{d\phi_a}{2\pi} \longrightarrow \oint_{|z_a|=1} \frac{dz_a}{2\pi i z_a}.
\end{equation}
Because the exponentiated kernels only depend on phase differences $\phi_a - \phi_b$, the system possesses an overall decoupled $U(1)$ symmetry. We gauge this away by fixing $z_3 = 1$ and defining $x = z_1$, $y = z_2$, reducing the integrals from 3D to 2D. We define the single-letter index $f(t, x, y)$. Rewriting the cosine kernels in terms of fugacities yields a simple finite polynomial function
\begin{equation}
    f(t, x, y) = \sum_{a=1}^3 \iota_{aa}(t) + \iota_{12}(t)\left(\frac{x}{y} + \frac{y}{x}\right) + \iota_{13}(t)\left(x + \frac{1}{x}\right) + \iota_{23}(t)\left(y + \frac{1}{y}\right).
\end{equation}
The full multi-particle index is
\begin{equation}
    \mathcal{I}(t, x, y) = \exp \left( \sum_{m=1}^{\infty} \frac{1}{m} f(t^m, x^m, y^m) \right).
\end{equation}
Computationally, we seek the index up to a finite order $\mathcal{O}(t^K)$. Because $f(t^m, x^m, y^m) \sim \mathcal{O}(t^{2m})$, we can safely truncate the sum at $m \le K/2$. By taking the Taylor series of the exponential with respect to $t$ \textit{before} integrating, we reduce the integrand to a finite polynomial. By residue theorem, $\oint \frac{dz}{2\pi i z} z^n = \delta_{n,0}$. Thus, the integration is exactly equivalent to extracting the constant term (the coefficient of $x^0 y^0$) from the expanded polynomial
\begin{equation}
    \mathcal{I}^{\text{BMN}}_{1,1,1;N_1,N_2,N_3}(t) = \oint \frac{dx}{2\pi i x} \oint \frac{dy}{2\pi i y} \mathcal{I}(t, x, y) = \Big[ \mathcal{I}(t, x, y) \Big]_{x^0, y^0}.
\end{equation}
Computationally, extracting polynomial coefficients is orders of magnitude faster than evaluating continuous symbolic integrals.
Finally, we isolate the $SU(N)$ index by factoring out the $U(1)$ contribution
\begin{equation}
    \mathcal{I}^{SU(N)}(t) = \frac{\mathcal{I}^{\text{BMN}}_{1,1,1;N_1,N_2,N_3}(t)}{\mathcal{I}^{U(1)}(t)}, \quad \text{where} \quad \mathcal{I}^{U(1)}(t) = \frac{(1+t^2)^3}{1-t^6}.
\end{equation}
The results for $N = N_1 + N_2+N_3$ are consistent with the character expansion analysis presented in \ref{tpbmn}.

\subsection*{$N=n_1N_1+n_2N_2$}
The fundamental bottleneck in evaluating the multi-partition BMN index directly from its matrix integral definition is the presence of continuous multidimensional integration over the gauge group Haar measures. We implemented an algorithmic strategy in Mathematica in using constant term extraction that maps the exact integrals into discrete, truncated polynomial algebra. 
\subsubsection*{Computational methodology: the constant term extraction algorithm}
For a double-partition sector with gauge group $U(n_1) \times U(n_2)$, we introduce the sets of variables $\{x_i\}_{i=1}^{n_1}$ and $\{y_i\}_{i=1}^{n_2}$. 

The matrix traces appearing in the index are then dynamically constructed as finite power sums of these variables
\begin{equation}
    \text{Tr}(U_1^m) = \sum_{i=1}^{n_1} x_i^m, \qquad \text{Tr}(U_1^{\dagger m}) = \sum_{i=1}^{n_1} x_i^{-m},
\end{equation}
with analogous expressions for the second node using the $y_i$ variables. The argument of the Plethystic exponential captures the un-exponentiated single-letter interactions between the nodes. We assemble this term, $f_m$, by multiplying the analytically derived kernels ($\iota_{ab}$) by the corresponding traces
\begin{equation}
\small
    f_m = \iota_{11}(m)\text{Tr}(U_1^{\dagger m})\text{Tr}(U_1^m) + \iota_{22}(m)\text{Tr}(U_2^{\dagger m})\text{Tr}(U_2^m) + \iota_{12}(m)\Big(\text{Tr}(U_1^{\dagger m})\text{Tr}(U_2^m) + \text{Tr}(U_2^{\dagger m})\text{Tr}(U_1^m)\Big).
\end{equation}
Because we only require the index up to a finite cutoff order $\mathcal{O}(t^K)$, we can truncate the infinite sum over $m$. Crucially, by applying a series expansion \textit{during} the exponentiation step, we force the computational kernel to aggressively drop higher-order cross-terms. This prevents the exponential from blowing up into massive intermediate expressions, keeping the integrand strictly as a finite multivariate Laurent polynomial
\begin{equation}
    \mathcal{Z}_{\text{PE}}(t, x_i, y_i) = \exp \left( \sum_{m=1}^{K} \frac{1}{m} f_m \right) + \mathcal{O}(t^{K+1}).
\end{equation}
To project the partition function onto the gauge-invariant sector, the Plethystic expansion must be integrated over the respective Haar measures. For a $U(n)$ node, the Haar measure introduces the Vandermonde determinant and a Weyl symmetry factor of $1/n!$
\begin{equation}
    \Delta_{U(n)}(z) = \frac{1}{n!} \prod_{1 \le i < j \le n} \left(1 - \frac{z_i}{z_j}\right)\left(1 - \frac{z_j}{z_i}\right).
\end{equation}
Furthermore, because the BMN index integrand depends strictly on the relative phases between different nodes, the system possesses an overall redundant $U(1)$ center-of-mass symmetry. We break this symmetry and eliminate one redundant integration by fixing the final variable of the second node to unity ($y_{n_2} = 1$). The full unintegrated polynomial is therefore
\begin{equation}
    \mathcal{P}(t, x_i, y_i) = \mathcal{Z}_{\text{PE}} \times \Delta_{U(n_1)}(x) \times \Delta_{U(n_2)}(y) \Big|_{y_{n_2} = 1}.
\end{equation}
By residue Theorem, integrating a phase over the unit circle Haar measure, $\oint \frac{dz}{2\pi i z} z^k$, is mathematically identical to extracting the constant term (where $k=0$) of the Laurent polynomial. We extract the zero-mode coefficient
\begin{equation}
    \mathcal{I}^{\text{BMN}}_{n_1,n_2;N_1,N_2}(t) = \Big[ \mathcal{P}(t, x_i, y_i) \Big]_{x_1^0 \dots x_{n_1}^0, \, y_1^0 \dots y_{n_2-1}^0}.
\end{equation}
This purely algebraic operation reduces the multidimensional continuous integration to a highly optimized coefficient search.
Finally, the resulting index corresponds to the $U(N)$ gauge group. To isolate the $SU(N)$ index, we factor out the $U(1)$ contribution
\begin{equation}
    \mathcal{I}^{SU(N)}(t) = \frac{\mathcal{I}^{\text{BMN}}_{n_1,n_2;N_1,N_2}(t)}{\mathcal{I}^{U(1)}(t)}, \quad \text{where} \quad \mathcal{I}^{U(1)}(t) = \frac{(1+t^2)^3}{1-t^6}.
\end{equation}
By matching the Taylor expansion of this division to the target order $\mathcal{O}(t^K)$, we obtain the final coefficients for the respective vacuum superselection sectors.
All results match with the results in \ref{dpbmn} using character expansion approach.

To compute the general multi-partition BMN index, first, we express the system using relative fugacities for all gauge nodes, fixing one variable to remove redundant overall phase symmetry. We then construct a finite Laurent polynomial by expanding the Plethystic exponential of the single-letter index up to the desired order in $t$, multiplied by the Vandermonde determinants from the Haar measures. Then, we simply extract the constant term (the coefficient where all fugacity powers are zero) from this polynomial. Finally, we divide by the universal $U(1)$ factor to isolate the physical $SU(N)$ index. This method transforms a difficult multidimensional integration problem into a fast, exact polynomial coefficient extraction.

\section{Analysis of the finite-$N$ BMN index data}
\label{datan}
In this section, we analyze the finite-$N$ data extracted from the BMN index in the single-, double-, triple-partition, and total sectors. Our goal is to quantify the growth of the coefficients $d_j$ at charges of order $j \sim N^2$ and to compare how this growth depends on the vacuum sector. To this end, we study the coefficient at $j=N^2$, plot $\log |d_{j=N^2}|$ as a function of $N^2$, and perform linear fits to test the $\mathcal{O}(N^2)$ scaling. We also examine $\log |d_{j=N^2}|/N^2$ as a function of $N$. This provides a quantitative summary of the finite-$N$ behavior of the BMN index and clarifies which sectors dominate the growth in the range of data available here.

\begin{figure}[H]
    \centering
    \includegraphics[width=0.8\linewidth]{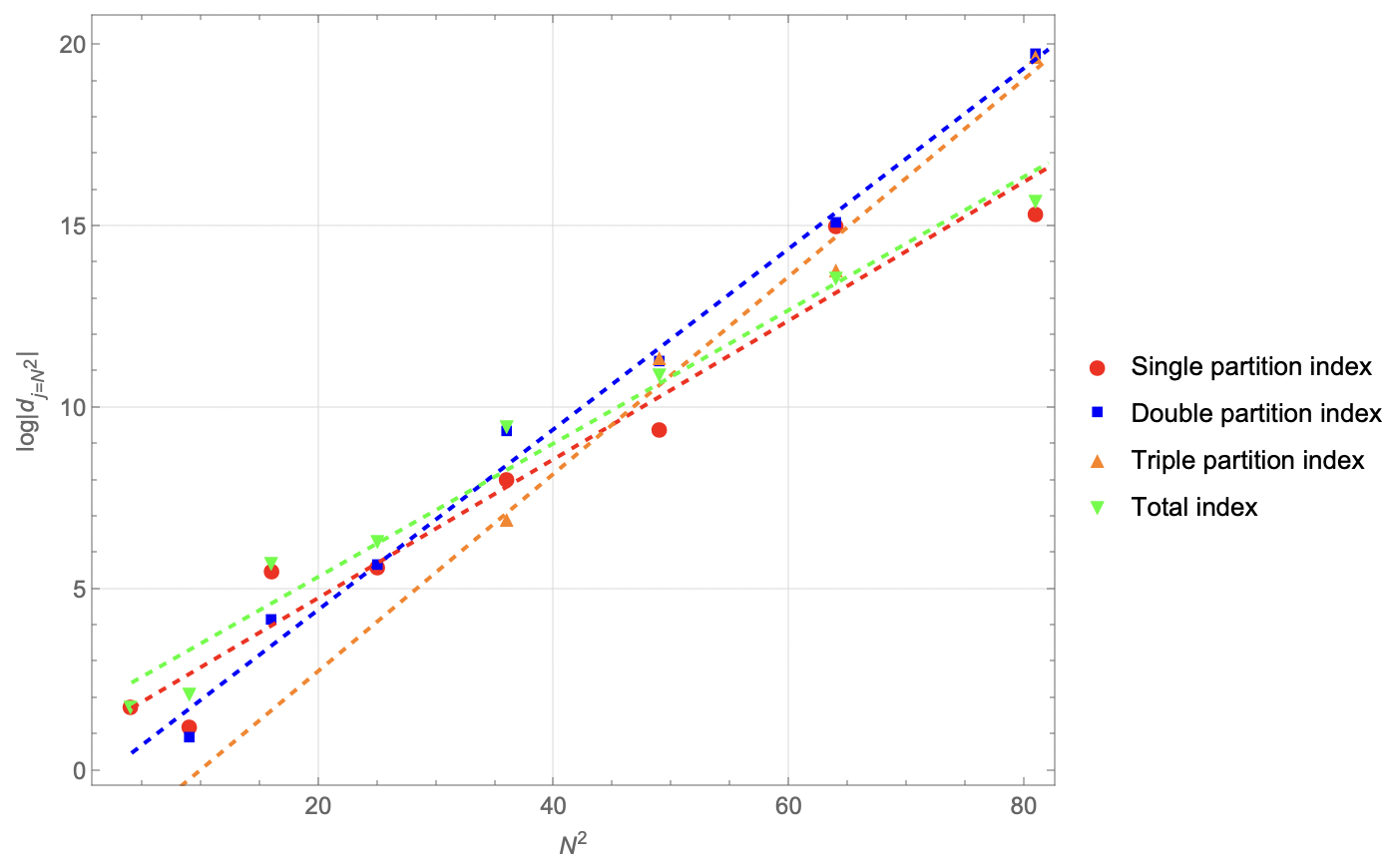}
    \caption{Linear fits of $\log |d_j|$ as a function of $N^2$ for the
    single-partition, double-partition, triple-partition, and total indices.}
    \label{fig:totalindex}
\end{figure}

Fitting $\log |d_j|$ as a linear function of $N^2$ for the single-partition,
double-partition, triple-partition, and total indices, we obtain
Fig.~\ref{fig:totalindex}. Using the data up to $N=9$, the slopes obtained
from the linear fits are summarized in Table.~\ref{tab:index_slopes}. Even a modest difference in these fitted slopes can give rise to a large difference in the magnitudes of the index coefficients, since $\log|d_j|\simeq cN^2$ implies $|d_j|\sim e^{cN^2}$. 
\begin{table}[H]
    \centering
    \begin{tabular}{|l|c|}
        \hline
        Index & Slope \\
        \hline
        Single partition & $0.19102$ \\
        Double partition & $0.24864$ \\
        Triple partition & $0.27172$ \\
        Total index & $0.18368$ \\
        \hline
    \end{tabular}
    \caption{Slopes of the linear fits of $\log |d_j|$ versus $N^2$.}
    \label{tab:index_slopes}
\end{table}

\begin{figure}[H]
    \centering
    \includegraphics[width=0.8\linewidth]{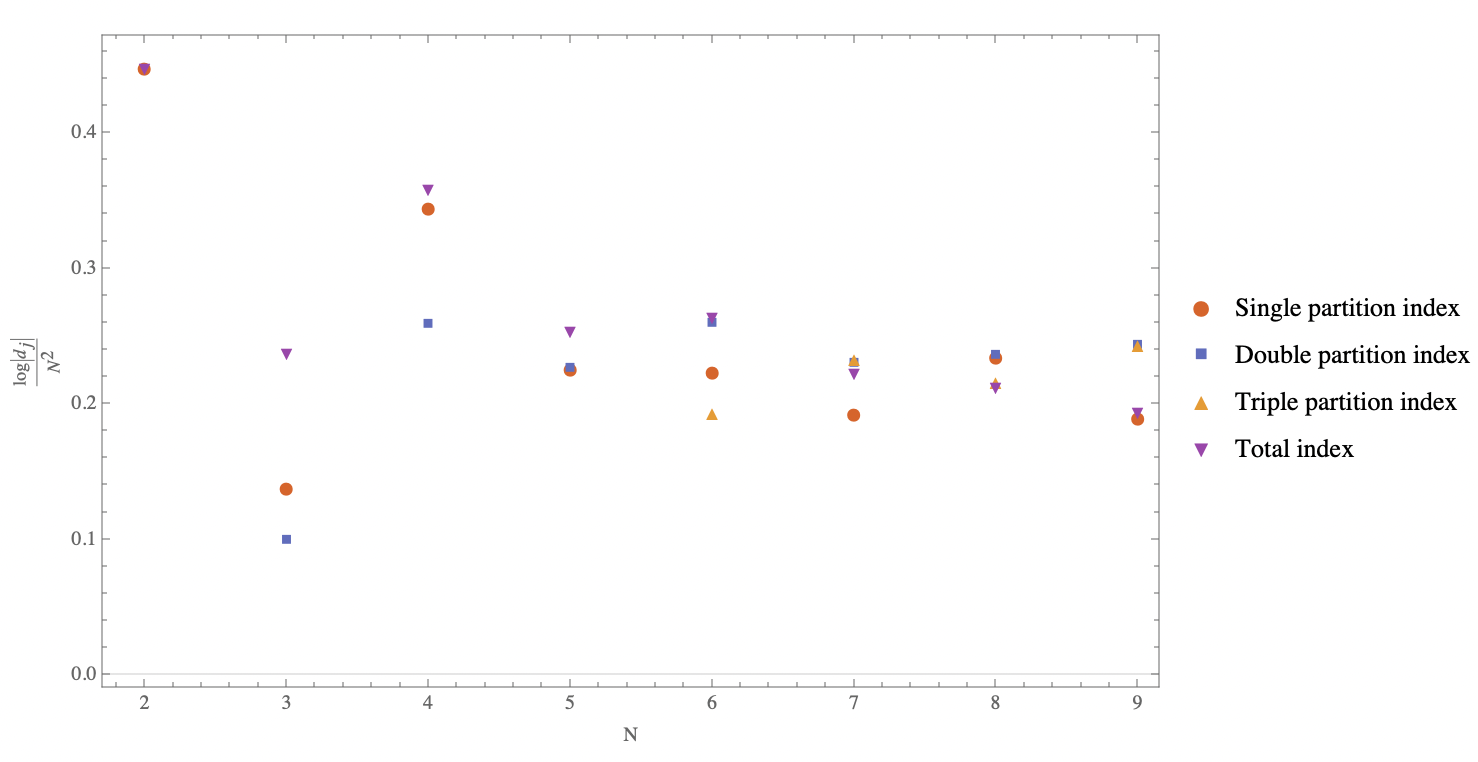}
    \caption{$\frac{\log |d_j|}{N^2}$ as a function of $N$ for the
    single-partition, double-partition, triple-partition, and total indices.}
    \label{fig:conv_totalindex}
\end{figure}

From the plot of $\frac{\log |d_j|}{N^2}$ versus $N$ in
Fig.~\ref{fig:conv_totalindex}, we list the values at $N=9$ below. 
\begin{table}[H]
    \centering
    \begin{tabular}{|l|c|}
        \hline
        Index & Value at $N=9$ of $\displaystyle\frac{\log |d_j|}{N^2}$ \\
        \hline
        Single partition & $0.189632$ \\
        Double partition & $0.243616$ \\
        Triple partition & $0.243454$ \\
        Total index & $0.194211$ \\
        \hline
    \end{tabular}
    \caption{Finite-$N$ values of
    $\frac{\log |d_j|}{N^2}$ at $N=9$.}
    \label{tab:conv_values}
\end{table}

The comparison between the sector sums and the total index reveals substantial
cancellations. For odd $N$, we use the average of the logarithms at
$j=N^2-1$ and $j=N^2+1$, as in the tables below. The normalized total-index
values for $N=6,7,8,9$ are approximately $0.2644$, $0.2231$, $0.2127$, and
$0.1942$, respectively. Thus the effective $N^2$ coefficient decreases over
the largest ranks available. At $N=9$, it is appreciably smaller than the
double- and triple-partition values, approximately $0.2436$ and $0.2435$.

This suppression can be seen directly from the signed coefficients. At
$N=9$, the contributions from the single-, double-, and triple-partition
classes, in that order, give
\begin{align}
 d^{\mathrm{tot}}_{80}
 &=-4\,283\,184+345\,173\,460-334\,491\,153=6\,399\,123,\\
 d^{\mathrm{tot}}_{82}
 &=5\,127\,519-399\,702\,471+401\,782\,191=7\,207\,239.
\end{align}
The large double- and triple-partition contributions have opposite signs
and nearly cancel at both charge levels. Their individual magnitudes
therefore do not determine the growth of the total index. In particular,
the positive slopes in Table~\ref{tab:index_slopes} summarize a finite range
and should not be interpreted as nonzero asymptotic $N^2$ coefficients.

To quantify the cancellation, let $d^{[K]}_{N,j}$ denote the signed sum of
coefficients over all vacuum sectors with $K$ distinct block dimensions,
so that $d^{\mathrm{tot}}_{N,j}=\sum_K d^{[K]}_{N,j}$. We define
\begin{equation}
 R_{N,j}=\frac{|d^{\mathrm{tot}}_{N,j}|}
 {\sum_K |d^{[K]}_{N,j}|},\qquad 0\leq R_{N,j}\leq1.
 \label{eq:class-cancellation-ratio}
\end{equation}
A smaller ratio indicates stronger cancellation between these partition
classes. The denominator already includes cancellations within each class. For $N\leq9$, the single-,
double-, and triple-partition classes exhaust the vacuum sectors. Using
the same near-$N^2$ charge prescription as above, we set
\begin{equation}
 \overline R_N=\begin{cases}
 R_{N,N^2}, & N\text{ even},\\[2pt]
 \sqrt{R_{N,N^2-1}R_{N,N^2+1}}, & N\text{ odd},
 \end{cases}
 \qquad \Delta_N=-\frac{\log\overline R_N}{N^2}.
 \label{eq:cancellation-measures}
\end{equation}
Here $\Delta_N$ measures the reduction in the normalized logarithm caused
by summing the signed class contributions, relative to summing their
absolute values; for odd $N$, the logarithms are averaged over the two
neighboring charges. Table~\ref{tab:cancellation-measures} quantifies the cancellation between different partition classes at charges near $N^2$. 
\begin{table}[H]
 \centering
 \begin{tabular}{|c|c|c|}
  \hline
  $N$ & $\overline R_N$ & $\Delta_N$ \\
  \hline
  $6$ & $0.866463$ & $0.003982$ \\
  $7$ & $0.294927$ & $0.024919$ \\
  $8$ & $0.101401$ & $0.035761$ \\
  $9$ & $0.009143$ & $0.057960$ \\
  \hline
 \end{tabular}
 \caption{Cancellation between partition classes at charges near $N^2$.
 Smaller $\overline R_N$ and larger $\Delta_N$ indicate stronger cancellation.}
 \label{tab:cancellation-measures}
\end{table}
The decrease of $\overline R_N$ and increase of $\Delta_N$ show that the
cancellation becomes stronger as $N$ increases from $6$ to $9$. At $N=9$,
$R_{9,80}\simeq0.00936$ and $R_{9,82}\simeq0.00894$: less than one percent
of the sum of the class magnitudes survives at either charge. This growing
suppression accompanies the decrease of the effective $N^2$ coefficient
of the total index noted above. The trend is a statement about the sampled
charges and ranks, and need not be monotonic at every charge or smaller $N$. The absence of $\mathcal{O}(N^2)$ entropy growth in the total index at fixed
$j/N^2$ is emphasized in
\cite{Komatsu:2026holography}. Their analysis of the proposed grand-canonical
expression for the fully summed BMN index instead gives $\mathcal{O}(N)$
entropy in this regime. 

We now evaluate the sector-wise dominant contributions, namely the single-partition index with the largest degeneracy, the double-partition index with the largest degeneracy, and the triple-partition index with the largest degeneracy.

For $N=8$ and $N=10$, the single partition with the largest degeneracy is not the trivial vacuum ($n_1=N, N_1=1$), unlike all other cases in the Table \ref{tab:singlepartition_dominance}.
\begin{table}[H]
\centering
\renewcommand{\arraystretch}{1} 
\begin{tabular}{|c|c|c|}
\hline
$N$ & Largest degeneracy& 
$\begin{cases}
\log |d_{j=N^2}| & \text{for even } N \\
\frac{1}{2}\Big(\log |d_{N^2-1}| + \log |d_{N^2+1}|\Big) & \text{for odd } N
\end{cases}$ \\
\hline
$2$ &  $\mathcal{I}_{n_1=2; N_1=1}^{\mathrm{BMN}}(t)$ & $1.79176$  \\
$3$ & $\mathcal{I}_{n_1=3; N_1=1}^{\mathrm{BMN}}(t)$ & $1.64792$ \\
$4$ & $\mathcal{I}_{n_1=4; N_1=1}^{\mathrm{BMN}}(t)$ & $4.90527$ \\
$5$ & $\mathcal{I}_{n_1=5; N_1=1}^{\mathrm{BMN}}(t)$ & $5.65369$ \\
$6$ &  $\mathcal{I}_{n_1=6; N_1=1}^{\mathrm{BMN}}(t)$ & $8.06558$  \\
$7$ & $\mathcal{I}_{n_1=7; N_1=1}^{\mathrm{BMN}}(t)$ & $9.42936$ \\
$8$ & $\mathcal{I}_{n_1=4; N_1=2}^{\mathrm{BMN}}(t)$  & $14.7956$ \\
$9$ & $\mathcal{I}_{n_1=9; N_1=1}^{\mathrm{BMN}}(t)$  & $16.4541$ \\ 
$10$ & $\mathcal{I}_{n_1=5; N_1=2}^{\mathrm{BMN}}(t)$  & $23.412$ \\
\hline
\end{tabular}
\caption{Single partition index with largest degeneracy.}
\label{tab:singlepartition_dominance}
\end{table}
Instead, the dominant entropy sectors are $\mathcal{I}_{n_1=4; N_1=2}^{\mathrm{BMN}}(t)$ for $N=8$ and $\mathcal{I}_{n_1=5; N_1=2}^{\mathrm{BMN}}(t)$ for $N=10$.

We list the dominant double-partition sectors for $N=3$ to $9$ in
Table~\ref{tab:doublepartition_dominance} and the dominant
triple-partition sectors for $N=6$ to $9$ in
Table~\ref{tab:triplepartition_dominance}.

\begin{table}[H]
\centering
\renewcommand{\arraystretch}{1}
\begin{tabular}{|c|c|c|}
\hline
$N$ & Largest degeneracy &
$\begin{cases}
\log |d_{j=N^2}|, & \text{for even } N, \\[2pt]
\frac{1}{2}\Bigl(
\log |d_{N^2-1}|+\log |d_{N^2+1}|
\Bigr), & \text{for odd } N
\end{cases}$
\\
\hline
3 &
$\mathcal{I}_{\mathrm{BMN}}^{
n_1=1,n_2=1;\,N_1=1,N_2=2}(t)$
& $0.89588$ \\

4 &
$\mathcal{I}_{\mathrm{BMN}}^{
n_1=2,n_2=1;\,N_1=1,N_2=2}(t)$
& $3.98898$ \\

5 &
$\mathcal{I}_{\mathrm{BMN}}^{
n_1=1,n_2=1;\,N_1=2,N_2=3}(t)$
& $4.48113$ \\

6 &
$\mathcal{I}_{\mathrm{BMN}}^{
n_1=2,n_2=2;\,N_1=1,N_2=2}(t)$
& $8.99082$ \\

7 &
$\mathcal{I}_{\mathrm{BMN}}^{
n_1=1,n_2=3;\,N_1=1,N_2=2}(t)$
& $11.62910$ \\

8 &
$\mathcal{I}_{\mathrm{BMN}}^{
n_1=2,n_2=3;\,N_1=1,N_2=2}(t)$
& $15.03520$ \\

9 &
$\mathcal{I}_{\mathrm{BMN}}^{
n_1=1,n_2=4;\,N_1=1,N_2=2}(t)$
& $19.97990$ \\
\hline
\end{tabular}
\caption{Double-partition indices with the largest individual-sector
degeneracy.}
\label{tab:doublepartition_dominance}
\end{table}

\begin{table}[H]
\centering
\renewcommand{\arraystretch}{1}
\begin{tabular}{|c|c|c|}
\hline
$N$ & Largest degeneracy &
$\begin{cases}
\log |d_{j=N^2}|, & \text{for even } N, \\[2pt]
\frac{1}{2}\Bigl(
\log |d_{N^2-1}|+\log |d_{N^2+1}|
\Bigr), & \text{for odd } N
\end{cases}$
\\
\hline
6 &
$\mathcal{I}_{\mathrm{BMN}}^{
n_1=1,n_2=1,n_3=1;\,
N_1=1,N_2=2,N_3=3}(t)$
& $6.95655$ \\

7 &
$\mathcal{I}_{\mathrm{BMN}}^{
n_1=2,n_2=1,n_3=1;\,
N_1=1,N_2=2,N_3=3}(t)$
& $11.32320$ \\

8 &
$\mathcal{I}_{\mathrm{BMN}}^{
n_1=3,n_2=1,n_3=1;\,
N_1=1,N_2=2,N_3=3}(t)$
& $15.45020$ \\

9 &
$\mathcal{I}_{\mathrm{BMN}}^{
n_1=4,n_2=1,n_3=1;\,
N_1=1,N_2=2,N_3=3}(t)$
& $19.81480$ \\
\hline
\end{tabular}
\caption{Triple-partition indices with the largest individual-sector
degeneracy.}
\label{tab:triplepartition_dominance}
\end{table}
The \textit{intra-sector notion of dominance} is summarized in Table~\ref{tab:singlepartition_dominance}, Table~\ref{tab:doublepartition_dominance}, and Table~\ref{tab:triplepartition_dominance}, which identify, within each partition sector, the individual sector with the largest value of $\log |d_j|$. Comparing the dominant representatives across all sectors, without performing any sector sum, reveals a non-monotonic sequence: the leading contribution comes from a single-partition sector for $N=2,3,4,5$, a double-partition sector for $N=6,7$, and a triple-partition sector at $N=8$, before returning to a double-partition sector at $N=9$. In the analysis of the BMN index at finite $N$, a striking feature emerges as $N$ increases: the dominant configuration in entropy transitions from a \emph{single-partition} sector to a \emph{double-partition} sector. The transition occurs at $N = 6$. 
The Table \ref{tab:BMN_dominance_switch} illustrate the dominance switching.
\begin{table}[H]
\centering
\begin{tabular}{|c|c|c|c|}
\hline
$N$ & Largest degeneracy & $\#$ of partitions &
$\begin{cases}
\log |d_{j=N^2}| & \text{for even } N \\[2pt]
\frac{1}{2}\Bigl(\log |d_{N^2-1}|+\log |d_{N^2+1}|\Bigr)
& \text{for odd } N
\end{cases}$ \\
\hline
$2$ &
$\mathcal{I}_{n_1=2;\,N_1=1}^{\mathrm{BMN}}(t)$
& $1$ & $1.79176$ \\

$3$ &
$\mathcal{I}_{n_1=3;\,N_1=1}^{\mathrm{BMN}}(t)$
& $1$ & $1.64792$ \\

$4$ &
$\mathcal{I}_{n_1=4;\,N_1=1}^{\mathrm{BMN}}(t)$
& $1$ & $4.90527$ \\

$5$ &
$\mathcal{I}_{n_1=5;\,N_1=1}^{\mathrm{BMN}}(t)$
& $1$ & $5.65369$ \\

$6$ &
$\mathcal{I}_{n_1=2,n_2=2;\,N_1=1,N_2=2}^{\mathrm{BMN}}(t)$
& $2$ & $8.99082$ \\

$7$ &
$\mathcal{I}_{n_1=1,n_2=3;\,N_1=1,N_2=2}^{\mathrm{BMN}}(t)$
& $2$ & $11.62910$ \\

$8$ &
$\mathcal{I}_{\mathrm{BMN}}^{
n_1=3,n_2=1,n_3=1;\,
N_1=1,N_2=2,N_3=3}(t)$
& $3$ & $15.45020$ \\

$9$ &
$\mathcal{I}_{n_1=1,n_2=4;\,N_1=1,N_2=2}^{\mathrm{BMN}}(t)$
& $2$ & $19.97990$ \\
\hline
\end{tabular}
\caption{Largest-degeneracy sectors in the BMN index for different
values of $N$. The switch from single- to double-partition dominance
occurs at $N=6$.}
\label{tab:BMN_dominance_switch}
\end{table}
We can now compare the total entropy across different partition sectors by first adding up all indices belonging to each sector (single, double, triple) and then identifying which sector produces the largest combined result. For small $N$ ($N=2,3,4, 5$), the dominant contribution in entropy at the sector‑summed level comes from the single‑partition sector. Starting at $N=6$, the double‑partition sector takes over and becomes dominant, as shown in Table \ref{tab:bmn_degeneracy}. Moreover, the data suggest that this trend continues as $N$ increases. In particular, by $N=9$ the double-partition entropy, $19.73289$, and the triple-partition entropy, $19.71977$, are already extremely close, with a difference of only $0.01312$. Sectors with more partitions---triple, quadruple, and so on---may become entropically dominant at larger $N$. However, the observed sequence is non-monotonic, so systematic growth in the number of partitions of the dominant sector is not yet established and requires investigation at larger $N$. The first switch from single- to double-partition dominance occurs at $N=6$ at the level of \textit{individual indices}. At the \textit{sector-summed level}, Table~\ref{tab:BMN_dominance_switch} shows that the initial switch occurs already at $N=5$, followed by a non-monotonic sequence: the summed double-partition contribution dominates for $N=5,6$, the triple-partition contribution at $N=7$, and the double-partition contribution again for $N=8,9$. The two levels therefore exhibit different dominance patterns: at $N=7$, an individual double-partition sector dominates while the summed triple-partition contribution is larger; at $N=8$, the reverse occurs. Thus, within the available data, double-partition dominance holds at both levels only for $N=6,9$.
\begin{table}[H]
\centering
\begin{tabular}{|c|c|c|c|c|}
\hline
$N$ & Single $\log |d_j|$ & Double $\log |d_j|$ &
Triple $\log |d_j|$ & Total $\log |d_j|$ \\
\hline
2 & 1.79176 & -- & -- & 1.79176 \\
\hline
3 & 1.24250 & 0.89588 & -- & 2.13833 \\
\hline
4 & 5.51750 & 4.14313 & -- & 5.74300 \\
\hline
5 & 5.63810 & 5.66653 & -- & 6.35199 \\
\hline
6 & 8.05706 & 9.35184 & 6.95655 & 9.51974 \\
\hline
7 & 9.42908 & 11.27553 & 11.41545 & 10.93107 \\
\hline
8 & 15.04570 & 15.10386 & 13.83871 & 13.61526 \\
\hline
9 & 15.36020 & 19.73289 & 19.71977 & 15.73113 \\
\hline
\end{tabular}
\caption{Multi-partition index degeneracy and total-index degeneracy.}
\label{tab:bmn_degeneracy}
\end{table}


\section{D2 dressed black-hole interpretation}
\label{sec:dressed-bh}

In this section we ask what bulk interpretation could explain the
$\mathcal{O}(N^2)$ entropy growth found in the BMN index.  The finite-$N$ data
suggest that dominant contributions can come from nontrivial vacuum sectors.
It is therefore natural to consider bulk states that retain black-hole degrees
of freedom as well as large fuzzy-sphere, or giant-graviton, degrees of
freedom.  We first fix the BMN scaling variables needed to compare these brane
scales with the relevant supergravity scales.

We use the dimensionful BMN 't Hooft coupling and the dimensionless effective
coupling at the BMN scale,
\begin{equation}
\label{eq:bmn_parameters}
\lambda=\frac{N R^3}{4\pi^2\ell_P^6},
\qquad
g_{\rm eff}^2(\mu)\equiv \frac{\lambda}{\mu^3}.
\end{equation}
Here $R$ is the DLCQ radius appearing in \eqref{coupling}.  The ratio
$g_{\rm eff}^2(\mu)$ is the parameter denoted by $g$ in
\cite{Costa:2014wya}; it should not be confused with the coupling $g$ in
\eqref{coupling}, since $g_{\rm eff}^2(\mu)=N g^2/(4\pi^2)$.

The D0 radial coordinate has dimensions of energy.  The radial scale
associated with a boundary quantum-mechanical energy scale follows from the
standard D0-brane UV/IR relation
\begin{equation}
\label{eq:d0_uv_ir}
U\sim \lambda^{1/5}\omega^{2/5}.
\end{equation}
Here $U$ is the D0 near-horizon radial coordinate in energy units, and $\omega$
is the boundary quantum-mechanical energy scale.\footnote{One direct derivation
uses the near-horizon D0 metric, written as
$ds^2=\alpha'G_{\mu\nu}dx^\mu dx^\nu$, with
$G_{tt}\sim -U^{7/2}/\lambda^{1/2}$ and
$G_{UU}\sim \lambda^{1/2}U^{-7/2}$.  A fluctuation localized at radial scale
$U$, with $\Delta U\sim U$, has proper radial size
$\Delta s\sim \sqrt{\alpha'G_{UU}}\,\Delta U
\sim \sqrt{\alpha'}\,\lambda^{1/4}U^{-3/4}$.  The corresponding local energy is
$E_{\rm loc}\sim 1/\Delta s
\sim U^{3/4}/(\sqrt{\alpha'}\,\lambda^{1/4})$.  Redshifting by
$\omega\sim \sqrt{-G_{tt}}\,E_{\rm loc}$ gives
$\omega\sim U^{5/2}/\lambda^{1/2}$, equivalently \eqref{eq:d0_uv_ir}.  For a
thermal D0 black hole, applying the same redshift relation at $U=U_0$ gives
$T\sim U_0^{5/2}/\lambda^{1/2}$, in agreement with the usual Euclidean
regularity result.}  More generally, \eqref{eq:d0_uv_ir} is the D0
scaling relation used in the Itzhaki--Maldacena--Sonnenschein--Yankielowicz
analysis \cite{Itzhaki:1998dd}.\footnote{In the notation of
\cite{Itzhaki:1998dd}, the translation to the variables used here is
$\lambda=N g_{YM}^2$ and $g_{\rm eff}^2(U)=\lambda/U^3$.  At the radial scale
corresponding to the BMN scale $\mu$ through \eqref{eq:d0_uv_ir}, this becomes
$g_{\rm eff}^2(U)\sim [g_{\rm eff}^2(\mu)]^{2/5}$.}
For the BMN branch considered here, we identify the boundary scale with the BMN
mass parameter, $\omega\sim\mu$, assuming no additional parametrically large or
small charge ratios.  The same D0 scaling then translates the validity regions
of \cite{Itzhaki:1998dd} into the BMN
windows shown in Table~\ref{tab:bmn-validity-windows}.
\begin{table}[H]
\centering
\small
\renewcommand{\arraystretch}{1.3}
\begin{tabular}{@{}>{\raggedright\arraybackslash}p{0.30\textwidth}
>{\raggedright\arraybackslash}p{0.35\textwidth}
>{\raggedright\arraybackslash}p{0.25\textwidth}@{}}
\hline
Region & $\mu$ range & $g_{\rm eff}^2(\mu)$ range \\
\hline
(a) perturbative matrix QM &
$\lambda^{1/3}\ll \mu$ &
$g_{\rm eff}^2(\mu)\ll 1$ \\
(b) type-IIA D0 supergravity &
$\lambda^{1/3}N^{-10/21}\ll \mu\ll \lambda^{1/3}$ &
$1\ll g_{\rm eff}^2(\mu)\ll N^{10/7}$ \\
(c) 11d boosted black string &
$\lambda^{1/3}N^{-5/9}\ll \mu\ll
\lambda^{1/3}N^{-10/21}$ &
$N^{10/7}\ll g_{\rm eff}^2(\mu)\ll N^{5/3}$ \\
(d1) semiclassical localized 11d black hole &
$\lambda^{1/3}N^{-25/21}\ll \mu\ll
\lambda^{1/3}N^{-5/9}$ &
$N^{5/3}\ll g_{\rm eff}^2(\mu)\ll N^{25/7}$ \\
(d2) 11d Planckian / graviton regime &
$\mu\ll \lambda^{1/3}N^{-25/21}$ &
$N^{25/7}\ll g_{\rm eff}^2(\mu)$ \\
\hline
\end{tabular}
\caption{BMN validity windows obtained from $\omega\sim\mu$ and
$g_{\rm eff}^2(\mu)=\lambda/\mu^3$.}
\label{tab:bmn-validity-windows}
\end{table}
The boosted-black-string label in Table~\ref{tab:bmn-validity-windows} refers
to the near-extremal black-object continuation of the D0 solution.\footnote{In
the strict extremal limit, the same D0 charge uplifts to the eleven-dimensional
M-wave carrying KK momentum along the M-theory circle.  Equivalently, the
uplifted near-extremal metric is a uniform boosted black string; taking
$r_0\to0$, the boost $\alpha\to\infty$, and
$r_0^7\sinh^2\alpha$ fixed gives the M-wave metric.}  Regions (d1) and (d2)
refine the IMSY matrix-black-hole region: the localized eleven-dimensional
black hole is semiclassical only for
$N^{5/3}\ll g_{\rm eff}^2(\mu)\ll N^{25/7}$, while for
$N^{25/7}\ll g_{\rm eff}^2(\mu)$ its Schwarzschild radius is Planckian and the gravitational
description gives way to graviton-like physics.  The comparison below uses the
controlled type-IIA window of region (b).

We next isolate the sector-dependent brane scale.  At large $N$, the dominant
sector of the protected index need not be the trivial vacuum sector.  We write
a dominant vacuum sector as $N=\sum_i n_i N_i$ and define $N_{\max}$ to be the
largest irreducible block size within that sector,
\begin{equation}
\begin{split}
N_{\max}=\max_i N_i .
\end{split}
\end{equation}
The index then poses a simple large-$N$ question: whether this $N_{\max}$ in
the dominant sector remains order one or instead grows without bound as
$N\to\infty$.

The largest fuzzy-sphere component is controlled by $N_{\max}$.  These fuzzy
spheres are the plane-wave giant gravitons, equivalently spherical M2-branes,
of \cite{Berenstein:2002jq}.  In the type-IIA description, the same fuzzy
sphere is a spherical D2-brane carrying D0-brane charge.  We compare this brane
scale with the inferred black-hole scale using the same D0 radial energy
coordinate $U$.  In this coordinate, the largest fuzzy-sphere component has the
scale\footnote{In the BMN vacuum,
$X^i=(\mu\ell_P^3/3R)J^i=(\mu\alpha'/3)J^i$, using
$\ell_P^3/R=\alpha'$.  Since the D0 radial energy coordinate is defined by
$r=2\pi\alpha' U$, the fuzzy-sphere coordinate is
$U_{\rm D2}^i=X^i/(2\pi\alpha')
=(\mu/6\pi)J^i$.  For an irreducible block of dimension $m$,
$\sum_i(J^i)^2=(m^2-1){\bf 1}_m/4$, so
$U_{\rm D2}(m)=\sqrt{m^{-1}\Tr\sum_i(U_{\rm D2}^i)^2}
=\mu\sqrt{m^2-1}/(12\pi)\sim \mu m$.  Setting $m=N_{\max}$ gives
\eqref{eq:UD2_scale}.}
\begin{equation}
\label{eq:UD2_scale}
U_{\rm D2}\sim \mu N_{\max}.
\end{equation}

In region (b), the controlled type-IIA description gives the relevant
black-hole scale.  Setting $\omega\sim\mu$ in
\eqref{eq:d0_uv_ir} gives the radial horizon scale
\begin{equation}
\label{eq:Uh_scale}
U_h\sim \lambda^{1/5}\mu^{2/5}
=\mu\bigl[g_{\rm eff}^2(\mu)\bigr]^{1/5}.
\end{equation}
The scale in \eqref{eq:Uh_scale} follows from the D0 redshift/UV/IR relation
of \cite{Itzhaki:1998dd}; the thermal BMN black holes of \cite{Costa:2014wya}
provide useful background but are not an input to this scaling estimate.

Because $U_{\rm D2}$ and $U_h$ are written in the same radial coordinate, the
relevant dimensionless comparison parameter is
\begin{equation}
\label{eq:chi_max_definition}
\begin{split}
\chi_{\max}\equiv \frac{U_{\rm D2}}{U_h}
\sim
N_{\max}\bigl[g_{\rm eff}^2(\mu)\bigr]^{-1/5}.
\end{split}
\end{equation}
If $\chi_{\max}<1$, the D2 component lies inside the type-IIA black-hole
horizon and is hidden behind it.  Conversely, the D2 component lies at or
outside the type-IIA black-hole horizon when $\chi_{\max}\gtrsim 1$, or
equivalently when
\begin{equation}
\label{eq:d2_outside_iia}
g_{\rm eff}^2(\mu)\lesssim N_{\max}^5 .
\end{equation}
Since $1\ll g_{\rm eff}^2(\mu)\ll N^{10/7}$ in the type-IIA window, a
sufficient condition for \eqref{eq:d2_outside_iia} to hold throughout that
window is
\[
N_{\max}\gtrsim N^{2/7}.
\]

When the dominant BMN index sector satisfies \eqref{eq:d2_outside_iia} for the
relevant value of $g_{\rm eff}^2(\mu)$, the
bulk description should not be a pure D0 black hole.  The
protected states should also retain macroscopic type-IIA spherical D2-brane
degrees of freedom outside the horizon, giving a D2 dressed black-hole picture:
a D0 black-hole sector accompanied by large D2-brane degrees of freedom.  This
picture suggests a possible reason why, to our knowledge, no pure BPS black
object in eleven-dimensional supergravity asymptotic to the plane wave is
currently known, whether localized as a black hole or extended along the
M-theory circle as a black string.  It should, however, be read heuristically:
the evidence is finite-$N$, and the radial scale $U_h$ is inferred from
strong-coupling BMN scaling rather than from an explicit BPS solution.

This picture is structurally similar to the recently proposed dual dressed
black holes (DDBH) in $AdS_5\times S^5$, where a core
black hole is dressed by large dual giant gravitons, namely D3-branes wrapping
an $S^3$ in the $AdS_5$ factor
\cite{Choi:2024ddbh,Choi:2025sgg}; related brane-fused black-hole cohomologies
were constructed in \cite{deMelloKoch:2024pcs}.

\section{Conclusion}
\label{sec:conclusion}

We computed the finite-$N$ BMN Witten index as a fugacity expansion, including
the contributions from all partition-labeled supersymmetric vacua.  In the
trivial vacuum sector, the coefficients at charges
$j\sim\mathcal{O}(N^2)$ already exhibit $\mathcal{O}(N^2)$ entropy growth
\cite{Chang:2024lkw}, suggesting protected plane-wave black-hole degrees of
freedom.  The main question addressed here was whether the full index, after
summing over all vacuum sectors, is reduced by cancellations.  Our
explicit computations up to $N=9$ show cancellations between sectors that
reduce the effective $N^2$ coefficient of the total index entropy. The absence
of $\mathcal{O}(N^2)$ entropy growth in the total index at fixed $j/N^2$ is
emphasized in \cite{Komatsu:2026holography}; our finite-$N$ data do not by
themselves establish the large-$N$ scaling.

The finite-$N$ data also reveal a nontrivial sectoral organization.  At order
$t^{N^2}$, the dominant sector is not fixed once and for all:
different partition sectors dominate over different ranges of $N$, and the
dominant sector jumps as the rank increases.  We call this phenomenon
\emph{dominance switching}.  It indicates that the finite-$N$ index is not
governed smoothly by a single vacuum sector.

One possible implication of this sector dependence is the D2 dressed black-hole
interpretation discussed in section~\ref{sec:dressed-bh}.  In the controlled
type-IIA window, the comparison of
$U_{\rm D2}\sim\mu N_{\max}$ with
$U_h\sim\mu\bigl[g_{\rm eff}^2(\mu)\bigr]^{1/5}$ shows that the spherical
D2 component lies at or outside the D0 horizon when
$g_{\rm eff}^2(\mu)\lesssim N_{\max}^5$.
This comparison motivates interpreting the relevant protected sector as
retaining macroscopic D2 degrees of freedom, rather than as being described by
a pure D0 black hole.  This makes it important to know whether $N_{\max}$ in the
dominant sector grows without bound.  A natural next step is to push the
multi-partition computation beyond
$N=9$ using the scalable tools employed here: the symmetric-group character
expansion, the residue-over-trees formula, and constant-term extraction.  A
larger dataset would test whether dominance switching settles into an
asymptotic pattern.

\paragraph{Further directions.}
Two complementary directions are worth emphasizing.  First, a related but
logically distinct problem is to isolate the genuinely finite-$N$ part of the
BMN protected spectrum.  In finite-$N$ $\mathcal{N}=4$ SYM, BPS operators can be
separated into monotone operators, which remain BPS for all $N$, and fortuitous
operators, which become BPS only at special finite values of $N$
\cite{Chang:2024zqi}.  Fortuity has been explored in several related systems
\cite{Chang:2022mjp,Choi:2022caq,Choi:2023znd,Chang:2023zqk,Budzik:2023vtr,
Choi:2023vdm,Chang:2024zqi,Chang:2024lxt,Chang:2025rqy,deMelloKoch:2025ngs,
deMelloKoch:2025cec,Chen:2025sum,Kim:2025vup,Chang:2025wgo,Belin:2025hsg,
Choi:2025pqr,Behan:2025hbx,Chang:2026scr}.  In the BMN model, one would like to define an
analogous black-hole index, at least at the level of protected indices, by
subtracting the monotone, graviton-like contribution from the full finite-$N$
protected index.  The sector-summed Witten index computed here is a necessary
input for such a subtraction, but the subtraction itself would require an
independent characterization of the BMN monotone contribution.  Related charge
constraints in the $U(N)$-symmetric BMN BPS sector were recently discussed in
\cite{Zigdon:2025bmn}, suggesting another possible handle on protected BMN
black-hole sectors.  A successful monotone subtraction would give a sharper way
to locate BMN fortuitous sectors before attempting the explicit construction of
the corresponding operators.  Second, it would be useful to derive the BMN
index directly by supersymmetric localization.  A
natural starting point is the four-dimensional construction of the
$\mathcal{N}=4$ SYM index on $S^1\times S^3$, where the index is realized as a
Scherk--Schwarz-deformed path integral and computed by localization
\cite{Nawata:2011un}.  Since BMN matrix quantum mechanics arises from
$\mathcal{N}=4$ SYM on $\mathbb{R}\times S^3$ by the consistent truncation to
the lowest Kaluza--Klein modes on $S^3$ \cite[section 2]{Kim:2003rza}, reducing
this localization construction on $S^3$ may lead to a localization derivation
of the BMN index itself.

\subsection*{Acknowledgements}
We thank Yoav Zigdon for helpful comments on the manuscript.
CC is supported by NSFC Grant No.~12575075. SD is supported by the Shuimu Tsinghua Scholar Program of Tsinghua University and the Beijing Natural Science Foundation of China Grant No.~IS25035. KL is supported by the NSFC special fund for theoretical physics No. 12447108 and the National Key Research and Development Program of China No. 2020YFA0713000.



\appendix

\section{Partition notation and a basic exponential identity}
\label{partnot}
Let $P$ be a partition of an integer, 
\begin{equation}
|P| = \sum_{j=1}^m j k_j.
\end{equation}
We regard $j$ as an element of a partition $P$ and $k_j$ as the multiplicity of $j$ in $P$. If $\lambda \vdash |P|$, this means that $\lambda$ is a partition of the integer $|P|$. The length of a partition $P$ is defined as
\begin{equation}
\ell(P) = \sum_{j=1}^m k_j.
\end{equation}
We also define the combinatorial factor
\begin{equation}
    z_P = \prod_{j=1}^m k_j ! \, j^{k_j}.
\end{equation}
Let $\{a_j\}_{j=1}^\infty$ be a sequence. For a partition $P$, we define
\begin{equation}
    a_P = \prod_{j=1}^m (a_j)^{k_j}.
\end{equation}
We claim the following identity
\begin{equation}
    \exp\!\left( \sum_{j=1}^\infty \frac{1}{j} a_j \right) 
    = \sum_P \frac{1}{z_P} a_P,
\end{equation}
where the sum is taken over all partitions $P$ of all integers.

\begin{proof}
Starting from the exponential, we rewrite
\begin{equation}
    \exp\!\left( \sum_{j=1}^\infty \frac{1}{j} a_j \right) 
    = \prod_{j=1}^\infty \exp\!\left( \frac{1}{j} a_j \right).
\end{equation}
Each factor expands as
\begin{equation}
    \exp\!\left( \frac{1}{j} a_j \right) 
    = \sum_{n=0}^\infty \frac{1}{n!} \left( \frac{1}{j} a_j \right)^n.
\end{equation}
Multiplying over all $j$, one sees that the coefficient of $a_P$ is precisely $\frac{1}{z_P}$
\begin{equation}
\exp\left( \sum_{j=1}^{\infty} \frac{1}{j} a_j \right) = \prod_{j=1}^{\infty} \sum_{n=0}^{\infty} \frac{1}{n!} \left( \frac{1}{j} a_j \right)^n.
\end{equation}
\end{proof}
Note that if $P=\emptyset$, then we always define $a_P = z_P = 1$.

\section{$SU(N)$ BMN index}
\label{sunbmn}
In this Appendix \ref{sunbmn}, we write down the $SU(N)$ BMN index for different partition sectors using character expansion approach.
\subsection{Single partition $SU(N)$ BMN index}
\label{spbmn}
In this Appendix \ref{spbmn}, we write down the single partition $SU(N)$ BMN index first for the $SU(N)$ BMN index in the trivial vacuum sector and then $SU(N)$ BMN index other than the trivial vacuum sector.
\subsubsection*{$SU(N)$ BMN index in the trivial vacuum sector}
We now write down the $SU(N)$ index in the trivial vacuum sector 
\begin{equation}
\label{b1}
\begin{split}
\mathcal{I}_{n_1=2; N_1=1}^{\mathrm{BMN}}(t) &= 1 + 6t^4+\mathcal{O}(t^{6}).
\end{split}
\end{equation}
\begin{equation}
\label{b2}
\begin{split}
\mathcal{I}_{n_1=3; N_1=1}^{\mathrm{BMN}}(t) &= 1 + 6 t^4 + 2 t^6 + 9 t^8 + 3 t^{10} + \mathcal{O}(t^{12}).
\end{split}
\end{equation}
\begin{equation}
\label{b3}
\begin{split}
\mathcal{I}_{n_1=4; N_1=1}^{\mathrm{BMN}}(t) &= 1 + 6 t^4 + 2 t^6 + 24 t^8 - 6 t^{10} + 60 t^{12} - 48 t^{14} + 135 t^{16} + \mathcal{O}(t^{18}).
\end{split}
\end{equation}
\begin{equation}
\label{b4}
\begin{split}
\mathcal{I}_{n_1=5; N_1=1}^{\mathrm{BMN}}(t) &= 1 + 6 t^4 + 2 t^6 + 24 t^8 + 15 t^{10} + 52 t^{12} + 39 t^{14} + 72 t^{16} + 74 t^{18} + 93 t^{20} \\
&+ 198 t^{22} + 236 t^{24} + 345 t^{26} + \mathcal{O}(t^{28}).
\end{split}
\end{equation}
\begin{equation}
\label{b5}
\begin{split}
\mathcal{I}_{n_1=6; N_1=1}^{\mathrm{BMN}}(t) &=1 + 6 t^4 + 2 t^6 + 24 t^8 + 15 t^{10} + 80 t^{12} + 33 t^{14} + 198 t^{16} + 22 t^{18} \\
&+ 399 t^{20} - 57 t^{22} + 795 t^{24} - 69 t^{26} + 1719 t^{28} - 144 t^{30} + 2523 t^{32} \\
&- 1494 t^{34} + 3183 t^{36} + \mathcal{O}(t^{38}).
\end{split}
\end{equation}
\begin{equation}
\label{b6}
\begin{split}
\mathcal{I}_{n_1=7; N_1=1}^{\mathrm{BMN}}(t) &=1 + 6 t^4 + 2 t^6 + 24 t^8 + 15 t^{10} + 80 t^{12} + 69 t^{14} + 195 t^{16} + 195 t^{18} \\
&+ 369 t^{20} + 405 t^{22} + 584 t^{24} + 762 t^{26} + 1026 t^{28} + 1720 t^{30} + 2316 t^{32} \\
&+ 3294 t^{34} + 3133 t^{36} + 2724 t^{38} + 1875 t^{40} + 3683 t^{42} + 10323 t^{44} \\
&+ 19407 t^{46} + 23630 t^{48} + 6558 t^{50} + \mathcal{O}(t^{52}).
\end{split}
\end{equation}
\begin{equation}
\begin{split}
\mathcal{I}_{n_1=8; N_1=1}^{\mathrm{BMN}}(t) &=1 + 6 t^4 + 2 t^6 + 24 t^8 + 15 t^{10} + 80 t^{12} + 69 t^{14} + 240 t^{16} + 196 t^{18} \\
&+ 597 t^{20} + 408 t^{22} + 1245 t^{24} + 663 t^{26} + 2319 t^{28} + 1066 t^{30} + 4443 t^{32} \\
&+ 2517 t^{34} + 9141 t^{36} + 5286 t^{38} + 14340 t^{40} + 3516 t^{42} + 14469 t^{44} - 204 t^{46} \\
&+ 31201 t^{48} + 32784 t^{50} + 91161 t^{52} + 54997 t^{54} + 30507 t^{56} - 138132 t^{58} \\
&- 107038 t^{60} + 71505 t^{62} + 723276 t^{64} + \mathcal{O}(t^{66}).
\end{split}
\end{equation}
\begin{equation}
\begin{split}
\mathcal{I}_{n_1=9; N_1=1}^{\mathrm{BMN}}(t) &=1 + 6 t^4 + 2 t^6 + 24 t^8 + 15 t^{10} + 80 t^{12} + 69 t^{14} + 240 t^{16} + 251 t^{18} \\
&+ 603 t^{20} + 699 t^{22} + 1292 t^{24} + 1566 t^{26} + 2400 t^{28} + 3011 t^{30} + 4146 t^{32} \\
&+ 5643 t^{34} + 7761 t^{36} + 11739 t^{38} + 16293 t^{40} + 22877 t^{42} + 26601 t^{44} \\
&+ 28290 t^{46} + 24276 t^{48} + 23895 t^{50} + 40194 t^{52} + 91396 t^{54} + 180120 t^{56} \\
&+ 252972 t^{58} + 216079 t^{60} - 45402 t^{62} - 430887 t^{64} - 525488 t^{66} + 373200 t^{68} \\
&+ 2440779 t^{70} + 4239379 t^{72} + 2494311 t^{74} - 5336595 t^{76} - 15340945 t^{78} \\
&- 13899420 t^{80} + 14087100 t^{82} + \mathcal{O}(t^{84}).
\end{split}
\end{equation}
\begin{equation}
\begin{split}
\mathcal{I}_{n_1=10; N_1=1}^{\mathrm{BMN}}(t) &=1 + 6 t^4 + 2 t^6 + 24 t^8 + 15 t^{10} + 80 t^{12} + 69 t^{14} + 240 t^{16} + 251 t^{18} \\
&+ 669 t^{20} + 711 t^{22} + 1654 t^{24} + 1668 t^{26} + 3588 t^{28} + 3340 t^{30} + 6933 t^{32} \\
&+ 6021 t^{34} + 12622 t^{36} + 11010 t^{38} + 23769 t^{40} + 22923 t^{42} + 47214 t^{44} \\
&+ 46062 t^{46} + 79819 t^{48} + 61821 t^{50} + 91011 t^{52} + 46887 t^{54} + 112302 t^{56} \\
&+ 139821 t^{58} + 397892 t^{60} + 576783 t^{62} + 840483 t^{64} + 481755 t^{66} - 268599 t^{68} \\
&- 1674861 t^{70} - 1697536 t^{72} + 995007 t^{74} + 8352888 t^{76} + 15403333 t^{78} \\
&+ 13034091 t^{80} - 12113328 t^{82} - 54039473 t^{84} - 75712605 t^{86} - 7013664 t^{88} \\
&+ 180078995 t^{90} + 371326038 t^{92} + 256877472 t^{94} - 439844576 t^{96} \\
&- 1456084758 t^{98} - 1565169936 t^{100} + \mathcal{O}(t^{102}).
\end{split}
\end{equation}

\subsubsection*{$SU(N)$ single partition index other than trivial vacuum sector }
We now write down the $SU(N)$ index other than trivial vacuum sector in $N=n_1 N_1$ case.

\textbf{$SU(N)$ index ($n_1 = 1$, $N_1 = 2$):}
\begin{equation}
\begin{split}
\mathcal{I}_{n_1=1; N_1=2}^{\mathrm{BMN}}(t) = &  1 +\mathcal{O}(t^{6}).
\end{split}
\end{equation}
\textbf{$SU(N)$ index ($n_1 = 1$, $N_1 = 3$):}
\begin{equation}
\begin{split}
\mathcal{I}_{n_1=1; N_1=3}^{\mathrm{BMN}}(t) = &  1 - t^6 + 3 t^8 - 3 t^{10}+\mathcal{O}(t^{12}).
\end{split}
\end{equation}
\textbf{$SU(N)$ index ($n_1 = 1$, $N_1 = 4$):}
\begin{equation}
\begin{split}
\mathcal{I}_{n_1=1; N_1=4}^{\mathrm{BMN}}(t) = &  1 - t^6 + 3 t^8 - 3 t^{10} + 6 t^{16}+\mathcal{O}(t^{18}).
\end{split}
\end{equation}
\textbf{$SU(N)$ index ($n_1 = 2$, $N_1 = 2$):}
\begin{equation}
\begin{split}
\mathcal{I}_{n_1=2; N_1=2}^{\mathrm{BMN}}(t) = &  1 + 6 t^4 - 9 t^6 + 18 t^8 - 30 t^{10} + 64 t^{12} - 90 t^{14} + 108 t^{16}+\mathcal{O}(t^{18}).
\end{split}
\end{equation}
\textbf{$SU(N)$ index ($n_1 = 1$, $N_1 = 5$):}
\begin{equation}
\begin{split}
\mathcal{I}_{n_1=1; N_1=5}^{\mathrm{BMN}}(t) = 1 - t^6 + 3 t^8 - 3 t^{10} + 6 t^{16} - 9 t^{18} + 3 t^{20} + 3 t^{22} + t^{24} - 12 t^{26}+\mathcal{O}(t^{28}).
\end{split}
\end{equation}
\textbf{$SU(N)$ index ($n_1 = 1$, $N_1 = 6$):}
\begin{equation}
\begin{split}
\mathcal{I}_{n_1=1; N_1=6}^{\mathrm{BMN}}(t) = &1 - t^6 + 3 t^8 - 3 t^{10} + 6 t^{16} - 9 t^{18} + 3 t^{20} + 3 t^{22} + t^{24} \\
&- 12 t^{26} + 15 t^{28} - 2 t^{30} - 9 t^{32} + 20 t^{36}+\mathcal{O}(t^{38}).
\end{split}
\end{equation}
\textbf{$SU(N)$ index ($n_1 = 2$, $N_1 = 3$):}
\begin{equation}
\begin{split}
\mathcal{I}_{n_1=2; N_1=3}^{\mathrm{BMN}}(t) = &  1 + 6 t^4 - 9 t^6 + 18 t^8 - 30 t^{10} + 63 t^{12} - 90 t^{14} + 108 t^{16} \\
&- 154 t^{18} + 264 t^{20} - 324 t^{22} + 300 t^{24} - 393 t^{26} + 732 t^{28} \\
&- 927 t^{30} + 657 t^{32} - 594 t^{34} + 1571 t^{36}+\mathcal{O}(t^{38}).
\end{split}
\end{equation}
\textbf{$SU(N)$ index ($n_1 = 3$, $N_1 = 2$):}
\begin{equation}
\begin{split}
\mathcal{I}_{n_1=3; N_1=2}^{\mathrm{BMN}}(t) = &  1 + 6 t^4 + t^6 + 9 t^8 - 3 t^{10} + 12 t^{12} + 12 t^{14} + 39 t^{16} \\
&- 29 t^{18} - 27 t^{20} - 24 t^{22} + 263 t^{24} + 30 t^{26} - 333 t^{28} \\
&- 511 t^{30} + 1068 t^{32} + 1230 t^{34} - 1618 t^{36}+\mathcal{O}(t^{38}).
\end{split}
\end{equation}
\textbf{$SU(N)$ index ($n_1 = 1$, $N_1 = 7$):}
\begin{equation}
\begin{split}
\mathcal{I}_{n_1=1; N_1=7}^{\mathrm{BMN}}(t) =& 1 - t^6 + 3 t^8 - 3 t^{10} + 6 t^{16} - 9 t^{18} + 3 t^{20} + 3 t^{22} + t^{24} \\
&- 12 t^{26} + 15 t^{28} - 2 t^{30} - 9 t^{32} + 19 t^{36} - 27 t^{38} + 9 t^{40} \\
&+ 20 t^{42} - 15 t^{44} - 27 t^{46} + 52 t^{48} - 18 t^{50}+ \mathcal{O}(t^{52}).
\end{split}
\end{equation}
\textbf{$SU(N)$ index ($n_1 = 1$, $N_1 = 8$):}
\begin{equation}
\begin{split}
\mathcal{I}_{n_1=1; N_1=8}^{\mathrm{BMN}}(t) =& 1 - t^6 + 3 t^8 - 3 t^{10} + 6 t^{16} - 9 t^{18} + 3 t^{20} + 3 t^{22} + t^{24} \\
&- 12 t^{26} + 15 t^{28} - 2 t^{30} - 9 t^{32} + 19 t^{36} - 27 t^{38} + 9 t^{40} \\
&+ 19 t^{42} - 12 t^{44} - 30 t^{46} + 54 t^{48} - 24 t^{50} - 36 t^{52} \\
&+ 43 t^{54} + 36 t^{56} - 105 t^{58} + 65 t^{60} + 51 t^{62} - 99 t^{64}+ \mathcal{O}(t^{66}).
\end{split}
\end{equation}
\textbf{$SU(N)$ index ($n_1 = 2$, $N_1 = 4$):}
\begin{equation}
\begin{split}
\mathcal{I}_{n_1=2; N_1=4}^{\mathrm{BMN}}(t) &= 1 + 6 t^4 - 9 t^6 + 18 t^8 - 30 t^{10} + 63 t^{12} - 90 t^{14} + 108 t^{16} \\
&- 155 t^{18} + 264 t^{20} - 324 t^{22} + 310 t^{24} - 405 t^{26} + 726 t^{28} \\
&- 936 t^{30} + 735 t^{32} - 651 t^{34} + 1463 t^{36} - 2385 t^{38} + 1767 t^{40} \\
&- 398 t^{42} + 1578 t^{44} - 5271 t^{46} + 5548 t^{48} + 96 t^{50} - 2259 t^{52} \\
&- 7376 t^{54} + 16695 t^{56} - 6042 t^{58} - 13918 t^{60} + 5040 t^{62} + 34044 t^{64}+ \mathcal{O}(t^{66}).
\end{split}
\end{equation}
\textbf{$SU(N)$ index ($n_1 = 4$, $N_1 = 2$):}
\begin{equation}
\begin{split}
\mathcal{I}_{n_1=4; N_1=2}^{\mathrm{BMN}}(t) &= 1 + 6 t^4 + t^6 + 24 t^8 - 12 t^{10} + 58 t^{12} - 57 t^{14} + 156 t^{16} \\
&- 116 t^{18} + 354 t^{20} - 393 t^{22} + 401 t^{24} - 660 t^{26} + 1503 t^{28} \\
&- 451 t^{30} + 714 t^{32} - 4326 t^{34} + 2915 t^{36} + 4620 t^{38} + 8202 t^{40} \\
&- 21606 t^{42} - 17859 t^{44} + 26352 t^{46} + 86647 t^{48} - 47706 t^{50} \\
&- 214914 t^{52} - 17160 t^{54} + 572751 t^{56} + 238956 t^{58} - 1248916 t^{60} \\
&- 1113066 t^{62} + 2664642 t^{64}+ \mathcal{O}(t^{66}).
\end{split}
\end{equation}
\textbf{$SU(N)$ index ($n_1 = 1$, $N_1 = 9$):}
\begin{equation}
\begin{split}
\mathcal{I}_{n_1=1; N_1=9}^{\mathrm{BMN}}(t) &= 1 - t^6 + 3 t^8 - 3 t^{10} + 6 t^{16} - 9 t^{18} + 3 t^{20} + 3 t^{22} + t^{24} \\
&- 12 t^{26} + 15 t^{28} - 2 t^{30} - 9 t^{32} + 19 t^{36} - 27 t^{38} + 9 t^{40} \\
&+ 19 t^{42} - 12 t^{44} - 30 t^{46} + 53 t^{48} - 21 t^{50} - 39 t^{52} + 45 t^{54} \\
&+ 30 t^{56} - 90 t^{58} + 46 t^{60} + 63 t^{62} - 108 t^{64} + 8 t^{66} + 138 t^{68} \\
&- 129 t^{70} - 64 t^{72} + 216 t^{74} - 108 t^{76} - 186 t^{78} + 300 t^{80} - 9 t^{82}+ \mathcal{O}(t^{84}).
\end{split}
\end{equation}
\textbf{$SU(N)$ index ($n_1 = 3$, $N_1 = 3$):}
\begin{equation}
\begin{split}
\mathcal{I}_{n_1=3; N_1=3}^{\mathrm{BMN}}(t) &=1 + 6 t^4 + t^6 + 9 t^8 - 3 t^{10} + 11 t^{12} + 12 t^{14} + 33 t^{16} \\
&- 19 t^{18} - 30 t^{20} + 9 t^{22} + 197 t^{24} + 18 t^{26} - 354 t^{28} \\
&- 184 t^{30} + 960 t^{32} + 630 t^{34} - 2209 t^{36} - 1635 t^{38} + 5157 t^{40} \\
&+ 4301 t^{42} - 11991 t^{44} - 9582 t^{46} + 27606 t^{48} + 20079 t^{50} \\
&- 62982 t^{52} - 38360 t^{54} + 142218 t^{56} + 64581 t^{58} - 313249 t^{60} \\
&- 87102 t^{62} + 672645 t^{64} + 56522 t^{66} - 1398387 t^{68} + 176637 t^{70} \\
&+ 2795578 t^{72} - 995943 t^{74} - 5333265 t^{76} + 3292418 t^{78} \\
&+ 9615936 t^{80} - 8959572 t^{82}+ \mathcal{O}(t^{84}).
\end{split}
\end{equation}
\textbf{$SU(N)$ index ($n_1 = 1$, $N_1 = 10$):}
\begin{equation}
 \begin{split}
 \mathcal{I}_{n_1=1; N_1=10}^{\mathrm{BMN}}(t)&= 1 - t^6 + 3 t^8 - 3 t^{10} + 6 t^{16} - 9 t^{18} + 3 t^{20} + 3 t^{22} + t^{24} \\
        & - 12 t^{26} + 15 t^{28} - 2 t^{30} - 9 t^{32} + 19 t^{36} - 27 t^{38} + 9 t^{40} \\
        & + 19 t^{42} - 12 t^{44} - 30 t^{46} + 53 t^{48} - 21 t^{50} - 39 t^{52} \\
        & + 44 t^{54} + 33 t^{56} - 93 t^{58} + 48 t^{60} + 57 t^{62} - 93 t^{64} \\
        & - 11 t^{66} + 150 t^{68} - 138 t^{70} - 37 t^{72} + 168 t^{74} - 69 t^{76} \\
        & - 196 t^{78} + 309 t^{80} - 60 t^{82} - 287 t^{84} + 276 t^{86} + 162 t^{88} \\
        & - 545 t^{90} + 318 t^{92} + 360 t^{94} - 644 t^{96} + 54 t^{98} + 813 t^{100}+\mathcal{O}(t^{102}).
 \end{split}
 \end{equation}
\textbf{$SU(N)$ index ($n_1 = 2$, $N_1 = 5$):}
 \begin{equation}
 \begin{split}
 \mathcal{I}_{n_1=2; N_1=5}^{\mathrm{BMN}}(t)&= 1 + 6 t^4 - 9 t^6 + 18 t^8 - 30 t^{10} + 63 t^{12} - 90 t^{14} + 108 t^{16} \\
        & - 155 t^{18} + 264 t^{20} - 324 t^{22} + 309 t^{24} - 405 t^{26} + 726 t^{28} \\
        & - 926 t^{30} + 723 t^{32} - 657 t^{34} + 1454 t^{36} - 2307 t^{38} + 1710 t^{40} \\
        & - 504 t^{42} + 1629 t^{44} - 4920 t^{46} + 5262 t^{48} - 516 t^{50} \\
        & - 1584 t^{52} - 6184 t^{54} + 14754 t^{56} - 7533 t^{58} - 9719 t^{60} \\
        & + 6492 t^{62} + 25080 t^{64} - 39505 t^{66} - 2154 t^{68} + 49272 t^{70} \\
        & - 8917 t^{72} - 93822 t^{74} + 94242 t^{76} + 68733 t^{78} - 169644 t^{80} \\
        & - 27837 t^{82} + 314750 t^{84} - 188727 t^{86} - 340524 t^{88} + 482278 t^{90} \\
        & + 252942 t^{92} - 952587 t^{94} + 276239 t^{96} + 1240224 t^{98} \\
        & - 1199643 t^{100}+ \mathcal{O}(t^{102}).
\end{split}
 \end{equation}
\textbf{$SU(N)$ index ($n_1 = 5$, $N_1 = 2$):}
\begin{equation}
 \begin{split}
 \mathcal{I}_{n_1=5; N_1=2}^{\mathrm{BMN}}(t) &=1 + 6 t^4 + t^6 + 24 t^8 + 9 t^{10} + 50 t^{12} + 15 t^{14} + 78 t^{16} + 48 t^{18} + 165 t^{20} \\
&+ 159 t^{22} + 307 t^{24} - 36 t^{26} - 27 t^{28} - 107 t^{30} + 1704 t^{32} + 2040 t^{34} \\
&+ 395 t^{36} - 6567 t^{38} - 5046 t^{40} + 12453 t^{42} + 34707 t^{44} - 1404 t^{46} - 96929 t^{48} \\
&- 94539 t^{50} + 201693 t^{52} + 451571 t^{54} - 165591 t^{56} - 1422096 t^{58} - 666768 t^{60} \\
&+ 3424923 t^{62} + 4419750 t^{64} - 6190139 t^{66} - 16232442 t^{68} + 5811177 t^{70} + 47015688 t^{72} \\
&+ 13588929 t^{74} - 115135479 t^{76} - 101061698 t^{78} + 241559943 t^{80} + 386289432 t^{82} \\
&- 416812659 t^{84} - 1177439610 t^{86} + 487315872 t^{88} + 3142761116 t^{90} + 171947556 t^{92} \\
&- 7610986839 t^{94} - 3494172510 t^{96} + 16978209045 t^{98} + 14712720957 t^{100}
+ \mathcal{O}(t^{102}).
 \end{split}
 \end{equation}



We now write down the $SU(N)$ BMN index in the $N=n_1 N_1$ i.e., single partition sector 
\begin{equation}
\begin{split}
\mathcal{I}_{N=2}^{\mathrm{BMN~SP}}(t)&=\mathcal{I}_{n_1=1; N_1=2}^{\mathrm{BMN}}(t) +\mathcal{I}_{n_1=2; N_1=1}^{\mathrm{BMN}}(t)\\
&= 2 + 6t^4+\mathcal{O}(t^{6}).
\end{split}
\end{equation}
\begin{equation}
\begin{split}
\mathcal{I}_{N=3}^{\mathrm{BMN~SP}}(t) &=\mathcal{I}_{n_1=1; N_1=3}^{\mathrm{BMN}}(t)+\mathcal{I}_{n_1=3; N_1=1}^{\mathrm{BMN}}(t)  \\&= 2 + 6 t^4 + t^6 + 12 t^8 + \mathcal{O}\left(t^{12}\right).
\end{split}
\end{equation}
\begin{equation}
\begin{split}
\mathcal{I}_{N=4}^{\mathrm{BMN~SP}}(t) &=\mathcal{I}_{n_1=1; N_1=4}^{\mathrm{BMN}}(t)+\mathcal{I}_{n_1=2; N_1=2}^{\mathrm{BMN}}(t)+\mathcal{I}_{n_1=4; N_1=1}^{\mathrm{BMN}}(t)\\
&= 3 + 12 t^4 - 8 t^6 + 45 t^8 - 39 t^{10} + 124 t^{12} - 138 t^{14} + 249 t^{16} + O\left(t^{18}\right).
\end{split}
\end{equation}
\begin{equation}
\begin{split}
\mathcal{I}_{N=5}^{\mathrm{BMN~SP}}(t) &= \mathcal{I}_{n_1=1; N_1=5}^{\mathrm{BMN}}(t)+\mathcal{I}_{n_1=5; N_1=1}^{\mathrm{BMN}}(t)  \\&=2 + 6 t^4 + t^6 + 27 t^8 + 12 t^{10} + 52 t^{12} + 39 t^{14} + 78 t^{16} + 65 t^{18} \\
&+ 96 t^{20} + 201 t^{22} + 237 t^{24} + 333 t^{26} + \mathcal{O}(t^{28}).
\end{split}
\end{equation}
\begin{equation}
\begin{split}
\mathcal{I}_{N=6}^{\mathrm{BMN~SP}}(t) &= \mathcal{I}_{n_1=1; N_1=6}^{\mathrm{BMN}}(t)+\mathcal{I}_{n_1=2; N_1=3}^{\mathrm{BMN}}(t) +\mathcal{I}_{n_1=3; N_1=2}^{\mathrm{BMN}}(t) +\mathcal{I}_{n_1=6; N_1=1}^{\mathrm{BMN}}(t)  \\&=4 + 18 t^4 - 7 t^6 + 54 t^8 - 21 t^{10} + 155 t^{12} - 45 t^{14} + 351 t^{16} - 170 t^{18} \\
&+ 639 t^{20} - 402 t^{22} + 1359 t^{24} - 444 t^{26} + 2133 t^{28} - 1584 t^{30} + 4239 t^{32} \\
&- 858 t^{34} + 3156 t^{36} + \mathcal{O}(t^{38}).
\end{split}
\end{equation}
\begin{equation}
\begin{split}
\mathcal{I}_{N=7}^{\mathrm{BMN~SP}}(t) &= \mathcal{I}_{n_1=1; N_1=7}^{\mathrm{BMN}}(t)+\mathcal{I}_{n_1=7; N_1=1}^{\mathrm{BMN}}(t)  \\&=2 + 6 t^4 + t^6 + 27 t^8 + 12 t^{10} + 80 t^{12} + 69 t^{14} + 201 t^{16} + 186 t^{18} \\
&+ 372 t^{20} + 408 t^{22} + 585 t^{24} + 750 t^{26} + 1041 t^{28} + 1718 t^{30} + 2307 t^{32} \\
&+ 3294 t^{34} + 3152 t^{36} + 2697 t^{38} + 1884 t^{40} + 3703 t^{42} + 10308 t^{44} + 19380 t^{46} \\
&+ 23682 t^{48} + 6540 t^{50} + \mathcal{O}(t^{52}).
\end{split}
\end{equation}
\begin{equation}
\begin{split}
\mathcal{I}_{N=8}^{\mathrm{BMN~SP}}(t) &= \mathcal{I}_{n_1=1; N_1=8}^{\mathrm{BMN}}(t)+\mathcal{I}_{n_1=2; N_1=4}^{\mathrm{BMN}}(t)+\mathcal{I}_{n_1=4; N_1=2}^{\mathrm{BMN}}(t)+\mathcal{I}_{n_1=8; N_1=1}^{\mathrm{BMN}}(t)  \\&=4 + 18 t^4 - 7 t^6 + 69 t^8 - 30 t^{10} + 201 t^{12} - 78 t^{14} + 510 t^{16} - 84 t^{18} \\
&+ 1218 t^{20} - 306 t^{22} + 1957 t^{24} - 414 t^{26} + 4563 t^{28} - 323 t^{30} + 5883 t^{32} \\
&- 2460 t^{34} + 13538 t^{36} + 7494 t^{38} + 24318 t^{40} - 18469 t^{42} - 1824 t^{44} + 20847 t^{46} \\
&+ 123450 t^{48} - 14850 t^{50} - 126048 t^{52} + 30504 t^{54} + 619989 t^{56} + 94677 t^{58} \\
&- 1369807 t^{60} - 1036470 t^{62} + 3421863 t^{64} + \mathcal{O}(t^{66}).
\end{split}
\end{equation}
\begin{equation}
\begin{split}
\mathcal{I}_{N=9}^{\mathrm{BMN~SP}}(t) &= \mathcal{I}_{n_1=1; N_1=9}^{\mathrm{BMN}}(t)+\mathcal{I}_{n_1=3; N_1=3}^{\mathrm{BMN}}(t)+\mathcal{I}_{n_1=9; N_1=1}^{\mathrm{BMN}}(t)\\
&=3 + 12 t^4 + 2 t^6 + 36 t^8 + 9 t^{10} + 91 t^{12} + 81 t^{14} + 279 t^{16} + 223 t^{18} \\
&+ 576 t^{20} + 711 t^{22} + 1490 t^{24} + 1572 t^{26} + 2061 t^{28} + 2825 t^{30} + 5097 t^{32} \\
&+ 6273 t^{34} + 5571 t^{36} + 10077 t^{38} + 21459 t^{40} + 27197 t^{42} + 14598 t^{44} + 18678 t^{46} \\
&+ 51935 t^{48} + 43953 t^{50} - 22827 t^{52} + 53081 t^{54} + 322368 t^{56} + 317463 t^{58} \\
&- 97124 t^{60} - 132441 t^{62} + 241650 t^{64} - 468958 t^{66} - 1025049 t^{68} + 2617287 t^{70} \\
&+ 7034893 t^{72} + 1498584 t^{74} - 10669968 t^{76} - 12048713 t^{78} - 4283184 t^{80} \\
&+ 5127519 t^{82} + \mathcal{O}(t^{84}).
\end{split}
\end{equation}
\begin{equation}
\begin{split}
\mathcal{I}_{N=10}^{\mathrm{BMN~SP}}(t) &= \mathcal{I}_{n_1=1; N_1=10}^{\mathrm{BMN}}(t)+\mathcal{I}_{n_1=2; N_1=5}^{\mathrm{BMN}}(t)+\mathcal{I}_{n_1=5; N_1=2}^{\mathrm{BMN}}(t)+\mathcal{I}_{n_1=10; N_1=1}^{\mathrm{BMN}}(t)\\&=4 + 18 t^4 - 7 t^6 + 69 t^8 - 9 t^{10} + 193 t^{12} - 6 t^{14} + 432 t^{16} + 135 t^{18} \\
&+ 1101 t^{20} + 549 t^{22} + 2271 t^{24} + 1215 t^{26} + 4302 t^{28} + 2305 t^{30} + 9351 t^{32} \\
&+ 7404 t^{34} + 14490 t^{36} + 2109 t^{38} + 20442 t^{40} + 34891 t^{42} + 83538 t^{44} + 39708 t^{46} \\
&- 11795 t^{48} - 33255 t^{50} + 291081 t^{52} + 492318 t^{54} - 38502 t^{56} - 1289901 t^{58} \\
&- 278547 t^{60} + 4008255 t^{62} + 5285220 t^{64} - 5747900 t^{66} - 16503045 t^{68} + 4185450 t^{70} \\
&+ 45309198 t^{72} + 14490282 t^{74} - 106688418 t^{76} - 85589828 t^{78} + 254424699 t^{80} \\
&+ 374148207 t^{82} - 470537669 t^{84} - 1253340666 t^{86} + 479961846 t^{88} + 3323321844 t^{90} \\
&+ 543526854 t^{92} - 7355061594 t^{94} - 3933741491 t^{96} + 15523364565 t^{98} \\
&+ 13146352191 t^{100} + \mathcal{O}(t^{102}).
\end{split}
\end{equation}

We have for single partitions
\begin{equation}
\begin{aligned}
\mathcal{I}_{N=2}^{\mathrm{BMN~SP}}(t): &\quad d_4 = 6 \\
\mathcal{I}_{N=3}^{\mathrm{BMN~SP}}(t): &\quad d_8 = 12, \quad d_{10} = 0 \\
\mathcal{I}_{N=4}^{\mathrm{BMN~SP}}(t): &\quad d_{16} = 249 \\
\mathcal{I}_{N=5}^{\mathrm{BMN~SP}}(t): &\quad d_{24} = 237, \quad d_{26} = 333 \\
\mathcal{I}_{N=6}^{\mathrm{BMN~SP}}(t): &\quad d_{36} = 3156 \\
\mathcal{I}_{N=7}^{\mathrm{BMN~SP}}(t): &\quad d_{48} = 23682, \quad d_{50} = 6540 \\
\mathcal{I}_{N=8}^{\mathrm{BMN~SP}}(t): &\quad d_{64} = 3421863 \\
\mathcal{I}_{N=9}^{\mathrm{BMN~SP}}(t): &\quad d_{80} = -4283184, \quad d_{82} = 5127519 \\
\mathcal{I}_{N=10}^{\mathrm{BMN~SP}}(t): &\quad d_{100} = 13146352191.
\end{aligned}
\end{equation}
We list all the coefficients for the term $d_{j=N^2}$ for $N=n_1 N_1$ even. For $N=n_1 N_1$ odd we list all the coefficients for the term $d_{N^2-1}$ and $d_{N^2+1}$.
Now, we compute 
\begin{equation}
\begin{split}
&\text{log} |d_{j=N^2}|~\text{for even}~N\\
&\frac{1}{2}\Bigg(\text{log} |d_{N^2-1}|+\text{log} |d_{N^2+1}|\Bigg)~\text{for odd}~N.
\end{split}
\end{equation}
We have Table \ref{spindex}
\begin{table}[h]
    \centering
    \label{tab:results}
    \begin{tabular}{|c|c|}
        \hline
        $N$ & $\text{log}|d_{j=N^2}|$ \\
        \hline
        2 & 1.79176 \\
        3 & 1.2425 \\
        4 & 5.5175 \\
        5 & 5.6381 \\
        6 & 8.05706 \\
        7 & 9.42908 \\
        8 & 15.0457 \\
        9 & 15.3602 \\
        10 & 23.2994 \\
        \hline
    \end{tabular}
\caption{Results for $\log |d_{j=N^2}|$ and $N$ single partition index.}
\label{spindex}
\end{table}

\subsection*{Trivial vacuum sector}
\begin{figure}[H]
	\centering
	\includegraphics[width=0.8\linewidth]{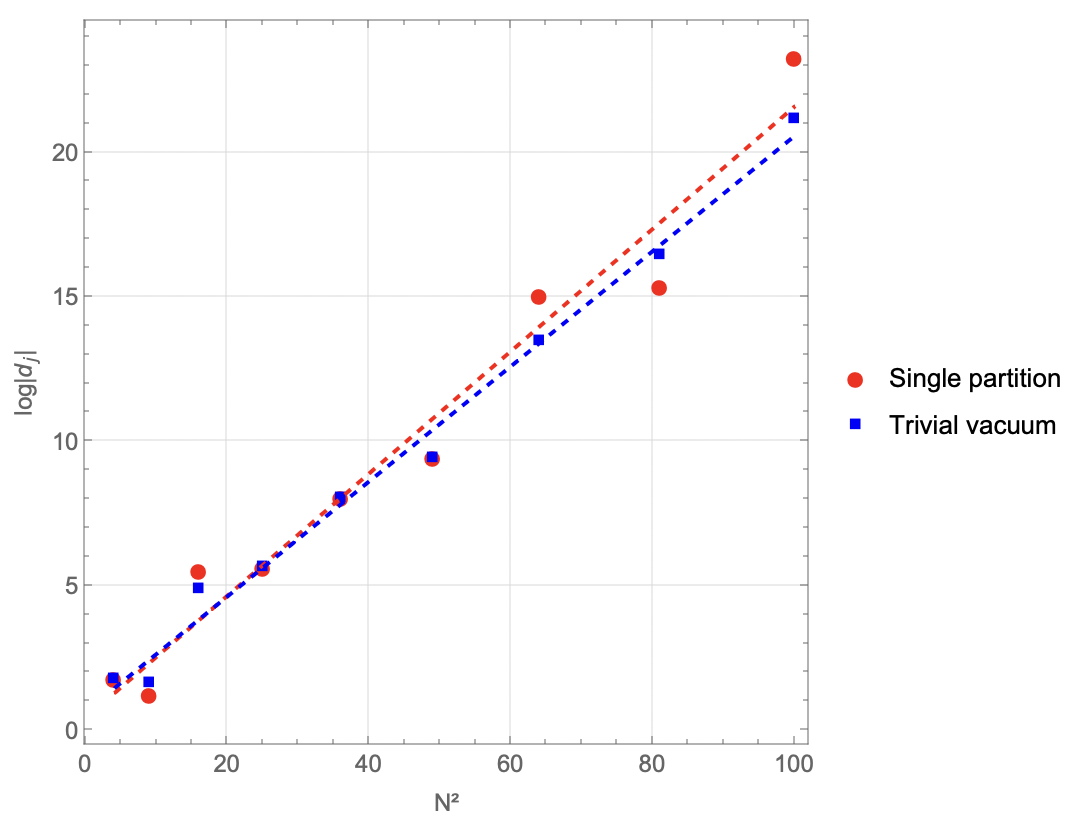}
	\caption{$\log |d_{j=N^2}|~$ with $N^2$.}
	\label{fig:plotrev}
\end{figure}

We have
\begin{equation}
\begin{split}
\mathcal{I}_{n_1=2; N_1=1}^{\mathrm{BMN}}(t) &: \quad d_4 = 6 \\
\mathcal{I}_{n_1=3; N_1=1}^{\mathrm{BMN}}(t) &: \quad d_8 = 9, \quad d_{10} = 3 \\
\mathcal{I}_{n_1=4; N_1=1}^{\mathrm{BMN}}(t) &: \quad d_{16} = 135 \\
\mathcal{I}_{n_1=5; N_1=1}^{\mathrm{BMN}}(t) &: \quad d_{24} = 236, \quad d_{26} = 345 \\
\mathcal{I}_{n_1=6; N_1=1}^{\mathrm{BMN}}(t) &: \quad d_{36} = 3183 \\
\mathcal{I}_{n_1=7; N_1=1}^{\mathrm{BMN}}(t) &: \quad d_{48} = 23630, \quad d_{50} = 6558 \\
\mathcal{I}_{n_1=8; N_1=1}^{\mathrm{BMN}}(t) &: \quad d_{64} = 723276 \\
\mathcal{I}_{n_1=9; N_1=1}^{\mathrm{BMN}}(t) &: \quad d_{80} = -13899420, \quad d_{82} = 14087100 \\
\mathcal{I}_{n_1=10; N_1=1}^{\mathrm{BMN}}(t) &: \quad d_{100} = -1565169936
\end{split}
\end{equation}
We have Table \ref{tvindex}
\begin{table}[H]
    \centering
    \label{tab:results_trvac}
    \begin{tabular}{|c|c|}
        \hline
        $N$ & $\log |d_{j=N^2}|$ \\
        \hline
        2 & 1.79176 \\
        3 & 1.64792 \\
        4 & 4.90527 \\
        5 & 5.65369 \\
        6 & 8.06558 \\
        7 & 9.42936 \\
        8 & 13.4915 \\
        9 & 16.4541 \\
        10 & 21.1713 \\
        \hline
    \end{tabular}
 \caption{Results for $\log |d_{j=N^2}|$ and $N$ trivial vacuum sector index.}
\label{tvindex}
\end{table}
Fitting $\log |d_j|$ with $N^2$ we get Fig.~.\ref{fig:plotrev}.

\begin{figure}[H]
	\centering
	\includegraphics[width=0.8\linewidth]{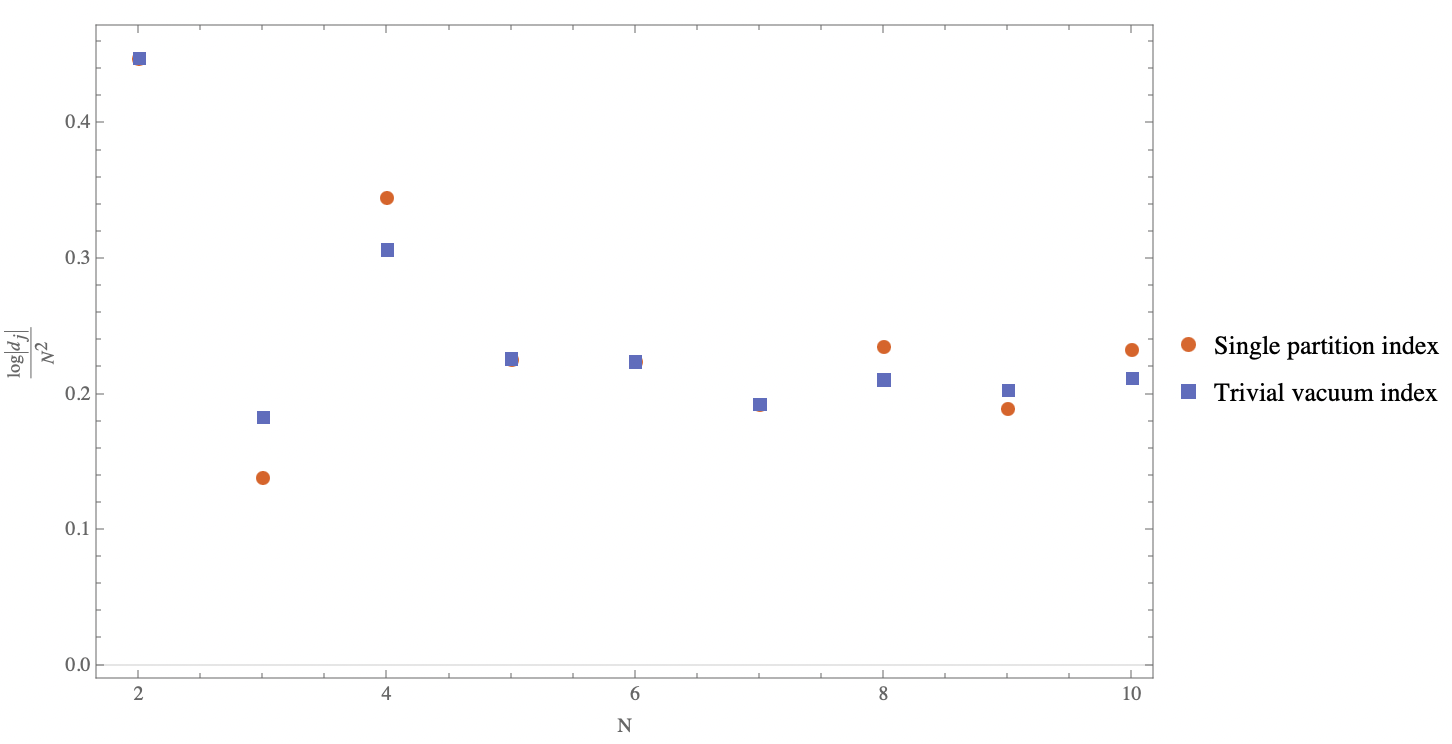}
	\caption{$\frac{\log |d_{j=N^2}|}{N^2}~$ with $N$ single partition index, and trivial vacuum index.}
	\label{fig:trspindex}
\end{figure}

The slopes we get by linear fit are given in Table \ref{slopessvtv}
\begin{table}[H]
    \centering
\begin{tabular}{|l|c|}
\hline
Index & Slope \\
\hline
Single partition & $0.211642$ \\
Trivial vacuum & $0.199249$ \\
\hline
\end{tabular}
\caption{Slopes of $\log |d_{j=N^2}|$ vs $N^2$ plots.}
\label{slopessvtv}
\end{table}
From the plot of $\frac{\log |d_j|}{N^2}$ versus $N$ (from $N=2$ to $N=10$) in Fig.~\ref{fig:trspindex}, we observe that $\frac{\log |d_j|}{N^2}$ appears to converge to the following values
\begin{table}[H]
\centering
\begin{tabular}{|l|c|}
\hline
Index & Convergent value of $\frac{\log |d_j|}{N^2}$ \\
\hline
Single partition & 0.232994 \\
Trivial vacuum & 0.211713 \\
\hline
\end{tabular}
\caption{Convergent values of $\frac{\log |d_j|}{N^2}$ for the single-partition and trivial-vacuum sectors.}
\label{tab:trsp_conv_values}
\end{table}

\subsection{Double partition $SU(N)$ BMN index}
\label{dpbmn}
We fix $ N $, and then enumerate all BMN indices in the sector corresponding to $ N = n_1 N_1 + n_2 N_2 $, i.e., the double partition sector.


\paragraph{$N=3.$}
\begin{equation}
\begin{split}
\mathcal{I}_{n_1=1, n_2=1; N_1=1, N_2=2}^{\mathrm{BMN}}(t)
&=1 + 3\,t^{2} + 3\,t^{4} + 2\,t^{6} + 6\,t^{10}
+\mathcal{O}(t^{12}).
\end{split}
\end{equation}
\begin{equation}
\begin{split}
\mathcal{I}_{N=3}^{\mathrm{BMN~DP}}(t)
&=\mathcal{I}_{n_1=1, n_2=1; N_1=1, N_2=2}^{\mathrm{BMN}}(t)\\
&=1 + 3\,t^{2} + 3\,t^{4} + 2\,t^{6} + 6\,t^{10}
+\mathcal{O}(t^{12}).
\end{split}
\end{equation}

\paragraph{$N=4.$}
\begin{equation}
\label{2112}
\begin{split}
\mathcal{I}_{n_1=2, n_2=1; N_1=1, N_2=2}^{\mathrm{BMN}}(t)
&=1 + 3\,t^{2} + 9\,t^{4} + 12\,t^{6} + 15\,t^{8}
+ 9\,t^{10} + 24\,t^{12} \\
&\quad + 36\,t^{14} + 54\,t^{16}
+\mathcal{O}(t^{18}).
\end{split}
\end{equation}
\begin{equation}
\begin{split}
\mathcal{I}_{n_1=1, n_2=1; N_1=1, N_2=3}^{\mathrm{BMN}}(t)
&=1 + 3\,t^{2} + 3\,t^{4} + t^{6} + 3\,t^{8}
+ 6\,t^{10} + t^{12} \\
&\quad - 6\,t^{14} + 9\,t^{16}
+\mathcal{O}(t^{18}).
\end{split}
\end{equation}
The largest individual-sector contribution at $j=N^2=16$ comes from
\[
\mathcal{I}_{n_1=2, n_2=1; N_1=1, N_2=2}^{\mathrm{BMN}}(t),
\]
with
\[
d_{16}=54,
\qquad
\log |d_{16}|=3.98898.
\]
Adding the two double-partition sectors gives
\begin{equation}
\begin{split}
\mathcal{I}_{N=4}^{\mathrm{BMN~DP}}(t)
&=2 + 6\,t^{2} + 12\,t^{4} + 13\,t^{6} + 18\,t^{8}
+ 15\,t^{10} + 25\,t^{12} \\
&\quad + 30\,t^{14} + 63\,t^{16}
+\mathcal{O}(t^{18}).
\end{split}
\end{equation}

\paragraph{$N=5.$}
\begin{equation}
\begin{split}
\mathcal{I}_{n_1=1, n_2=2; N_1=1, N_2=2}^{\mathrm{BMN}}(t)
&=1 + 3\,t^{2} + 9\,t^{4} + 12\,t^{6} + 12\,t^{8}
+ 9\,t^{10} + 22\,t^{12} \\
&\quad + 54\,t^{14} + 48\,t^{16} - 33\,t^{18}
- 54\,t^{20} + 117\,t^{22} \\
&\quad + 309\,t^{24} - 6\,t^{26}
+\mathcal{O}(t^{28}).
\end{split}
\end{equation}
\begin{equation}
\begin{split}
\mathcal{I}_{n_1=3, n_2=1; N_1=1, N_2=2}^{\mathrm{BMN}}(t)
&=1 + 3\,t^{2} + 9\,t^{4} + 22\,t^{6} + 36\,t^{8}
+ 51\,t^{10} + 54\,t^{12} \\
&\quad + 75\,t^{14} + 117\,t^{16} + 203\,t^{18}
+ 246\,t^{20} + 105\,t^{22} \\
&\quad + 33\,t^{24} + 204\,t^{26}
+\mathcal{O}(t^{28}).
\end{split}
\end{equation}
\begin{equation}
\begin{split}
\mathcal{I}_{n_1=2, n_2=1; N_1=1, N_2=3}^{\mathrm{BMN}}(t)
&=1 + 3\,t^{2} + 9\,t^{4} + 11\,t^{6} + 15\,t^{8}
+ 12\,t^{10} + 31\,t^{12} \\
&\quad + 24\,t^{14} + 24\,t^{16} + 3\,t^{18}
+ 84\,t^{20} + 66\,t^{22} \\
&\quad - 13\,t^{24} - 69\,t^{26}
+\mathcal{O}(t^{28}).
\end{split}
\end{equation}
\begin{equation}
\begin{split}
\mathcal{I}_{n_1=1, n_2=1; N_1=1, N_2=4}^{\mathrm{BMN}}(t)
&=1 + 3\,t^{2} + 3\,t^{4} + t^{6} + 3\,t^{8}
+ 6\,t^{10} - 3\,t^{14} \\
&\quad + 9\,t^{16} + 10\,t^{18} - 12\,t^{20}
+ 28\,t^{24} - 9\,t^{26}
+\mathcal{O}(t^{28}).
\end{split}
\end{equation}
\begin{equation}
\begin{split}
\mathcal{I}_{n_1=1, n_2=1; N_1=2, N_2=3}^{\mathrm{BMN}}(t)
&=1 + 3\,t^{2} + 3\,t^{4} + t^{6} + 9\,t^{10}
+ 9\,t^{12} - 15\,t^{14} \\
&\quad - 9\,t^{16} + 35\,t^{18} + 30\,t^{20}
- 63\,t^{22} - 51\,t^{24} \\
&\quad + 153\,t^{26}
+\mathcal{O}(t^{28}).
\end{split}
\end{equation}
The largest individual-sector contribution comes from
\[
\mathcal{I}_{n_1=1, n_2=1; N_1=2, N_2=3}^{\mathrm{BMN}}(t),
\]
for which
\[
d_{24}=-51,
\qquad
d_{26}=153,
\]
and therefore
\[
\frac{1}{2}\Big(\log |d_{24}|+\log |d_{26}|\Big)=4.48113.
\]
Adding all five double-partition sectors gives
\begin{equation}
\begin{split}
\mathcal{I}_{N=5}^{\mathrm{BMN~DP}}(t)
&=5 + 15\,t^{2} + 33\,t^{4} + 47\,t^{6} + 66\,t^{8}
+ 87\,t^{10} + 116\,t^{12} \\
&\quad + 135\,t^{14} + 189\,t^{16} + 218\,t^{18}
+ 294\,t^{20} + 225\,t^{22} \\
&\quad + 306\,t^{24} + 273\,t^{26}
+\mathcal{O}(t^{28}).
\end{split}
\end{equation}
\paragraph{$N=6.$}
\begin{equation}
\begin{split}
\mathcal{I}_{n_1=2, n_2=2; N_1=1, N_2=2}^{\mathrm{BMN}}(t)
&=1 + 3\,t^{2} + 15\,t^{4} + 22\,t^{6} + 63\,t^{8}
+ 33\,t^{10} + 128\,t^{12} \\
&\quad + 3\,t^{14} + 372\,t^{16} + 58\,t^{18}
+ 663\,t^{20} - 573\,t^{22} \\
&\quad + 764\,t^{24} - 300\,t^{26} + 3084\,t^{28}
- 1098\,t^{30} + 540\,t^{32} \\
&\quad - 4644\,t^{34} + 8029\,t^{36}
+\mathcal{O}(t^{38}).
\end{split}
\end{equation}
\begin{equation}
\begin{split}
\mathcal{I}_{n_1=4, n_2=1; N_1=1, N_2=2}^{\mathrm{BMN}}(t)
&=1 + 3\,t^{2} + 9\,t^{4} + 22\,t^{6} + 51\,t^{8}
+ 87\,t^{10} + 138\,t^{12} \\
&\quad + 174\,t^{14} + 237\,t^{16} + 311\,t^{18}
+ 525\,t^{20} + 804\,t^{22} \\
&\quad + 1046\,t^{24} + 744\,t^{26} + 423\,t^{28}
+ 720\,t^{30} + 3021\,t^{32} \\
&\quad + 4878\,t^{34} + 3230\,t^{36}
+\mathcal{O}(t^{38}).
\end{split}
\end{equation}
\begin{equation}
\begin{split}
\mathcal{I}_{n_1=3, n_2=1; N_1=1, N_2=3}^{\mathrm{BMN}}(t)
&=1 + 3\,t^{2} + 9\,t^{4} + 21\,t^{6} + 36\,t^{8}
+ 48\,t^{10} + 61\,t^{12} \\
&\quad + 84\,t^{14} + 123\,t^{16} + 152\,t^{18}
+ 162\,t^{20} + 159\,t^{22} \\
&\quad + 282\,t^{24} + 420\,t^{26} + 285\,t^{28}
+ 462\,t^{32} + 1362\,t^{34} \\
&\quad + 612\,t^{36}
+\mathcal{O}(t^{38}).
\end{split}
\end{equation}
\begin{equation}
\begin{split}
\mathcal{I}_{n_1=2, n_2=1; N_1=1, N_2=4}^{\mathrm{BMN}}(t)
&=1 + 3\,t^{2} + 9\,t^{4} + 11\,t^{6} + 15\,t^{8}
+ 12\,t^{10} + 30\,t^{12} \\
&\quad + 24\,t^{14} + 27\,t^{16} + 11\,t^{18}
+ 69\,t^{20} + 39\,t^{22} \\
&\quad + 14\,t^{24} + 12\,t^{26} + 168\,t^{28}
+ 13\,t^{30} - 93\,t^{32} \\
&\quad + 111\,t^{34} + 429\,t^{36}
+\mathcal{O}(t^{38}).
\end{split}
\end{equation}
\begin{equation}
\begin{split}
\mathcal{I}_{n_1=1, n_2=1; N_1=1, N_2=5}^{\mathrm{BMN}}(t)
&=1 + 3\,t^{2} + 3\,t^{4} + t^{6} + 3\,t^{8}
+ 6\,t^{10} - 3\,t^{14} \\
&\quad + 9\,t^{16} + 9\,t^{18} - 9\,t^{20}
+ 20\,t^{24} - 9\,t^{26} \\
&\quad - 15\,t^{28} + 36\,t^{30} + 9\,t^{32}
- 54\,t^{34} + 35\,t^{36} \\
&\quad + \mathcal{O}(t^{38}).
\end{split}
\end{equation}
\begin{equation}
\begin{split}
\mathcal{I}_{n_1=1, n_2=1; N_1=2, N_2=4}^{\mathrm{BMN}}(t)
&=1 + 3\,t^{2} + 3\,t^{4} + 3\,t^{8} + 9\,t^{10}
+ t^{12} - 15\,t^{14} \\
&\quad + 12\,t^{16} + 36\,t^{18} - 24\,t^{20}
- 63\,t^{22} + 91\,t^{24} \\
&\quad + 90\,t^{26} - 198\,t^{28} - 110\,t^{30}
+ 456\,t^{32} + 15\,t^{34} \\
&\quad - 815\,t^{36}
+\mathcal{O}(t^{38}).
\end{split}
\end{equation}
The largest individual-sector contribution comes from
\[
\mathcal{I}_{n_1=2, n_2=2; N_1=1, N_2=2}^{\mathrm{BMN}}(t),
\]
with
\[
d_{36}=8029,
\qquad
\log |d_{36}|=8.99082.
\]
Adding all six double-partition sectors gives
\begin{equation}
\begin{split}
\mathcal{I}_{N=6}^{\mathrm{BMN~DP}}(t)
&=6 + 18\,t^{2} + 48\,t^{4} + 77\,t^{6} + 171\,t^{8}
+ 195\,t^{10} + 358\,t^{12} \\
&\quad + 267\,t^{14} + 780\,t^{16} + 577\,t^{18}
+ 1386\,t^{20} + 366\,t^{22} \\
&\quad + 2217\,t^{24} + 957\,t^{26} + 3747\,t^{28}
- 439\,t^{30} + 4395\,t^{32} \\
&\quad + 1668\,t^{34} + 11520\,t^{36}
+\mathcal{O}(t^{38}).
\end{split}
\end{equation}

\paragraph{$N=7.$}

\begin{equation}
\begin{split}
\mathcal{I}_{n_1=1,n_2=1;\,N_1=1,N_2=6}^{\mathrm{BMN}}(t)
&=1+3t^2+3t^4+t^6+3t^8+6t^{10}-3t^{14}
+9t^{16}+9t^{18}-9t^{20} \\
&\quad +19t^{24}-6t^{26}-15t^{28}+28t^{30}
+9t^{32}-33t^{34}+26t^{36}+30t^{38} \\
&\quad -69t^{40}+19t^{42}+111t^{44}-105t^{46}
-79t^{48}+231t^{50}
+\mathcal{O}(t^{52}).
\end{split}
\end{equation}

\begin{equation}
\begin{split}
\mathcal{I}_{n_1=1,n_2=1;\,N_1=2,N_2=5}^{\mathrm{BMN}}(t)
&=1+3t^2+3t^4+3t^8+9t^{10}-12t^{14}
+12t^{16}+28t^{18}-24t^{20}-33t^{22} \\
&\quad +65t^{24}+36t^{26}-126t^{28}+15t^{30}
+213t^{32}-132t^{34}-270t^{36}+396t^{38} \\
&\quad +198t^{40}-797t^{42}+198t^{44}+1248t^{46}
-1115t^{48}-1416t^{50}
+\mathcal{O}(t^{52}).
\end{split}
\end{equation}

\begin{equation}
\begin{split}
\mathcal{I}_{n_1=2,n_2=1;\,N_1=1,N_2=5}^{\mathrm{BMN}}(t)
&=1+3t^2+9t^4+11t^6+15t^8+12t^{10}
+30t^{12}+24t^{14}+27t^{16}+10t^{18} \\
&\quad +69t^{20}+42t^{22}+22t^{24}-3t^{26}
+141t^{28}+41t^{30}-15t^{32}+48t^{34} \\
&\quad +248t^{36}-93t^{38}-36t^{40}+298t^{42}
+144t^{44}-429t^{46}+459t^{48}+672t^{50} \\
&\quad +\mathcal{O}(t^{52}).
\end{split}
\end{equation}

\begin{equation}
\begin{split}
\mathcal{I}_{n_1=1,n_2=1;\,N_1=3,N_2=4}^{\mathrm{BMN}}(t)
&=1+3t^2+3t^4+t^6+9t^{10}+8t^{12}
-15t^{14}-6t^{16}+36t^{18}+21t^{20} \\
&\quad -69t^{22}-16t^{24}+156t^{26}-30t^{28}
-287t^{30}+183t^{32}+477t^{34}-499t^{36} \\
&\quad -669t^{38}+1188t^{40}+695t^{42}-2448t^{44}
-117t^{46}+4346t^{48}-1779t^{50} \\
&\quad +\mathcal{O}(t^{52}).
\end{split}
\end{equation}

\begin{equation}
\begin{split}
\mathcal{I}_{n_1=3,n_2=1;\,N_1=1,N_2=4}^{\mathrm{BMN}}(t)
&=1+3t^2+9t^4+21t^6+36t^8+48t^{10}
+60t^{12}+84t^{14}+120t^{16}+160t^{18} \\
&\quad +171t^{20}+165t^{22}+224t^{24}+369t^{26}
+366t^{28}+231t^{30}+366t^{32}+738t^{34} \\
&\quad +554t^{36}+117t^{38}+810t^{40}+1543t^{42}
-162t^{44}-663t^{46}+2809t^{48}+3477t^{50} \\
&\quad +\mathcal{O}(t^{52}).
\end{split}
\end{equation}

\begin{equation}
\begin{split}
\mathcal{I}_{n_1=1,n_2=2;\,N_1=1,N_2=3}^{\mathrm{BMN}}(t)
&=1+3t^2+9t^4+11t^6+12t^8+12t^{10}
+30t^{12}+39t^{14}+6t^{16}-17t^{18} \\
&\quad +90t^{20}+156t^{22}-78t^{24}-240t^{26}
+354t^{28}+699t^{30}-768t^{32}-1359t^{34} \\
&\quad +2101t^{36}+2964t^{38}-4737t^{40}-5392t^{42}
+11214t^{44}+9420t^{46} \\
&\quad -24924t^{48}-13872t^{50}
+\mathcal{O}(t^{52}).
\end{split}
\end{equation}

\begin{equation}
\begin{split}
\mathcal{I}_{n_1=2,n_2=1;\,N_1=2,N_2=3}^{\mathrm{BMN}}(t)
&=1+3t^2+9t^4+11t^6+12t^8+6t^{10}
+30t^{12}+51t^{14}+39t^{16}-71t^{18} \\
&\quad -15t^{20}+240t^{22}+290t^{24}-447t^{26}
-633t^{28}+1068t^{30}+1923t^{32}-2175t^{34} \\
&\quad -4566t^{36}+4728t^{38}+10881t^{40}-9708t^{42}
-23835t^{44}+21000t^{46} \\
&\quad +50626t^{48}-44895t^{50}
+\mathcal{O}(t^{52}).
\end{split}
\end{equation}

\begin{equation}
\begin{split}
\mathcal{I}_{n_1=4,n_2=1;\,N_1=1,N_2=3}^{\mathrm{BMN}}(t)
&=1+3t^2+9t^4+21t^6+51t^8+84t^{10}
+135t^{12}+177t^{14}+261t^{16}+322t^{18} \\
&\quad +504t^{20}+660t^{22}+864t^{24}+810t^{26}
+1131t^{28}+1617t^{30}+2139t^{32}+1566t^{34} \\
&\quad +1599t^{36}+3351t^{38}+5892t^{40}+3293t^{42}
-2430t^{44}+2484t^{46} \\
&\quad +21288t^{48}+18864t^{50}
+\mathcal{O}(t^{52}).
\end{split}
\end{equation}

\begin{equation}
\begin{split}
\mathcal{I}_{n_1=1,n_2=3;\,N_1=1,N_2=2}^{\mathrm{BMN}}(t)
&=1+3t^2+9t^4+22t^6+33t^8+45t^{10}
+44t^{12}+69t^{14}+150t^{16}+245t^{18} \\
&\quad +204t^{20}-147t^{22}-163t^{24}+723t^{26}
+1950t^{28}+552t^{30}-3936t^{32}-3117t^{34} \\
&\quad +8741t^{36}+15966t^{38}-10299t^{40}-44244t^{42}
+903t^{44}+106893t^{46} \\
&\quad +58872t^{48}-214272t^{50}
+\mathcal{O}(t^{52}).
\end{split}
\end{equation}

\begin{equation}
\begin{split}
\mathcal{I}_{n_1=3,n_2=2;\,N_1=1,N_2=2}^{\mathrm{BMN}}(t)
&=1+3t^2+15t^4+32t^6+84t^8+135t^{10}
+222t^{12}+276t^{14}+381t^{16}+632t^{18} \\
&\quad +1050t^{20}+1521t^{22}+1403t^{24}+666t^{26}
+687t^{28}+3433t^{30}+8223t^{32}+7206t^{34} \\
&\quad -3432t^{36}-10803t^{38}+8268t^{40}+43919t^{42}
+34470t^{44}-41565t^{46} \\
&\quad -78833t^{48}+38163t^{50}
+\mathcal{O}(t^{52}).
\end{split}
\end{equation}

\begin{equation}
\begin{split}
\mathcal{I}_{n_1=5,n_2=1;\,N_1=1,N_2=2}^{\mathrm{BMN}}(t)
&=1+3t^2+9t^4+22t^6+51t^8+108t^{10}
+193t^{12}+315t^{14}+447t^{16}+625t^{18} \\
&\quad +822t^{20}+1200t^{22}+1858t^{24}+2865t^{26}
+3744t^{28}+3504t^{30}+2520t^{32}+2622t^{34} \\
&\quad +8378t^{36}+18180t^{38}+22536t^{40}+6485t^{42}
-21570t^{44}-13818t^{46} \\
&\quad +66223t^{48}+152421t^{50}
+\mathcal{O}(t^{52}).
\end{split}
\end{equation}
Among all double-partition sectors for $SU(7)$, the dominant
contribution to the BMN index comes from
$\mathcal{I}_{n_1=1,n_2=3;\,N_1=1,N_2=2}^{\mathrm{BMN}}(t)$. For this sector,
$d_{48}=58872$ and $d_{50}=-214272$, giving
\begin{equation}
\frac{1}{2}
\left(
\log |d_{48}|+\log |d_{50}|
\right)=11.6291.
\nonumber
\end{equation}
Adding all $11$ double-partition sectors for $N=7$, we obtain
\begin{equation}
\begin{split}
\mathcal{I}_{N=7}^{\mathrm{BMN~DP}}(t)
&=11+33t^2+87t^4+153t^6+300t^8+474t^{10}
+752t^{12}+1005t^{14} \\
&\quad +1446t^{16}+1979t^{18}+2883t^{20}
+3735t^{22}+4488t^{24}+4929t^{26} \\
&\quad +7569t^{28}+10901t^{30}+10857t^{32}
+5841t^{34}+12880t^{36}+34167t^{38} \\
&\quad +34632t^{40}-3889t^{42}-3081t^{44}
+84348t^{46}+99672t^{48}-62406t^{50} \\
&\quad +\mathcal{O}(t^{52}).
\end{split}
\end{equation}
\paragraph{$N=8.$}

\begin{equation}
\small
\begin{split}
\mathcal{I}_{n_1=1,n_2=1;\,N_1=1,N_2=7}^{\mathrm{BMN}}(t)
&=1+3t^2+3t^4+t^6+3t^8+6t^{10}-3t^{14}
+9t^{16}+9t^{18}-9t^{20} \\
&\quad +19t^{24}-6t^{26}-15t^{28}+27t^{30}
+12t^{32}-33t^{34}+18t^{36}+30t^{38}-48t^{40} \\
&\quad +10t^{42}+75t^{44}-69t^{46}-35t^{48}
+147t^{50}-78t^{52}-162t^{54}+258t^{56} \\
&\quad +54t^{58}-404t^{60}+246t^{62}+387t^{64}
+\mathcal{O}(t^{66}).
\end{split}
\end{equation}

\begin{equation}
\small
\begin{split}
\mathcal{I}_{n_1=1,n_2=1;\,N_1=2,N_2=6}^{\mathrm{BMN}}(t)
&=1+3t^2+3t^4+3t^8+9t^{10}-12t^{14}
+12t^{16}+27t^{18}-21t^{20}-33t^{22} \\
&\quad +57t^{24}+36t^{26}-96t^{28}-11t^{30}
+159t^{32}-51t^{34}-172t^{36}+153t^{38} \\
&\quad +123t^{40}-269t^{42}+102t^{44}+258t^{46}
-445t^{48}+60t^{50}+864t^{52}-969t^{54} \\
&\quad -897t^{56}+2469t^{58}+37t^{60}-4314t^{62}
+2592t^{64}+\mathcal{O}(t^{66}).
\end{split}
\end{equation}

\begin{equation}
\small
\begin{split}
\mathcal{I}_{n_1=2,n_2=1;\,N_1=1,N_2=6}^{\mathrm{BMN}}(t)
&=1+3t^2+9t^4+11t^6+15t^8+12t^{10}
+30t^{12}+24t^{14}+27t^{16}+10t^{18} \\
&\quad +69t^{20}+42t^{22}+21t^{24}-3t^{26}
+144t^{28}+49t^{30}-30t^{32}+21t^{34} \\
&\quad +276t^{36}-15t^{38}-99t^{40}+118t^{42}
+342t^{44}-150t^{46}+69t^{48}+111t^{50} \\
&\quad +129t^{52}+115t^{54}+474t^{56}-897t^{58}
+181t^{60}+1845t^{62}-384t^{64}
+\mathcal{O}(t^{66}).
\end{split}
\end{equation}

\begin{equation}
\small
\begin{split}
\mathcal{I}_{n_1=1,n_2=1;\,N_1=3,N_2=5}^{\mathrm{BMN}}(t)
&=1+3t^2+3t^4+3t^8+9t^{10}-15t^{14}
+15t^{16}+37t^{18}-33t^{20}-60t^{22} \\
&\quad +109t^{24}+75t^{26}-246t^{28}-39t^{30}
+531t^{32}-171t^{34}-908t^{36}+762t^{38} \\
&\quad +1332t^{40}-2119t^{42}-1320t^{44}+4512t^{46}
+52t^{48}-7995t^{50}+4224t^{52} \\
&\quad +11469t^{54}-13425t^{56}-12165t^{58}
+29331t^{60}+4014t^{62}-50517t^{64} \\
&\quad +\mathcal{O}(t^{66}).
\end{split}
\end{equation}

\begin{equation}
\small
\begin{split}
\mathcal{I}_{n_1=3,n_2=1;\,N_1=1,N_2=5}^{\mathrm{BMN}}(t)
&=1+3t^2+9t^4+21t^6+36t^8+48t^{10}
+60t^{12}+84t^{14}+120t^{16}+159t^{18} \\
&\quad +171t^{20}+162t^{22}+232t^{24}+378t^{26}
+372t^{28}+174t^{30}+315t^{32}+819t^{34} \\
&\quad +777t^{36}+57t^{38}+210t^{40}+1450t^{42}
+1458t^{44}-42t^{46}-18t^{48}+1938t^{50} \\
&\quad +2193t^{52}+678t^{54}+324t^{56}+234t^{58}
+1645t^{60}+6309t^{62}+2988t^{64} \\
&\quad +\mathcal{O}(t^{66}).
\end{split}
\end{equation}

\begin{equation}
\small
\begin{split}
\mathcal{I}_{n_1=2,n_2=1;\,N_1=2,N_2=4}^{\mathrm{BMN}}(t)
&=1+3t^2+9t^4+10t^6+12t^8+9t^{10}
+38t^{12}+36t^{14}+6t^{16}-45t^{18} \\
&\quad +120t^{20}+186t^{22}-111t^{24}-402t^{26}
+543t^{28}+1018t^{30}-1152t^{32}-2283t^{34} \\
&\quad +3368t^{36}+4884t^{38}-8142t^{40}-9634t^{42}
+20673t^{44}+17103t^{46}-48358t^{48} \\
&\quad -26613t^{50}+110253t^{52}+29824t^{54}
-236226t^{56}-2487t^{58}+483538t^{60} \\
&\quad -123366t^{62}-930297t^{64}
+\mathcal{O}(t^{66}).
\end{split}
\end{equation}

\begin{equation}
\small
\begin{split}
\mathcal{I}_{n_1=4,n_2=1;\,N_1=1,N_2=4}^{\mathrm{BMN}}(t)
&=1+3t^2+9t^4+21t^6+51t^8+84t^{10}
+134t^{12}+177t^{14}+258t^{16}+320t^{18} \\
&\quad +507t^{20}+690t^{22}+868t^{24}+786t^{26}
+1008t^{28}+1503t^{30}+2268t^{32}+2205t^{34} \\
&\quad +1851t^{36}+1800t^{38}+3906t^{40}+5548t^{42}
+4605t^{44}+1545t^{46}+4455t^{48} \\
&\quad +9273t^{50}+9303t^{52}+4635t^{54}+8904t^{56}
+8223t^{58}+1762t^{60}+16821t^{62} \\
&\quad +49206t^{64}+\mathcal{O}(t^{66}).
\end{split}
\end{equation}

\begin{equation}
\small
\begin{split}
\mathcal{I}_{n_1=1,n_2=2;\,N_1=2,N_2=3}^{\mathrm{BMN}}(t)
&=1+3t^2+9t^4+11t^6+12t^8+6t^{10}
+30t^{12}+48t^{14}+39t^{16}-72t^{18} \\
&\quad +6t^{20}+228t^{22}+263t^{24}-501t^{26}
-465t^{28}+1158t^{30}+1653t^{32}-2841t^{34} \\
&\quad -3585t^{36}+6333t^{38}+9168t^{40}-14994t^{42}
-19119t^{44}+33840t^{46}+40752t^{48} \\
&\quad -78210t^{50}-77934t^{52}+173563t^{54}
+143370t^{56}-381138t^{58}-233446t^{60} \\
&\quad +808179t^{62}+333345t^{64}
+\mathcal{O}(t^{66}).
\end{split}
\end{equation}

\begin{equation}
\small
\begin{split}
\mathcal{I}_{n_1=2,n_2=2;\,N_1=1,N_2=3}^{\mathrm{BMN}}(t)
&=1+3t^2+15t^4+21t^6+60t^8+30t^{10}
+142t^{12}+33t^{14}+354t^{16}-91t^{18} \\
&\quad +525t^{20}-234t^{22}+1434t^{24}-609t^{26}
+1029t^{28}-1543t^{30}+5067t^{32}-1269t^{34} \\
&\quad -974t^{36}-7062t^{38}+19638t^{40}+3739t^{42}
-21648t^{44}-32166t^{46}+84817t^{48} \\
&\quad +42846t^{50}-147435t^{52}-137694t^{54}
+395277t^{56}+224769t^{58}-794635t^{60} \\
&\quad -499281t^{62}+1834701t^{64}
+\mathcal{O}(t^{66}).
\end{split}
\end{equation}

\begin{equation}
\small
\begin{split}
\mathcal{I}_{n_1=5,n_2=1;\,N_1=1,N_2=3}^{\mathrm{BMN}}(t)
&=1+3t^2+9t^4+21t^6+51t^8+105t^{10}
+190t^{12}+303t^{14}+456t^{16}+642t^{18} \\
&\quad +867t^{20}+1188t^{22}+1770t^{24}+2529t^{26}
+3225t^{28}+3573t^{30}+4218t^{32}+5610t^{34} \\
&\quad +7933t^{36}+9924t^{38}+10638t^{40}+10471t^{42}
+14022t^{44}+22581t^{46}+25064t^{48} \\
&\quad +10995t^{50}+10680t^{52}+62181t^{54}
+104466t^{56}-1398t^{58}-155981t^{60} \\
&\quad +41301t^{62}+565959t^{64}
+\mathcal{O}(t^{66}).
\end{split}
\end{equation}

\begin{equation}
\small
\begin{split}
\mathcal{I}_{n_1=2,n_2=3;\,N_1=1,N_2=2}^{\mathrm{BMN}}(t)
&=1+3t^2+15t^4+32t^6+84t^8+129t^{10}
+214t^{12}+246t^{14}+390t^{16}+622t^{18} \\
&\quad +1179t^{20}+1461t^{22}+1167t^{24}-138t^{26}
+747t^{28}+5373t^{30}+11487t^{32}+4461t^{34} \\
&\quad -16031t^{36}-20808t^{38}+35640t^{40}+101244t^{42}
+25368t^{44}-216009t^{46}-228334t^{48} \\
&\quad +343104t^{50}+856812t^{52}-88623t^{54}
-1965891t^{56}-1255290t^{58}+3406662t^{60} \\
&\quad +5142297t^{62}-3386178t^{64}
+\mathcal{O}(t^{66}).
\end{split}
\end{equation}

\begin{equation}
\small
\begin{split}
\mathcal{I}_{n_1=4,n_2=2;\,N_1=1,N_2=2}^{\mathrm{BMN}}(t)
&=1+3t^2+15t^4+32t^6+99t^8+171t^{10}
+396t^{12}+501t^{14}+960t^{16}+933t^{18} \\
&\quad +2061t^{20}+2172t^{22}+5186t^{24}+4830t^{26}
+8145t^{28}+2354t^{30}+7464t^{32}+8685t^{34} \\
&\quad +37434t^{36}+35310t^{38}+22317t^{40}-53170t^{42}
-15171t^{44}+114156t^{46}+324945t^{48} \\
&\quad +96444t^{50}-405285t^{52}-610859t^{54}
+607311t^{56}+1766652t^{58}+509186t^{60} \\
&\quad -3094629t^{62}-2121636t^{64}
+\mathcal{O}(t^{66}).
\end{split}
\end{equation}

\begin{equation}
\small
\begin{split}
\mathcal{I}_{n_1=6,n_2=1;\,N_1=1,N_2=2}^{\mathrm{BMN}}(t)
&=1+3t^2+9t^4+22t^6+51t^8+108t^{10}
+221t^{12}+393t^{14}+660t^{16}+988t^{18} \\
&\quad +1443t^{20}+1956t^{22}+2740t^{24}+3918t^{26}
+6033t^{28}+8989t^{30}+12150t^{32}+12936t^{34} \\
&\quad +11253t^{36}+9942t^{38}+20982t^{40}+50025t^{42}
+85890t^{44}+79923t^{46}-6011t^{48} \\
&\quad -116427t^{50}-42594t^{52}+372282t^{54}
+832881t^{56}+470820t^{58}-1112154t^{60} \\
&\quad -2314026t^{62}+73026t^{64}
+\mathcal{O}(t^{66}).
\end{split}
\end{equation}
Among all double-partition sectors for $SU(8)$, the dominant
contribution to the BMN index comes from
$\mathcal{I}_{n_1=2,n_2=3;\,N_1=1,N_2=2}^{\mathrm{BMN}}(t)$.
For this sector, the coefficient at order $t^{64}$ is
$d_{64}=-3386178$. It therefore has the largest absolute
degeneracy,
\begin{equation}
|d_{64}|=3386178,
\qquad
\log |d_{64}|
=15.0352.
\nonumber
\end{equation}
Adding all $13$ double-partition sectors for $N=8$, we obtain
\begin{equation}
\begin{split}
\mathcal{I}_{N=8}^{\mathrm{BMN~DP}}(t)
&=13+39t^2+117t^4+203t^6+480t^8+726t^{10}
+1455t^{12}+1815t^{14} \\
&\quad +3306t^{16}+3539t^{18}+6885t^{20}
+7758t^{22}+13755t^{24}+10893t^{26} \\
&\quad +20424t^{28}+22625t^{30}+44142t^{32}
+28089t^{34}+41240t^{36}+41310t^{38} \\
&\quad +115665t^{40}+92419t^{42}+95277t^{44}
+25482t^{46}+196953t^{48}+275673t^{50} \\
&\quad +321132t^{52}-183560t^{54}-123174t^{56}
+819846t^{58}+2135722t^{60} \\
&\quad -14604t^{62}-3626808t^{64}
+\mathcal{O}(t^{66}).
\end{split}
\end{equation}
\paragraph{$N=9.$}
\begin{equation}
\small
\begin{split}
\mathcal{I}_{n_1=1,n_2=1;\,N_1=1,N_2=8}^{\mathrm{BMN}}(t)
&=1 + 3 t^2 + 3 t^{4} + t^{6} + 3 t^{8} + 6 t^{10}
- 3 t^{14} + 9 t^{16} + 9 t^{18} - 9 t^{20} \\
&\quad + 19 t^{24} - 6 t^{26} - 15 t^{28} + 27 t^{30}
+ 12 t^{32} - 33 t^{34} + 17 t^{36} + 33 t^{38} \\
&\quad - 48 t^{40} + 2 t^{42} + 75 t^{44} - 48 t^{46}
- 44 t^{48} + 111 t^{50} - 42 t^{52} \\
&\quad - 118 t^{54} + 174 t^{56} + 21 t^{58} - 251 t^{60}
+ 213 t^{62} + 186 t^{64} - 475 t^{66} \\
&\quad + 135 t^{68} + 576 t^{70} - 641 t^{72} - 255 t^{74}
+ 1080 t^{76} - 540 t^{78} - 1011 t^{80} \\
&\quad + 1548 t^{82} + \mathcal{O}(t^{84}).
\end{split}
\end{equation}

\begin{equation}
\small
\begin{split}
\mathcal{I}_{n_1=1,n_2=1;\,N_1=2,N_2=7}^{\mathrm{BMN}}(t)
&=1 + 3 t^2 + 3 t^{4} + 3 t^{8} + 9 t^{10}
- 12 t^{14} + 12 t^{16} + 27 t^{18} - 21 t^{20} \\
&\quad - 33 t^{22} + 56 t^{24} + 39 t^{26} - 96 t^{28}
- 19 t^{30} + 159 t^{32} - 21 t^{34} \\
&\quad - 198 t^{36} + 99 t^{38} + 204 t^{40} - 171 t^{42}
- 141 t^{44} + 192 t^{46} + 56 t^{48} \\
&\quad - 36 t^{50} - 54 t^{52} - 316 t^{54} + 363 t^{56}
+ 780 t^{58} - 1251 t^{60} - 939 t^{62} \\
&\quad + 2853 t^{64} + 170 t^{66} - 4812 t^{68} + 2223 t^{70}
+ 6124 t^{72} - 6405 t^{74} \\
&\quad - 5103 t^{76} + 11296 t^{78} + 327 t^{80}
- 14226 t^{82} + \mathcal{O}(t^{84}).
\end{split}
\end{equation}

\begin{equation}
\small
\begin{split}
\mathcal{I}_{n_1=2,n_2=1;\,N_1=1,N_2=7}^{\mathrm{BMN}}(t)
&=1 + 3 t^2 + 9 t^{4} + 11 t^{6} + 15 t^{8}
+ 12 t^{10} + 30 t^{12} + 24 t^{14} + 27 t^{16} \\
&\quad + 10 t^{18} + 69 t^{20} + 42 t^{22} + 21 t^{24}
- 3 t^{26} + 144 t^{28} + 48 t^{30} \\
&\quad - 30 t^{32} + 24 t^{34} + 284 t^{36} - 30 t^{38}
- 126 t^{40} + 146 t^{42} + 420 t^{44} \\
&\quad - 213 t^{46} - 111 t^{48} + 309 t^{50} + 408 t^{52}
- 274 t^{54} - 90 t^{56} + 150 t^{58} \\
&\quad + 700 t^{60} + 132 t^{62} - 1029 t^{64} - 58 t^{66}
+ 2616 t^{68} - 504 t^{70} - 3864 t^{72} \\
&\quad + 2625 t^{74} + 5493 t^{76} - 5988 t^{78}
- 4872 t^{80} + 10656 t^{82} + \mathcal{O}(t^{84}).
\end{split}
\end{equation}

\begin{equation}
\small
\begin{split}
\mathcal{I}_{n_1=1,n_2=1;\,N_1=3,N_2=6}^{\mathrm{BMN}}(t)
&=1 + 3 t^2 + 3 t^{4} + 3 t^{8} + 9 t^{10}
- t^{12} - 12 t^{14} + 15 t^{16} + 29 t^{18} \\
&\quad - 33 t^{20} - 30 t^{22} + 83 t^{24} + 21 t^{26}
- 165 t^{28} + 68 t^{30} + 270 t^{32} \\
&\quad - 264 t^{34} - 307 t^{36} + 666 t^{38} + 108 t^{40}
- 1261 t^{42} + 675 t^{44} + 1773 t^{46} \\
&\quad - 2317 t^{48} - 1545 t^{50} + 5076 t^{52} - 781 t^{54}
- 8190 t^{56} + 6798 t^{58} \\
&\quad + 9362 t^{60} - 17487 t^{62} - 4053 t^{64} + 31505 t^{66}
- 14280 t^{68} - 41613 t^{70} \\
&\quad + 50444 t^{72} + 33270 t^{74} - 101895 t^{76}
+ 15074 t^{78} + 149364 t^{80} - 123201 t^{82}
+ \mathcal{O}(t^{84}).
\end{split}
\end{equation}

\begin{equation}
\small
\begin{split}
\mathcal{I}_{n_1=3,n_2=1;\,N_1=1,N_2=6}^{\mathrm{BMN}}(t)
&=1 + 3 t^2 + 9 t^{4} + 21 t^{6} + 36 t^{8}
+ 48 t^{10} + 60 t^{12} + 84 t^{14} + 120 t^{16} \\
&\quad + 159 t^{18} + 171 t^{20} + 162 t^{22} + 231 t^{24}
+ 378 t^{26} + 369 t^{28} + 182 t^{30} \\
&\quad + 324 t^{32} + 825 t^{34} + 720 t^{36} + 6 t^{38}
+ 291 t^{40} + 1674 t^{42} + 1398 t^{44} \\
&\quad - 642 t^{46} - 119 t^{48} + 3270 t^{50} + 2838 t^{52}
- 2185 t^{54} - 1542 t^{56} \\
&\quad + 6015 t^{58} + 5913 t^{60} - 4596 t^{62} - 5649 t^{64}
+ 9146 t^{66} + 13971 t^{68} \\
&\quad - 6165 t^{70} - 18974 t^{72} + 10374 t^{74} + 39225 t^{76}
- 6711 t^{78} - 60750 t^{80} \\
&\quad + 16626 t^{82} + \mathcal{O}(t^{84}).
\end{split}
\end{equation}

\begin{equation}
\small
\begin{split}
\mathcal{I}_{n_1=1,n_2=1;\,N_1=4,N_2=5}^{\mathrm{BMN}}(t)
&=1 + 3 t^2 + 3 t^{4} + t^{6} + 9 t^{10}
+ 8 t^{12} - 15 t^{14} - 6 t^{16} + 35 t^{18} \\
&\quad + 21 t^{20} - 66 t^{22} - 15 t^{24} + 147 t^{26}
- 36 t^{28} - 251 t^{30} + 186 t^{32} \\
&\quad + 390 t^{34} - 500 t^{36} - 456 t^{38} + 1140 t^{40}
+ 216 t^{42} - 2166 t^{44} + 786 t^{46} \\
&\quad + 3422 t^{48} - 3237 t^{50} - 4044 t^{52} + 7871 t^{54}
+ 2250 t^{56} - 14622 t^{58} \\
&\quad + 4792 t^{60} + 21585 t^{62} - 20724 t^{64} - 23007 t^{66}
+ 47862 t^{68} + 8130 t^{70} \\
&\quad - 82501 t^{72} + 37773 t^{74} + 108726 t^{76}
- 127683 t^{78} - 92019 t^{80} + 258714 t^{82}
+ \mathcal{O}(t^{84}).
\end{split}
\end{equation}

\begin{equation}
\small
\begin{split}
\mathcal{I}_{n_1=2,n_2=1;\,N_1=2,N_2=5}^{\mathrm{BMN}}(t)
&=1 + 3 t^2 + 9 t^{4} + 10 t^{6} + 12 t^{8}
+ 9 t^{10} + 37 t^{12} + 36 t^{14} + 9 t^{16} \\
&\quad - 36 t^{18} + 102 t^{20} + 156 t^{22} - 67 t^{24}
- 270 t^{26} + 369 t^{28} + 615 t^{30} \\
&\quad - 639 t^{32} - 1056 t^{34} + 1780 t^{36} + 1722 t^{38}
- 3630 t^{40} - 1990 t^{42} \\
&\quad + 7950 t^{44} + 705 t^{46} - 15033 t^{48} + 5643 t^{50}
+ 26427 t^{52} - 23517 t^{54} \\
&\quad - 39024 t^{56} + 65670 t^{58} + 43474 t^{60} - 149379 t^{62}
- 11277 t^{64} + 294645 t^{66} \\
&\quad - 117078 t^{68} - 504981 t^{70} + 452715 t^{72}
+ 732996 t^{74} - 1163223 t^{76} \\
&\quad - 806446 t^{78} + 2450727 t^{80} + 326301 t^{82}
+ \mathcal{O}(t^{84}).
\end{split}
\end{equation}

\begin{equation}
\small
\begin{split}
\mathcal{I}_{n_1=4,n_2=1;\,N_1=1,N_2=5}^{\mathrm{BMN}}(t)
&=1 + 3 t^2 + 9 t^{4} + 21 t^{6} + 51 t^{8}
+ 84 t^{10} + 134 t^{12} + 177 t^{14} + 258 t^{16} \\
&\quad + 319 t^{18} + 507 t^{20} + 687 t^{22} + 866 t^{24}
+ 789 t^{26} + 1038 t^{28} + 1508 t^{30} \\
&\quad + 2244 t^{32} + 2088 t^{34} + 1729 t^{36} + 1926 t^{38}
+ 4566 t^{40} + 5876 t^{42} \\
&\quad + 3078 t^{44} - 558 t^{46} + 6083 t^{48} + 15855 t^{50}
+ 10152 t^{52} - 8900 t^{54} \\
&\quad - 2856 t^{56} + 32391 t^{58} + 38989 t^{60} - 16152 t^{62}
- 44406 t^{64} + 35027 t^{66} \\
&\quad + 114723 t^{68} + 21564 t^{70} - 136026 t^{72} - 54033 t^{74}
+ 229680 t^{76} + 230776 t^{78} \\
&\quad - 218010 t^{80} - 427608 t^{82} + \mathcal{O}(t^{84}).
\end{split}
\end{equation}

\begin{equation}
\small
\begin{split}
\mathcal{I}_{n_1=1,n_2=2;\,N_1=1,N_2=4}^{\mathrm{BMN}}(t)
&=1 + 3 t^2 + 9 t^{4} + 11 t^{6} + 12 t^{8}
+ 12 t^{10} + 29 t^{12} + 39 t^{14} + 9 t^{16} \\
&\quad - 8 t^{18} + 72 t^{20} + 117 t^{22} - 43 t^{24}
- 96 t^{26} + 270 t^{28} + 266 t^{30} \\
&\quad - 522 t^{32} - 192 t^{34} + 1414 t^{36} + 54 t^{38}
- 2748 t^{40} + 1243 t^{42} + 5418 t^{44} \\
&\quad - 4584 t^{46} - 8902 t^{48} + 13284 t^{50} + 12225 t^{52}
- 31091 t^{54} - 9300 t^{56} \\
&\quad + 64512 t^{58} - 11999 t^{60} - 116661 t^{62} + 78099 t^{64}
+ 181045 t^{66} - 231354 t^{68} \\
&\quad - 220113 t^{70} + 530744 t^{72} + 137205 t^{74}
- 1023702 t^{76} + 267878 t^{78} \\
&\quad + 1683324 t^{80} - 1325865 t^{82} + \mathcal{O}(t^{84}).
\end{split}
\end{equation}

\begin{equation}
\small
\begin{split}
\mathcal{I}_{n_1=5,n_2=1;\,N_1=1,N_2=4}^{\mathrm{BMN}}(t)
&=1 + 3 t^2 + 9 t^{4} + 21 t^{6} + 51 t^{8}
+ 105 t^{10} + 189 t^{12} + 303 t^{14} + 453 t^{16} \\
&\quad + 640 t^{18} + 855 t^{20} + 1203 t^{22} + 1790 t^{24}
+ 2577 t^{26} + 3195 t^{28} \\
&\quad + 3527 t^{30} + 3915 t^{32} + 5256 t^{34} + 8165 t^{36}
+ 11664 t^{38} + 12018 t^{40} \\
&\quad + 7796 t^{42} + 7002 t^{44} + 20271 t^{46} + 39266 t^{48}
+ 35844 t^{50} + 4359 t^{52} \\
&\quad - 10477 t^{54} + 41592 t^{56} + 122511 t^{58} + 109549 t^{60}
- 37887 t^{62} - 121791 t^{64} \\
&\quad + 83326 t^{66} + 391926 t^{68} + 300864 t^{70} - 232303 t^{72}
- 497337 t^{74} + 92775 t^{76} \\
&\quad + 1105352 t^{78} + 1186749 t^{80} - 697848 t^{82}
+ \mathcal{O}(t^{84}).
\end{split}
\end{equation}

\begin{equation}
\small
\begin{split}
\mathcal{I}_{n_1=3,n_2=2;\,N_1=1,N_2=3}^{\mathrm{BMN}}(t)
&=1 + 3 t^2 + 15 t^{4} + 31 t^{6} + 81 t^{8}
+ 126 t^{10} + 218 t^{12} + 291 t^{14} + 450 t^{16} \\
&\quad + 665 t^{18} + 858 t^{20} + 972 t^{22} + 1205 t^{24}
+ 1974 t^{26} + 2913 t^{28} + 2396 t^{30} \\
&\quad + 1170 t^{32} + 3087 t^{34} + 9624 t^{36} + 9540 t^{38}
- 6624 t^{40} - 10568 t^{42} \\
&\quad + 34668 t^{44} + 65379 t^{46} - 41180 t^{48}
- 150354 t^{50} + 102093 t^{52} + 444518 t^{54} \\
&\quad - 114450 t^{56} - 1068429 t^{58} + 143475 t^{60}
+ 2654580 t^{62} + 98136 t^{64} \\
&\quad - 6164882 t^{66} - 815346 t^{68} + 14242368 t^{70}
+ 3107026 t^{72} - 31968423 t^{74} \\
&\quad - 8631735 t^{76} + 70929950 t^{78} + 21421746 t^{80}
- 154699392 t^{82} + \mathcal{O}(t^{84}).
\end{split}
\end{equation}

\begin{equation}
\small
\begin{split}
\mathcal{I}_{n_1=3,n_2=1;\,N_1=2,N_2=3}^{\mathrm{BMN}}(t)
&=1 + 3 t^2 + 9 t^{4} + 21 t^{6} + 33 t^{8}
+ 42 t^{10} + 42 t^{12} + 75 t^{14} + 150 t^{16} \\
&\quad + 230 t^{18} + 138 t^{20} - 165 t^{22} - 5 t^{24}
+ 1068 t^{26} + 1680 t^{28} - 989 t^{30} \\
&\quad - 4470 t^{32} + 1371 t^{34} + 14479 t^{36} + 5859 t^{38}
- 35274 t^{40} - 30899 t^{42} \\
&\quad + 82905 t^{44} + 116970 t^{46} - 168225 t^{48}
- 356637 t^{50} + 314928 t^{52} + 993251 t^{54} \\
&\quad - 506061 t^{56} - 2561571 t^{58} + 632254 t^{60}
+ 6284934 t^{62} - 261567 t^{64} \\
&\quad - 14805680 t^{66} - 1787931 t^{68} + 33824334 t^{70}
+ 8542948 t^{72} - 75328278 t^{74} \\
&\quad - 26997420 t^{76} + 164343520 t^{78} + 72456876 t^{80}
- 352417971 t^{82} + \mathcal{O}(t^{84}).
\end{split}
\end{equation}

\begin{equation}
\small
\begin{split}
\mathcal{I}_{n_1=6,n_2=1;\,N_1=1,N_2=3}^{\mathrm{BMN}}(t)
&=1 + 3 t^2 + 9 t^{4} + 21 t^{6} + 51 t^{8}
+ 105 t^{10} + 218 t^{12} + 381 t^{14} + 648 t^{16} \\
&\quad + 978 t^{18} + 1473 t^{20} + 1992 t^{22} + 2799 t^{24}
+ 3849 t^{26} + 5784 t^{28} \\
&\quad + 8083 t^{30} + 10905 t^{32} + 12798 t^{34} + 14984 t^{36}
+ 17691 t^{38} + 24660 t^{40} \\
&\quad + 35589 t^{42} + 48489 t^{44} + 51450 t^{46} + 44309 t^{48}
+ 46632 t^{50} + 87024 t^{52} \\
&\quad + 138218 t^{54} + 134565 t^{56} + 84609 t^{58} + 124746 t^{60}
+ 291225 t^{62} + 299454 t^{64} \\
&\quad - 71340 t^{66} - 214620 t^{68} + 898077 t^{70} + 2380804 t^{72}
+ 619239 t^{74} \\
&\quad - 4987680 t^{76} - 4913589 t^{78} + 10333356 t^{80}
+ 22778742 t^{82} + \mathcal{O}(t^{84}).
\end{split}
\end{equation}

\begin{equation}
\small
\begin{split}
\mathcal{I}_{n_1=1,n_2=4;\,N_1=1,N_2=2}^{\mathrm{BMN}}(t)
&=1 + 3 t^2 + 9 t^{4} + 22 t^{6} + 48 t^{8}
+ 81 t^{10} + 118 t^{12} + 147 t^{14} + 201 t^{16} \\
&\quad + 342 t^{18} + 621 t^{20} + 888 t^{22} + 671 t^{24}
- 186 t^{26} - 234 t^{28} + 2806 t^{30} \\
&\quad + 7818 t^{32} + 5184 t^{34} - 12540 t^{36} - 23976 t^{38}
+ 17550 t^{40} + 97746 t^{42} \\
&\quad + 57243 t^{44} - 214476 t^{46} - 351613 t^{48}
+ 297384 t^{50} + 1249500 t^{52} \\
&\quad + 242568 t^{54} - 3168966 t^{56} - 3070053 t^{58}
+ 6132353 t^{60} + 12456831 t^{62} \\
&\quad - 7251927 t^{64} - 36691095 t^{66} - 5168865 t^{68}
+ 88738329 t^{70} + 65999337 t^{72} \\
&\quad - 179397945 t^{74} - 260453964 t^{76} + 290271939 t^{78}
+ 771306303 t^{80} - 293176299 t^{82} \\
&\quad + \mathcal{O}(t^{84}).
\end{split}
\end{equation}

\begin{equation}
\small
\begin{split}
\mathcal{I}_{n_1=3,n_2=3;\,N_1=1,N_2=2}^{\mathrm{BMN}}(t)
&=1 + 3 t^2 + 15 t^{4} + 42 t^{6} + 105 t^{8}
+ 231 t^{10} + 408 t^{12} + 675 t^{14} \\
&\quad + 948 t^{16} + 1373 t^{18} + 2127 t^{20} + 3645 t^{22}
+ 5804 t^{24} + 7116 t^{26} \\
&\quad + 6003 t^{28} + 3083 t^{30} + 7488 t^{32} + 27141 t^{34}
+ 52805 t^{36} + 39762 t^{38} \\
&\quad - 41616 t^{40} - 101430 t^{42} + 57087 t^{44}
+ 440787 t^{46} + 485273 t^{48} - 439575 t^{50} \\
&\quad - 1563429 t^{52} - 426560 t^{54} + 3523104 t^{56}
+ 4733538 t^{58} - 3210030 t^{60} \\
&\quad - 13141149 t^{62} - 4440150 t^{64} + 23702063 t^{66}
+ 28344744 t^{68} - 21912339 t^{70} \\
&\quad - 64499264 t^{72} - 8252244 t^{74} + 81913014 t^{76}
+ 48466577 t^{78} - 17589354 t^{80} \\
&\quad + 75647319 t^{82} + \mathcal{O}(t^{84}).
\end{split}
\end{equation}

\begin{equation}
\small
\begin{split}
\mathcal{I}_{n_1=5,n_2=2;\,N_1=1,N_2=2}^{\mathrm{BMN}}(t)
&=1 + 3 t^2 + 15 t^{4} + 32 t^{6} + 99 t^{8}
+ 192 t^{10} + 451 t^{12} + 768 t^{14} \\
&\quad + 1377 t^{16} + 2013 t^{18} + 3006 t^{20} + 4272 t^{22}
+ 6386 t^{24} + 10044 t^{26} \\
&\quad + 15351 t^{28} + 21697 t^{30} + 24705 t^{32} + 22152 t^{34}
+ 20315 t^{36} + 40476 t^{38} \\
&\quad + 101934 t^{40} + 171115 t^{42} + 150066 t^{44}
- 41694 t^{46} - 237251 t^{48} + 21477 t^{50} \\
&\quad + 978951 t^{52} + 1742849 t^{54} + 399759 t^{56}
- 3202290 t^{58} - 4306523 t^{60} \\
&\quad + 3367593 t^{62} + 15065193 t^{64} + 9687620 t^{66}
- 23509332 t^{68} - 43685763 t^{70} \\
&\quad + 15687760 t^{72} + 116851893 t^{74} + 62514927 t^{76}
- 225207839 t^{78} - 321293121 t^{80} \\
&\quad + 319972134 t^{82} + \mathcal{O}(t^{84}).
\end{split}
\end{equation}

\begin{equation}
\small
\begin{split}
\mathcal{I}_{n_1=7,n_2=1;\,N_1=1,N_2=2}^{\mathrm{BMN}}(t)
&=1 + 3 t^2 + 9 t^{4} + 22 t^{6} + 51 t^{8}
+ 108 t^{10} + 221 t^{12} + 429 t^{14} + 765 t^{16} \\
&\quad + 1288 t^{18} + 2001 t^{20} + 2997 t^{22} + 4225 t^{24}
+ 5907 t^{26} + 8205 t^{28} \\
&\quad + 11899 t^{30} + 17721 t^{32} + 26223 t^{34} + 35752 t^{36}
+ 41616 t^{38} + 40809 t^{40} \\
&\quad + 36782 t^{42} + 53127 t^{44} + 117471 t^{46} + 237130 t^{48}
+ 330264 t^{50} + 235767 t^{52} \\
&\quad - 143124 t^{54} - 521685 t^{56} - 97044 t^{58}
+ 1749582 t^{60} + 3931350 t^{62} \\
&\quad + 2835285 t^{64} - 4456060 t^{66} - 13099848 t^{68}
- 7583478 t^{70} + 23179224 t^{72} \\
&\quad + 55283694 t^{74} + 24566034 t^{76} - 100057394 t^{78}
- 196556175 t^{80} - 15832101 t^{82} \\
&\quad + \mathcal{O}(t^{84}).
\end{split}
\end{equation}

Among all double-partition sectors for $SU(9)$, the dominant
contribution to the BMN index comes from
$\mathcal{I}_{n_1=1,n_2=4;\,N_1=1,N_2=2}^{\mathrm{BMN}}(t)$.
For this sector,
$d_{80}=771306303$ and $d_{82}=-293176299$, and hence
\begin{equation}
\begin{split}
\frac{1}{2}\left(\log|d_{80}|+\log|d_{82}|\right)
&= 19.9799.
\end{split}
\nonumber
\end{equation}

Adding all $17$ double-partition indices for $N=9$, we obtain
\begin{equation}
\small
\begin{split}
\mathcal{I}_{N=9}^{\mathrm{BMN~DP}}(t)
&=17 + 51 t^2 + 147 t^{4} + 288 t^{6} + 654 t^{8}
+ 1188 t^{10} + 2162 t^{12} + 3387 t^{14} \\
&\quad + 5445 t^{16} + 8073 t^{18} + 11958 t^{20}
+ 16839 t^{22} + 24026 t^{24} + 33348 t^{26} \\
&\quad + 44775 t^{28} + 54946 t^{30} + 71256 t^{32}
+ 104973 t^{34} + 148523 t^{36} + 146652 t^{38} \\
&\quad + 113214 t^{40} + 211866 t^{42} + 507294 t^{44}
+ 553569 t^{46} - 9256 t^{48} - 181311 t^{50} \\
&\quad + 1462179 t^{52} + 2921932 t^{54} - 270357 t^{56}
- 4897014 t^{58} + 1465135 t^{60} \\
&\quad + 15524193 t^{62} + 6216633 t^{64} - 28188050 t^{66}
- 16047489 t^{68} + 64081509 t^{70} \\
&\quad + 54963553 t^{72} - 121795851 t^{74} - 133893768 t^{76}
+ 244516172 t^{78} + 345173460 t^{80} \\
&\quad - 399702471 t^{82} + \mathcal{O}(t^{84}).
\end{split}
\end{equation}

\subsection{Triple partition $SU(N)$ BMN index}
\label{tpbmn}
We fix $ N $, and then enumerate all BMN indices in the sector corresponding to $ N = n_1 N_1 + n_2 N_2 +n_3N_3$, i.e., the triple partition sector.
\paragraph{$N=6.$}
\begin{equation}
\begin{split}
\mathcal{I}_{n_1=1, n_2=1, n_3=1; N_1=1, N_2=2, N_3=3}^{\mathrm{BMN}}(t)
&=1 + 6\,t^2 + 15\,t^4 + 22\,t^6 + 21\,t^8 + 24\,t^{10} + 58\,t^{12} + 81\,t^{14} \\
&\quad + 24\,t^{16} - 41\,t^{18} + 99\,t^{20} + 342\,t^{22} + 62\,t^{24} - 498\,t^{26} \\
&\quad + 15\,t^{28} + 1379\,t^{30} + 501\,t^{32} - 2649\,t^{34} - 1050\,t^{36} \\
&\quad + \mathcal{O}(t^{38}).
\end{split}
\end{equation}
We have
\begin{equation}
\begin{split}
\mathcal{I}_{N=6}^{\mathrm{BMN~TP}}(t)
&=1 + 6\,t^2 + 15\,t^4 + 22\,t^6 + 21\,t^8 + 24\,t^{10} + 58\,t^{12} + 81\,t^{14} \\
&\quad + 24\,t^{16} - 41\,t^{18} + 99\,t^{20} + 342\,t^{22} + 62\,t^{24} - 498\,t^{26} \\
&\quad + 15\,t^{28} + 1379\,t^{30} + 501\,t^{32} - 2649\,t^{34} - 1050\,t^{36} \\
&\quad + \mathcal{O}(t^{38}).
\end{split}
\end{equation}

\paragraph{$N=7.$}
\begin{equation}
\begin{split}
\mathcal{I}_{n_1=2, n_2=1, n_3=1; N_1=1, N_2=2, N_3=3}^{\mathrm{BMN}}(t)
&=1 + 6\,t^2 + 21\,t^4 + 50\,t^6 + 84\,t^8 + 114\,t^{10} + 147\,t^{12} \\
&\quad + 240\,t^{14} + 390\,t^{16} + 430\,t^{18} + 273\,t^{20} + 288\,t^{22} \\
&\quad + 1029\,t^{24} + 1680\,t^{26} + 417\,t^{28} - 1318\,t^{30} + 1431\,t^{32} \\
&\quad + 7278\,t^{34} + 2958\,t^{36} - 12090\,t^{38} - 6009\,t^{40} \\
&\quad + 30662\,t^{42} + 28152\,t^{44} - 52746\,t^{46} - 66501\,t^{48} \\
&\quad + 102885\,t^{50} + \mathcal{O}(t^{52}).
\end{split}
\end{equation}

\begin{equation}
\begin{split}
\mathcal{I}_{n_1=1, n_2=1, n_3=1; N_1=1, N_2=2, N_3=4}^{\mathrm{BMN}}(t)
&=1 + 6\,t^2 + 15\,t^4 + 21\,t^6 + 21\,t^8 + 30\,t^{10} + 57\,t^{12} \\
&\quad + 60\,t^{14} + 18\,t^{16} + 30\,t^{18} + 135\,t^{20} + 126\,t^{22} \\
&\quad - 56\,t^{24} + 36\,t^{26} + 393\,t^{28} + 81\,t^{30} - 516\,t^{32} \\
&\quad + 381\,t^{34} + 1334\,t^{36} - 822\,t^{38} - 2022\,t^{40} \\
&\quad + 2756\,t^{42} + 3594\,t^{44} - 5955\,t^{46} - 4634\,t^{48} \\
&\quad + 12792\,t^{50} + \mathcal{O}(t^{52}).
\end{split}
\end{equation}
The dominant individual triple-partition sector is
\begin{equation}
\mathcal{I}_{n_1=2, n_2=1, n_3=1; N_1=1, N_2=2, N_3=3}^{\mathrm{BMN}}(t),
\nonumber
\end{equation}
for which
\begin{equation}
d_{48}=-66501,
\qquad
d_{50}=102885.
\nonumber
\end{equation}
Therefore,
\begin{equation}
\frac{1}{2}\Big(\log |d_{48}|+\log |d_{50}|\Big)=11.3232.
\nonumber
\end{equation}

Adding the two triple-partition sectors for $N=7$, we obtain
\begin{equation}
\begin{split}
\mathcal{I}_{N=7}^{\mathrm{BMN~TP}}(t)
&=2 + 12\,t^2 + 36\,t^4 + 71\,t^6 + 105\,t^8 + 144\,t^{10} + 204\,t^{12} \\
&\quad + 300\,t^{14} + 408\,t^{16} + 460\,t^{18} + 408\,t^{20} + 414\,t^{22} \\
&\quad + 973\,t^{24} + 1716\,t^{26} + 810\,t^{28} - 1237\,t^{30} + 915\,t^{32} \\
&\quad + 7659\,t^{34} + 4292\,t^{36} - 12912\,t^{38} - 8031\,t^{40} \\
&\quad + 33418\,t^{42} + 31746\,t^{44} - 58701\,t^{46} - 71135\,t^{48} \\
&\quad + 115677\,t^{50} + \mathcal{O}(t^{52}).
\end{split}
\end{equation}

\paragraph{$N=8.$}

\begin{equation}
\begin{split}
\mathcal{I}_{n_1=1,n_2=1,n_3=1;\,N_1=1,N_2=2,N_3=5}^{\mathrm{BMN}}(t)
&=1+6t^2+15t^4+21t^6+21t^8+30t^{10}
+56t^{12}+60t^{14} \\
&\quad +24t^{16}+29t^{18}+114t^{20}+129t^{22}
+16t^{24}+18t^{26}+168t^{28} \\
&\quad +153t^{30}+66t^{32}+159t^{34}-10t^{36}
-138t^{38}+789t^{40}+794t^{42} \\
&\quad -1713t^{44}-963t^{46}+4339t^{48}+1308t^{50}
-7764t^{52}+604t^{54} \\
&\quad +13212t^{56}-5208t^{58}-17621t^{60}
+16119t^{62}+17916t^{64}
+\mathcal{O}(t^{66}).
\end{split}
\end{equation}

\begin{equation}
\begin{split}
\mathcal{I}_{n_1=1,n_2=1,n_3=1;\,N_1=1,N_2=3,N_3=4}^{\mathrm{BMN}}(t)
&=1+6t^2+15t^4+21t^6+21t^8+30t^{10}
+58t^{12}+57t^{14} \\
&\quad +6t^{16}+38t^{18}+174t^{20}+111t^{22}
-159t^{24}+72t^{26}+666t^{28} \\
&\quad -54t^{30}-1182t^{32}+804t^{34}+2744t^{36}
-2037t^{38}-4749t^{40} \\
&\quad +6217t^{42}+8328t^{44}-15102t^{46}-11459t^{48}
+34509t^{50}+11586t^{52} \\
&\quad -70674t^{54}+3117t^{56}+134994t^{58}
-54012t^{60}-233883t^{62} \\
&\quad +186351t^{64}+\mathcal{O}(t^{66}).
\end{split}
\end{equation}

\begin{equation}
\begin{split}
\mathcal{I}_{n_1=2,n_2=1,n_3=1;\,N_1=1,N_2=2,N_3=4}^{\mathrm{BMN}}(t)
&=1+6t^2+21t^4+49t^6+84t^8+114t^{10}
+154t^{12}+237t^{14} \\
&\quad +360t^{16}+408t^{18}+372t^{20}+456t^{22}
+799t^{24}+933t^{26}+732t^{28} \\
&\quad +943t^{30}+1761t^{32}+1020t^{34}-265t^{36}
+2787t^{38}+6990t^{40} \\
&\quad -1214t^{42}-10458t^{44}+9888t^{46}+31678t^{48}
-15375t^{50}-56742t^{52} \\
&\quad +46963t^{54}+118968t^{56}-104334t^{58}
-200742t^{60}+265443t^{62} \\
&\quad +342993t^{64}+\mathcal{O}(t^{66}).
\end{split}
\end{equation}

\begin{equation}
\begin{split}
\mathcal{I}_{n_1=1,n_2=2,n_3=1;\,N_1=1,N_2=2,N_3=3}^{\mathrm{BMN}}(t)
&=1+6t^2+21t^4+50t^6+84t^8+105t^{10}
+129t^{12}+231t^{14} \\
&\quad +438t^{16}+510t^{18}+162t^{20}-117t^{22}
+972t^{24}+2868t^{26}+1425t^{28} \\
&\quad -4320t^{30}-3327t^{32}+12924t^{34}
+19216t^{36}-19443t^{38}-52761t^{40} \\
&\quad +29319t^{42}+145956t^{44}-4338t^{46}
-337387t^{48}-106209t^{50} \\
&\quad +741333t^{52}+497300t^{54}-1478205t^{56}
-1550172t^{58}+2750383t^{60} \\
&\quad +4157148t^{62}-4651662t^{64}
+\mathcal{O}(t^{66}).
\end{split}
\end{equation}

\begin{equation}
\begin{split}
\mathcal{I}_{n_1=3,n_2=1,n_3=1;\,N_1=1,N_2=2,N_3=3}^{\mathrm{BMN}}(t)
&=1+6t^2+21t^4+60t^6+135t^8+249t^{10}
+386t^{12}+537t^{14} \\
&\quad +798t^{16}+1249t^{18}+1830t^{20}+2175t^{22}
+2120t^{24}+2634t^{26} \\
&\quad +4977t^{28}+7753t^{30}+5673t^{32}-159t^{34}
+5330t^{36}+28791t^{38} \\
&\quad +31650t^{40}-26022t^{42}-59085t^{44}
+78708t^{46}+241627t^{48} \\
&\quad -20406t^{50}-539106t^{52}-116091t^{54}
+1313133t^{56}+877788t^{58} \\
&\quad -2621034t^{60}-2809425t^{62}+5127870t^{64}
+\mathcal{O}(t^{66}).
\end{split}
\end{equation}

Among all triple-partition sectors for $SU(8)$, the dominant
contribution to the BMN index comes from
$\mathcal{I}_{n_1=3,n_2=1,n_3=1;\,N_1=1,N_2=2,N_3=3}^{\mathrm{BMN}}(t)$.
For this sector, the coefficient at order $t^{64}$ is
$d_{64}=5127870$, giving
\begin{equation}
\log|d_{64}|=15.4502.
\nonumber
\end{equation}

Adding all five triple-partition indices for $N=8$, we obtain
\begin{equation}
\begin{split}
\mathcal{I}_{N=8}^{\mathrm{BMN~TP}}(t)
&=5+30t^2+93t^4+201t^6+345t^8+528t^{10}
+783t^{12}+1122t^{14} \\
&\quad +1626t^{16}+2234t^{18}+2652t^{20}
+2754t^{22}+3748t^{24}+6525t^{26} \\
&\quad +7968t^{28}+4475t^{30}+2991t^{32}
+14748t^{34}+27015t^{36}+9960t^{38} \\
&\quad -18081t^{40}+9094t^{42}+83028t^{44}
+68193t^{46}-71202t^{48}-106173t^{50} \\
&\quad +149307t^{52}+358102t^{54}-29775t^{56}
-646932t^{58}-143026t^{60} \\
&\quad +1395402t^{62}+1023468t^{64}
+\mathcal{O}(t^{66}).
\end{split}
\end{equation}

\paragraph{$N=9.$}

\begin{equation}
\begin{split}
\mathcal{I}_{
\substack{
n_1=1,n_2=1,n_3=1;\\
N_1=1,N_2=2,N_3=6
}}^{\mathrm{BMN}}(t)
&=1+6t^2+15t^4+21t^6+21t^8 \\
&\quad +30t^{10}+56t^{12}+60t^{14}+24t^{16} \\
&\quad +28t^{18}+114t^{20}+135t^{22}+15t^{24} \\
&\quad -3t^{26}+171t^{28}+225t^{30}+48t^{32} \\
&\quad -57t^{34}+63t^{36}+390t^{38}+558t^{40} \\
&\quad -360t^{42}-1017t^{44}+1251t^{46}+2564t^{48} \\
&\quad -2337t^{50}-3921t^{52}+5589t^{54}+5928t^{56} \\
&\quad -10005t^{58}-6100t^{60}+16791t^{62}+4230t^{64} \\
&\quad -22288t^{66}+2904t^{68}+23190t^{70}-12096t^{72} \\
&\quad -10068t^{74}+17796t^{76}-23326t^{78}-411t^{80} \\
&\quad +80796t^{82}+\mathcal{O}(t^{84}).
\end{split}
\end{equation}

\begin{equation}
\begin{split}
\mathcal{I}_{
\substack{
n_1=1,n_2=1,n_3=1;\\
N_1=1,N_2=3,N_3=5
}}^{\mathrm{BMN}}(t)
&=1+6t^2+15t^4+20t^6+21t^8 \\
&\quad +36t^{10}+57t^{12}+36t^{14}+9t^{16} \\
&\quad +93t^{18}+165t^{20}-24t^{22}-96t^{24} \\
&\quad +375t^{26}+366t^{28}-662t^{30}-207t^{32} \\
&\quad +1866t^{34}+97t^{36}-3450t^{38}+1776t^{40} \\
&\quad +6673t^{42}-6171t^{44}-10092t^{46}+17292t^{48} \\
&\quad +12300t^{50}-38517t^{52}-4670t^{54}+76032t^{56} \\
&\quad -27225t^{58}-128409t^{60}+113310t^{62}+181122t^{64} \\
&\quad -293958t^{66}-180831t^{68}+616062t^{70}+14900t^{72} \\
&\quad -1091460t^{74}+521886t^{76}+1629707t^{78} \\
&\quad -1719975t^{80}-1907106t^{82}
+\mathcal{O}(t^{84}).
\end{split}
\end{equation}

\begin{equation}
\begin{split}
\mathcal{I}_{
\substack{
n_1=2,n_2=1,n_3=1;\\
N_1=1,N_2=2,N_3=5
}}^{\mathrm{BMN}}(t)
&=1+6t^2+21t^4+49t^6+84t^8 \\
&\quad +114t^{10}+153t^{12}+237t^{14}+360t^{16} \\
&\quad +415t^{18}+369t^{20}+435t^{22}+778t^{24} \\
&\quad +1035t^{26}+825t^{28}+689t^{30}+1224t^{32} \\
&\quad +1647t^{34}+1386t^{36}+1803t^{38}+2304t^{40} \\
&\quad +407t^{42}+495t^{44}+7644t^{46}+7401t^{48} \\
&\quad -9885t^{50}-7725t^{52}+29986t^{54}+23928t^{56} \\
&\quad -46179t^{58}-26786t^{60}+89178t^{62}+37212t^{64} \\
&\quad -123853t^{66}+3957t^{68}+171909t^{70}-103839t^{72} \\
&\quad -144687t^{74}+382614t^{76}-17085t^{78} \\
&\quad -888240t^{80}+577224t^{82}
+\mathcal{O}(t^{84}).
\end{split}
\end{equation}

\begin{equation}
\begin{split}
\mathcal{I}_{
\substack{
n_1=1,n_2=1,n_3=1;\\
N_1=2,N_2=3,N_3=4
}}^{\mathrm{BMN}}(t)
&=1+6t^2+15t^4+21t^6+18t^8 \\
&\quad +24t^{10}+65t^{12}+81t^{14}-9t^{16} \\
&\quad -62t^{18}+201t^{20}+417t^{22}-237t^{24} \\
&\quad -783t^{26}+861t^{28}+2250t^{30}-1842t^{32} \\
&\quad -5031t^{34}+5020t^{36}+11670t^{38}-12084t^{40} \\
&\quad -24414t^{42}+30555t^{44}+49647t^{46}-73844t^{48} \\
&\quad -93666t^{50}+173637t^{52}+164354t^{54}-391608t^{56} \\
&\quad -254967t^{58}+850928t^{60}+322482t^{62}-1769280t^{64} \\
&\quad -216469t^{66}+3515436t^{68}-456594t^{70} \\
&\quad -6644784t^{72}+2605944t^{74}+11890386t^{76} \\
&\quad -8117892t^{78}-19955280t^{80}+20625837t^{82} \\
&\quad +\mathcal{O}(t^{84}).
\end{split}
\end{equation}

\begin{equation}
\begin{split}
\mathcal{I}_{
\substack{
n_1=2,n_2=1,n_3=1;\\
N_1=1,N_2=3,N_3=4
}}^{\mathrm{BMN}}(t)
&=1+6t^2+21t^4+49t^6+81t^8 \\
&\quad +114t^{10}+162t^{12}+255t^{14}+333t^{16} \\
&\quad +318t^{18}+363t^{20}+687t^{22}+950t^{24} \\
&\quad +480t^{26}+210t^{28}+1867t^{30}+2889t^{32} \\
&\quad -900t^{34}-2749t^{36}+7233t^{38}+11646t^{40} \\
&\quad -11979t^{42}-18897t^{44}+37398t^{46}+45458t^{48} \\
&\quad -82641t^{50}-78243t^{52}+205003t^{54}+142422t^{56} \\
&\quad -456369t^{58}-194452t^{60}+1016844t^{62}+189060t^{64} \\
&\quad -2130413t^{66}+142110t^{68}+4300392t^{70} \\
&\quad -1363442t^{72}-8188500t^{74}+4905237t^{76} \\
&\quad +14718239t^{78}-13619928t^{80}-24479973t^{82} \\
&\quad +\mathcal{O}(t^{84}).
\end{split}
\end{equation}

\begin{equation}
\begin{split}
\mathcal{I}_{
\substack{
n_1=1,n_2=2,n_3=1;\\
N_1=1,N_2=2,N_3=4
}}^{\mathrm{BMN}}(t)
&=1+6t^2+21t^4+49t^6+81t^8 \\
&\quad +105t^{10}+143t^{12}+249t^{14}+393t^{16} \\
&\quad +389t^{18}+213t^{20}+372t^{22}+1147t^{24} \\
&\quad +1350t^{26}-108t^{28}-491t^{30}+3300t^{32} \\
&\quad +4986t^{34}-3405t^{36}-7038t^{38}+13179t^{40} \\
&\quad +21594t^{42}-25491t^{44}-41352t^{46}+70389t^{48} \\
&\quad +99075t^{50}-159780t^{52}-204142t^{54}+382764t^{56} \\
&\quad +433662t^{58}-854479t^{60}-847032t^{62}+1893480t^{64} \\
&\quad +1602792t^{66}-4028649t^{68}-2800776t^{70} \\
&\quad +8361102t^{72}+4517544t^{74}-16775673t^{76} \\
&\quad -6354201t^{78}+32649159t^{80}+6982590t^{82} \\
&\quad +\mathcal{O}(t^{84}).
\end{split}
\end{equation}

\begin{equation}
\begin{split}
\mathcal{I}_{
\substack{
n_1=3,n_2=1,n_3=1;\\
N_1=1,N_2=2,N_3=4
}}^{\mathrm{BMN}}(t)
&=1+6t^2+21t^4+59t^6+135t^8 \\
&\quad +249t^{10}+383t^{12}+543t^{14}+795t^{16} \\
&\quad +1215t^{18}+1776t^{20}+2256t^{22}+2434t^{24} \\
&\quad +2694t^{26}+4020t^{28}+6233t^{30}+7032t^{32} \\
&\quad +6015t^{34}+7246t^{36}+11325t^{38}+11418t^{40} \\
&\quad +9691t^{42}+21405t^{44}+33771t^{46}+4148t^{48} \\
&\quad -31521t^{50}+47070t^{52}+169182t^{54}+23211t^{56} \\
&\quad -280491t^{58}-8523t^{60}+684726t^{62}+188241t^{64} \\
&\quad -1226486t^{66}-149445t^{68}+2659893t^{70}+68647t^{72} \\
&\quad -5437845t^{74}+1076190t^{76}+12016506t^{78} \\
&\quad -3848106t^{80}-25579299t^{82}
+\mathcal{O}(t^{84}).
\end{split}
\end{equation}

\begin{equation}
\begin{split}
\mathcal{I}_{
\substack{
n_1=1,n_2=1,n_3=2;\\
N_1=1,N_2=2,N_3=3
}}^{\mathrm{BMN}}(t)
&=1+6t^2+21t^4+50t^6+81t^8 \\
&\quad +105t^{10}+136t^{12}+249t^{14}+423t^{16} \\
&\quad +392t^{18}+99t^{20}+195t^{22}+1529t^{24} \\
&\quad +2244t^{26}-873t^{28}-3807t^{30}+3903t^{32} \\
&\quad +15810t^{34}-920t^{36}-38091t^{38}-4458t^{40} \\
&\quad +102152t^{42}+45387t^{44}-237897t^{46}-157599t^{48} \\
&\quad +557745t^{50}+490770t^{52}-1237025t^{54}-1335564t^{56} \\
&\quad +2703303t^{58}+3422201t^{60}-5763204t^{62} \\
&\quad -8255832t^{64}+12164833t^{66}+19145031t^{68} \\
&\quad -25392786t^{70}-42840933t^{72}+52739937t^{74} \\
&\quad +93080973t^{76}-109123431t^{78}-196818909t^{80} \\
&\quad +225450327t^{82}+\mathcal{O}(t^{84}).
\end{split}
\end{equation}

\begin{equation}
\begin{split}
\mathcal{I}_{
\substack{
n_1=2,n_2=2,n_3=1;\\
N_1=1,N_2=2,N_3=3
}}^{\mathrm{BMN}}(t)
&=1+6t^2+27t^4+78t^6+183t^8 \\
&\quad +324t^{10}+498t^{12}+684t^{14}+1101t^{16} \\
&\quad +1859t^{18}+2832t^{20}+2883t^{22}+1987t^{24} \\
&\quad +2631t^{26}+9081t^{28}+15996t^{30}+7992t^{32} \\
&\quad -15555t^{34}-6934t^{36}+70017t^{38}+112833t^{40} \\
&\quad -58914t^{42}-288435t^{44}+10653t^{46}+812752t^{48} \\
&\quad +576324t^{50}-1566156t^{52}-2257755t^{54} \\
&\quad +2547903t^{56}+6933744t^{58}-1940966t^{60} \\
&\quad -16900008t^{62}-4541469t^{64}+35683642t^{66} \\
&\quad +29002806t^{68}-62937834t^{70}-95447335t^{72} \\
&\quad +86418729t^{74}+249612129t^{76}-53987983t^{78} \\
&\quad -559438059t^{80}-178714413t^{82}
+\mathcal{O}(t^{84}).
\end{split}
\end{equation}

\begin{equation}
\begin{split}
\mathcal{I}_{
\substack{
n_1=4,n_2=1,n_3=1;\\
N_1=1,N_2=2,N_3=3
}}^{\mathrm{BMN}}(t)
&=1+6t^2+21t^4+60t^6+150t^8 \\
&\quad +330t^{10}+623t^{12}+1026t^{14}+1539t^{16} \\
&\quad +2262t^{18}+3381t^{20}+5085t^{22}+7347t^{24} \\
&\quad +9540t^{26}+10989t^{28}+12866t^{30}+19236t^{32} \\
&\quad +30876t^{34}+36328t^{36}+24708t^{38}+19221t^{40} \\
&\quad +70857t^{42}+152310t^{44}+96714t^{46}-146382t^{48} \\
&\quad -144255t^{50}+587208t^{52}+1196069t^{54}-301680t^{56} \\
&\quad -2984052t^{58}-811635t^{60}+7811322t^{62}+7855437t^{64} \\
&\quad -14200988t^{66}-27416118t^{68}+20510004t^{70} \\
&\quad +79843349t^{72}-6704943t^{74}-195142839t^{76} \\
&\quad -77543812t^{78}+429148596t^{80}+378746208t^{82} \\
&\quad +\mathcal{O}(t^{84}).
\end{split}
\end{equation}

Among all triple-partition sectors for $SU(9)$, the dominant
contribution to the BMN index comes from
\begin{equation}
\mathcal{I}_{
\substack{
n_1=4,n_2=1,n_3=1;\\
N_1=1,N_2=2,N_3=3
}}^{\mathrm{BMN}}(t).
\nonumber
\end{equation}
For this sector,
\begin{equation}
d_{80}=429148596,
\qquad
d_{82}=378746208.
\nonumber
\end{equation}
Therefore,
\begin{equation}
\begin{split}
\frac{1}{2}
\left(
\log|d_{80}|+\log|d_{82}|
\right)
&=19.8148.
\end{split}
\nonumber
\end{equation}

Adding all ten triple-partition indices for $N=9$, we obtain
\begin{equation}
\begin{split}
\mathcal{I}_{N=9}^{\mathrm{BMN~TP}}(t)
&=10+60t^2+198t^4+456t^6+855t^8 \\
&\quad +1431t^{10}+2276t^{12}+3420t^{14}+4968t^{16} \\
&\quad +6909t^{18}+9513t^{20}+12441t^{22}+15854t^{24} \\
&\quad +19563t^{26}+25542t^{28}+35166t^{30}+43575t^{32} \\
&\quad +39657t^{34}+36132t^{36}+78567t^{38}+156393t^{40} \\
&\quad +115707t^{42}-89859t^{44}-52263t^{46}+582179t^{48} \\
&\quad +881139t^{50}-555657t^{52}-1933409t^{54}+1173336t^{56} \\
&\quad +6011421t^{58}+301779t^{60}-13455591t^{62} \\
&\quad -4217799t^{64}+31236812t^{66}+20037201t^{68} \\
&\quad -63306540t^{70}-58124431t^{72}+124704651t^{74} \\
&\quad +149568699t^{76}-226803278t^{78}-334491153t^{80} \\
&\quad +401782191t^{82}+\mathcal{O}(t^{84}).
\end{split}
\end{equation}
\subsection{Total $SU(N)$ BMN index}
\label{totalbmnin}
For $N=2,3,4,5,6,7,8,9$, we obtain the following full
$SU(N)$ BMN indices.

\begin{equation}
\begin{split}
\mathcal{I}_{N=2}^{\mathrm{BMN}}(t)
&=
\mathcal{I}_{N=2}^{\mathrm{BMN~SP}}(t) \\
&=
2+6t^4+\mathcal{O}(t^5).
\end{split}
\end{equation}

\begin{equation}
\begin{split}
\mathcal{I}_{N=3}^{\mathrm{BMN}}(t)
&=
\mathcal{I}_{N=3}^{\mathrm{BMN~SP}}(t)
+\mathcal{I}_{N=3}^{\mathrm{BMN~DP}}(t) \\
&=
3+3t^2+9t^4+3t^6+12t^8+6t^{10}
+\mathcal{O}(t^{11}).
\end{split}
\end{equation}

\begin{equation}
\begin{split}
\mathcal{I}_{N=4}^{\mathrm{BMN}}(t)
&=
\mathcal{I}_{N=4}^{\mathrm{BMN~SP}}(t)
+\mathcal{I}_{N=4}^{\mathrm{BMN~DP}}(t) \\
&=
5+6t^2+24t^4+5t^6+63t^8-24t^{10} \\
&\quad
+149t^{12}-108t^{14}+312t^{16}
+\mathcal{O}(t^{17}).
\end{split}
\end{equation}

\begin{equation}
\begin{split}
\mathcal{I}_{N=5}^{\mathrm{BMN}}(t)
&=
\mathcal{I}_{N=5}^{\mathrm{BMN~SP}}(t)
+\mathcal{I}_{N=5}^{\mathrm{BMN~DP}}(t) \\
&=
7+15t^2+39t^4+48t^6+93t^8+99t^{10} \\
&\quad
+168t^{12}+174t^{14}+267t^{16}+283t^{18} \\
&\quad
+390t^{20}+426t^{22}+543t^{24}+606t^{26}
+\mathcal{O}(t^{27}).
\end{split}
\end{equation}

\begin{equation}
\begin{split}
\mathcal{I}_{N=6}^{\mathrm{BMN}}(t)
&=
\mathcal{I}_{N=6}^{\mathrm{BMN~SP}}(t)
+\mathcal{I}_{N=6}^{\mathrm{BMN~DP}}(t) \\
&\quad
+\mathcal{I}_{N=6}^{\mathrm{BMN~TP}}(t) \\
&=
11+24t^2+81t^4+92t^6+246t^8+198t^{10} \\
&\quad
+571t^{12}+303t^{14}+1155t^{16}+366t^{18} \\
&\quad
+2124t^{20}+306t^{22}+3638t^{24}+15t^{26} \\
&\quad
+5895t^{28}-644t^{30}+9135t^{32}-1839t^{34} \\
&\quad
+13626t^{36}+\mathcal{O}(t^{37}).
\end{split}
\end{equation}

\begin{equation}
\begin{split}
\mathcal{I}_{N=7}^{\mathrm{BMN}}(t)
&=
\mathcal{I}_{N=7}^{\mathrm{BMN~SP}}(t)
+\mathcal{I}_{N=7}^{\mathrm{BMN~DP}}(t) \\
&\quad
+\mathcal{I}_{N=7}^{\mathrm{BMN~TP}}(t) \\
&=
15+45t^2+129t^4+225t^6+432t^8+630t^{10} \\
&\quad
+1036t^{12}+1374t^{14}+2055t^{16}+2625t^{18} \\
&\quad
+3663t^{20}+4557t^{22}+6046t^{24}+7395t^{26} \\
&\quad
+9420t^{28}+11382t^{30}+14079t^{32}+16794t^{34} \\
&\quad
+20324t^{36}+23952t^{38}+28485t^{40}+33232t^{42} \\
&\quad
+38973t^{44}+45027t^{46}+52219t^{48}+59811t^{50} \\
&\quad
+\mathcal{O}(t^{51}).
\end{split}
\end{equation}

\begin{equation}
\begin{split}
\mathcal{I}_{N=8}^{\mathrm{BMN}}(t)
&=
\mathcal{I}_{N=8}^{\mathrm{BMN~SP}}(t)
+\mathcal{I}_{N=8}^{\mathrm{BMN~DP}}(t) \\
&\quad
+\mathcal{I}_{N=8}^{\mathrm{BMN~TP}}(t) \\
&=
22+69t^2+228t^4+397t^6+894t^8+1224t^{10} \\
&\quad
+2439t^{12}+2859t^{14}+5442t^{16}+5689t^{18} \\
&\quad
+10755t^{20}+10206t^{22}+19460t^{24}+17004t^{26} \\
&\quad
+32955t^{28}+26777t^{30}+53016t^{32}+40377t^{34} \\
&\quad
+81793t^{36}+58764t^{38}+121902t^{40}+83044t^{42} \\
&\quad
+176481t^{44}+114522t^{46}+249201t^{48}+154650t^{50} \\
&\quad
+344391t^{52}+205046t^{54}+467040t^{56}+267591t^{58} \\
&\quad
+622889t^{60}+344328t^{62}+818523t^{64}
+\mathcal{O}(t^{65}).
\end{split}
\end{equation}

\begin{equation}
\begin{split}
\mathcal{I}_{N=9}^{\mathrm{BMN}}(t)
&=
\mathcal{I}_{N=9}^{\mathrm{BMN~SP}}(t)
+\mathcal{I}_{N=9}^{\mathrm{BMN~DP}}(t) \\
&\quad
+\mathcal{I}_{N=9}^{\mathrm{BMN~TP}}(t) \\
&=
30+111t^2+357t^4+746t^6+1545t^8+2628t^{10} \\
&\quad
+4529t^{12}+6888t^{14}+10692t^{16}+15205t^{18} \\
&\quad
+22047t^{20}+29991t^{22}+41370t^{24}+54483t^{26} \\
&\quad
+72378t^{28}+92937t^{30}+119928t^{32}+150903t^{34} \\
&\quad
+190226t^{36}+235296t^{38}+291066t^{40}+354770t^{42} \\
&\quad
+432033t^{44}+519984t^{46}+624858t^{48}+743781t^{50} \\
&\quad
+883695t^{52}+1041604t^{54}+1225347t^{56}
+1431870t^{58} \\
&\quad
+1669790t^{60}+1936161t^{62}+2240484t^{64}
+2579804t^{66} \\
&\quad
+2964663t^{68}+3392256t^{70}+3874015t^{72}
+4407384t^{74} \\
&\quad
+5004963t^{76}+5664181t^{78}+6399123t^{80}
+7207239t^{82} \\
&\quad
+\mathcal{O}(t^{83}).
\end{split}
\end{equation}
We have for double partitions
\begin{equation}
\begin{aligned}
\mathcal{I}_{N=3}^{\mathrm{BMN~DP}}(t): &\quad d_8 = 0, \quad d_{10} = 6 \\
\mathcal{I}_{N=4}^{\mathrm{BMN~DP}}(t): &\quad d_{16} = 63 \\
\mathcal{I}_{N=5}^{\mathrm{BMN~DP}}(t): &\quad d_{24} = 306, \quad d_{26} = 273 \\
\mathcal{I}_{N=6}^{\mathrm{BMN~DP}}(t): &\quad d_{36} = 11520 \\
\mathcal{I}_{N=7}^{\mathrm{BMN~DP}}(t): &\quad d_{48} = 99672, \quad d_{50} = -62406 \\
\mathcal{I}_{N=8}^{\mathrm{BMN~DP}}(t): &\quad d_{64} = -3626808 \\
\mathcal{I}_{N=9}^{\mathrm{BMN~DP}}(t): &\quad d_{80} = 345173460, \quad d_{82} = -399702471.
\end{aligned}
\end{equation}

We have Table \ref{dpindextab}
\begin{table}[H]
    \centering
    \label{tab:results}
    \begin{tabular}{|c|c|}
        \hline
        $N$ & $\text{log}|d_{j=N^2}|$ \\
        \hline
        3 & 0.89588 \\
        4 & 4.14313 \\
        5 & 5.66653 \\
        6 & 9.35184 \\
        7 & 11.27553 \\
        8 & 15.10386 \\
        9 & 19.73289 \\
        \hline
    \end{tabular}
\caption{Results for $\log |d_{j=N^2}|$ vs $N$ double partition index.}
\label{dpindextab}
\end{table}
We have for triple partitions
\begin{equation}
\begin{aligned}
\mathcal{I}_{N=6}^{\mathrm{BMN~TP}}(t): &\quad d_{36} = -1050 \\
\mathcal{I}_{N=7}^{\mathrm{BMN~TP}}(t): &\quad d_{48} = -71135, \quad d_{50} = 115677 \\
\mathcal{I}_{N=8}^{\mathrm{BMN~TP}}(t): &\quad d_{64} = 1023468 \\
\mathcal{I}_{N=9}^{\mathrm{BMN~TP}}(t): &\quad d_{80} = -334491153, \quad d_{82} = 401782191~.
\end{aligned}
\end{equation}

We have Table \ref{tptab}
\begin{table}[H]
    \centering
    \label{tab:results}
    \begin{tabular}{|c|c|}
        \hline
        $N$ & $\text{log}|d_{j=N^2}|$ \\
        \hline
        6 & 6.95655 \\
        7 & 11.41545 \\
        8 & 13.83871 \\
        9 & 19.71977 \\
        \hline
    \end{tabular}
\caption{Results for $\log |d_{j=N^2}|$ and $N$ triple partition index.}
\label{tptab}
\end{table}
We have the total BMN index  
\begin{equation}
\begin{aligned}
\mathcal{I}_{N=2}^{\mathrm{BMN}}(t): &\quad d_4 = 6 \\
\mathcal{I}_{N=3}^{\mathrm{BMN}}(t): &\quad d_8 = 12, \quad d_{10} = 6 \\
\mathcal{I}_{N=4}^{\mathrm{BMN}}(t): &\quad d_{16} = 312 \\
\mathcal{I}_{N=5}^{\mathrm{BMN}}(t): &\quad d_{24} = 543, \quad d_{26} = 606 \\
\mathcal{I}_{N=6}^{\mathrm{BMN}}(t): &\quad d_{36} = 13626 \\
\mathcal{I}_{N=7}^{\mathrm{BMN}}(t): &\quad d_{48} = 52219, \quad d_{50} = 59811 \\
\mathcal{I}_{N=8}^{\mathrm{BMN}}(t): &\quad d_{64} = 818523 \\
\mathcal{I}_{N=9}^{\mathrm{BMN}}(t): &\quad d_{80} = 6399123, \quad d_{82} = 7207239 .
\end{aligned}
\end{equation}
We have Table \ref{totaltab}
\begin{table}[H]
    \centering
    \begin{tabular}{|c|c|}
        \hline
        $N$ & $\text{log}|d_{j=N^2}|$ \\
        \hline
        2 & 1.79176 \\
        3 & 2.13833 \\
        4 & 5.74300 \\
        5 & 6.35199 \\
        6 & 9.51974 \\
        7 & 10.93107 \\
        8 & 13.61526 \\
        9 & 15.73113 \\
        \hline
    \end{tabular}
\caption{Results for $\log |d_{j=N^2}|$ and $N$ total BMN index.}
\label{totaltab}
\end{table}

\section{Closed-form $SU(N)$ BMN index single partition}
\label{closedform}
\subsection{Closed-form $SU(N)$ BMN index trivial vacuum}
\label{cf}
In \ref{cf}, we write down the closed-form $SU(N)$ BMN index for the trivial vacuum from $N=2$ to $N=7$ by evaluating residues as a sum over trees in section \ref{treesres} 
\begin{equation}
\begin{split}
\mathcal{I}_{n_1=2;N_1=1}^{\mathrm{BMN}~SU(N)}(t)=& \left(1 - t^{6}\right) \left(1 + 3 t^{2} + 12 t^{4} + 20 t^{6} + 42 t^{8} + 48 t^{10} + 75 t^{12} + 66 t^{14} + 81 t^{16} \right. \\
& \left. + 55 t^{18} + 54 t^{20} + 27 t^{22} + 19 t^{24} + 6 t^{26} + 3 t^{28}\right) \Bigg/ \\
& \left( \left(1 + t^{2}\right)^{3} \left(1 + t^{4}\right)^{3} \left(-1 + t^{6}\right)^{2} \left(1 + t^{6}\right) \right).
\end{split}
\end{equation}
\begin{equation}
\small
\begin{split}
\mathcal{I}_{n_1=3;N_1=1}^{\mathrm{BMN}~SU(N)}(t)=\bigg( & \left(1 - t^{6}\right) \left(-1 - 7 t^{2} - 35 t^{4} - 135 t^{6} - 433 t^{8} - 1200 t^{10} - 2945 t^{12} - 6529 t^{14} - 13\,276 t^{16} \right. \\
& \left. - 25\,056 t^{18} - 44\,296 t^{20} - 73\,861 t^{22} - 116\,796 t^{24} - 175\,918 t^{26} - 253\,296 t^{28} - 349\,675 t^{30} \right. \\
& \left. - 463\,933 t^{32} - 592\,731 t^{34} - 730\,438 t^{36} - 869\,391 t^{38} - 1\,000\,510 t^{40} - 1\,114\,205 t^{42} \right. \\
& \left. - 1\,201\,489 t^{44} - 1\,255\,095 t^{46} - 1\,270\,406 t^{48} - 1\,246\,087 t^{50} - 1\,184\,255 t^{52} - 1\,090\,197 t^{54} \right. \\
& \left. - 971\,672 t^{56} - 837\,897 t^{58} - 698\,439 t^{60} - 562\,137 t^{62} - 436\,232 t^{64} - 325\,841 t^{66} - 233\,773 t^{68} \right. \\
& \left. - 160\,691 t^{70} - 105\,508 t^{72} - 65\,920 t^{74} - 39\,006 t^{76} - 21\,732 t^{78} - 11\,316 t^{80} - 5\,454 t^{82} \right. \\
& \left. - 2\,400 t^{84} - 945 t^{86} - 324 t^{88} - 93 t^{90} - 21 t^{92} - 3 t^{94}\right) \bigg) \Bigg/ \\
& \left( \left(-1 + t^{2}\right)^{3} \left(1 + t^{2}\right)^{3} \left(1 + t^{4}\right)^{3} \left(1 - t^{2} + t^{4}\right) \left(1 + t^{2} + t^{4}\right)^{3} \right. \\
& \left. \left(1 + t^{2} + t^{4} + t^{6} + t^{8}\right)^{2} \left(1 + t^{6} + t^{12}\right) \left(1 + t^{2} + t^{4} + t^{6} + t^{8} + t^{10} + t^{12}\right)^{3} \right).
\end{split}
\end{equation}

\begin{equation}
\small
\begin{split}
&\mathcal{I}_{n_1=4;N_1=1}^{\mathrm{BMN}~SU(N)}(t)\\
=\bigg( & -1 - 8 t^2 - 50 t^4 - 234 t^6 - 955 t^8 - 3\,392 t^{10} - 10\,965 t^{12} - 32\,367 t^{14} - 88\,970 t^{16} - \\
& 228\,640 t^{18} - 555\,287 t^{20} - 1\,279\,096 t^{22} - 2\,813\,610 t^{24} - 5\,927\,835 t^{26} - 12\,017\,153 t^{28} - \\
& 23\,497\,180 t^{30} - 44\,459\,087 t^{32} - 81\,556\,349 t^{34} - 145\,397\,160 t^{36} - 252\,296\,386 t^{38} - \\
& 426\,893\,471 t^{40} - 705\,196\,298 t^{42} - 1\,138\,944\,252 t^{44} - 1\,800\,227\,375 t^{46} - 2\,787\,899\,485 t^{48} - \\
& 4\,233\,561\,491 t^{50} - 6\,309\,738\,732 t^{52} - 9\,236\,077\,766 t^{54} - 13\,288\,018\,130 t^{56} - 18\,800\,747\,478 t^{58} - \\
& 26\,175\,926\,188 t^{60} - 35\,879\,624\,540 t^{62} - 48\,443\,779\,958 t^{64} - 64\,453\,613\,400 t^{66} - 84\,540\,585\,792 t^{68} - \\
& 109\,355\,364\,992 t^{70} - 139\,550\,711\,647 t^{72} - 175\,738\,422\,768 t^{74} - 218\,464\,004\,869 t^{76} - \\
& 268\,150\,037\,661 t^{78} - 325\,069\,125\,365 t^{80} - 389\,280\,866\,300 t^{82} - 460\,614\,039\,472 t^{84} - \\
& 538\,608\,182\,497 t^{86} - 622\,518\,401\,149 t^{88} - 711\,273\,598\,353 t^{90} - 803\,516\,014\,117 t^{92} - \\
& 897\,585\,150\,836 t^{94} - 991\,598\,494\,213 t^{96} - 1\,083\,463\,207\,795 t^{98} - 1\,170\,995\,364\,759 t^{100} - \\
& 1\,251\,952\,235\,640 t^{102} - 1\,324\,177\,252\,229 t^{104} - 1\,385\,637\,143\,482 t^{106} - 1\,434\,572\,062\,240 t^{108} - \\
& 1\,469\,517\,134\,830 t^{110} - 1\,489\,434\,631\,239 t^{112} - 1\,493\,700\,851\,309 t^{114} - 1\,482\,201\,227\,095 t^{116} - \\
& 1\,455\,271\,010\,157 t^{118} - 1\,413\,743\,808\,907 t^{120} - 1\,358\,846\,156\,136 t^{122} - 1\,292\,202\,220\,703 t^{124} - \\
& 1\,215\,694\,359\,573 t^{126} - 1\,131\,437\,663\,120 t^{128} - 1\,041\,627\,178\,293 t^{130} - 948\,502\,144\,291 t^{132} - \\
& 854\,203\,118\,497 t^{134} - 760\,745\,610\,741 t^{136} - 669\,906\,426\,625 t^{138} - 583\,220\,570\,020 t^{140} - \\
& 501\,907\,508\,998 t^{142} - 426\,896\,198\,695 t^{144} - 358\,791\,800\,697 t^{146} - 297\,924\,735\,567 t^{148} - \\
& 244\,349\,863\,587 t^{150} - 197\,909\,345\,772 t^{152} - 158\,251\,586\,540 t^{154} - 124\,895\,691\,097 t^{156} - \\
& 97\,257\,111\,835 t^{158} - 74\,703\,346\,303 t^{160} - 56\,576\,262\,101 t^{162} - 42\,232\,909\,561 t^{164} - \\
& 31\,059\,134\,481 t^{166} - 22\,494\,109\,780 t^{168} - 16\,034\,277\,748 t^{170} - 11\,243\,914\,274 t^{172} - \\
& 7\,751\,473\,307 t^{174} - 5\,250\,438\,908 t^{176} - 3\,491\,403\,388 t^{178} - 2\,277\,684\,685 t^{180} - \\
& 1\,456\,272\,926 t^{182} - 911\,761\,002 t^{184} - 558\,296\,127 t^{186} - 334\,000\,685 t^{188} - 194\,909\,526 t^{190} - \\
& 110\,808\,374 t^{192} - 61\,242\,223 t^{194} - 32\,853\,691 t^{196} - 17\,058\,060 t^{198} - 8\,555\,090 t^{200} - \\
& 4\,127\,698 t^{202} - 1\,911\,069 t^{204} - 843\,898 t^{206} - 354\,262 t^{208} - 139\,996 t^{210} - 51\,858 t^{212} - \\
& 17\,690 t^{214} - 5\,526 t^{216} - 1\,523 t^{218} - 368 t^{220} - 71 t^{222} - 11 t^{224} - t^{226} \bigg) \Bigg/ \\
& \bigg( (-1 + t^2)^3 (1 + t^2)^5 (1 + t^4)^9 (1 - t^2 + t^4)^2 (1 + t^2 + t^4)^3 (1 + t^8)^5 (1 - t^4 + t^8) \cdot \\
& (1 - t^2 + t^4 - t^6 + t^8)^3 (1 + t^2 + t^4 + t^6 + t^8)^5 (1 + t^6 + t^{12}) (1 + t^2 + t^4 + t^6 + t^8 + t^{10} + t^{12})^3 \bigg).
\end{split}
\end{equation}

\begin{equation}
\mathcal{I}_{n_1=5;N_1=1}^{\mathrm{BMN}~SU(N)}(t)=\frac{\mathcal{N}(t)}{\mathcal{D}(t)},
\end{equation}
where $\mathcal{N}(t) = \sum_{i=1}^{8} \mathcal{N}_i(t)$ with the numerator split into eight parts for readability
\begin{equation}
\small
\begin{split}
\mathcal{N}_1(t) = &  1 + 13 t^{2} + 108 t^{4} + 681 t^{6} + {} \\
& 3\,585 t^{8} + 16\,453 t^{10} + 67\,778 t^{12} + 255\,331 t^{14} + {} \\
& 891\,703 t^{16} + 2\,915\,996 t^{18} + 8\,998\,994 t^{20} + 26\,370\,705 t^{22} + {} \\
& 73\,749\,803 t^{24} + 197\,664\,754 t^{26} + 509\,526\,213 t^{28} + {} \\
& 1\,267\,048\,769 t^{30} + 3\,047\,595\,360 t^{32} + 7\,106\,653\,589 t^{34} + {} \\
& 16\,099\,323\,116 t^{36} + 35\,495\,894\,883 t^{38} + 76\,293\,518\,041 t^{40} + {} \\
& 160\,094\,468\,869 t^{42} + 328\,415\,135\,707 t^{44} + 659\,407\,297\,596 t^{46} + {} \\
& 1\,297\,319\,251\,160 t^{48} + 2\,503\,456\,059\,507 t^{50} + 4\,742\,790\,427\,573 t^{52} + {} \\
& 8\,828\,705\,017\,192 t^{54} + 16\,161\,003\,615\,598 t^{56} ,
\end{split}
\end{equation}

\begin{equation}
\small
\begin{split}
\mathcal{N}_2(t) = &  29\,111\,254\,966\,641 t^{58} + 51\,637\,399\,311\,835 t^{60} + {} \\
& 90\,249\,957\,344\,550 t^{62} + 155\,509\,917\,630\,928 t^{64} + {} \\
& 264\,319\,674\,562\,495 t^{66} + 443\,381\,226\,949\,456 t^{68} + {} \\
& 734\,350\,254\,176\,334 t^{70} + 1\,201\,423\,022\,279\,323 t^{72} + {} \\
& 1\,942\,370\,250\,420\,173 t^{74} + 3\,104\,395\,699\,482\,503 t^{76} + {} \\
& 4\,906\,667\,602\,865\,891 t^{78} + 7\,671\,970\,856\,208\,621 t^{80} + {} \\
& 11\,870\,681\,712\,805\,923 t^{82} + 18\,181\,200\,180\,741\,079 t^{84} + {} \\
& 27\,572\,113\,565\,609\,705 t^{86} + 41\,412\,730\,541\,684\,905 t^{88} + {} \\
& 61\,620\,236\,831\,527\,806 t^{90} + 90\,853\,591\,174\,446\,016 t^{92} + {} \\
& 132\,766\,402\,184\,256\,518 t^{94} + 192\,333\,385\,347\,741\,375 t^{96} + {} \\
& 276\,267\,556\,366\,913\,542 t^{98} + 393\,548\,008\,079\,607\,970 t^{100} + {} \\
& 556\,080\,848\,230\,436\,499 t^{102} + 779\,518\,514\,011\,536\,155 t^{104} + {} \\
& 1\,084\,265\,057\,040\,345\,037 t^{106} + 1\,496\,696\,898\,525\,957\,431 t^{108} + {} \\
& 2\,050\,629\,736\,394\,823\,563 t^{110} + 2\,789\,062\,452\,616\,054\,854 t^{112} ,
\end{split}
\end{equation}

\begin{equation}
\small
\begin{split}
\mathcal{N}_3(t) = &  3\,766\,227\,694\,790\,290\,759 t^{114} + 5\,049\,975\,942\,180\,292\,912 t^{116} + {} \\
& 6\,724\,514\,953\,164\,451\,157 t^{118} + 8\,893\,519\,176\,791\,097\,617 t^{120} + {} \\
& 11\,683\,613\,673\,928\,321\,535 t^{122} + 15\,248\,224\,068\,555\,561\,346 t^{124} + {} \\
& 19\,771\,767\,858\,563\,497\,316 t^{126} + 25\,474\,142\,998\,439\,682\,512 t^{128} + {} \\
& 32\,615\,447\,115\,022\,454\,371 t^{130} + 41\,500\,835\,306\,778\,272\,558 t^{132} + {} \\
& 52\,485\,396\,691\,806\,658\,824 t^{134} + 65\,978\,900\,427\,518\,343\,082 t^{136} + {} \\
& 82\,450\,231\,787\,747\,652\,556 t^{138} + 102\,431\,309\,259\,392\,151\,312 t^{140} + {} \\
& 126\,520\,245\,951\,627\,785\,331 t^{142} + 155\,383\,494\,541\,905\,868\,602 t^{144} + {} \\
& 189\,756\,696\,317\,014\,342\,990 t^{146} + 230\,443\,943\,500\,934\,096\,380 t^{148} + {} \\
& 278\,315\,161\,901\,493\,092\,401 t^{150} + 334\,301\,329\,777\,265\,322\,082 t^{152} + {} \\
& 399\,387\,270\,351\,159\,257\,198 t^{154} + 474\,601\,790\,891\,755\,561\,161 t^{156} + {} \\
& 561\,004\,991\,628\,028\,645\,580 t^{158} + 659\,672\,633\,290\,073\,489\,446 t^{160} + {} \\
& 771\,677\,532\,456\,962\,689\,220 t^{162} + 898\,068\,048\,086\,117\,810\,491 t^{164} + {} \\
& 1\,039\,843\,828\,745\,353\,092\,313 t^{166} + 1\,197\,929\,105\,503\,009\,670\,786 t^{168} + {} \\
& 1\,373\,143\,936\,688\,892\,783\,286 t^{170} ,
\end{split}
\end{equation}

\begin{equation}
\small
\begin{split}
\mathcal{N}_4(t) = &  1\,566\,173\,933\,621\,681\,575\,251 t^{172} + 1\,777\,539\,116\,085\,067\,060\,841 t^{174} + {} \\
& 2\,007\,562\,657\,538\,033\,290\,421 t^{176} + 2\,256\,340\,377\,210\,624\,989\,628 t^{178} + {} \\
& 2\,523\,711\,913\,785\,323\,119\,944 t^{180} + 2\,809\,234\,567\,953\,149\,083\,447 t^{182} + {} \\
& 3\,112\,160\,823\,942\,794\,991\,462 t^{184} + 3\,431\,420\,549\,134\,786\,507\,118 t^{186} + {} \\
& 3\,765\,608\,823\,154\,382\,876\,906 t^{188} + 4\,112\,980\,261\,768\,224\,736\,941 t^{190} + {} \\
& 4\,471\,450\,576\,390\,694\,908\,380 t^{192} + 4\,838\,605\,948\,588\,528\,422\,345 t^{194} + {} \\
& 5\,211\,720\,603\,928\,281\,331\,912 t^{196} + 5\,587\,782\,745\,811\,761\,232\,806 t^{198} + {} \\
& 5\,963\,528\,764\,158\,914\,905\,381 t^{200} + 6\,335\,485\,373\,898\,398\,690\,616 t^{202} + {} \\
& 6\,700\,019\,073\,317\,447\,224\,343 t^{204} + 7\,053\,392\,052\,288\,079\,804\,223 t^{206} + {} \\
& 7\,391\,823\,435\,484\,130\,466\,480 t^{208} + 7\,711\,554\,526\,111\,660\,314\,441 t^{210} + {} \\
& 8\,008\,916\,531\,027\,918\,219\,168 t^{212} + 8\,280\,399\,107\,040\,770\,120\,783 t^{214} + {} \\
& 8\,522\,717\,977\,834\,640\,500\,842 t^{216} + 8\,732\,879\,836\,699\,639\,704\,244 t^{218} + {} \\
& 8\,908\,242\,775\,249\,035\,220\,725 t^{220} + 9\,046\,570\,563\,441\,551\,888\,209 t^{222} + {} \\
& 9\,146\,079\,249\,866\,155\,606\,925 t^{224} + 9\,205\,474\,749\,333\,054\,525\,422 t^{226} ,
\end{split}
\end{equation}

\begin{equation}
\small
\begin{split}
\mathcal{N}_5(t) = &  9\,223\,980\,330\,962\,585\,289\,307 t^{228} + 9\,201\,353\,205\,711\,758\,614\,644 t^{230} + {} \\
& 9\,137\,889\,727\,438\,519\,111\,648 t^{232} + 9\,034\,419\,054\,692\,765\,950\,770 t^{234} + {} \\
& 8\,892\,285\,459\,167\,444\,218\,380 t^{236} + 8\,713\,319\,798\,616\,430\,570\,209 t^{238} + {} \\
& 8\,499\,800\,984\,828\,465\,514\,010 t^{240} + 8\,254\,408\,559\,558\,757\,443\,842 t^{242} + {} \\
& 7\,980\,167\,733\,123\,592\,038\,781 t^{244} + 7\,680\,388\,433\,384\,199\,670\,228 t^{246} + {} \\
& 7\,358\,600\,050\,917\,046\,608\,072 t^{248} + 7\,018\,483\,645\,457\,648\,264\,688 t^{250} + {} \\
& 6\,663\,803\,397\,841\,052\,619\,877 t^{252} + 6\,298\,339\,051\,724\,779\,650\,658 t^{254} + {} \\
& 5\,925\,820\,993\,789\,744\,684\,697 t^{256} + 5\,549\,869\,475\,404\,605\,964\,003 t^{258} + {} \\
& 5\,173\,939\,290\,241\,823\,644\,615 t^{260} + 4\,801\,270\,999\,789\,091\,347\,866 t^{262} + {} \\
& 4\,434\,849\,551\,816\,982\,022\,259 t^{264} + 4\,077\,370\,875\,851\,588\,485\,343 t^{266} + {} \\
& 3\,731\,216\,774\,802\,507\,848\,054 t^{268} + 3\,398\,438\,172\,938\,777\,137\,752 t^{270} + {} \\
& 3\,080\,746\,536\,398\,724\,142\,957 t^{272} + 2\,779\,513\,061\,212\,954\,546\,127 t^{274} + {} \\
& 2\,495\,775\,031\,835\,207\,813\,600 t^{276} + 2\,230\,248\,595\,223\,631\,958\,439 t^{278} + {} \\
& 1\,983\,347\,074\,699\,459\,161\,329 t^{280} + 1\,755\,203\,865\,518\,477\,642\,468 t^{282} ,
\end{split}
\end{equation}

\begin{equation}
\begin{split}
\mathcal{N}_6(t) = &  1\,545\,698\,910\,067\,523\,223\,233 t^{284} + 1\,354\,487\,743\,086\,419\,324\,900 t^{286} + {} \\
& 1\,181\,032\,123\,260\,771\,323\,251 t^{288} + 1\,024\,631\,322\,824\,812\,429\,614 t^{290} + {} \\
& 884\,453\,226\,570\,107\,870\,317 t^{292} + 759\,564\,490\,495\,165\,332\,067 t^{294} + {} \\
& 648\,959\,122\,661\,463\,073\,970 t^{296} + 551\,584\,969\,103\,167\,870\,303 t^{298} + {} \\
& 466\,367\,710\,618\,513\,481\,981 t^{300} + 392\,232\,097\,181\,391\,603\,939 t^{302} + {} \\
& 328\,120\,261\,420\,347\,864\,733 t^{304} + 273\,007\,057\,729\,280\,058\,957 t^{306} + {} \\
& 225\,912\,466\,512\,092\,716\,262 t^{308} + 185\,911\,182\,050\,151\,144\,161 t^{310} + {} \\
& 152\,139\,566\,534\,603\,812\,408 t^{312} + 123\,800\,201\,673\,204\,628\,809 t^{314} + {} \\
& 100\,164\,303\,354\,633\,402\,412 t^{316} + 80\,572\,285\,061\,727\,103\,705 t^{318} + {} \\
& 64\,432\,763\,419\,531\,966\,191 t^{320} + 51\,220\,296\,096\,893\,410\,306 t^{322} + {} \\
& 40\,472\,130\,088\,059\,680\,390 t^{324} + 31\,784\,219\,095\,485\,481\,378 t^{326} + {} \\
& 24\,806\,744\,208\,479\,493\,119 t^{328} + 19\,239\,344\,115\,724\,881\,591 t^{330} + {} \\
& 14\,826\,231\,331\,186\,322\,582 t^{332} + 11\,351\,340\,772\,702\,138\,103 t^{334} + {} \\
& 8\,633\,627\,695\,707\,983\,258 t^{336} + 6\,522\,604\,389\,323\,827\,670 t^{338} + {} \\
& 4\,894\,179\,882\,475\,085\,612 t^{340} ,
\end{split}
\end{equation}

\begin{equation}
\small
\begin{split}
\mathcal{N}_7(t) = &  3\,646\,844\,647\,136\,996\,143 t^{342} + 2\,698\,223\,179\,298\,202\,270 t^{344} + {} \\
& 1\,982\,001\,460\,912\,787\,418 t^{346} + 1\,445\,223\,584\,763\,143\,182 t^{348} + {} \\
& 1\,045\,942\,070\,035\,229\,028 t^{350} + 751\,199\,337\,219\,596\,160 t^{352} + {} \\
& 535\,313\,120\,430\,901\,126 t^{354} + 378\,435\,919\,401\,444\,490 t^{356} + {} \\
& 265\,357\,572\,597\,524\,874 t^{358} + 184\,520\,319\,509\,749\,426 t^{360} + {} \\
& 127\,216\,991\,794\,051\,438 t^{362} + 86\,944\,941\,515\,106\,377 t^{364} + {} \\
& 58\,890\,731\,786\,729\,960 t^{366} + 39\,523\,273\,847\,723\,019 t^{368} + {} \\
& 26\,275\,829\,511\,209\,091 t^{370} + 17\,299\,982\,231\,011\,606 t^{372} + {} \\
& 11\,277\,221\,973\,651\,705 t^{374} + 7\,276\,127\,260\,343\,560 t^{376} + {} \\
& 4\,645\,226\,084\,792\,987 t^{378} + 2\,933\,460\,187\,731\,486 t^{380} + {} \\
& 1\,831\,764\,230\,523\,320 t^{382} + 1\,130\,613\,915\,136\,896 t^{384} + {} \\
& 689\,513\,801\,309\,576 t^{386} + 415\,309\,753\,015\,550 t^{388} + {} \\
& 246\,947\,961\,813\,879 t^{390} + 144\,887\,925\,639\,393 t^{392} + {} \\
& 83\,835\,089\,818\,087 t^{394} + 47\,812\,593\,448\,709 t^{396} ,
\end{split}
\end{equation}

\begin{equation}
\begin{split}
\mathcal{N}_8(t) = &  26\,860\,746\,294\,305 t^{398} + 14\,854\,838\,933\,527 t^{400} + {} \\
& 8\,081\,322\,854\,965 t^{402} + 4\,321\,404\,440\,822 t^{404} + {} \\
& 2\,269\,488\,173\,780 t^{406} + 1\,169\,470\,651\,287 t^{408} + {} \\
& 590\,700\,580\,143 t^{410} + 292\,129\,316\,283 t^{412} + {} \\
& 141\,277\,656\,699 t^{414} + 66\,721\,072\,357 t^{416} + {} \\
& 30\,723\,733\,177 t^{418} + 13\,770\,632\,701 t^{420} + {} \\
& 5\,995\,868\,513 t^{422} + 2\,530\,464\,688 t^{424} + {} \\
& 1\,032\,493\,475 t^{426} + 406\,093\,816 t^{428} + {} \\
& 153\,430\,139 t^{430} + 55\,456\,543 t^{432} + {} \\
& 19\,080\,852 t^{434} + 6\,211\,498 t^{436} + {} \\
& 1\,898\,480 t^{438} + 539\,359 t^{440} + {} \\
& 140\,507 t^{442} + 32\,918 t^{444} + {} \\
& 6\,731 t^{446} + 1\,143 t^{448} + {} \\
& 146 t^{450} + 11 t^{452} .
\end{split}
\end{equation}

and the denominator $\mathcal{D}(t)$ is given by

\begin{equation}
\small
\begin{split}
\mathcal{D}(t) = & \bigg( (-1 + t^{2})^{4} (1 + t^{2})^{2} (1 + t^{4})^{9} (1 - t^{2} + t^{4})^{2} \cdot {} \\
& (1 + t^{2} + t^{4})^{4} (1 + t^{8})^{5} (1 - t^{4} + t^{8}) \cdot {} \\
& (1 - t^{2} + t^{4} - t^{6} + t^{8})^{3} (1 + t^{2} + t^{4} + t^{6} + t^{8})^{6} \cdot {} \\
& (1 + t^{6} + t^{12})^{4} (1 + t^{2} + t^{4} + t^{6} + t^{8} + t^{10} + t^{12})^{3} \cdot {} \\
& (1 - t^{2} + t^{6} - t^{8} + t^{10} - t^{14} + t^{16}) \cdot {} \\
& (1 + t^{2} + t^{4} + t^{6} + t^{8} + t^{10} + t^{12} + t^{14} + t^{16} + t^{18} + t^{20})^{5} \cdot {} \\
& (1 + t^{2} + t^{4} + t^{6} + t^{8} + t^{10} + t^{12} + t^{14} + t^{16} + t^{18} + t^{20} + t^{22} + t^{24})^{3} \bigg).
\end{split}
\end{equation}

\begin{equation}
\mathcal{I}_{n_1=6;N_1=1}^{\mathrm{BMN}~SU(N)}(t) = \frac{\sum_{i=1}^{14} \mathcal{N}_i(t)}{\mathcal{D}(t)},
\end{equation}
\begin{equation}
\small
\begin{split}
\mathcal{N}_1(t) &= 1 + 14t^{2} + 131t^{4} + 933t^{6} + 5586t^{8} + 29209t^{10} + 137407t^{12} \\
&\quad + 591867t^{14} + 2366809t^{16} + 8873484t^{18} + 31437815t^{20} + 105903439t^{22} \\
&\quad + 340938954t^{24} + 1053322967t^{26} + 3133987331t^{28} + 9007032700t^{30} \\
&\quad + 25069008620t^{32} + 67723161394t^{34} + 177925551132t^{36} + 455404485078t^{38} \\
&\quad + 1137337674446t^{40} + 2775364364048t^{42} + 6625756655656t^{44} + 15492846695638t^{46} \\
&\quad + 35518834488935t^{48} + 79915477359829t^{50} + 176614802049048t^{52} \\
&\quad + 383704387459823t^{54} + 820092626762600t^{56} + 1725539358080799t^{58}
\end{split}
\end{equation}
\begin{equation}
\small
\begin{split}
\mathcal{N}_2(t) &= 3576521160159606t^{60} + 7306842000489090t^{62} + 14722222038199083t^{64} \\
&\quad + 29269666964549024t^{66} + 57447974838563368t^{68} + 111363504369997993t^{70} \\
&\quad + 213308922794752106t^{72} + 403876787177771641t^{74} + 756185192285970181t^{76} \\
&\quad + 1400564788621897454t^{78} + 2566975624085297324t^{80} + 4657185859491067175t^{82} \\
&\quad + 8366447104451220147t^{84} + 14886745657548896030t^{86} + 26243321848441456197t^{88} \\
&\quad + 45847047732526720719t^{90} + 79393530373543474229t^{92} + 136314633130610344925t^{94} \\
&\quad + 232103171922329752049t^{96} + 392007130681419532588t^{98} + 656855678076374252450t^{100} \\
&\quad + 1092179090960590507004t^{102} + 1802381487716031459301t^{104} + 2952609811865083332076t^{106} \\
&\quad + 4802264467731501775149t^{108} + 7756000395839626508215t^{110} + 12440831509599677742505t^{112} \\
&\quad + 19821939308205142334843t^{114} + 31375502996091591602935t^{116} + 49345011109213111830060t^{118}
\end{split}
\end{equation}
\begin{equation}
\small
\begin{split}
\mathcal{N}_3(t) &= 77119041108339053370671t^{120} + 119784709961009844027711t^{122} \\
&\quad + 184933676316070974996613t^{124} + 283829099216170572593653t^{126} \\
&\quad + 433085523902471623911567t^{128} + 657073523309119645683934t^{130} \\
&\quad + 991342697887257941863157t^{132} + 1487467720428868095616094t^{134} \\
&\quad + 2219872171045531909538318t^{136} + 3295386496955430275451620t^{138} \\
&\quad + 4866565771010098570006447t^{140} + 7150150836603321732481655t^{142} \\
&\quad + 10452529497733654599230020t^{144} + 15204676374120905041837656t^{146} \\
&\quad + 22009863448434068945964019t^{148} + 31708491554517648723060069t^{150} \\
&\quad + 45465762628956174286575155t^{152} + 64889675931158862639436726t^{154} \\
&\quad + 92189090212544422741315559t^{156} + 130384472367205669817157344t^{158} \\
&\quad + 183587602807961716913424477t^{160} + 257371111780205988215168990t^{162} \\
&\quad + 359254499132746362078478879t^{164} + 499340505544963984724670591t^{166} \\
&\quad + 691144667734011740982009351t^{168} + 952671971301460005270196983t^{170} \\
&\quad + 1307808142895302150160572220t^{172} + 1788109798334543285058858014t^{174} \\
&\quad + 2435097962249936105947919521t^{176} + 3303184058211756089166675518t^{178}
\end{split}
\end{equation}
\begin{equation}
\small
\begin{split}
\mathcal{N}_4(t) &= 4463387086517565513345654323t^{180} + 6008036204662665457811230878t^{182} \\
&\quad + 8056695245886722861846353065t^{184} + 10763595896033435949514973459t^{186} \\
&\quad + 14326925437692072515874018375t^{188} + 19000384394611580727018495265t^{190} \\
&\quad + 25107510382385625848732075794t^{192} + 33059358373979853462971503210t^{194} \\
&\quad + 43376235846880517635519142758t^{196} + 56714315334838053369712064237t^{198} \\
&\quad + 73898088182963571583124488762t^{200} + 95959783154192183051119992197t^{202} \\
&\quad + 124187053187599246907632699232t^{204} + 160180434102155538416891772750t^{206} \\
&\quad + 205922301139697575193081757222t^{208} + 263859293352186109379011909948t^{210} \\
&\quad + 337000441900593556571474744878t^{212} + 429033525707314521048266751603t^{214} \\
&\quad + 544462485251034666386291946967t^{216} + 688769050438442695192529951587t^{218} \\
&\quad + 868602078285881990020810241298t^{220} + 1091998446336969708971201505395t^{222} \\
&\quad + 1368639702817630480625684544029t^{224} + 1710149027586104471922021008743t^{226} \\
&\quad + 2130433400558255279086905965066t^{228} + 2646076196441348647898025276852t^{230} \\
&\quad + 3276785714552876938898516802774t^{232} + 4045905396737108618797063892555t^{234} \\
&\quad + 4980991669656826732102581815960t^{236} + 6114465453032993083179417961269t^{238}
\end{split}
\end{equation}
\begin{equation}
\small
\begin{split}
\mathcal{N}_5(t) &= 7484343384097534216491033182091t^{240} + 9135054700499246289531579912730t^{242} \\
&\quad + 11118349477803593350256059925024t^{244} + 13494303511262604431255683560116t^{246} \\
&\quad + 16332424541884481641053063413471t^{248} + 19712863731144157896960249803201t^{250} \\
&\quad + 23727735264638914873412065487639t^{252} + 28482545691518152107591938348682t^{254} \\
&\quad + 34097733064511865865636127652863t^{256} + 40710314118586279720695762403129t^{258} \\
&\quad + 48475635602169846496940117530050t^{260} + 57569223445739484276627615813245t^{262} \\
&\quad + 68188720716305167511641334291172t^{264} + 80555902267696653332635815785309t^{266} \\
&\quad + 94918750668032386374454217942222t^{268} + 111553574388535905185389818592882t^{270} \\
&\quad + 130767145402768801305460390241999t^{272} + 152898829313507533890783249763597t^{274} \\
&\quad + 178322676947995135853746376661356t^{276} + 207449442104534886952572429874576t^{278} \\
&\quad + 240728485869230675484059594130335t^{280} + 278649523737270224943285095976301t^{282} \\
&\quad + 321744167765515847542758829497305t^{284} + 370587212259319509046423344701279t^{286} \\
&\quad + 425797608172268789797127892763814t^{288} + 488039068596020180981826960395322t^{290} \\
&\quad + 558020245566749063098003223238529t^{292} + 636494417046042109280055004539286t^{294} \\
&\quad + 724258622478312122773241139552082t^{296} + 822152185911830518838247784945914t^{298}
\end{split}
\end{equation}
\begin{equation}
\small
\begin{split}
\mathcal{N}_6(t) &= 931054567417461795877405538533187t^{300} + 1051882486558916353516293536913465t^{302} \\
&\quad + 1185586266057360289193995730504570t^{304} + 1333145349629706338429252365661374t^{306} \\
&\quad + 1495562955319811706031429120685831t^{308} + 1673859834514803539029063553650668t^{310} \\
&\quad + 1869067117244774907898844907824720t^{312} + 2082218236269851517370517674236772t^{314} \\
&\quad + 2314339935795018986389166622181631t^{316} + 2566442385312810157833274994410936t^{318} \\
&\quad + 2839508434910269161993463708528360t^{320} + 3134482065202592753551735879725342t^{322} \\
&\quad + 3452256102644987954089639147582330t^{324} + 3793659289061629581630495288256659t^{326} \\
&\quad + 4159442812515599360811185864705470t^{328} + 4550266424793510658825180547311853t^{330} \\
&\quad + 4966684288433203063451128726452668t^{332} + 5409130713000852931374672316141290t^{334} \\
&\quad + 5877905955829234732144639519298018t^{336} + 6373162276259019727388494925682299t^{338} \\
&\quad + 6894890444179570429523694873547049t^{340} + 7442906912955535183052873330276166t^{342} \\
&\quad + 8016841873282980610345875334028455t^{344} + 8616128407807079906420440670751167t^{346} \\
&\quad + 9239992966157203938185341255814175t^{348} + 9887447376169885918741921150598515t^{350} \\
&\quad + 10557282599291012584431561883711319t^{352} + 11248064426359032371687209775282284t^{354} \\
&\quad + 11958131294129837991399599993178492t^{356} + 12685594383050772036814402703972874t^{358}
\end{split}
\end{equation}
\begin{equation}
\small
\begin{split}
\mathcal{N}_7(t) &= 13428340133050823560128528183314466t^{360} + 14184035286698378864219759147060094t^{362} \\
&\quad + 14950134538286676060075404664935013t^{364} + 15723890833625544676737665006448074t^{366} \\
&\quad + 16502368329013372670216512529114580t^{368} + 17282457979577896390710856440499414t^{370} \\
&\quad + 18060895687518321538251848510889378t^{372} + 18834282900420761331586778911682045t^{374} \\
&\quad + 19599109509465382016247968840467469t^{376} + 20351778857738608364973930322379649t^{378} \\
&\quad + 21088634630765002836186352260366922t^{380} + 21805989365537692719817076837461448t^{382} \\
&\quad + 22500154281493435789569166193820478t^{384} + 23167470107753790396890121792309608t^{386} \\
&\quad + 23804338556192216635947393848713291t^{388} + 24407254070075567528071246826075000t^{390} \\
&\quad + 24972835463673267868631194417673449t^{392} + 25497857059738510651350228112999179t^{394} \\
&\quad + 25979278929446714272774492971815962t^{396} + 26414275843414171263711030957917537t^{398} \\
&\quad + 26800264552878848842201329342936931t^{400} + 27134929036942682514637096141506477t^{402} \\
&\quad + 27416243374760079311808135092438572t^{404} + 27642491930393654508453771431000395t^{406} \\
&\quad + 27812286572307930090335992861716498t^{408} + 27924580688583382226389217703867314t^{410} \\
&\quad + 27978679802253437628107591650908160t^{412} + 27974248637952260348169049264109970t^{414} \\
&\quad + 27911314540494508187170349454216197t^{416} + 27790267197215215170105022147372160t^{418}
\end{split}
\end{equation}
\begin{equation}
\small
\begin{split}
\mathcal{N}_8(t) &= 27611854667966593861264848333668844t^{420} + 27377175778667604829514471966420308t^{422} \\
&\quad + 27087668985302536154792192770905125t^{424} + 26745097864358355363204169820967529t^{426} \\
&\quad + 26351533432010732161620890030186328t^{428} + 25909333537107704178064184336083213t^{430} \\
&\quad + 25421119611427448071890422944373051t^{432} + 24889751094162892919310191248082588t^{434} \\
&\quad + 24318297875576369663908894099390080t^{436} + 23710011126851470245830088578574883t^{438} \\
&\quad + 23068292899048835703982433025152411t^{440} + 22396664883575647687509377070026442t^{442} \\
&\quad + 21698736729661786985902410903509890t^{444} + 20978174311081469383464152622719471t^{446} \\
&\quad + 20238668324972943041281962879444944t^{448} + 19483903590410160743591560237500055t^{450} \\
&\quad + 18717529393794790219486170782987203t^{452} + 17943131202682956732897162241938494t^{454} \\
&\quad + 17164204039937574844575560764790950t^{456} + 16384127776766961609293867289938441t^{458} \\
&\quad + 15606144566985622344264061357043901t^{460} + 14833338606456813092897037652203307t^{462} \\
&\quad + 14068618361907165584596610488787012t^{464} + 13314701372897326342149001977409425t^{466} \\
&\quad + 12574101690427808659950710716324284t^{468} + 11849119976161278809050974298264455t^{470} \\
&\quad + 11141836248210462908301543793868886t^{472} + 10454105223474999528177060068441698t^{474} \\
&\quad + 9787554173143017424594807572983218t^{476} + 9143583177659108776798584880842524t^{478}
\end{split}
\end{equation}
\begin{equation}
\small
\begin{split}
\mathcal{N}_9(t) &= 8523367640572775456207339326663983t^{480} + 7927862897506849692065490751376359t^{482} \\
&\quad + 7357810737223037095894534565911511t^{484} + 6813747636523301852105198246538496t^{486} \\
&\quad + 6296014499538807500035936174043517t^{488} + 5804767684769454073792615645937777t^{490} \\
&\quad + 5339991099919673308285645085754519t^{492} + 4901509144935686059837905510450109t^{494} \\
&\quad + 4489000287433202955172496620087796t^{496} + 4102011061610137790663529597910807t^{498} \\
&\quad + 3739970291424457237940187974988670t^{500} + 3402203350910968037521515014927314t^{502} \\
&\quad + 3087946288621320638024171307734963t^{504} + 2796359658897203437106284967641233t^{506} \\
&\quad + 2526541919625509988282650105470228t^{508} + 2277542273881683852876983602029848t^{510} \\
&\quad + 2048372851064776918011030036448782t^{512} + 1838020141408465644870573993882175t^{514} \\
&\quad + 1645455615788125816681735889417811t^{516} + 1469645480239397601001776773852977t^{518} \\
&\quad + 1309559531299199402209245235634409t^{520} + 1164179093955014580136694236007064t^{522} \\
&\quad + 1032504038461620459963054824815560t^{524} + 913558885415339326026859209492556t^{526} \\
&\quad + 806398020162960022057138164512494t^{528} + 710110047802004382174006221546653t^{530} \\
&\quad + 623821328672960089103037785365074t^{532} + 546698741357002293024939196311898t^{534} \\
&\quad + 477951725808535980124502701882539t^{536} + 416833663430100337119659005908293t^{538}
\end{split}
\end{equation}
\begin{equation}
\small
\begin{split}
\mathcal{N}_{10}(t) &= 362642653719078755687129653443272t^{540} + 314721748680170024608109460190589t^{542} \\
&\quad + 272458706617391697869479088422265t^{544} + 235285326316940767891007792120216t^{546} \\
&\quad + 202676421136142488997915737933465t^{548} + 174148490255210566327761247732878t^{550} \\
&\quad + 149258141458482927765250150502950t^{552} + 127600316417824475523389273989305t^{554} \\
&\quad + 108806365675156835823671399829774t^{556} + 92542016478218896003568774817236t^{558} \\
&\quad + 78505272419331078563387881696809t^{560} + 66424279556575323715511804899770t^{562} \\
&\quad + 56055189444949379237311243138869t^{564} + 47180045344857283333318341234809t^{566} \\
&\quad + 39604713868473995177378773371012t^{568} + 33156880521480349416626499860304t^{570} \\
&\quad + 27684124037973262153266718198501t^{572} + 23052081119315865751626832547155t^{574} \\
&\quad + 19142710193147627668786646814809t^{576} + 15852660117967814228887323085221t^{578} \\
&\quad + 13091747375276107696465190913117t^{580} + 10781543212218942980638083047640t^{582} \\
&\quad + 8854070414492065948038135733458t^{584} + 7250607888739558908388697975758t^{586} \\
&\quad + 5920599999113270789891939912067t^{588} + 4820666614545536675767606563547t^{590} \\
&\quad + 3913709060312065025199545896568t^{592} + 3168106607167363182591943164913t^{594} \\
&\quad + 2556997750822738595814299307961t^{596} + 2057640311020734730799584172282t^{598}
\end{split}
\end{equation}
\begin{equation}
\small
\begin{split}
\mathcal{N}_{11}(t) &= 1650844290756264319989441042217t^{600} + 1320471461100718056192153371398t^{602} \\
&\quad + 1052995755688814978728417737130t^{604} + 837118752821774741666856599101t^{606} \\
&\quad + 663434775578155765391636449477t^{608} + 524140436298491084834401736676t^{610} \\
&\quad + 412783778088062670334456194877t^{612} + 324048511080358410622740115342t^{614} \\
&\quad + 253569195335478986648515361524t^{616} + 197773577232169663043147234688t^{618} \\
&\quad + 153748635366434667114308054167t^{620} + 119127229979646811516169378157t^{622} \\
&\quad + 91992572724875995733415935545t^{624} + 70798038153232218518494873664t^{626} \\
&\quad + 54300122625612166391884303370t^{628} + 41502619205414803028080985811t^{630} \\
&\quad + 31610317923792158298841439762t^{632} + 23990759649853667881843013028t^{634} \\
&\quad + 18142769111772521540854729230t^{636} + 13670669215929990932787798467t^{638} \\
&\quad + 10263235773814695864993850470t^{640} + 7676590324776917819321331269t^{642} \\
&\quad + 5720350306483598382316242295t^{644} + 4246461803034337555961687950t^{646} \\
&\quad + 3140231936857905319553344557t^{648} + 2313157086726321264455010924t^{650} \\
&\quad + 1697210883358037891753686988t^{652} + 1240313659029657193415776410t^{654} \\
&\quad + 902753927171035992792725169t^{656} + 654373667513207304860460816t^{658}
\end{split}
\end{equation}
\begin{equation}
\small
\begin{split}
\mathcal{N}_{12}(t) &= 472363719471226405599369383t^{660} + 339544369260150065965780279t^{662} \\
&\quad + 243030085301369184262048739t^{664} + 173197048229809156246353507t^{666} \\
&\quad + 122888283859017144148744361t^{668} + 86804404681601120746525918t^{670} \\
&\quad + 61038686903960898297801818t^{672} + 42723875873136601683905779t^{674} \\
&\quad + 29765081659361434120467207t^{676} + 20638702419076779787072972t^{678} \\
&\quad + 14241751840060151171107318t^{680} + 9779482458339536288986790t^{682} \\
&\quad + 6681966791157586885895932t^{684} + 4542469936173745141600763t^{686} \\
&\quad + 3072141112854176701977986t^{688} + 2066865925941586483606654t^{690} \\
&\quad + 1383135674852367978209664t^{692} + 920569107449497403195658t^{694} \\
&\quad + 609317138173537168730459t^{696} + 401033252477232830458254t^{698} \\
&\quad + 262434095447892606104579t^{700} + 170731670303716731130821t^{702} \\
&\quad + 110410669284066165803610t^{704} + 70967297420181613275381t^{706} \\
&\quad + 45331572345303085440028t^{708} + 28772749489407988773166t^{710} \\
&\quad + 18144307907474394074939t^{712} + 11366213007986817772849t^{714} \\
&\quad + 7072007756237073341533t^{716} + 4369719964721992559135t^{718}
\end{split}
\end{equation}
\begin{equation}
\small
\begin{split}
\mathcal{N}_{13}(t) &= 2680882668338132199128t^{720} + 1632832080154222253883t^{722} \\
&\quad + 987114554423453604791t^{724} + 592209105816468611001t^{726} \\
&\quad + 352516479286075696040t^{728} + 208157368463555921162t^{730} \\
&\quad + 121904285912524926679t^{732} + 70788557904392570230t^{734} \\
&\quad + 40749387578588609953t^{736} + 23247948491581079716t^{738} \\
&\quad + 13141334444839528588t^{740} + 7358089286463832256t^{742} \\
&\quad + 4079754886448561232t^{744} + 2239297866992321823t^{746} \\
&\quad + 1216342387348422257t^{748} + 653604119279225842t^{750} \\
&\quad + 347318648031221994t^{752} + 182442080617318828t^{754} \\
&\quad + 94694247156572978t^{756} + 48543346844580876t^{758} \\
&\quad + 24566080994348077t^{760} + 12266477702584790t^{762} \\
&\quad + 6040087991175823t^{764} + 2931228452687289t^{766} \\
&\quad + 1401079547738756t^{768} + 659149889096119t^{770} \\
&\quad + 304993546316844t^{772} + 138685559425815t^{774} \\
&\quad + 61918987477947t^{776} + 27117586843351t^{778}
\end{split}
\end{equation}
\begin{equation}
\small
\begin{split}
\mathcal{N}_{14}(t) &= 11637387429615t^{780} + 4888033612959t^{782} + 2006929236867t^{784} \\
&\quad + 804328971413t^{786} + 314162619846t^{788} + 119379047412t^{790} \\
&\quad + 44044718603t^{792} + 15742613364t^{794} + 5437129346t^{796} \\
&\quad + 1809246325t^{798} + 578084366t^{800} + 176650616t^{802} \\
&\quad + 51383253t^{804} + 14143976t^{806} + 3657851t^{808} \\
&\quad + 879923t^{810} + 194211t^{812} + 38396t^{814} \\
&\quad + 6524t^{816} + 846t^{818} + 51t^{820} \\
&\quad - 14t^{822} - 5t^{824} - t^{826}.
\end{split}
\end{equation}
$\mathcal{D}(t)$ is defined as
\begin{equation}
\small
\begin{split}
\mathcal{D}(t) &= (-1 + t^2)^6(1 + t^2)^3(1 + t^4)^{19}(1 - t^2 + t^4)^5(1 + t^2 + t^4)^6(1 + t^8)^8 \\
&\quad \times (1 - t^4 + t^8)^7(1 - t^2 + t^4 - t^6 + t^8)^4(1 + t^2 + t^4 + t^6 + t^8)^6 \\
&\quad \times (1 - t^6 + t^{12})(1 + t^6 + t^{12})^7(1 - t^2 + t^4 - t^6 + t^8 - t^{10} + t^{12})^5 \\
&\quad \times (1 + t^2 + t^4 + t^6 + t^8 + t^{10} + t^{12})^9(1 + t^{16})^3 \\
&\quad \times (1 - t^2 + t^6 - t^8 + t^{10} - t^{14} + t^{16}) \\
&\quad \times (1 + t^2 + t^4 + t^6 + t^8 + t^{10} + t^{12} + t^{14} + t^{16} + t^{18} + t^{20})^5 \\
&\quad \times (1 + t^2 + t^4 + t^6 + t^8 + t^{10} + t^{12} + t^{14} + t^{16} + t^{18} + t^{20} + t^{22} + t^{24})^3.
\end{split}
\end{equation}

The numerator of $\mathcal{I}_{n_1=7; N_1=1}^{\mathrm{BMN}~SU(N)}(t)$ is a polynomial in $t$ of degree $1422$ with integer coefficients. Because its full expression is extremely long, we write only the first few and last few terms, indicating the omitted part by $\cdots$.

\begin{equation}
\begin{split}
\mathcal{I}_{n_1=7; N_1=1}^{\mathrm{BMN}~SU(N)}(t)=
\frac{ \Bigl( -1 - 21t^2 - 263t^4 - 2460t^6 - 18902t^8 + \cdots - 45t^{1420} - 3t^{1422} \Bigr) }{%
\begin{aligned}
&(-1 + t^2)^7 (1 + t^4)^{19} (1 - t^2 + t^4)^6 (1 + t^2 + t^4)^7 (1 + t^8)^8 \\
&\times (1 - t^4 + t^8)^7 (1 - t^2 + t^4 - t^6 + t^8)^6 (1 + t^2 + t^4 + t^6 + t^8)^{12} \\
&\times (1 - t^6 + t^{12}) (1 + t^6 + t^{12})^8 (1 - t^2 + t^4 - t^6 + t^8 - t^{10} + t^{12})^5 \\
&\times (1 + t^2 + t^4 + t^6 + t^8 + t^{10} + t^{12})^9 (1 + t^{16})^3 \\
&\times (1 - t^2 + t^6 - t^8 + t^{10} - t^{14} + t^{16})^7 \\
&\times (1 + t^2 + t^4 + t^6 + t^8 + t^{10} + t^{12} + t^{14} + t^{16} + t^{18} + t^{20})^8 \\
&\times (1 - t^2 + t^6 - t^8 + t^{12} - t^{16} + t^{18} - t^{22} + t^{24}) \\
&\times (1 + t^2 + t^4 + t^6 + t^8 + t^{10} + t^{12} + t^{14} + t^{16} + t^{18} + t^{20} + t^{22} + t^{24})^6 \\
&\times (1 + t^2 + t^4 + t^6 + t^8 + t^{10} + t^{12} + t^{14} + t^{16} + t^{18} + t^{20} + t^{22} + t^{24} \\
&\qquad + t^{26} + t^{28} + t^{30} + t^{32})^5 \\
&\times (1 + t^2 + t^4 + t^6 + t^8 + t^{10} + t^{12} + t^{14} + t^{16} + t^{18} + t^{20} + t^{22} + t^{24} \\
&\qquad + t^{26} + t^{28} + t^{30} + t^{32} + t^{34} + t^{36})^3
\end{aligned}
}.
\end{split}
\end{equation}

Expanding in series
\begin{equation}
\small
\begin{split}
\mathcal{I}_{n_1=2; N_1=1}^{\mathrm{BMN}~SU(N)}(t) &= 1 + 6t^4+\mathcal{O}(t^{6}).
\end{split}
\end{equation}
\begin{equation}
\small
\begin{split}
\mathcal{I}_{n_1=3; N_1=1}^{\mathrm{BMN}~SU(N)}(t) &= 1 + 6 t^4 + 2 t^6 + 9 t^8 + 3 t^{10} + \mathcal{O}(t^{12}).
\end{split}
\end{equation}
\begin{equation}
\small
\begin{split}
\mathcal{I}_{n_1=4; N_1=1}^{\mathrm{BMN}~SU(N)}(t) &= 1 + 6 t^4 + 2 t^6 + 24 t^8 - 6 t^{10} + 60 t^{12} - 48 t^{14} + 135 t^{16} + \mathcal{O}(t^{18}).
\end{split}
\end{equation}
\begin{equation}
\small
\begin{split}
\mathcal{I}_{n_1=5; N_1=1}^{\mathrm{BMN}~SU(N)}(t) &= 1 + 6 t^4 + 2 t^6 + 24 t^8 + 15 t^{10} + 52 t^{12} + 39 t^{14} + 72 t^{16} + 74 t^{18} + 93 t^{20} \\
&+ 198 t^{22} + 236 t^{24} + 345 t^{26} + \mathcal{O}(t^{28}).
\end{split}
\end{equation}
\begin{equation}
\small
\begin{split}
\mathcal{I}_{n_1=6; N_1=1}^{\mathrm{BMN}~SU(N)}(t) &=1 + 6 t^4 + 2 t^6 + 24 t^8 + 15 t^{10} + 80 t^{12} + 33 t^{14} + 198 t^{16} + 22 t^{18} \\
&+ 399 t^{20} - 57 t^{22} + 795 t^{24} - 69 t^{26} + 1719 t^{28} - 144 t^{30} + 2523 t^{32} \\
&- 1494 t^{34} + 3183 t^{36} + \mathcal{O}(t^{38}).
\end{split}
\end{equation}
\begin{equation}
\small 
\begin{split}
\mathcal{I}_{n_1=7; N_1=1}^{\mathrm{BMN}~SU(N)}(t) &=1 + 6 t^4 + 2 t^6 + 24 t^8 + 15 t^{10} + 80 t^{12} + 69 t^{14} + 195 t^{16} + 195 t^{18} \\
&+ 369 t^{20} + 405 t^{22} + 584 t^{24} + 762 t^{26} + 1026 t^{28} + 1720 t^{30} + 2316 t^{32} \\
&+ 3294 t^{34} + 3133 t^{36} + 2724 t^{38} + 1875 t^{40} + 3683 t^{42} + 10323 t^{44} \\
&+ 19407 t^{46} + 23630 t^{48} + 6558 t^{50} + \mathcal{O}(t^{52}).
\end{split}
\end{equation}
By expanding the closed-form trivial vacuum indices in fugacity, we verified that the series matches the results obtained via character expansion noted in \eqref{b1}-\eqref{b6} in Appendix \ref{spbmn}.

\subsection{Closed-form $SU(N)$ single partition BMN index besides trivial vacuum}
\label{single-partitionclosedform}
In this Appendix \ref{single-partitionclosedform}, we list the closed-form results for single partition sector $SU(N)$ BMN indices other than trivial vacuum sector indices 
\begin{equation}
\begin{split}
\mathcal{I}_{n_1=2;N_1=2}^{\mathrm{BMN}~SU(N)}(t)&=\big( (1 - t^6) \, (1 + 3\,t^2 + 20\,t^4 + 44\,t^6 + 174\,t^8 + 325\,t^{10} + 990\,t^{12} + 1680\,t^{14} + 4279\,t^{16} \\
&\quad + 6806\,t^{18} + 15113\,t^{20} + 22932\,t^{22} + 45588\,t^{24} + 66604\,t^{26} + 120816\,t^{28} \\
&\quad + 170782\,t^{30} + 286816\,t^{32} + 393307\,t^{34} + 618449\,t^{36} + 824001\,t^{38} \\
&\quad + 1223718\,t^{40} + 1585932\,t^{42} + 2239503\,t^{44} + 2825612\,t^{46} + 3814061\,t^{48} \\
&\quad + 4688362\,t^{50} + 6074539\,t^{52} + 7279364\,t^{54} + 9083013\,t^{56} + 10616902\,t^{58} \\
&\quad + 12790958\,t^{60} + 14590541\,t^{62} + 17006961\,t^{64} + 18939937\,t^{66} \\
&\quad + 21392686\,t^{68} + 23267720\,t^{70} + 25496799\,t^{72} + 27091567\,t^{74} \\
&\quad + 28825679\,t^{76} + 29928153\,t^{78} + 30937050\,t^{80} + 31389999\,t^{82} \\
&\quad + 31533040\,t^{84} + 31269341\,t^{86} + 30526488\,t^{88} + 29584734\,t^{90} \\
&\quad + 28061030\,t^{92} + 26576638\,t^{94} + 24478768\,t^{96} + 22653231\,t^{98} \\
&\quad + 20245119\,t^{100} + 18302795\,t^{102} + 15852945\,t^{104} + 13997595\,t^{106} \\
&\quad + 11732206\,t^{108} + 10114426\,t^{110} + 8187080\,t^{112} + 6889192\,t^{114} \\
&\quad + 5371500\,t^{116} + 4410295\,t^{118} + 3301389\,t^{120} + 2644008\,t^{122} \\
&\quad + 1892143\,t^{124} + 1477720\,t^{126} + 1005500\,t^{128} + 765586\,t^{130} \\
&\quad + 491844\,t^{132} + 365047\,t^{134} + 219399\,t^{136} + 158716\,t^{138} + 88162\,t^{140} \\
&\quad + 62148\,t^{142} + 31394\,t^{144} + 21544\,t^{146} + 9684\,t^{148} + 6449\,t^{150} + 2502\,t^{152} \\
&\quad + 1605\,t^{154} + 513\,t^{156} + 312\,t^{158} + 75\,t^{160} + 42\,t^{162} + 6\,t^{164} + 3\,t^{166}) \big) \\
&\Big/ \Big( (-1 + t^2)^2 (1 + t^2)^5 (1 + t^4)^9 (1 + t^8)^5 (1 - t^4 + t^8) (1 + t^4 + t^8)^2 \\
&\quad \cdot (1 + t^4 + t^8 + t^{10} + t^{12} + t^{16} + t^{20})^3 \Big).
\end{split}
\end{equation}

\begin{equation}
\begin{split}
\mathcal{I}_{n_1=2;N_1=3}^{\mathrm{BMN}~SU(N)}(t) = \frac{\mathcal{N}(t)}{\mathcal{D}(t)},
\end{split}
\end{equation}
where $\mathcal{N}(t) = (1 - t^6) \sum_{i=1}^{8} \mathcal{N}_i(t)$ with the numerator parts defined as
\begin{equation}
\small
\begin{split}
\mathcal{N}_1(t) &=  1 + 12 t^2 + 92 t^4 + 524 t^6 + 2460 t^8 + 9943 t^{10} + 35769 t^{12} + 116888 t^{14} + 352569 t^{16} + 992908 t^{18} \\
&\quad + 2634931 t^{20} + 6636508 t^{22} + 15958513 t^{24} + 36815081 t^{26} + 81810976 t^{28} + 175727882 t^{30} \\
&\quad + 365927544 t^{32} + 740588844 t^{34} + 1459986030 t^{36} + 2808989675 t^{38} + 5283522948 t^{40} \\
&\quad + 9730288510 t^{42} + 17568800924 t^{44} + 31138401630 t^{46} + 54232441916 t^{48} + 92908442556 t^{50} \\
&\quad + 156700012470 t^{52} + 260405441457 t^{54} + 426694286910 t^{56} + 689859284082 t^{58} + 1101154211436 t^{60} ,
\end{split}
\end{equation}

\begin{equation}
\small
\begin{split}
\mathcal{N}_2(t) &=  1736301458655 t^{62} + 2705936992427 t^{64} + 4169987020852 t^{66} + 6357255003954 t^{68} \\
&\quad + 9591839105081 t^{70} + 14328418505966 t^{72} + 21198936040547 t^{74} + 31073789305585 t^{76} \\
&\quad + 45141306262033 t^{78} + 65010054930589 t^{80} + 92839386310249 t^{82} + 131504571982224 t^{84} \\
&\quad + 184803912587829 t^{86} + 257716305237653 t^{88} + 356718868111049 t^{90} + 490175385279538 t^{92} \\
&\quad + 668807402973262 t^{94} + 906260855114053 t^{96} + 1219781901865309 t^{98} + 1631016337049964 t^{100} \\
&\quad + 2166947155091789 t^{102} + 2860984825687796 t^{104} + 3754224100324427 t^{106} + 4896880014392617 t^{108} \\
&\quad + 6349913630427960 t^{110} + 8186855367710819 t^{112} + 10495829825801974 t^{114} + 13381781400063945 t^{116} \\
&\quad + 16968893898384976 t^{118} + 21403190628681348 t^{120},
\end{split}
\end{equation}

\begin{equation}
\small
\begin{split}
\mathcal{N}_3(t) &= 26855293052895206 t^{122} + 33523307272072569 t^{124} + 41635797124844742 t^{126} \\
&\quad + 51454792167512981 t^{128} + 63278766801685870 t^{130} + 77445515483775318 t^{132} \\
&\quad + 94334836493770569 t^{134} + 114370925951349437 t^{136} + 138024372400306127 t^{138} \\
&\quad + 165813633850993350 t^{140} + 198305870832316881 t^{142} + 236117005069661876 t^{144} \\
&\quad + 279910870244545851 t^{146} + 330397324076792585 t^{148} + 388329195060755277 t^{150} \\
&\quad + 454497948674495996 t^{152} + 529727970859394540 t^{154} + 614869388110740421 t^{156} \\
&\quad + 710789366188406132 t^{158} + 818361861386635138 t^{160} + 938455830258258454 t^{162} \\
&\quad + 1071921944983848830 t^{164} + 1219577900944217989 t^{166} + 1382192451147880527 t^{168} \\
&\quad + 1560468345498869485 t^{170} + 1755024403681229218 t^{172} + 1966376992857302001 t^{174} \\
&\quad + 2194921229495927800 t^{176} + 2440912260240643364 t^{178} + 2704447016060684335 t^{180} ,
\end{split}
\end{equation}

\begin{equation}
\small
\begin{split}
\mathcal{N}_4(t) &=  2985446856393454522 t^{182} + 3283641544424894347 t^{184} + 3598554998141112846 t^{186} \\
&\quad + 3929493266118091215 t^{188} + 4275535157047921265 t^{190} + 4635525932833094951 t^{192} \\
&\quad + 5008074429427360685 t^{194} + 5391553926138770269 t^{196} + 5784107013460302104 t^{198} \\
&\quad + 6183654644051553043 t^{200} + 6587909459779825414 t^{202} + 6994393405851590049 t^{204} \\
&\quad + 7400459537348806572 t^{206} + 7803317833586965918 t^{208} + 8200064725595679490 t^{210} \\
&\quad + 8587715954567011729 t^{212} + 8963242275707257060 t^{214} + 9323607448207805026 t^{216} \\
&\quad + 9665807866599996452 t^{218} + 9986913139443978172 t^{220} + 10284106863707111436 t^{222} \\
&\quad + 10554726826833619260 t^{224} + 10796303845889080888 t^{226} + 11006598474915857434 t^{228} \\
&\quad + 11183634827216174108 t^{230} + 11325730819981081536 t^{232} + 11431524201159381736 t^{234} \\
&\quad + 11499993814322607080 t^{236} + 11530475640808454274 t^{238} + 11522673281338761597 t^{240} ,
\end{split}
\end{equation}

\begin{equation}
\small
\begin{split}
\mathcal{N}_5(t) &=  11476662644123341845 t^{242} + 11392890744923940408 t^{244} + 11272168638168726136 t^{246} \\
&\quad + 11115658638743993440 t^{248} + 10924856102882439435 t^{250} + 10701566165437605786 t^{252} \\
&\quad + 10447875921800540186 t^{254} + 10166122647665158124 t^{256} + 9858858712273964843 t^{258} \\
&\quad + 9528813912568458260 t^{260} + 9178855982454803988 t^{262} + 8811950064898037057 t^{264} \\
&\quad + 8431117923029671403 t^{266} + 8039397661140382249 t^{268} + 7639804678508532305 t^{270} \\
&\quad + 7235294538824166897 t^{272} + 6828728359719439721 t^{274} + 6422841259510858710 t^{276} \\
&\quad + 6020214298897957113 t^{278} + 5623250270934872723 t^{280} + 5234153583129163352 t^{282} \\
&\quad + 4854914385335485418 t^{284} + 4487296988852378972 t^{286} + 4132832537032608556 t^{288} \\
&\quad + 3792815790836554041 t^{290} + 3468305821874792412 t^{292} + 3160130327885062606 t^{294} \\
&\quad + 2868893235371271962 t^{296} + 2594985200707285917 t^{298} + 2338596594736171953 t^{300} ,
\end{split}
\end{equation}

\begin{equation}
\small
\begin{split}
\mathcal{N}_6(t) &=  2099732528282045670 t^{302} + 1878229473752617362 t^{304} + 1673773034788270856 t^{306} \\
&\quad + 1485916435008000832 t^{308} + 1314099314177686878 t^{310} + 1157666455788244922 t^{312} \\
&\quad + 1015886102264428107 t^{314} + 887967560314593214 t^{316} + 773077839862409509 t^{318} \\
&\quad + 670357120718293995 t^{320} + 578932884670608820 t^{322} + 497932600163843304 t^{324} \\
&\quad + 426494887606163065 t^{326} + 363779137090112196 t^{328} + 308973584642162359 t^{330} \\
&\quad + 261301888078873270 t^{332} + 220028268995135375 t^{334} + 184461312125336326 t^{336} \\
&\quad + 153956529064107295 t^{338} + 127917807762379867 t^{340} + 105797875738725820 t^{342} \\
&\quad + 87097910179303378 t^{344} + 71366426832620451 t^{346} + 58197577481277031 t^{348} \\
&\quad + 47228978698265033 t^{350} + 38139187304689855 t^{352} + 30644927223235667 t^{354} \\
&\quad + 24498162157712665 t^{356} + 19483096182957093 t^{358} + 15413173022682686 t^{360} ,
\end{split}
\end{equation}

\begin{equation}
\small
\begin{split}
\mathcal{N}_7(t) &=  12128132497128355 t^{362} + 9491171776634749 t^{364} + 7386248041732957 t^{366} \\
&\quad + 5715549777272030 t^{368} + 4397154877382531 t^{370} + 3362886374877889 t^{372} \\
&\quad + 2556369831379569 t^{374} + 1931291209015171 t^{376} + 1449849486579154 t^{378} \\
&\quad + 1081395060013257 t^{380} + 801242343320451 t^{382} + 589643419659540 t^{384} \\
&\quad + 430908495026591 t^{386} + 312658582611377 t^{388} + 225195835115940 t^{390} \\
&\quad + 160977445666538 t^{392} + 114179707138675 t^{394} + 80339789070912 t^{396} \\
&\quad + 56063806121189 t^{398} + 38790905593961 t^{400} + 26604215243313 t^{402} \\
&\quad + 18080634893330 t^{404} + 12172507065884 t^{406} + 8115215963332 t^{408} \\
&\quad + 5355665685155 t^{410} + 3497420358159 t^{412} + 2259007052028 t^{414} \\
&\quad + 1442522010313 t^{416} + 910218781952 t^{418} + 567221735551 t^{420} ,
\end{split}
\end{equation}

\begin{equation}
\small
\begin{split}
\mathcal{N}_8(t) &=  348889997091 t^{422} + 211677594149 t^{424} + 126592607699 t^{426} \\
&\quad + 74568665623 t^{428} + 43226827300 t^{430} + 24637487192 t^{432} + 13792384270 t^{434} \\
&\quad + 7575074904 t^{436} + 4076487574 t^{438} + 2146435955 t^{440} + 1104038388 t^{442} \\
&\quad + 553727016 t^{444} + 270241085 t^{446} + 128032624 t^{448} + 58722961 t^{450} \\
&\quad + 25991209 t^{452} + 11059433 t^{454} + 4503878 t^{456} + 1745892 t^{458} \\
&\quad + 639973 t^{460} + 219964 t^{462} + 70149 t^{464} + 20456 t^{466} + 5352 t^{468} \\
&\quad + 1218 t^{470} + 231 t^{472} + 33 t^{474} + 3 t^{476},
\end{split}
\end{equation}

\begin{equation}
\small
\begin{split}
\mathcal{D}(t) &= (1 + t^2)^4 (1 + t^4)^3 (1 - t^2 + t^4) (1 - t^2 + t^4 - t^6 + t^8)^3 (1 - t^6 + t^{12}) \\
&\quad \times (1 + t^2 + t^4 + t^6 + t^8 + t^{10} + t^{12})^9 (1 + t^{16})^3 (-1 + t^{18})^2 \\
&\quad \times (1 + t^2 + t^4 + t^6 + t^8 + t^{10} + t^{12} + t^{14} + t^{16} + t^{18} + t^{20} + t^{22} + t^{24})^3 \\
&\quad \times (1 + t^4 + t^8 + t^{12} + t^{14} + t^{16} + t^{18} + t^{20} + t^{24} + t^{28} + t^{32})^5.
\end{split}
\end{equation}
\begin{equation}
\mathcal{I}_{n_1=2;N_1=4}^{\mathrm{BMN}~SU(N)}(t)=\frac{\mathcal{N}(t)}{\mathcal{D}(t)},
\end{equation}

\begin{equation}
\mathcal{N}(t) = (1 - t^6) \sum_{i=1}^{8} \mathcal{N}_i(t),
\end{equation}

\begin{equation}
\small
\begin{split}
\mathcal{N}_1(t) &= 1 + 8 t^2 + 51 t^4 + 232 t^6 + 928 t^8 + 3191 t^{10} \\
&\quad + 10113 t^{12} + 29337 t^{14} + 80316 t^{16} + 207088 t^{18} \\
&\quad + 510899 t^{20} + 1205926 t^{22} + 2748205 t^{24} + 6050753 t^{26} \\
&\quad + 12943155 t^{28} + 26923485 t^{30} + 54662601 t^{32} + 108421914 t^{34} \\
&\quad + 210632531 t^{36} + 401134636 t^{38} + 750267713 t^{40} + 1379244567 t^{42} \\
&\quad + 2495558929 t^{44} + 4447280344 t^{46} + 7814226608 t^{48} + 13545789722 t^{50} \\
&\quad + 23185525695 t^{52} + 39206055776 t^{54} + 65540507400 t^{56} + 108364642946 t^{58} \\
&\quad + 177309002644 t^{60} + 287220478712 t^{62} + 460835071894 t^{64} + 732615377171 t^{66} \\
&\quad + 1154462289330 t^{68} + 1803820886255 t^{70} + 2795535977157 t^{72} + 4298490062887 t^{74} \\
&\quad + 6559550580590 t^{76} + 9936817851231 t^{78} + 14946776653108 t^{80} + 22329087009955 t^{82} \\
&\quad + 33137217550889 t^{84} + 48861750333971 t^{86} + 71600676750962 t^{88} + 104288868918399 t^{90} \\
&\quad + 151011232492677 t^{92} + 217420788282214 t^{94} + 311302833559933 t^{96} + 443321437897801 t^{98} \\
&\quad + 628016128553185 t^{100} + 885109360374703 t^{102} + 1241234713502055 t^{104} + 1732185110978604 t^{106}
\end{split}
\end{equation}

\begin{equation}
\small
\begin{split}
\mathcal{N}_2(t) &= 2405856200886045 t^{108} + 3326044546085274 t^{110} + 4577375091723645 t^{112} \\
&\quad + 6271609939476561 t^{114} + 8555761746944934 t^{116} + 11622402670410600 t^{118} \\
&\quad + 15722811676263769 t^{120} + 21183556284276667 t^{122} + 28427470203261140 t^{124} \\
&\quad + 37999920727994208 t^{126} + 50601782740117312 t^{128} + 67130438192447004 t^{130} \\
&\quad + 88730858784849840 t^{132} + 116858687112370834 t^{134} + 153358262199616692 t^{136} \\
&\quad + 200558327681537492 t^{138} + 261389580923354078 t^{140} + 339527918071230936 t^{142} \\
&\quad + 439569163129404773 t^{144} + 567240625997603786 t^{146} + 729657435148848861 t^{148} \\
&\quad + 935630943990941655 t^{150} + 1196039968915594898 t^{152} + 1524274676457938507 t^{154} \\
&\quad + 1936767485983984449 t^{156} + 2453623992760475250 t^{158} + 3099372833584051852 t^{160} \\
&\quad + 3903851458604247278 t^{162} + 4903252388427167090 t^{164} + 6141351739624560443 t^{166} \\
&\quad + 7670951501358035223 t^{168} + 9555563088637965560 t^{170} + 11871371927903867026 t^{172} \\
&\quad + 14709517282401079056 t^{174} + 18178736796267098161 t^{176} + 22408417534841124731 t^{178} \\
&\quad + 27552114189819664646 t^{180} + 33791584538743976331 t^{182} + 41341415408670281685 t^{184} \\
&\quad + 50454298016887128040 t^{186} + 61427039715976416213 t^{188} + 74607379844460186278 t^{190} \\
&\quad + 90401711338261881361 t^{192} + 109283784064083232714 t^{194} + 131804506483497931056 t^{196} \\
&\quad + 158602928471400397086 t^{198} + 190418536448665586760 t^{200} + 228104948052190067594 t^{202} \\
&\quad + 272645150703294337605 t^{204} + 325168371920730654961 t^{206} + 386968736420407321604 t^{208} \\
&\quad + 459525793137877561114 t^{210} + 544527073997889807909 t^{212} + 643892755747827616050 t^{214}
\end{split}
\end{equation}

\begin{equation}
\small
\begin{split}
\mathcal{N}_3(t) &= 759802588010208380310 t^{216} + 894725137997051649929 t^{218} \\
&\quad + 1051449509256324408964 t^{220} + 1233119552925800483666 t^{222} \\
&\quad + 1443270714277181370083 t^{224} + 1685869487941126288251 t^{226} \\
&\quad + 1965355599618069680662 t^{228} + 2286686827593604562672 t^{230} \\
&\quad + 2655386545154781103066 t^{232} + 3077593820618875864417 t^{234} \\
&\quad + 3560116106716055849364 t^{236} + 4110484261860516318569 t^{238} \\
&\quad + 4737009872753476347158 t^{240} + 5448844508792750952429 t^{242} \\
&\quad + 6256040803017377774087 t^{244} + 7169614860738318317051 t^{246} \\
&\quad + 8201609805000045123119 t^{248} + 9365159815186573438752 t^{250} \\
&\quad + 10674554374211209417840 t^{252} + 12145301923301815145963 t^{254} \\
&\quad + 13794192542074608376424 t^{256} + 15639358687608561901002 t^{258} \\
&\quad + 17700333513915045143453 t^{260} + 19998105638145593756773 t^{262} \\
&\quad + 22555169786968051249924 t^{264} + 25395572027359143327685 t^{266} \\
&\quad + 28544948943784987553164 t^{268} + 32030559317799502937784 t^{270} \\
&\quad + 35881307626307560917955 t^{272} + 40127757789763636776842 t^{274} \\
&\quad + 44802136476492748707637 t^{276} + 49938324303259798444289 t^{278} \\
&\quad + 55571834274030095800814 t^{280} + 61739775749716217749671 t^{282} \\
&\quad + 68480803382199702891321 t^{284} + 75835049311508280086540 t^{286} \\
&\quad + 83844038215072422201120 t^{288} + 92550583575726651578459 t^{290} \\
&\quad + 101998664984168429712092 t^{292} + 112233284978215318450125 t^{294} \\
&\quad + 123300305536759612252970 t^{296} + 135246262937655934887527 t^{298} \\
&\quad + 148118161475891842624000 t^{300} + 161963245029142778033892 t^{302} \\
&\quad + 176828747418600036252529 t^{304} + 192761620896585851706108 t^{306} \\
&\quad + 209808244225733707675034 t^{308} + 228014110083189193017135 t^{310} \\
&\quad + 247423493824248766023160 t^{312} + 268079103784568321860980 t^{314} \\
&\quad + 290021715760247141527078 t^{316} + 313289792315871443561696 t^{318}
\end{split}
\end{equation}

\begin{equation}
\small
\begin{split}
\mathcal{N}_4(t) &= 337919090177835662031225 t^{320} + 363942256836560273178522 t^{322} \\
&\quad + 391388420219927821900086 t^{324} + 420282773011519545845766 t^{326} \\
&\quad + 450646156038786692302354 t^{328} + 482494642702627832224708 t^{330} \\
&\quad + 515839129363182988632203 t^{332} + 550684933969689281476877 t^{334} \\
&\quad + 587031408234609898295369 t^{336} + 624871565846019596450772 t^{338} \\
&\quad + 664191732270025218813325 t^{340} + 704971218706930677466682 t^{342} \\
&\quad + 747182025844571134533106 t^{344} + 790788579884081677868713 t^{346} \\
&\quad + 835747506392527950441701 t^{348} + 882007444195522650165783 t^{350} \\
&\quad + 929508904581017030965180 t^{352} + 978184177583874072424399 t^{354} \\
&\quad + 1027957290139963674797113 t^{356} + 1078744017255954545662203 t^{358} \\
&\quad + 1130451950306658740763534 t^{360} + 1182980622814204463252741 t^{362} \\
&\quad + 1236221696964710249050180 t^{364} + 1290059210278107057231107 t^{366} \\
&\quad + 1344369884676833795470703 t^{368} + 1399023496315409418697923 t^{370} \\
&\quad + 1453883307288222762135701 t^{372} + 1508806556449623927377238 t^{374} \\
&\quad + 1563645009260709461710501 t^{376} + 1618245562742860566756797 t^{378} \\
&\quad + 1672450904225464523449504 t^{380} + 1726100218841183684764444 t^{382} \\
&\quad + 1779029943260056319039842 t^{384} + 1831074559572483170005269 t^{386} \\
&\quad + 1882067425703345072733244 t^{388} + 1931841635363073983409137 t^{390} \\
&\quad + 1980230902950669892721934 t^{392} + 2027070465701526196988777 t^{394} \\
&\quad + 2072197997719511608204153 t^{396} + 2115454527710065153836512 t^{398} \\
&\quad + 2156685354512065666618039 t^{400} + 2195740952042078214751571 t^{402} \\
&\quad + 2232477857472599309386978 t^{404} + 2266759534351874498941644 t^{406} \\
&\quad + 2298457204495378420418290 t^{408} + 2327450640758042709772796 t^{410} \\
&\quad + 2353628914815338522293036 t^{412} + 2376891092766924503735285 t^{414} \\
&\quad + 2397146873268658508130131 t^{416} + 2414317161994912846006619 t^{418} \\
&\quad + 2428334577969841550108478 t^{420} + 2439143886807421725333694 t^{422} \\
&\quad + 2446702357448816603273795 t^{424} + 2450980038877536299145156 t^{426}
\end{split}
\end{equation}

\begin{equation}
\small
\begin{split}
\mathcal{N}_5(t) &= 2451959954618067662943101 t^{428} + 2449638213084357831105757 t^{430} \\
&\quad + 2444024032909236075047413 t^{432} + 2435139682987121429939668 t^{434} \\
&\quad + 2423020337729930007799994 t^{436} + 2407713848945092070761421 t^{438} \\
&\quad + 2389280436180572794382483 t^{440} + 2367792298562955303769520 t^{442} \\
&\quad + 2343333151228650164091683 t^{444} + 2315997690865181073862696 t^{446} \\
&\quad + 2285890994566930595226255 t^{448} + 2253127857826788288255642 t^{450} \\
&\quad + 2217832076768738272732008 t^{452} + 2180135681510597624694129 t^{454} \\
&\quad + 2140178126417129066041874 t^{456} + 2098105444925783135866242 t^{458} \\
&\quad + 2054069375086335304814249 t^{460} + 2008226463991066754050268 t^{462} \\
&\quad + 1960737157329399401695441 t^{464} + 1911764882430433914657964 t^{466} \\
&\quad + 1861475130832155303236737 t^{468} + 1810034548626020223625856 t^{470} \\
&\quad + 1757610040147774999673695 t^{472} + 1704367892867169224787919 t^{474} \\
&\quad + 1650472928334900342357878 t^{476} + 1596087686394848038173400 t^{478} \\
&\quad + 1541371646602160333684866 t^{480} + 1486480493205106233374261 t^{482} \\
&\quad + 1431565426556175884363413 t^{484} + 1376772526305176765172311 t^{486} \\
&\quad + 1322242168061045446017082 t^{488} + 1268108497769887616511642 t^{490} \\
&\quad + 1214498964269696744605717 t^{492} + 1161533913120196550015856 t^{494} \\
&\quad + 1109326240950083873833943 t^{496} + 1057981112281693978098297 t^{498} \\
&\quad + 1007595736917409066970300 t^{500} + 958259208770007649881157 t^{502} \\
&\quad + 910052403169640342411721 t^{504} + 863047932555515233712494 t^{506} \\
&\quad + 817310156677711636020115 t^{508} + 772895246381299418418719 t^{510} \\
&\quad + 729851296363838988414483 t^{512} + 688218485305480678361025 t^{514} \\
&\quad + 648029278219583829741920 t^{516} + 609308668927542066225636 t^{518} \\
&\quad + 572074457161595403649708 t^{520} + 536337557885554639274749 t^{522} \\
&\quad + 502102337189932841462867 t^{524} + 469366972214325549141822 t^{526} \\
&\quad + 438123829491932659769173 t^{528} + 408359859195295298358777 t^{530} \\
&\quad + 380056999883445099122353 t^{532}
\end{split}
\end{equation}

\begin{equation}
\small
\begin{split}
\mathcal{N}_6(t) &= 353192591398990534354803 t^{534} + 327739790862862145595825 t^{536} \\
&\quad + 303667989702850097122653 t^{538} + 280943227124671750689679 t^{540} \\
&\quad + 259528598339311269473699 t^{542} + 239384653499009253209353 t^{544} \\
&\quad + 220469786093848721629717 t^{546} + 202740607356396752399655 t^{548} \\
&\quad + 186152305896213810895656 t^{550} + 170658989728350714251400 t^{552} \\
&\quad + 156214010392200805598578 t^{554} + 142770266935903178954060 t^{556} \\
&\quad + 130280489918104783059783 t^{558} + 118697503784362172199830 t^{560} \\
&\quad + 107974468186552076057634 t^{562} + 98065097136735346439494 t^{564} \\
&\quad + 88923856926336547163073 t^{566} + 80506142174247349694372 t^{568} \\
&\quad + 72768431232726879220993 t^{570} + 65668420715695844026537 t^{572} \\
&\quad + 59165140606087924130199 t^{574} + 53219050032169217781669 t^{576} \\
&\quad + 47792115325642186090943 t^{578} + 42847870700800112997619 t^{580} \\
&\quad + 38351463254417191223869 t^{582} + 34269682802339475140733 t^{584} \\
&\quad + 30570978275481933747006 t^{586} + 27225461301393871252243 t^{588} \\
&\quad + 24204898661027132633477 t^{590} + 21482694298787106565268 t^{592} \\
&\quad + 19033862495641985140728 t^{594} + 16834992886527352505659 t^{596} \\
&\quad + 14864208815280038650355 t^{598} + 13101119672829692577637 t^{600} \\
&\quad + 11526768568898272740194 t^{602} + 10123575918802945430608 t^{604} \\
&\quad + 8875280135997090399673 t^{606} + 7766875928657824336325 t^{608} \\
&\quad + 6784551223644155748592 t^{610} + 5915623122199625500993 t^{612} \\
&\quad + 5148473743464309196220 t^{614} + 4472486262865473197280 t^{616} \\
&\quad + 3877981840608177230435 t^{618} + 3356157652563539608693 t^{620} \\
&\quad + 2899026569252356808939 t^{622} + 2499358607473299602533 t^{624} \\
&\quad + 2150624565534158529556 t^{626} + 1846941889037714337705 t^{628} \\
&\quad + 1583023060348684743614 t^{630} + 1354126493131068716823 t^{632} \\
&\quad + 1156010125094286356890 t^{634} + 984887637618560644604 t^{636} \\
&\quad + 837387413196515500222 t^{638}
\end{split}
\end{equation}

\begin{equation}
\small
\begin{split}
\mathcal{N}_7(t) &= 710514119411536575120 t^{640} + 601612965132825373808 t^{642} \\
&\quad + 508336489713843508305 t^{644} + 428613881144285137601 t^{646} \\
&\quad + 360622666791781810505 t^{648} + 302762736686672870342 t^{650} \\
&\quad + 253632535088749409135 t^{652} + 212007356029984044938 t^{654} \\
&\quad + 176819578258437149124 t^{656} + 147140760706745196307 t^{658} \\
&\quad + 122165439500428347241 t^{660} + 101196540762653961644 t^{662} \\
&\quad + 83632260133433759237 t^{664} + 68954322262116329422 t^{666} \\
&\quad + 56717483973753261759 t^{668} + 46540197611634967609 t^{670} \\
&\quad + 38096312687099365796 t^{672} + 31107738434141261096 t^{674} \\
&\quad + 25337960467019822736 t^{676} + 20586341944293028652 t^{678} \\
&\quad + 16683117351522239570 t^{680} + 13485017923009959523 t^{682} \\
&\quad + 10871450995204099691 t^{684} + 8741181063116683549 t^{686} \\
&\quad + 7009447885326211331 t^{688} + 5605477813028925714 t^{690} \\
&\quad + 4470335379469777497 t^{692} + 3555079066984513136 t^{694} \\
&\quad + 2819178510956223792 t^{696} + 2229163955629029829 t^{698} \\
&\quad + 1757473970122108270 t^{700} + 1381478214913762501 t^{702} \\
&\quad + 1082648608126366350 t^{704} + 845860732287894164 t^{706} \\
&\quad + 658804877680046671 t^{708} + 511492737871179378 t^{710} \\
&\quad + 395844047864136094 t^{712} + 305342561673039187 t^{714} \\
&\quad + 234749555389155233 t^{716} + 179866938731454070 t^{718} \\
&\quad + 137341211974929572 t^{720} + 104502446433860971 t^{722} \\
&\quad + 79231877137181646 t^{724} + 59853891356932694 t^{726} \\
&\quad + 45047786540810836 t^{728} + 33776290448501315 t^{730} \\
&\quad + 25227551125371756 t^{732} + 18768484777106132 t^{734} \\
&\quad + 13907171227989259 t^{736} + 10262836793510416 t^{738} \\
&\quad + 7541826257413055 t^{740} + 5518570349253657 t^{742} \\
&\quad + 4020458834499834 t^{744}
\end{split}
\end{equation}

\begin{equation}
\small
\begin{split}
\mathcal{N}_8(t) &= 2915953995795000 t^{746} + 2105212149139493 t^{748} + 1512775181414998 t^{750} \\
&\quad + 1081847320302889 t^{752} + 769873624237794 t^{754} + 545104165572081 t^{756} \\
&\quad + 383963613127100 t^{758} + 269023414547918 t^{760} + 187464004218108 t^{762} \\
&\quad + 129898983355285 t^{764} + 89492306873283 t^{766} + 61288956174295 t^{768} \\
&\quad + 41717679117291 t^{770} + 28217257700878 t^{772} + 18961940840001 t^{774} \\
&\quad + 12656937654672 t^{776} + 8389951527147 t^{778} + 5521624267637 t^{780} \\
&\quad + 3607007163498 t^{782} + 2338169102030 t^{784} + 1503616721837 t^{786} \\
&\quad + 958931441581 t^{788} + 606311952919 t^{790} + 379923322605 t^{792} \\
&\quad + 235851770363 t^{794} + 144986904653 t^{796} + 88226183590 t^{798} \\
&\quad + 53114098813 t^{800} + 31621028508 t^{802} + 18604155434 t^{804} \\
&\quad + 10811763396 t^{806} + 6201285422 t^{808} + 3508504738 t^{810} \\
&\quad + 1956004685 t^{812} + 1073869892 t^{814} + 579811137 t^{816} \\
&\quad + 307660607 t^{818} + 160151471 t^{820} + 81722081 t^{822} \\
&\quad + 40777517 t^{824} + 19881586 t^{826} + 9437851 t^{828} \\
&\quad + 4359317 t^{830} + 1948577 t^{832} + 842722 t^{834} \\
&\quad + 349527 t^{836} + 139159 t^{838} + 52368 t^{840} + 18697 t^{842} \\
&\quad + 6146 t^{844} + 1881 t^{846} + 501 t^{848} + 120 t^{850} + 21 t^{852} + 3 t^{854}
\end{split}
\end{equation}

\begin{equation}
\small
\begin{split}
\mathcal{D}(t) &= (-1 + t^2)^2 (1 + t^4)^2 (1 + t^2 + t^4 + t^6 + t^8)^2 (1 + t^{16})^3 \\
&\quad \times (1 + t^8 + t^{16})^2 (1 + t^4 + t^8 + t^{10} + t^{12} + t^{16} + t^{20})^9 \\
&\quad \times (1 - t^2 + t^4 - t^6 + t^8 - t^{10} + t^{12} - t^{14} + t^{16} - t^{18} + t^{20})^3 \\
&\quad \times (1 + t^{24}) (1 + t^2 + t^4 + t^6 + t^8 + t^{10} + t^{12} + t^{14} + t^{16} + t^{18} + t^{20} + t^{22} + t^{24})^3 \\
&\quad \times (1 + t^2 + t^4 + t^6 + t^8 + t^{10} + t^{12} + t^{14} + t^{16} + t^{18} + t^{20} + t^{22} + t^{24} \\
&\quad \quad + t^{26} + t^{28} + t^{30} + t^{32} + t^{34} + t^{36})^3 \\
&\quad \times (1 + t^2 + t^8 + t^{10} + t^{14} + 2 t^{16} + t^{18} + t^{20} + 2 t^{22} + 2 t^{24} + t^{26} + t^{28} \\
&\quad \quad + 2 t^{30} + 2 t^{32} + t^{34} + t^{36} + 2 t^{38} + 2 t^{40} + t^{42} + t^{44} + 2 t^{46} + t^{48} \\
&\quad \quad + t^{52} + t^{54} + t^{60} + t^{62})^5~~.
\end{split}
\end{equation}
The numerator of $\mathcal{I}_{n_1=2; N_1=5}^{\mathrm{BMN}~SU(N)}(t)$ is a polynomial in $t$ of degree $1420$ with integer coefficients. Because its full expression is extremely long, we write only the first few and last few terms, indicating the omitted part by $\cdots$.
\begin{equation}
\small
\begin{split}
\mathcal{I}_{n_1=2; N_1=5}^{\mathrm{BMN}~SU(N)}(t) = 
\frac{ (1-t^6) \bigl( 1 + 10t^2 + 71t^4 + 372t^6 + 1653t^8 + \cdots + 27t^{1418} + 3t^{1420} \bigr) }{%
\begin{aligned}
&(1 + t^2)^4 (1 + t^4)^3 (1 - t^2 + t^4 - t^6 + t^8)^{11} \\
&\times (1 + t^2 + t^4 + t^6 + t^8 + t^{10} + t^{12})^{3} \\
&\times (1 + t^{16})^{3} (1 - t^4 + t^8 - t^{12} + t^{16})^{5} \\
&\times (1 - t^{10} + t^{20}) \\
&\times (1 - t^2 + t^4 - t^6 + t^8 - t^{10} + t^{12} - t^{14} + t^{16} - t^{18} + t^{20})^{3} \\
&\times (1 - t^4 + t^8 - t^{12} + t^{16} - t^{20} + t^{24})^{3} \\
&\times (1 - t^2 + t^4 - t^6 + t^8 - t^{10} + t^{12} - t^{14} + t^{16} - t^{18} + t^{20} - t^{22} + t^{24})^{5} \\
&\times (1 + t^2 + t^4 + t^6 + t^8 + t^{10} + t^{12} + t^{14} + t^{16} + t^{18} + t^{20} + t^{22} + t^{24})^{9} \\
&\times (-1 + t^{30})^{2} \\
&\times (1 + t^2 + t^4 + t^6 + t^8 + t^{10} + t^{12} + t^{14} + t^{16} + t^{18} + t^{20} + t^{22} + t^{24} \\
&\qquad\; + t^{26} + t^{28} + t^{30} + t^{32})^{5} \\
&\times (1 + t^2 + t^4 + t^6 + t^8 + t^{10} + t^{12} + t^{14} + t^{16} + t^{18} + t^{20} + t^{22} + t^{24} \\
&\qquad\; + t^{26} + t^{28} + t^{30} + t^{32} + t^{34} + t^{36})^{3} \\
&\times (1 + t^{10} + t^{20} + t^{30} + t^{40})^{3} \\
&\times (1 + t^2 + t^4 + t^6 + t^8 + t^{10} + t^{12} + t^{14} + t^{16} + t^{18} + t^{20} + t^{22} + t^{24} \\
&\qquad\; + t^{26} + t^{28} + t^{30} + t^{32} + t^{34} + t^{36} + t^{38} + t^{40} + t^{42} + t^{44})^{5}
\end{aligned}
}.
\end{split}
\end{equation}
We perform a series expansion in fugacity on the closed-form expressions for the single partition indices, which we derive using the residue method. The series shows perfect agreement with the character expansion approach noted in Appendix \ref{spbmn}.
\begin{equation}
\small
\mathcal{I}_{n_1=2;N_1=2}^{\mathrm{BMN}~SU(N)}(t)=  1 + 6 t^4 - 9 t^6 + 18 t^8 - 30 t^{10} + 64 t^{12} - 90 t^{14} + 108 t^{16}+\mathcal{O}(t^{18}),
\end{equation}
\begin{equation}
\small
\begin{split}
\mathcal{I}_{n_1=2; N_1=3}^{\mathrm{BMN}~SU(N)}(t) = &  1 + 6 t^4 - 9 t^6 + 18 t^8 - 30 t^{10} + 63 t^{12} - 90 t^{14} + 108 t^{16} \\
&- 154 t^{18} + 264 t^{20} - 324 t^{22} + 300 t^{24} - 393 t^{26} + 732 t^{28} \\
&- 927 t^{30} + 657 t^{32} - 594 t^{34} + 1571 t^{36}+\mathcal{O}(t^{38}),
\end{split}
\end{equation}
\begin{equation}
\small
\begin{split}
\mathcal{I}_{n_1=2; N_1=4}^{\mathrm{BMN}~SU(N)}(t) &= 1 + 6 t^4 - 9 t^6 + 18 t^8 - 30 t^{10} + 63 t^{12} - 90 t^{14} + 108 t^{16} \\
&- 155 t^{18} + 264 t^{20} - 324 t^{22} + 310 t^{24} - 405 t^{26} + 726 t^{28} \\
&- 936 t^{30} + 735 t^{32} - 651 t^{34} + 1463 t^{36} - 2385 t^{38} + 1767 t^{40} \\
&- 398 t^{42} + 1578 t^{44} - 5271 t^{46} + 5548 t^{48} + 96 t^{50} - 2259 t^{52} \\
&- 7376 t^{54} + 16695 t^{56} - 6042 t^{58} - 13918 t^{60} + 5040 t^{62} + 34044 t^{64}+ \mathcal{O}(t^{66}).
\end{split}
\end{equation}

\begin{equation}
\small
\begin{split}
\mathcal{I}_{n_1=2; N_1=5}^{\mathrm{BMN}~SU(N)}(t) &= 1 + 6 t^4 - 9 t^6 + 18 t^8 - 30 t^{10} + 63 t^{12} - 90 t^{14} + 108 t^{16} \\
        & - 155 t^{18} + 264 t^{20} - 324 t^{22} + 309 t^{24} - 405 t^{26} + 726 t^{28} \\
        & - 926 t^{30} + 723 t^{32} - 657 t^{34} + 1454 t^{36} - 2307 t^{38} + 1710 t^{40} \\
        & - 504 t^{42} + 1629 t^{44} - 4920 t^{46} + 5262 t^{48} - 516 t^{50} \\
        & - 1584 t^{52} - 6184 t^{54} + 14754 t^{56} - 7533 t^{58} - 9719 t^{60} \\
        & + 6492 t^{62} + 25080 t^{64} - 39505 t^{66} - 2154 t^{68} + 49272 t^{70} \\
        & - 8917 t^{72} - 93822 t^{74} + 94242 t^{76} + 68733 t^{78} - 169644 t^{80} \\
        & - 27837 t^{82} + 314750 t^{84} - 188727 t^{86} - 340524 t^{88} + 482278 t^{90} \\
        & + 252942 t^{92} - 952587 t^{94} + 276239 t^{96} + 1240224 t^{98} \\
        & - 1199643 t^{100}+ \mathcal{O}(t^{102}).
\end{split}
\end{equation}
\begin{equation}
\label{42ex}
\small
\begin{split}
\mathcal{I}_{n_1=4; N_1=2}^{\mathrm{BMN}~SU(N)}(t) &= \frac{\mathcal{N}(t)}{\mathcal{D}(t)}, \quad \text{where} \\[1em]
\mathcal{N}(t) &= 1 + 10t^2 + 96t^4 + 633t^6 + 3862t^8 + \dots + 120t^{1678} + 13t^{1680} + t^{1682}, \\[1em]
\mathcal{D}(t) &= (-1 + t^2)^4 (1 + t^2)^4 (1 + t^4)^{33} (1 - t^2 + t^4)^4 (1 + t^2 + t^4)^4 \\
&\quad \times (1 + t^8)^{17} (1 - t^4 + t^8)^{10} (1 - t^2 + t^4 - t^6 + t^8)^9 \\
&\quad \times (1 + t^2 + t^4 + t^6 + t^8)^2 (1 - t^6 + t^{12})^7 (1 + t^6 + t^{12})^4 \\
&\quad \times (1 - t^2 + t^4 - t^6 + t^8 - t^{10} + t^{12})^8 \\
&\quad \times (1 + t^2 + t^4 + t^6 + t^8 + t^{10} + t^{12})^{11} (1 + t^{16})^9 \\
&\quad \times (1 - t^8 + t^{16}) (1 - t^4 + t^8 - t^{12} + t^{16})^5 \\
&\quad \times (1 - t^2 + t^6 - t^8 + t^{10} - t^{14} + t^{16})^4 \\
&\quad \times \left( \begin{aligned} &1 - t^2 + t^4 - t^6 + t^8 - t^{10} \\ &+ t^{12} - t^{14} + t^{16} - t^{18} + t^{20} \end{aligned} \right)^3 \\
&\quad \times \left( \begin{aligned} &1 + t^2 + t^4 + t^6 + t^8 + t^{10} \\ &+ t^{12} + t^{14} + t^{16} + t^{18} + t^{20} \end{aligned} \right)^7 \\
&\quad \times \left( \begin{aligned} &1 + t^2 + t^4 + t^6 + t^8 + t^{10} + t^{12} \\ &+ t^{14} + t^{16} + t^{18} + t^{20} + t^{22} + t^{24} \end{aligned} \right)^3 \\
&\quad \times \left( \begin{aligned} &1 + t^2 + t^4 + t^6 + t^8 + t^{10} + t^{12} \\ &+ t^{14} + t^{16} + t^{18} + t^{20} + t^{22} + t^{24} \\ &+ t^{26} + t^{28} + t^{30} + t^{32} \end{aligned} \right)^5 \\
&\quad \times \left( \begin{aligned} &1 + t^2 + t^4 + t^6 + t^8 + t^{10} + t^{12} \\ &+ t^{14} + t^{16} + t^{18} + t^{20} + t^{22} + t^{24} \\ &+ t^{26} + t^{28} + t^{30} + t^{32} + t^{34} + t^{36} \end{aligned} \right)^3.
\end{split}
\end{equation}
Using sum over trees approach, we evaluated the single partition indices besides trivial vacuum sector $n_1 \geq 3$ in closed-form, one such example is expressed in \eqref{42ex}. 

\section{Single partition BMN index using iterative residue evaluation}
\label{itreseval}
In this Appendix \ref{itreseval}, we evaluate the single partition BMN index directly by iterative contour integration. The method is conceptually straightforward: one performs the contour integrals successively, identifying the poles at each stage and summing their residues. Its limitation is also clear: the pole structure is generated dynamically, so after each integration new singularities appear in the remaining variables. Consequently, while the method is effective for low-rank examples, it becomes increasingly inefficient for larger $n_1$, which motivates the residue-over-trees reorganization used in the main text section \ref{treesres}.

We specialize chemical potentials as
\begin{equation}
	\Delta=\Delta_1 = \Delta_2 = \Delta_3 , \quad e^{-\Delta} = t^2.
\end{equation}

The single-partition contribution to the BMN index is then given by
\begin{equation}
\mathcal{I}_{n_1; N_1}^{\mathrm{BMN}}(t)
= \int [dU]\,
\exp\left[
\sum_{m=1}^\infty \frac{1}{m}\,
\iota_{11}(t^m)\,
\mathrm{Tr}(U^{\dagger m})\,
\mathrm{Tr}(U^m)
\right],
\end{equation}
where \( U \in U(n_1) \) and \( [dU] \) denotes the Haar measure.

\vspace{1em}
\noindent
Using the eigenvalue decomposition
\begin{equation}
U = \mathrm{diag}(e^{i\theta_1}, \dots, e^{i\theta_{n_1}}),
\end{equation}
the Haar measure on \( U(n_1) \) takes the form
\begin{equation}
[dU] = \frac{1}{(2\pi)^{n_1} n_1!}
\prod_{j=1}^{n_1} d\theta_j
\prod_{1 \le j < k \le n_1} | e^{i\theta_j} - e^{i\theta_k} |^2.
\end{equation}
The traces in the exponent are
\begin{equation}
\mathrm{Tr}(U^m) = \sum_{j=1}^{n_1} e^{i m \theta_j}, 
\qquad
\mathrm{Tr}(U^{\dagger m}) = \sum_{k=1}^{n_1} e^{-i m \theta_k},
\end{equation}
so that
\begin{equation}
\mathrm{Tr}(U^{\dagger m})\,\mathrm{Tr}(U^m)
= \sum_{j,k=1}^{n_1} e^{i m (\theta_j - \theta_k)}.
\end{equation}

\noindent
Substituting this into the exponential gives
\begin{equation}
\exp\!\left[
\sum_{m=1}^\infty \frac{1}{m}\,
\iota_{11}(t^m)
\sum_{j,k=1}^{n_1} e^{i m (\theta_j - \theta_k)}
\right]
= 
\prod_{j,k=1}^{n_1}
\exp\!\left[
\sum_{m=1}^\infty \frac{1}{m}\,
\iota_{11}(t^m)\,
e^{i m (\theta_j - \theta_k)}
\right].
\end{equation}

\noindent
For convenience, we define
\begin{equation}
z = e^{i(\theta_j - \theta_k)}.
\end{equation}
Then, using the explicit form of \( \iota_{11}(t^m) \),
\begin{equation}
\exp\!\left[
\sum_{m=1}^\infty \frac{1}{m}
\Big(
t^{6N_1 m}
+ 3 \sum_{r=0}^{N_1-1} (t^{(6r+2)m} - t^{(6r+4)m})
\Big)
z^m
\right].
\end{equation}

\noindent
We now employ the standard identity
\begin{equation}
\sum_{m=1}^\infty \frac{x^m}{m} = -\ln(1-x)
\quad \Longrightarrow \quad
\exp\!\left[\sum_{m=1}^\infty \frac{x^m}{m}\right] = \frac{1}{1-x}.
\end{equation}
Applying this term by term gives
\begin{equation}
\begin{split}
&\exp\!\left[\sum_{m=1}^\infty \frac{1}{m}\,t^{6N_1 m} z^m\right]
\prod_{r=0}^{N_1-1}
\exp\!\left[3\sum_{m=1}^\infty \frac{1}{m}\,t^{(6r+2)m} z^m\right]
\exp\!\left[-3\sum_{m=1}^\infty \frac{1}{m}\,t^{(6r+4)m} z^m\right]
\\
&\quad =
\frac{1}{1 - t^{6N_1} z}
\prod_{r=0}^{N_1-1}
\frac{(1 - t^{6r+4} z)^3}{(1 - t^{6r+2} z)^3}.
\end{split}
\end{equation}

\noindent
Thus, for each pair \((j,k)\), the integrand contributes a factor
\begin{equation}
\frac{1}{1 - t^{6N_1} e^{i(\theta_j - \theta_k)}} 
\prod_{r=0}^{N_1-1}
\frac{(1 - t^{6r+4} e^{i(\theta_j - \theta_k)})^3}
     {(1 - t^{6r+2} e^{i(\theta_j - \theta_k)})^3}.
\end{equation}

\noindent
Collecting all factors, the single-partition BMN index becomes
\begin{equation}
\begin{split}
\mathcal{I}^{\text{BMN}}_{n_1;N_1}(t)
&= 
\frac{1}{(2\pi)^{n_1} n_1!} 
\int_{[0,2\pi]^{n_1}}
\prod_{j=1}^{n_1} d\theta_j
\prod_{1 \le j < k \le n_1} 
|e^{i\theta_j} - e^{i\theta_k}|^2
\\
&\quad \times
\prod_{1 \le j,k \le n_1}
\left[
\frac{1}{1 - t^{6N_1} e^{i(\theta_j - \theta_k)}} 
\prod_{r=0}^{N_1-1}
\frac{(1 - t^{6r+4} e^{i(\theta_j - \theta_k)})^3}
     {(1 - t^{6r+2} e^{i(\theta_j - \theta_k)})^3}
\right].
\end{split}
\end{equation}

\noindent
This is the final compact expression for the single-partition BMN index in terms of eigenvalue angles \( \theta_j \), valid for arbitrary \( n_1 \) and \( N_1 \).
Therefore, the single partition BMN index given by 
\begin{equation}
\begin{split}
\frac{1}{(2\pi)^{n_1} n_1!} \int_0^{2\pi} \prod_{j=1}^{n_1} \mathrm{d}\theta_j \prod_{1 \le k < l \le N} |e^{i\theta_k} - e^{i\theta_l}|^2 \exp\left( \sum_{n=1}^\infty \frac{1}{n} \iota_{11}(t^n) \sum_{j,k=1}^{n_1} e^{in(\theta_j - \theta_k)} \right).
\end{split}
\end{equation}
is simplified to the final compact expression
\begin{equation}
\begin{split}
\mathcal{I}^{\text{BMN}}_{n_1;N_1} 
&= \frac{1}{(2\pi)^{n_1} n_1!} 
\int_{[0,2\pi]^{n_1}} 
\prod_{j=1}^{n_1} d\theta_j 
\prod_{j<k} |e^{i\theta_j} - e^{i\theta_k}|^2 
\\
&\quad \times 
\prod_{1\leq j,k \leq n_1} 
\left[
\frac{1}{1 - t^{6N_1} e^{i(\theta_j - \theta_k)}} 
\prod_{r=0}^{N_1-1} 
\frac{(1 - t^{6r+4} e^{i(\theta_j - \theta_k)})^3}{(1 - t^{6r+2} e^{i(\theta_j - \theta_k)})^3}
\right].
\end{split}
\end{equation}

\noindent
Introducing the contour variables $z_j = e^{i\theta_j}$, and using $d\theta_j = \frac{dz_j}{i z_j}$, this integral takes the form
\begin{equation}
\mathcal{I}^{\text{BMN}}_{n_1;N_1} 
= \frac{1}{(2\pi i)^{n_1} n_1!} 
\oint_{|z_1|=1} \frac{dz_1}{z_1} 
\cdots 
\oint_{|z_{n_1}|=1} \frac{dz_{n_1}}{z_{n_1}}
\;
\Delta(z)
\prod_{j,k=1}^{n_1} F\!\left(\frac{z_j}{z_k}\right).
\label{eq:contour_form}
\end{equation}
where
\begin{equation}
\Delta(z) = \prod_{1\le j<k\le n_1} (z_j - z_k)\,(z_j^{-1} - z_k^{-1}),
\qquad
F(z) = \frac{1}{1-t^{6N_1} z} 
\prod_{r=0}^{N_1-1} 
\frac{(1 - t^{6r+4} z)^3}{(1 - t^{6r+2} z)^3}.
\end{equation}

\subsubsection*{Example: Case $n_1 = 2$}
For \(n_1 = 2\), the Vandermonde factor simplifies to
\begin{equation}
\Delta(z_1,z_2)
= (z_1 - z_2)\left(\frac{1}{z_1} - \frac{1}{z_2}\right)
= \frac{(z_1 - z_2)^2}{z_1 z_2}.
\end{equation}
The index then becomes
\begin{equation}
\mathcal{I}^{\text{BMN}}_{2;N_1} 
= \frac{1}{2(2\pi i)^2}
\oint_{|z_1|=1} \frac{dz_1}{z_1}
\oint_{|z_2|=1} \frac{dz_2}{z_2}
\,
\frac{(z_1 - z_2)^2}{z_1 z_2}
\, F\!\left(\frac{z_1}{z_2}\right)
F\!\left(\frac{z_2}{z_1}\right).
\label{eq:I2N1_contour}
\end{equation}
Since \(F(z_2/z_1)\) contains denominators of the form
\((1 - t^{6r+2} z_2/z_1)^3\) and \((1 - t^{6N_1} z_2/z_1)\),
the integrand has poles at
\begin{equation}
z_1 = 0,
\qquad
z_1 = t^{6r+2} z_2, \quad r = 0, 1, \ldots, N_1 - 1,
\qquad
z_1 = t^{6N_1} z_2.
\label{eq:poles_z1}
\end{equation}
Here,
\begin{itemize}
    \item \(z_1 = 0\) is a simple pole from the \(1/z_1^2\) prefactor,
    \item \(z_1 = t^{6r+2} z_2\) are triple poles, 
    \item \(z_1 = t^{6N_1} z_2\) is a simple pole.
\end{itemize}
Since \(0<|t|<1\), all these poles lie \emph{inside} the unit circle.
We define the first contour integral as
\begin{equation}
R(z_2)
= \frac{1}{2\pi i}
\oint_{|z_1|=1} \frac{dz_1}{z_1}
\,
\frac{(z_1 - z_2)^2}{z_1 z_2}
\, F\!\left(\frac{z_1}{z_2}\right)
F\!\left(\frac{z_2}{z_1}\right).
\end{equation}
After the \(z_1\)-integration, the remaining single contour is
\begin{equation}
\mathcal{I}^{\text{BMN}}_{2;N_1}
= \frac{1}{2(2\pi i)}
\oint_{|z_2|=1} \frac{dz_2}{z_2}
\, R(z_2)
= \frac{1}{2}\,
\operatorname{Res}_{z_2=0}
\!\left[
\frac{R(z_2)}{z_2}
\right].
\end{equation}
This final residue at \(z_2=0\) gives the complete index for given \(N_1\).
Putting all pieces together, the residue representation for \(n_1=2\) is
\begin{equation}
\begin{aligned}
\mathcal{I}^{\mathrm{BMN}}_{2;N_1}
&= 
\frac{1}{2}
\operatorname{Res}_{z_2=0}
\left[
\frac{1}{z_2}
\sum_{p \in P}
\operatorname{Res}_{z_1 = p}
\left(
\frac{(z_1 - z_2)^2}{z_1^2 z_2}
F\!\left(\frac{z_1}{z_2}\right)
F\!\left(\frac{z_2}{z_1}\right)
\right)
\right], \\[5pt]
P &= \{0\} \cup \{t^{6r+2}z_2\}_{r=0}^{N_1-1} \cup \{t^{6N_1}z_2\}, \\[4pt]
F(z) &= \frac{1}{1 - t^{6N_1} z}
\prod_{r=0}^{N_1-1}
\left(\frac{1 - t^{6r+4} z}{1 - t^{6r+2} z}\right)^3.
\end{aligned}
\label{eq:I2N1_residue}
\end{equation}
For $SU(N)$ index we divide the $U(N)$ index by the $U(1)$ part coming from $\mathcal{I}^{\mathrm{BMN}}_{n_1=1;N_1=1}(t)$
\begin{equation}
\mathcal{I}^{\mathrm{BMN}~U(1)}(t)
=\frac{(1+t^{2})^{3}}{1-t^{6}},
\end{equation}
and get
\begin{equation}
\small
\begin{split}
\mathcal{I}_{n_1=2;N_1=1}^{\mathrm{BMN}~SU(N)}(t)&=-\Big( \big(1 + 3\,t^2 + 12\,t^4 + 20\,t^6 + 42\,t^8 + 48\,t^{10} + 75\,t^{12} + 66\,t^{14} + 81\,t^{16} + 55\,t^{18} \\
&\quad + 54\,t^{20} + 27\,t^{22} + 19\,t^{24} + 6\,t^{26} + 3\,t^{28}\big) \Big/ \big( (1 + t^2)^4 (1 + t^4)^3 (1 - t^2 + t^4) (-1 + t^6) \big) \Big).
\end{split}
\end{equation}
Expanding in series 
\begin{equation}
\mathcal{I}_{n_1=2;N_1=1}^{\mathrm{BMN}~SU(N)}(t)=1+6t^4+\mathcal{O}\left(t^{6}\right).
\end{equation}
We list more closed-form BMN indices for the single-partition sector $n_1=2$ in Appendix \ref{single-partitionclosedform}. 
Evaluating iterative contour integrals becomes computationally prohibitive for a single partition index $N = n_1 N_1$ when $n_1 > 2$. The primary bottleneck lies in the dynamic generation of singularities. It is difficult to identify and classify all relevant poles a priori. As each successive integration is evaluated, the resulting expression gives entirely new poles. Attempting to systematically track, organize, and compute residues in this step-by-step iterative manner leads to massive computational inefficiency. To circumvent this combinatorial explosion, we utilize the \textit{sum over trees} formula for calculating residues.

\subsubsection*{BMN index in the irreducible vacuum sector}
We start from the contour representation of the single-partition BMN index
\[
\mathcal{I}^{\mathrm{BMN}}_{n_1;N_1}(t)
=
\frac{1}{(2\pi i)^{n_1} n_1!}
\oint_{|z_1|=1}\frac{dz_1}{z_1}\cdots
\oint_{|z_{n_1}|=1}\frac{dz_{n_1}}{z_{n_1}}\,
\Delta(z)\,
\prod_{j,k=1}^{n_1} F\!\left(\frac{z_j}{z_k}\right),
\]
with
\[
\Delta(z)=\prod_{1\le j<k\le n_1}(z_j-z_k)(z_j^{-1}-z_k^{-1}),
\qquad
F(z)=\frac{1}{1-t^{6N_1}z}\prod_{r=0}^{N_1-1}\frac{(1-t^{6r+4}z)^3}{(1-t^{6r+2}z)^3}.
\]
For \(n_1=1\) the Vandermonde \(\Delta(z)\) is trivial (empty product) and the integrand does not depend on the integration variable
\[
\prod_{j,k=1}^{1} F\!\left(\frac{z_j}{z_k}\right)=F(1).
\]
Hence the unit-circle integral simply returns the constant \(F(1)\). Therefore for \(n_1=1\) and \(N_1=N\) we obtain the closed form
\begin{equation}\label{eq:single_eigenvalue_generalN}
\mathcal{I}^{\mathrm{BMN}}_{1;N}(t)
=F(1)
=\frac{1}{1-t^{6N}}
\prod_{r=0}^{N-1}\frac{\big(1-t^{6r+4}\big)^3}{\big(1-t^{6r+2}\big)^3}.
\end{equation}
Putting \(N=1\) into \eqref{eq:single_eigenvalue_generalN} yields
\[
\mathcal{I}^{\mathrm{BMN}}_{1;1}(t)
=\frac{1}{1-t^{6}}\frac{(1-t^{4})^{3}}{(1-t^{2})^{3}}
=\frac{(1+t^{2})^{3}}{1-t^{6}}.
\]
By \textit{irreducible vacuum sector} we mean the contribution coming from \(n_1=1\) and \(N_1=N\)). Thus the formula \eqref{eq:single_eigenvalue_generalN} provides the $U(N)$ BMN index for the irreducible vacuum sector
\begin{equation}
\label{unirrvac}
\mathcal{I}^{\mathrm{BMN}}_{U(N),\,\text{irr.\ vac.}}(t)
=\mathcal{I}^{\mathrm{BMN}}_{1;N}(t)
=\frac{1}{1-t^{6N}}
\prod_{r=0}^{N-1}\frac{\big(1-t^{6r+4}\big)^3}{\big(1-t^{6r+2}\big)^3}.
\end{equation}
For $SU(N)$ index we divide the $U(N)$ index by the $U(1)$ part coming from $\mathcal{I}^{\mathrm{BMN}}_{1;1}(t)$
\begin{equation}
\mathcal{I}^{\mathrm{BMN}}_{1;1}(t)
=\frac{(1+t^{2})^{3}}{1-t^{6}}.
\end{equation}
From the formula of the $U(N)$ BMN index in the irreducible vacuum \(n_1 = 1,\; N_1 = N\) \eqref{unirrvac} and the \(U(1)\) factor, the \(SU(N)\) BMN index for the irreducible vacuum can be written as
\begin{equation}
\begin{split}
\mathcal I^{\rm BMN}_{SU(N),{\rm irr.\ vac.}}(t)
=
\frac{1-t^6}{1-t^{6N}}
\prod_{r=1}^{N-1}
\frac{\left(1-t^{6r+4}\right)^3}
{\left(1-t^{6r+2}\right)^3}.
\end{split}
\label{eq:irr-index-product}
\end{equation}
This can be written in plethystic-exponential form.  Using
\begin{equation}
\begin{split}
{\rm PE}\left[t^a\right]
=
\frac{1}{1-t^a},
\qquad
{\rm PE}\left[-t^a\right]
=
1-t^a,
\end{split}
\end{equation}
we obtain
\begin{equation}
\begin{split}
\mathcal I^{\rm BMN}_{SU(N),{\rm irr.\ vac.}}(t)
=
{\rm PE}
\left[
t^{6N}
-
t^6
+
3\sum_{r=1}^{N-1}
\left(
t^{6r+2}
-
t^{6r+4}
\right)
\right].
\end{split}
\label{eq:irr-index-pe}
\end{equation}
\subsubsection*{BMN index in the trivial vacuum sector}
Defining \(z_j = e^{i\theta_j}\) with \(d\theta_j = \frac{dz_j}{i z_j}\). Then, the $U(N)$ BMN index in the trivial vacuum sector is expressed as
\begin{equation}
\mathcal{I}^{\mathrm{BMN}}_{N;1}(t)
=
\frac{1}{(2\pi i)^{N} N!}
\oint_{|z_1|=1} \frac{dz_1}{z_1}
\cdots
\oint_{|z_{N}|=1} \frac{dz_{N}}{z_{N}}
\;
\Delta(z)\,
\prod_{j,k=1}^{N}
F\!\left( \frac{z_j}{z_k} \right),
\end{equation}
where
\begin{equation}
\Delta(z)
=
\prod_{1\le j<k\le N}
(z_j - z_k)(z_j^{-1} - z_k^{-1}),
\qquad
F(z)
=
\frac{(1 - t^{4} z)^3}{(1 - t^{2} z)^3(1 - t^{6} z)}.
\end{equation}
For $SU(N)$ index in the trivial vacuum sector, we divide the $U(N)$ index by the $U(1)$ part coming from $\mathcal{I}^{\mathrm{BMN}}_{n_1=1;N_1=1}(t)$
\begin{equation}
\mathcal{I}^{\mathrm{BMN}~U(1)}(t)
=\frac{(1+t^{2})^{3}}{1-t^{6}}.
\end{equation}

\section{Explicit computation in the $N_1 = 1, N_2 = 2, n_1 = 1, n_2 = 1$ sector}
\label{exbmn}
We consider the BMN index for the case $N_1 = 1, N_2 = 2, n_1 = 1, n_2 = 1$. The general formula is
\begin{equation}
\begin{split}
\mathcal{I}_{n_1,n_2;N_1,N_2}^{\mathrm{BMN}}(t) = \sum_{P_{11},P_{12},P_{21},P_{22}} &\frac{\iota_{11}(t)_{P_{11}}\,\iota_{12}(t)_{P_{12}}\,\iota_{21}(t)_{P_{21}}\,\iota_{22}(t)_{P_{22}}}{z_{P_{11}}z_{P_{12}}z_{P_{21}}z_{P_{22}}} \\
&\times \sum_{\substack{\lambda_1\vdash|P_{11}+P_{12}|\\ \ell(\lambda_1)\le n_1}} \chi^{\lambda_1}(P_{11}+P_{12})\,\chi^{\lambda_1}(P_{11}+P_{21})\\
&\times \sum_{\substack{\lambda_2\vdash|P_{22}+P_{12}|\\ \ell(\lambda_2)\le n_2}} \chi^{\lambda_2}(P_{22}+P_{12})\,\chi^{\lambda_2}(P_{22}+P_{21}),
\end{split}
\end{equation}
with the consistency condition $|P_{12}| = |P_{21}|$. Here the diagonal terms are
\begin{equation}
\iota_{aa}(t^m) = 1 - (1-t^{2m})^3\sum_{j=0}^{N_a-1}t^{6jm},\qquad a=1,2,
\end{equation}
and the off-diagonal term (symmetric) is
\begin{equation}
\iota_{12}(t^m)=\iota_{21}(t^m)= (1-t^{2m})^3\sum_{j=\frac12|N_1-N_2|}^{\frac12(N_1+N_2)-1}(-1)^{2jm+1}t^{6jm}.
\end{equation}
For the double-partition case with $n_1=n_2=1$, the index formula contains the two factors
\begin{equation}
\sum_{\substack{\lambda_1\vdash |P_{11}+P_{12}|\\ \ell(\lambda_1)\le 1}}
\chi^{\lambda_1}\bigl(P_{11}+P_{12}\bigr)\;
\chi^{\lambda_1}\bigl(P_{11}+P_{21}\bigr)~~,
\end{equation}
and an analogous one for $\lambda_2$. Here $P_{11}+P_{12}$ denotes the partition obtained by combining the parts of $P_{11}$ and $P_{12}$ and similarly for $P_{11}+P_{21}$. The consistency condition $|P_{12}|=|P_{21}|$ guarantees that
\begin{equation}
|P_{11}+P_{12}| = |P_{11}+P_{21}| \qquad\text{and}\qquad |P_{22}+P_{12}| = |P_{22}+P_{21}|,
\end{equation}
so both characters in each product are evaluated on partitions of the same integer $d$.
The condition $\ell(\lambda)\le 1$ means that $\lambda$ has at most one part. For a partition of a positive integer $d$, the only possibilities are the \textit{single-row} partition $\lambda=(d)$. The empty partition (which corresponds to $d=0$) is also allowed, and by convention $\chi^{\emptyset}(\emptyset)=1$. For $d>0$, the irreducible representation of the symmetric group $S_d$ labelled by $(d)$ is the \textit{trivial representation}. The character of the trivial representation of any finite group is $1$ on every conjugacy class. For $S_d$, the conjugacy classes are indexed by partitions of $d$. Hence for any partition $P$ of $d$,
\begin{equation}
\chi^{(d)}(P)=1.
\end{equation}
Because the sum over $\lambda_1$ runs only over partitions of $d_1=|P_{11}+P_{12}|$ with length $\le1$, the only contribution is from $\lambda_1=(d_1)$ (or the empty partition when $d_1=0$). Therefore
\begin{equation}
\sum_{\substack{\lambda_1\vdash d_1\\ \ell(\lambda_1)\le 1}}
\chi^{\lambda_1}(P_{11}+P_{12})\;\chi^{\lambda_1}(P_{11}+P_{21})
= \chi^{(d_1)}(P_{11}+P_{12})\;\chi^{(d_1)}(P_{11}+P_{21})=1\cdot 1=1.
\end{equation}
The same reasoning applies to the $\lambda_2$ sum. Hence both character sums equal $1$, regardless of the specific partitions $P_{11},P_{12},P_{21},P_{22}$.

We have
\begin{equation}
\mathcal{I}_{n_1=1,n_2=1;N_1,N_2}^{\mathrm{BMN}}(t) = \Bigl(\sum_{P_{11}}\frac{\iota_{11}(t)_{P_{11}}}{z_{P_{11}}}\Bigr)
\Bigl(\sum_{P_{22}}\frac{\iota_{22}(t)_{P_{22}}}{z_{P_{22}}}\Bigr)
\Bigl(\sum_{\substack{P_{12},P_{21}\\|P_{12}|=|P_{21}|}}\frac{\iota_{12}(t)_{P_{12}}\,\iota_{21}(t)_{P_{21}}}{z_{P_{12}}z_{P_{21}}}\Bigr).
\end{equation}
Using the exponential formula
\begin{equation}
\exp\!\left(\sum_{m\ge1}\frac{1}{m}\,a(t^m)\right)=\sum_{P}\frac{a_P}{z_P},
\end{equation}
we introduce the generating functions
\begin{equation}
\begin{split}
F_{11}(u) &= \exp\!\left(\sum_{m\ge1}\frac{1}{m}\,\iota_{11}(t^m)u^m\right), \\
F_{22}(u) &= \exp\!\left(\sum_{m\ge1}\frac{1}{m}\,\iota_{22}(t^m)u^m\right), \\
F_{12}(u) &= \exp\!\left(\sum_{m\ge1}\frac{1}{m}\,\iota_{12}(t^m)u^m\right).
\end{split}
\end{equation}
Then the first two factors are \(F_{11}(1)\) and \(F_{22}(1)\), while the constrained double sum becomes
\begin{equation}
\sum_{d\ge0}\bigl([u^d]F_{12}(u)\bigr)^2.
\end{equation}
We compute expansion up to $t^{10}$. We have
\begin{equation}
\iota_{11}(t^m)=3t^{2m}-3t^{4m}+t^{6m}.
\end{equation}
\begin{equation}
\iota_{22}(t^m)=3t^{2m}-3t^{4m}+3t^{8m}-3t^{10m}+t^{12m}.
\end{equation}
\begin{equation}
\begin{split}
\iota_{12}(t^m)
&=-(-1)^m t^{3m}+3(-1)^m t^{5m}
-3(-1)^m t^{7m}+(-1)^m t^{9m}.
\end{split}
\end{equation}
We have
\begin{equation}
\begin{split}
S_{11}=\sum_{m\ge1}\frac{1}{m}\iota_{11}(t^m)&=3\sum_{m\ge1}\frac{t^{2m}}{m}-3\sum_{m\ge1}\frac{t^{4m}}{m}+\sum_{m\ge1}\frac{t^{6m}}{m}.
\end{split}
\end{equation}
Expanding up to $t^{10}$
\begin{equation}
\begin{split}
S_{11}&=\bigl(3t^2-3t^4+t^6\bigr)+\frac12\bigl(3t^4-3t^8\bigr)+\frac13\bigl(3t^6\bigr)+\frac14\bigl(3t^8\bigr)+\frac15\bigl(3t^{10}\bigr)+\cdots\\
&=3t^2-\frac32t^4+2t^6-\frac34t^8+\frac35t^{10}+\mathcal{O}(t^{12}).
\end{split}
\end{equation}
Then $F_{11}(1)=\exp(S_{11})$. Expanding the exponential
\begin{equation}
\begin{split}
F_{11}(1)&=1+S_{11}+\frac12S_{11}^2+\frac16S_{11}^3+\frac1{24}S_{11}^4+\frac1{120}S_{11}^5+\cdots\\
&=1+3t^2+3t^4+2t^6+3t^8+3t^{10}+\mathcal{O}(t^{12}).
\end{split}
\end{equation}
We have
\begin{equation}
S_{22}=3\sum_{m\ge1}\frac{t^{2m}}{m}-3\sum_{m\ge1}\frac{t^{4m}}{m}+3\sum_{m\ge1}\frac{t^{8m}}{m}-3\sum_{m\ge1}\frac{t^{10m}}{m}+\sum_{m\ge1}\frac{t^{12m}}{m}.
\end{equation}
Up to $t^{10}$
\begin{equation}
\begin{split}
S_{22}&=\bigl(3t^2-3t^4+3t^8-3t^{10}\bigr)+\frac12\bigl(3t^4-3t^8\bigr)+\frac13\bigl(3t^6\bigr)+\frac14\bigl(3t^8\bigr)+\frac15\bigl(3t^{10}\bigr)+\cdots\\
&=3t^2-\frac32t^4+t^6+\frac94t^8-\frac{12}{5}t^{10}+\mathcal{O}(t^{12}).
\end{split}
\end{equation}
Exponentiating
\begin{equation}
F_{22}(1)=1+3t^2+3t^4+t^6+3t^8+6t^{10}+\mathcal{O}(t^{12}).
\end{equation}
We need the coefficient of $u^d$ in $F_{12}(u)$ for small $d$ because only $d=0$ and $d=1$ contribute up to $t^{10}$. We first compute
\begin{equation}
\begin{split}
S_{12}(u)
&=\sum_{m\ge1}\frac{1}{m}\iota_{12}(t^m)u^m \\
&=u\bigl(t^3-3t^5+3t^7-t^9\bigr)
+\frac{u^2}{2}\bigl(-t^6+3t^{10}\bigr)
+\frac{u^3}{3}t^9+\mathcal{O}(t^{11}).
\end{split}
\end{equation}
Then $F_{12}(u)=\exp(S_{12}(u))$. The coefficient of $u^1$ is 
\begin{equation}
A_1(t)=[u^1]F_{12}(u)=t^3-3t^5+3t^7-t^9.
\end{equation}
Hence
\begin{equation}
\bigl(A_1(t)\bigr)^2 = t^6-6t^8+15t^{10}+\mathcal{O}(t^{12}).
\end{equation}
The empty partition contributes $1$. Therefore the constrained double sum is
\begin{equation}
\sum_{d\ge0}\bigl([u^d]F_{12}(u)\bigr)^2 = 1 + \bigl(t^6-6t^8+15t^{10}\bigr) +\mathcal{O}(t^{12}).
\end{equation}
Multiplying the three factors,
\begin{equation}
\mathcal{I}_{n_1=1,n_2=1;N_1,N_2}^{\mathrm{BMN}}(t)=F_{11}(1)\,F_{22}(1)\,\Bigl(1+\bigl(t^6-6t^8+15t^{10}\bigr)\Bigr)+\mathcal{O}(t^{12}).
\end{equation}
We first compute $P=F_{11}(1)F_{22}(1)$
\begin{equation}
\begin{split}
P&=\bigl(1+3t^2+3t^4+2t^6+3t^8+3t^{10}\bigr)\bigl(1+3t^2+3t^4+t^6+3t^8+6t^{10}\bigr)\\
&=1+6t^2+15t^4+21t^6+24t^8+36t^{10}+\mathcal{O}(t^{12}).
\end{split}
\end{equation}
Then
\begin{equation}
\begin{split}
\mathcal{I}_{n_1=1,n_2=1;N_1,N_2}^{\mathrm{BMN}}(t)&=P + P\bigl(t^6-6t^8+15t^{10}\bigr)\\
&=P + \bigl(t^6+6t^8+15t^{10}\bigr) -6t^8-36t^{10}+15t^{10} +\mathcal{O}(t^{12})\\
&=P + t^6 -6t^{10} +\mathcal{O}(t^{12})\\
&=1+6t^2+15t^4+22t^6+24t^8+30t^{10}+\mathcal{O}(t^{12}).
\end{split}
\end{equation}
Thus the $U(N)$ BMN index for $N_1=1,\;N_2=2,\;n_1=1,\;n_2=1$ expanded to order $t^{10}$ is
\begin{equation}
\mathcal{I}_{n_1=1,n_2=1;N_1,N_2}^{\mathrm{BMN}}(t)=1+6t^2+15t^4+22t^6+24t^8+30t^{10}+\mathcal{O}(t^{12}).
\end{equation}
The $SU(N)$ BMN index for $N_1=1,\;N_2=2,\;n_1=1,\;n_2=1$ is
\begin{equation}
\mathcal{I}_{n_1=1,n_2=1;N_1,N_2}^{\mathrm{BMN}~SU(N)}(t)=1 + 3\,t^{2} + 3\,t^{4} + 2\,t^{6} + 6\,t^{10} +\mathcal{O}(t^{12}).
\end{equation}

\bibliography{References}

@article{Berenstein:2002jq,
    author = "Berenstein, David Eliecer and Maldacena, Juan Martin and Nastase, Horatiu Stefan",
    title = "{Strings in flat space and pp waves from N=4 superYang-Mills}",
    eprint = "hep-th/0202021",
    archivePrefix = "arXiv",
    doi = "10.1088/1126-6708/2002/04/013",
    journal = "JHEP",
    volume = "04",
    pages = "013",
    year = "2002"
}

@article{Maldacena:2002rb,
    author = "Maldacena, Juan Martin and Sheikh-Jabbari, Mohammad M. and Van Raamsdonk, Mark",
    title = "{Transverse five-branes in matrix theory}",
    eprint = "hep-th/0211139",
    archivePrefix = "arXiv",
    reportNumber = "SU-ITP-02-33",
    doi = "10.1088/1126-6708/2003/01/038",
    journal = "JHEP",
    volume = "01",
    pages = "038",
    year = "2003"
}

@article{Chang:2024lkw,
    author = "Chang, Chi-Ming",
    title = "{Witten index of BMN matrix quantum mechanics}",
    eprint = "2404.18442",
    archivePrefix = "arXiv",
    primaryClass = "hep-th",
    doi = "10.21468/SciPostPhys.19.6.147",
    journal = "SciPost Phys.",
    volume = "19",
    number = "6",
    pages = "147",
    year = "2025"
}

@article{Gadde:2025yoa,
    author = "Gadde, Abhijit and Lee, Eunwoo and Raj, Rajat and Tomar, Shivansh",
    title = "{Probing non-graviton spectra in $ \mathcal{N}=4 $ SYM via BMN truncation and S-duality}",
    eprint = "2506.13887",
    archivePrefix = "arXiv",
    primaryClass = "hep-th",
    doi = "10.1007/JHEP02(2026)026",
    journal = "JHEP",
    volume = "02",
    pages = "026",
    year = "2026"
}

@article{Imamura:2021ytr,
    author = "Imamura, Yosuke",
    title = "{Finite-N superconformal index via the AdS/CFT correspondence}",
    eprint = "2108.12090",
    archivePrefix = "arXiv",
    primaryClass = "hep-th",
    reportNumber = "TIT/HEP-686",
    doi = "10.1093/ptep/ptab141",
    journal = "PTEP",
    volume = "2021",
    number = "12",
    pages = "123B05",
    year = "2021"
}

@article{Dasgupta:2002hx,
    author = "Dasgupta, Keshav and Sheikh-Jabbari, Mohammad M. and Van Raamsdonk, Mark",
    title = "{Matrix perturbation theory for M theory on a PP wave}",
    eprint = "hep-th/0205185",
    archivePrefix = "arXiv",
    reportNumber = "SU-ITP-02-14",
    doi = "10.1088/1126-6708/2002/05/056",
    journal = "JHEP",
    volume = "05",
    pages = "056",
    year = "2002"
}

@article{Banks:1996vh,
    author = "Banks, Tom and Fischler, W. and Shenker, S. H. and Susskind, Leonard",
    title = "{M theory as a matrix model: A conjecture}",
    eprint = "hep-th/9610043",
    archivePrefix = "arXiv",
    reportNumber = "RU-96-95, SU-ITP-96-12, UTTG-13-96",
    doi = "10.1103/PhysRevD.55.5112",
    journal = "Phys. Rev. D",
    volume = "55",
    pages = "5112--5128",
    year = "1997"
}

@article{Taylor:2001vb,
    author = "Taylor, Washington",
    title = "{M(atrix) Theory: Matrix Quantum Mechanics as a Fundamental Theory}",
    eprint = "hep-th/0101126",
    archivePrefix = "arXiv",
    doi = "10.1103/RevModPhys.73.419",
    journal = "Rev. Mod. Phys.",
    volume = "73",
    pages = "419--462",
    year = "2001"
}

@article{Kim:2002if,
    author = "Kim, Nakwoo and Plefka, Jan",
    title = "{On the spectrum of PP wave matrix theory}",
    eprint = "hep-th/0207034",
    archivePrefix = "arXiv",
    reportNumber = "AEI-2002-047",
    doi = "10.1016/S0550-3213(02)00738-1",
    journal = "Nucl. Phys. B",
    volume = "643",
    pages = "31--48",
    year = "2002"
}

@article{Dasgupta:2002ru,
    author = "Dasgupta, Keshav and Sheikh-Jabbari, Mohammad M. and Van Raamsdonk, Mark",
    title = "{Protected multiplets of M theory on a plane wave}",
    eprint = "hep-th/0207050",
    archivePrefix = "arXiv",
    reportNumber = "SU-ITP-02-29",
    doi = "10.1088/1126-6708/2002/09/021",
    journal = "JHEP",
    volume = "09",
    pages = "021",
    year = "2002"
}

@article{Lin:2005nh,
    author = "Lin, Hai and Maldacena, Juan Martin",
    title = "{Fivebranes from gauge theory}",
    eprint = "hep-th/0509235",
    archivePrefix = "arXiv",
    reportNumber = "PUPT-2172",
    doi = "10.1103/PhysRevD.74.084014",
    journal = "Phys. Rev. D",
    volume = "74",
    pages = "084014",
    year = "2006"
}

@article{Kim:2003rza,
    author = "Kim, Nakwoo and Klose, Thomas and Plefka, Jan",
    title = "{Plane wave matrix theory from $\mathcal{N}=4$ super Yang-Mills on ${\mathbb R} \times S^3$}",
    eprint = "hep-th/0306054",
    archivePrefix = "arXiv",
    reportNumber = "AEI-2003-048",
    doi = "10.1016/j.nuclphysb.2003.08.019",
    journal = "Nucl. Phys. B",
    volume = "671",
    pages = "359--382",
    year = "2003"
}

@article{Itzhaki:1998dd,
    author = "Itzhaki, Nissan and Maldacena, Juan Martin and Sonnenschein, Jacob and Yankielowicz, Shimon",
    title = "{Supergravity and the large N limit of theories with sixteen supercharges}",
    eprint = "hep-th/9802042",
    archivePrefix = "arXiv",
    reportNumber = "TAUP-2474-98, HUTP-98/A003",
    doi = "10.1103/PhysRevD.58.046004",
    journal = "Phys. Rev. D",
    volume = "58",
    pages = "046004",
    year = "1998"
}

@article{Costa:2014wya,
    author = "Costa, Miguel S. and Greenspan, Lauren and Penedones, Joao and Santos, Jorge",
    title = "{Thermodynamics of the BMN matrix model at strong coupling}",
    eprint = "1411.5541",
    archivePrefix = "arXiv",
    primaryClass = "hep-th",
    doi = "10.1007/JHEP03(2015)069",
    journal = "JHEP",
    volume = "03",
    pages = "069",
    year = "2015"
}

@article{Ishiki:2006rt,
    author = "Ishiki, Goro and Takayama, Yastoshi and Tsuchiya, Asato",
    title = "{N=4 SYM on ${\mathbb R} \times S^3$ and theories with 16 supercharges}",
    eprint = "hep-th/0605163",
    archivePrefix = "arXiv",
    reportNumber = "OU-HET-560",
    doi = "10.1088/1126-6708/2006/10/007",
    journal = "JHEP",
    volume = "10",
    pages = "007",
    year = "2006"
}

@article{Ishiki:2006yr,
    author = "Ishiki, Goro and Shimasaki, Shinji and Takayama, Yastoshi and Tsuchiya, Asato",
    title = "{Embedding of theories with $SU(2|4)$ symmetry into the plane wave matrix model}",
    eprint = "hep-th/0610038",
    archivePrefix = "arXiv",
    reportNumber = "OU-HET-567",
    doi = "10.1088/1126-6708/2006/11/089",
    journal = "JHEP",
    volume = "11",
    pages = "089",
    year = "2006"
}

@article{Asano:2012zt,
    author = "Asano, Yuhma and Ishiki, Goro and Okada, Takashi and Shimasaki, Shinji",
    title = "{Exact results for perturbative partition functions of theories with $SU(2|4)$ symmetry}",
    eprint = "1211.0364",
    archivePrefix = "arXiv",
    primaryClass = "hep-th",
    reportNumber = "KUNS-2422",
    doi = "10.1007/JHEP02(2013)148",
    journal = "JHEP",
    volume = "02",
    pages = "148",
    year = "2013"
}

@article{Asano:2014vba,
    author = "Asano, Yuhma and Ishiki, Goro and Okada, Takashi and Shimasaki, Shinji",
    title = "{Emergent bubbling geometries in the plane wave matrix model}",
    eprint = "1401.5079",
    archivePrefix = "arXiv",
    primaryClass = "hep-th",
    doi = "10.1007/JHEP05(2014)075",
    journal = "JHEP",
    volume = "05",
    pages = "075",
    year = "2014"
}

@article{Asano:2017nxw,
    author = "Asano, Yuhma and Ishiki, Goro and Shimasaki, Shinji and Terashima, Seiji",
    title = "{Spherical transverse M5-branes from the plane wave matrix model}",
    eprint = "1711.07681",
    archivePrefix = "arXiv",
    primaryClass = "hep-th",
    reportNumber = "DIAS-STP-17-13, UTHEP-709, YITP-17-123",
    doi = "10.1007/JHEP02(2018)076",
    journal = "JHEP",
    volume = "02",
    pages = "076",
    year = "2018"
}

@article{Dolan:2007rq,
    author = "Dolan, F. A.",
    title = "{Counting BPS operators in N=4 SYM}",
    eprint = "0704.1038",
    archivePrefix = "arXiv",
    primaryClass = "hep-th",
    reportNumber = "DIAS-STP-07-05",
    doi = "10.1016/j.nuclphysb.2007.07.026",
    journal = "Nucl. Phys. B",
    volume = "790",
    pages = "432--464",
    year = "2008"
}

@article{Murthy:2020scj,
    author = "Murthy, Sameer",
    title = "{Growth of the $\frac {1} {16}$-BPS index in 4d $N=4$ supersymmetric Yang-Mills theory}",
    eprint = "2005.10843",
    archivePrefix = "arXiv",
    primaryClass = "hep-th",
    doi = "10.1103/PhysRevD.105.L021903",
    journal = "Phys. Rev. D",
    volume = "105",
    number = "2",
    pages = "L021903",
    year = "2022"
}

@article{Gaiotto:2021xce,
    author = "Gaiotto, Davide and Lee, Ji Hoon",
    title = "{The giant graviton expansion}",
    eprint = "2109.02545",
    archivePrefix = "arXiv",
    primaryClass = "hep-th",
    doi = "10.1007/JHEP08(2024)025",
    journal = "JHEP",
    volume = "08",
    pages = "025",
    year = "2024"
}

@article{Chang:2022mjp,
    author = "Chang, Chi-Ming and Lin, Ying-Hsuan",
    title = "{Words to describe a black hole}",
    eprint = "2209.06728",
    archivePrefix = "arXiv",
    primaryClass = "hep-th",
    doi = "10.1007/JHEP02(2023)109",
    journal = "JHEP",
    volume = "02",
    pages = "109",
    year = "2023"
}

@article{Choi:2022caq,
    author = "Choi, Sunjin and Kim, Seok and Lee, Eunwoo and Park, Jaemo",
    title = "{The shape of non-graviton operators for SU(2)}",
    eprint = "2209.12696",
    archivePrefix = "arXiv",
    primaryClass = "hep-th",
    reportNumber = "KIAS-P22052",
    doi = "10.1007/JHEP09(2024)029",
    journal = "JHEP",
    volume = "09",
    pages = "029",
    year = "2024"
}

@article{Choi:2023znd,
    author = "Choi, Sunjin and Kim, Seok and Lee, Eunwoo and Lee, Siyul and Park, Jaemo",
    title = "{Towards quantum black hole microstates}",
    eprint = "2304.10155",
    archivePrefix = "arXiv",
    primaryClass = "hep-th",
    doi = "10.1007/JHEP11(2023)175",
    journal = "JHEP",
    volume = "11",
    pages = "175",
    year = "2023",
    note = "[Erratum: JHEP 03, 091 (2025)]"
}

@article{Chang:2023zqk,
    author = "Chang, Chi-Ming and Feng, Li and Lin, Ying-Hsuan and Tao, Yi-Xiao",
    title = "{Decoding stringy near-supersymmetric black holes}",
    eprint = "2306.04673",
    archivePrefix = "arXiv",
    primaryClass = "hep-th",
    doi = "10.21468/SciPostPhys.16.4.109",
    journal = "SciPost Phys.",
    volume = "16",
    number = "4",
    pages = "109",
    year = "2024"
}

@article{Budzik:2023vtr,
    author = "Budzik, Kasia and Murali, Harish and Vieira, Pedro",
    title = "{Following Black Hole States}",
    eprint = "2306.04693",
    archivePrefix = "arXiv",
    primaryClass = "hep-th",
    month = "6",
    year = "2023"
}

@article{Choi:2023vdm,
    author = "Choi, Jaehyeok and Choi, Sunjin and Kim, Seok and Lee, Jehyun and Lee, Siyul",
    title = "{Finite N black hole cohomologies}",
    eprint = "2312.16443",
    archivePrefix = "arXiv",
    primaryClass = "hep-th",
    reportNumber = "SNUTP23-002, KIAS-P23070, LCTP-23-20, SNUTP23-002; KIAS-P23070; LCTP-23-20;",
    doi = "10.1007/JHEP12(2024)029",
    journal = "JHEP",
    volume = "12",
    pages = "029",
    year = "2024"
}

@article{Chang:2024zqi,
    author = "Chang, Chi-Ming and Lin, Ying-Hsuan",
    title = "{Holographic covering and the fortuity of black holes}",
    eprint = "2402.10129",
    archivePrefix = "arXiv",
    primaryClass = "hep-th",
    month = "2",
    year = "2024"
}

@article{Chang:2024lxt,
    author = "Chang, Chi-Ming and Chen, Yiming and Sia, Bik Soon and Yang, Zhenbin",
    title = "{Fortuity in SYK models}",
    eprint = "2412.06902",
    archivePrefix = "arXiv",
    primaryClass = "hep-th",
    doi = "10.1007/JHEP08(2025)003",
    journal = "JHEP",
    volume = "08",
    pages = "003",
    year = "2025"
}

@article{deMelloKoch:2024pcs,
    author = "de Mello Koch, Robert and Kim, Minkyoo and Kim, Seok and Lee, Jehyun and Lee, Siyul",
    title = "{Brane-fused black hole operators}",
    eprint = "2412.08695",
    archivePrefix = "arXiv",
    primaryClass = "hep-th",
    reportNumber = "SNUTP24-004",
    doi = "10.1007/JHEP07(2025)216",
    journal = "JHEP",
    volume = "07",
    pages = "216",
    year = "2025"
}

@article{Chang:2025rqy,
    author = "Chang, Chi-Ming and Lin, Ying-Hsuan and Zhang, Haoyu",
    title = "{Fortuity in the D1-D5 system}",
    eprint = "2501.05448",
    archivePrefix = "arXiv",
    primaryClass = "hep-th",
    month = "1",
    year = "2025"
}

@article{deMelloKoch:2025ngs,
    author = "de Mello Koch, Robert and Jevicki, Antal",
    title = "{Structure of loop space at finite N}",
    eprint = "2503.20097",
    archivePrefix = "arXiv",
    primaryClass = "hep-th",
    doi = "10.1007/JHEP06(2025)011",
    journal = "JHEP",
    volume = "06",
    pages = "011",
    year = "2025"
}

@article{deMelloKoch:2025cec,
    author = "de Mello Koch, Robert and Ghosh, Animik and Van Zyl, Hendrik J. R.",
    title = "{Bosonic fortuity in vector models}",
    eprint = "2504.14181",
    archivePrefix = "arXiv",
    primaryClass = "hep-th",
    doi = "10.1007/JHEP06(2025)246",
    journal = "JHEP",
    volume = "06",
    pages = "246",
    year = "2025"
}

@article{Chang:2025wgo,
    author = "Chang, Chi-Ming and Zhang, Haoyu",
    title = "{Fortuity and R-charge concentration in the D1-D5 CFT}",
    eprint = "2511.23294",
    archivePrefix = "arXiv",
    primaryClass = "hep-th",
    month = "11",
    year = "2025"
}

@article{Kim:2025vup,
    author = "Kim, Seok and Lee, Jehyun and Lee, Siyul and Oh, Hyunwoo",
    title = "{BPS phases and fortuity in higher spin holography}",
    eprint = "2511.03105",
    archivePrefix = "arXiv",
    primaryClass = "hep-th",
    month = "11",
    year = "2025"
}

@article{Chen:2025sum,
    author = "Chen, Yiming",
    title = "{Fortuity with a single matrix}",
    eprint = "2511.00790",
    archivePrefix = "arXiv",
    primaryClass = "hep-th",
    month = "11",
    year = "2025"
}

@article{Belin:2025hsg,
    author = "Belin, Alexandre and Singh, Palash and Vadala, Rita and Zaffaroni, Alberto",
    title = "{Fortuity in ABJM}",
    eprint = "2512.04146",
    archivePrefix = "arXiv",
    primaryClass = "hep-th",
    month = "12",
    year = "2025"
}

@article{Behan:2025hbx,
    author = "Behan, Connor and de Gioia, Leonardo Pipolo",
    title = "{Two roads to fortuity in ABJM theory}",
    eprint = "2512.23603",
    archivePrefix = "arXiv",
    primaryClass = "hep-th",
    month = "12",
    year = "2025"
}

@article{Choi:2025pqr,
    author = "Choi, Jaehyeok and Kim, Seunggyu",
    title = "{Fortuity and relevant deformation}",
    eprint = "2512.12674",
    archivePrefix = "arXiv",
    primaryClass = "hep-th",
    month = "12",
    year = "2025"
}

@article{Choi:2024ddbh,
    author = "Choi, Sunjin and Jain, Diksha and Kim, Seok and Krishna, Vineeth and Lee, Eunwoo and Minwalla, Shiraz and Patel, Chintan",
    title = "{Dual Dressed Black Holes as the end point of the Charged Superradiant instability in ${\cal N} = 4$ Yang Mills}",
    eprint = "2409.18178",
    archivePrefix = "arXiv",
    primaryClass = "hep-th",
    doi = "10.21468/SciPostPhys.18.4.137",
    journal = "SciPost Phys.",
    volume = "18",
    number = "4",
    pages = "137",
    year = "2025"
}

@article{Choi:2025sgg,
    author = "Choi, Sunjin and Jain, Diksha and Kim, Seok and Krishna, Vineeth and Kwon, Goojin and Lee, Eunwoo and Minwalla, Shiraz and Patel, Chintan",
    title = "{Supersymmetric Grey Galaxies, Dual Dressed Black Holes and the Superconformal Index}",
    eprint = "2501.17217",
    archivePrefix = "arXiv",
    primaryClass = "hep-th",
    doi = "10.21468/SciPostPhys.19.3.072",
    journal = "SciPost Phys.",
    volume = "19",
    number = "3",
    pages = "072",
    year = "2025"
}

@article{Nawata:2011un,
    author = "Nawata, Satoshi",
    title = "{Localization of ${\cal N}=4$ Superconformal Field Theory on $S^1 \times S^3$ and Index}",
    eprint = "1104.4470",
    archivePrefix = "arXiv",
    primaryClass = "hep-th",
    doi = "10.1007/JHEP11(2011)144",
    journal = "JHEP",
    volume = "11",
    pages = "144",
    year = "2011"
}

@article{Zigdon:2025bmn,
    author = "Zigdon, Yoav",
    title = "{A Charge Constraint in BMN}",
    eprint = "2506.19924",
    archivePrefix = "arXiv",
    primaryClass = "hep-th",
    month = "6",
    year = "2025"
}

@article{Chang:2026scr,
    author = "Chang, Chi-Ming",
    title = "{Super-Chevalley Restriction and Relative Lie Algebra Cohomology over the 2|3 Algebra}",
    eprint = "2604.23549",
    archivePrefix = "arXiv",
    primaryClass = "math.RT",
    month = "4",
    year = "2026"
}

@article{vanLeuven:2025gwr,
    author = "van Leuven, Sam and Mathieson, Kayleigh and Roy, Pratik",
    title = "{Residue sums for superconformal indices}",
    eprint = "2511.10732",
    archivePrefix = "arXiv",
    primaryClass = "hep-th",
    doi = "10.1007/JHEP05(2026)204",
    journal = "JHEP",
    volume = "05",
    pages = "204",
    year = "2026"
}

@article{Lee:2026xoo,
    author = "Lee, Eunwoo",
    title = "{BMN-like Matrix Models}",
    eprint = "2602.22163",
    archivePrefix = "arXiv",
    primaryClass = "hep-th",
    month = "2",
    year = "2026"
}

@article{Gaikwad:2025doublegauge,
    author = "Gaikwad, Adwait and Kibe, Tanay and van Leuven, Sam and Mathieson, Kayleigh",
    title = "{To gauge or to double gauge? Matrix models, global symmetry, and black hole cohomologies}",
    eprint = "2512.02103",
    archivePrefix = "arXiv",
    primaryClass = "hep-th",
    month = "12",
    year = "2025"
}

@article{Komatsu:2026holography,
    author = "Komatsu, Shota and Lee, Eunwoo and Patel, Chintan",
    title = "{Matrix Theory from Holography}",
    eprint = "2609.13386v2",
    archivePrefix = "arXiv",
    primaryClass = "hep-th",
    month = "9",
    year = "2026",
    note = "arXiv version 2, 22 September 2026"
}
\bibliographystyle{JHEP}
\end{document}